\documentclass[12pt,twoside]{report}

\usepackage{psfig}
\usepackage[dvips]{graphics,psfrag}

\newcommand{\EV}{~\mbox{eV}}
\newcommand{\KEV}{~\mbox{keV}}
\newcommand{\MEV}{~\mbox{MeV}}
\newcommand{\GEV}{~\mbox{GeV}}
\newcommand{\TEV}{~\mbox{TeV}}
\newcommand{\gsim}{ \mathop{}_{\textstyle \sim}^{\textstyle >} }
\newcommand{\lsim}{ \mathop{}_{\textstyle \sim}^{\textstyle <} }
\newcommand{\vev}[1]{ \left\langle {#1} \right\rangle }
\newcommand{\sneu}{\widetilde{N_1}}
\newcommand{\mnu}[1]{ m_{\nu {#1}} }
\def\Frac#1#2{{\displaystyle\frac{#1}{#2}}}

\def\SEC#1{Sec.~\ref{#1}}
\def\FIG#1{Fig.~\ref{#1}}
\def\EQ#1{Eq.~(\ref{#1})}
\def\EQS#1{Eqs.~(\ref{#1})}
\def\footnoterule{%
\kern 5pt
\hrule width .4\columnwidth
\kern 10pt}

\begin{document}%%%%%%%%%%%%%
%%%%%%%%%%%%%%%%%%%%%%%%%%%%%
%%%%%%%%%%%%%%%%%%%%%%%%%%%%%
%%%%%%%%%%%%%%%%%%%%%%%%%%%%%
\baselineskip 0.6cm

%%%%%%%%%%%%%%%%%%%%%%%%%%%%%%%%%%%%%%%%%%%%%%%%%%%%%%%%%%%%%%%%%%%
%\include{Title-Contents}%%%%%%%%%%%%%%%%%%%%%%%%%%%%%%%%%%%%%%%%%%%
%%%%%%%%%%%%%%%%%%%%%%%%%%%%%%%%%%%%%%%%%%%%%%%%%%%%%%%%%%%%%%%%%%%
%%%%%%%%%%%%%%%%%%%%%%%%%%%%%%%%%%%%%%%%%%%%%%%%%%%%%%%%%%%%%%%%%%%
\begin{titlepage}%%%%%%%%%%%%%%%%%%%%%%%%%%%%%%%%%%%%%%%%%%%%%%%%%%
\begin{center}%%%%%%%%%%%%%%%%%%%%%%%%%%%%%%%%%%%%%%%%%%%%%%%%%%%%%
%%%%%%%%%%%%%%%%%%%%%%%%%%%%%%%%%%%%%%%%%%%%%%%%%%%%%%%%%%%%%%%%%%%
~

\begin{flushright}
% UT-???            \\
% DESY-???     \\
 hep-ph/0212305
\end{flushright} 
\vskip 1cm

{\Large{\bf Cosmological Baryon Asymmetry and Neutrinos: }}
\vskip 0.1cm

{\large{\bf Baryogenesis via Leptogenesis in Supersymmetric Theories }}

\vskip 4cm 

{\large Koichi Hamaguchi}
\vskip 0.4cm

{\it Deutsches Elektronen-Synchrotron DESY, D-22603, Hamburg, Germany}
\vskip 3cm

{\it Dr.thesis submitted to}
\\
{\it  Department of Physics, University of Tokyo}
\\
{\it January 2002}

%%%%%%%%%%%%%%%%%%%%%%%%%%%%%%%%%%%%%%%%%%%%%%%%%%%%%%%%%%%%%%%%%%%
\end{center}%%%%%%%%%%%%%%%%%%%%%%%%%%%%%%%%%%%%%%%%%%%%%%%%%%%%%%%
\end{titlepage}%%%%%%%%%%%%%%%%%%%%%%%%%%%%%%%%%%%%%%%%%%%%%%%%%%%%
%%%%%%%%%%%%%%%%%%%%%%%%%%%%%%%%%%%%%%%%%%%%%%%%%%%%%%%%%%%%%%%%%%%
\newpage
~
\thispagestyle{empty}

\newpage 
%%%%%%%%%%%%%%%%%%%%%%%%%%%%%%%%%%%%%%%%%%%%%%%%%%%%%%%%%%%%%%%%%%%
%\begin{titlepage}%%%%%%%%%%%%%%%%%%%%%%%%%%%%%%%%%%%%%%%%%%%%%%%%%%
\begin{center}%%%%%%%%%%%%%%%%%%%%%%%%%%%%%%%%%%%%%%%%%%%%%%%%%%%%%
%%%%%%%%%%%%%%%%%%%%%%%%%%%%%%%%%%%%%%%%%%%%%%%%%%%%%%%%%%%%%%%%%%%
~
\vskip 3cm

{\Large Ph.D thesis}
\vskip 0.5cm

{\Large{\bf Cosmological Baryon Asymmetry and Neutrinos: }}
\vskip 0.1cm

{\large{\bf Baryogenesis via Leptogenesis in Supersymmetric Theories }}

\vskip 4cm 

{\large Koichi Hamaguchi}
\vskip 0.4cm

{\it Department of Physics, University of Tokyo, Tokyo 113-0033, Japan}
\vskip 3cm

%%%%%%%%%%%%%%%%%%%%%%%%%%%%%%%%%%%%%%%%%%%%%%%%%%%%%%%%%%%%%%%%%%%
\end{center}%%%%%%%%%%%%%%%%%%%%%%%%%%%%%%%%%%%%%%%%%%%%%%%%%%%%%%%
\begin{flushright}
Submitted to Department of Physics, University of Tokyo: {\it January 2002}
\\
Accepted: {\it February 2002}
\\
Revised to submit to the eprint arXiv: {\it December 2002}
\end{flushright}
%\end{titlepage}%%%%%%%%%%%%%%%%%%%%%%%%%%%%%%%%%%%%%%%%%%%%%%%%%%%%
%%%%%%%%%%%%%%%%%%%%%%%%%%%%%%%%%%%%%%%%%%%%%%%%%%%%%%%%%%%%%%%%%%%
\thispagestyle{empty}

\pagenumbering{roman}
\setcounter{page}{0}
%%%%%%%%%%%%%%%%%%%%%%%%%%%%%%%%%%%%%%%%%%%%%%%%%%%%%%%%%%%%%%%%%%%
\tableofcontents%%%%%%%%%%%%%%%%%%%%%%%%%%%%%%%%%%%%%%%%%%%%%%%%%%%
%%%%%%%%%%%%%%%%%%%%%%%%%%%%%%%%%%%%%%%%%%%%%%%%%%%%%%%%%%%%%%%%%%%
\newpage
\pagenumbering{arabic}
\setcounter{page}{1}

%%%%%%%%%%%%%%%%%%%%%%%%%%%%%%
%\include{Chap-Intro}%%%%%%%%%%
%%%%%%%%%%%%%%%%%%%%%%%%%%%%%%%%%%%%%%%%%%%%%%%%%%%%%%%%%%%%%%%%%%%
%%%%%%%%%%%%%%%%%%%%%%%%%%%%%%%%%%%%%%%%%%%%%%%%%%%%%%%%%%%%%%%%%%%
%%%%%%%%%%%%%%%%%%%%%%%%%%%%%%%%%%%%%%%%%%%%%%%%%%%%%%%%%%%%%%%%%%%
\chapter{Introduction}%%%%%%%%%%%%%%%%%%%%%%%%%%%%%%%%%%%%%%%%%%%%%
%%%%%%%%%%%%%%%%%%%%%%%%%%%%%%%%%%%%%%%%%%%%%%%%%%%%%%%%%%%%%%%%%%%
%%%%%%%%%%%%%%%%%%%%%%%%%%%%%%%%%%%%%%%%%%%%%%%%%%%%%%%%%%%%%%%%%%%
%%%%%%%%%%%%%%%%%%%%%%%%%%%%%%%%%%%%%%%%%%%%%%%%%%%%%%%%%%%%%%%%%%%

%%%%%%%%%%%%%%%%%%%%%%%%%%%%%%%%%%%%%%%%%%%%%%%%%%%%%%%%%%%%%%%%%%%
\section{Overview}%%%%%%%%%%%%%%%%%%%%%%%%%%%%%%%%%%%%%%%%%%%%%%%%%
%%%%%%%%%%%%%%%%%%%%%%%%%%%%%%%%%%%%%%%%%%%%%%%%%%%%%%%%%%%%%%%%%%%
\label{SEC-1-1}

The origin of baryon asymmetry (matter-antimatter asymmetry) in the
present universe is one of the fundamental puzzles in particle physics
as well as in cosmology. Assuming an inflationary phase in the early
universe, any initial baryon asymmetry is diluted and becomes
essentially zero during the inflation. Therefore, the observed baryon
asymmetry should be generated dynamically after the inflation. Such a
dynamical generation of baryon asymmetry (baryogenesis) is possible if
the following three conditions are satisfied; (i) baryon-number
violation, (ii) $C$- and $CP$-violations, and (iii) departure from
thermal equilibrium, which are known as the Sakharov's three
conditions~\cite{Sakharov}.

The original idea of the baryogenesis in the grand unified theory
(GUT)~\cite{GUT-baryo-1} satisfies all of these
conditions~\cite{GUT-baryo-2}, and provides an elegant interplay between
particle physics and cosmology. Actually, the GUT~\cite{GUT} unifies
baryons and leptons in the same gauge multiplets and predicts the
existence of baryon number violation, the delayed decay of superheavy
GUT particle satisfies the out-of-equilibrium condition, and its
$CP$-violating asymmetric decay into baryons and antibaryons can offer a
net baryon asymmetry.

On the other hand, the electroweak gauge theory itself violates the
baryon asymmetry by a quantum anomaly~\cite{tHooft}. This effect is
highly suppressed by a large exponential factor at zero temperature, and
hence the stability of the proton is practically guaranteed. However, it
was found that in the early universe at high temperatures above the
electroweak scale, the baryon-number violating interaction is not
suppressed and even be rapid enough to be in thermal
equilibrium~\cite{sphaleron}. An important point here is that this
baryon number violating (``sphaleron'') process violates lepton ($L$)
number as well as baryon ($B$) number, and it conserves a linear
combination of them, $B-L$. Therefore, if the baryon asymmetry is
produced in $(B-L)$-conserving processes, like those in the ${\rm
SU}(5)$ GUT, it would be washed out by the sphaleron effect before the
electroweak phase transition, and hence no baryon asymmetry can remain
until the stage of big-bang nucleosynthesis to generate the light
nuclei.

The existence of the sphaleron effect opens a new possibility of
baryogenesis, the generation of baryon asymmetry at electroweak phase
transition~\cite{sphaleron,CKN}. This mechanism does not use any
$(B-L)$-violation, and the baryon asymmetry is generated by the
sphaleron-induced $(B+L)$-violating process. (Hence, same amount of
baryons and leptons are produced.) As other baryogenesis scenarios have,
this mechanism also has a double-edged behavior, i.e., the produced
baryon (and lepton) asymmetry tends to be washed out afterwards by the
sphaleron process itself. In order to avoid the erasure of the produced
baryon asymmetry, the electroweak phase transition must be strongly
first-order. However, in the case of the minimal standard model with one
Higgs doublet, it turns out that, for Higgs masses allowed by current
experimental bounds, the electroweak transition is too weakly first
order or just a smooth transition, which excludes the electroweak
baryogenesis in the standard model. Although the extension of the
standard model to the supersymmetric version may cure this difficulty,
there remains only a small parameter range~\cite{EW-MSSM-rev}.
\vspace{0.8em}

Therefore, it seems natural to consider that the baryon asymmetry was
generated before the electroweak phase transition.  As long as we
consider a baryogenesis above the electroweak scale, some
$(B-L)$-violating interaction is mandatory, since the sphaleron process
washes out any baryon asymmetry in the universe unless there exists
nonzero $B-L$ asymmetry. (Namely, the first one of the Sakharov's
conditions, (i) $B$-violation, should be replaced with (i)'
$(B-L)$-violation.)  This means, on the other hand, as first suggested
by Fukugita and Yanagida~\cite{FY}, that the {\it lepton}-number
violation is enough for baryogenesis and explicit baryon-number
violation is not necessarily required. Actually, there now exists an
implication of lepton-number ($L$) violation, whereas there has been
discovered no evidence for baryon-number ($B$) violations: that is, the
neutrino oscillation.

Neutrino oscillation~\cite{SK-Atm,Solar},\footnote{Very recently the
KamLAND experiment has announced the first results, which exclude all
oscillation solutions but the 'Large Mixing Angle' solution to the
solar neutrino problem~\cite{KamLAND}.} especially the strong evidence
for the atmospheric neutrino oscillation reported in 1998 by the
Super-Kamiokande Collaboration~\cite{SK-Atm}, is one of the greatest
discoveries in particle physics after the success of the standard
model.  The experimental data strongly suggest that the neutrinos have
small but finite masses, impelling the standard model of particle
physics to be modified.

Such tiny masses of light neutrinos are naturally understood if we
introduce heavy right-handed Majorana neutrinos to the standard model,
in terms of so-called seesaw mechanism~\cite{seesaw}. Because the
right-handed neutrinos $N$ are singlets under the standard gauge
symmetries, they can have Majorana masses as well as Yukawa couplings
to the left-handed leptons ($l$) and Higgs ($\varphi$)
doublets:\footnote{Here, we have omitted the family- and ${\rm
SU}(2)_L$-indices for simplicity.}
%%%
\begin{eqnarray}
 \label{EQ-lag-seesaw}
  {\cal L} = -\frac{1}{2}M\, N^T N 
  +
  \left(
   - h\,N^T l\,\varphi + {\rm H.c.}
   \right)
   \,,
\end{eqnarray}
%%%
where we have taken the $N$ to be the Majorana mass eigenstates.  Then
by integrating out the heavy right-handed neutrinos, we obtain the small
Majorana masses for light neutrinos, $m_{\nu} = h^2\vev{\varphi}^2/M$.

A crucial observation in the lagrangian \EQ{EQ-lag-seesaw} is that the
Majorana mass term $M N^T N$ represents nothing but a lepton number
($L$) violation, actually a $(B-L)$-violation. In fact, the heavy
right-handed neutrinos have two distinct decay channels into leptons and
anti-leptons, and produce $B-L$ asymmetry if the Yukawa couplings $h$
violate the $CP$ and if the decay is out of thermal equilibrium. The
produced lepton asymmetry is partially converted into baryon asymmetry
via the aforementioned sphaleron effect, which explains the baryon
asymmetry in the present universe. This is the original idea of the
leptogenesis~\cite{FY}.

We should also note that if we assume a gauged ${\rm U}(1)_{B-L}$
symmetry, the existence of three generations of the right-handed
neutrinos are automatically required in order to cancel the gauge
anomaly.  As well known, the ${\rm U}(1)_{B-L}$ symmetry is the unique
extra ${\rm U}(1)$ symmetry which can be gauged consistently with the
standard model. Furthermore, the breaking of this ${\rm U}(1)_{B-L}$
symmetry naturally provides large Majorana masses to the right-handed
neutrinos, which leads to the tiny neutrino masses via the seesaw
mechanism. It should be noticed that the ${\rm U}(1)_{B-L}$ symmetry is
consistent with the ${\rm SU}(5)$ GUT, and is embedded in larger GUT
groups such as ${\rm SO}(10)$.
\vspace{0.8em}

Meantime, the supersymmetry (SUSY)~\cite{SUSY} has been attracting wide
interests in particle physics as one of the best candidates for new
physics beyond the standard model. It protects the huge hierarchy
between the electroweak scale and unification (or Planck) scale against
the quadratically divergent radiative corrections, and the particle
contents of the minimal SUSY standard model (MSSM) lead to a beautiful
unification of the three gauge couplings of the standard model at the
scale $\simeq 2\times 10^{16}\GEV$~\cite{unification-theory}, which
strongly suggests the SUSY GUT~\cite{SUSY-GUT}. The MSSM also gives a
natural framework to break the electroweak symmetry
radiatively~\cite{SUSY-rad-breaking}.  On the other hand, from the
viewpoint of cosmology, SUSY provides an ideal dark matter candidate,
the lightest SUSY particle~\cite{SUSY-DM}.  It also protects the
flatness of the inflaton potential against the radiative corrections,
which is inevitable for successful inflation.

However, SUSY also causes a cosmological difficulty if the reheating
temperature of the inflation is too high, that is, the cosmological
gravitino problems~\cite{Gprob,Gprob-GMSB}. After the end of the
inflation, the gravitinos are produced by scattering processes of
particles from the thermal bath, and its abundance is proportional to
the reheating temperature.  Because the gravitino's interaction is
suppressed by the gravitational scale, it has a very long lifetime
unless it is completely stable, and its decay during or after the
big-bang nucleosynthesis (BBN) epoch ($t\sim 1$--$100$ sec) might spoil
the success of the BBN~\cite{Gprob,Gprob-heavy}.  On the other hand, if
the gravitino is completely stable, its present mass density must be
below the critical density of the present universe~\cite{Gprob-GMSB}. In
both cases, the primordial gravitino abundance should be low enough, and
hence the reheating temperature of the inflation is severely constrained
from above, depending on the gravitino mass.
\vspace{0.8em}

In this thesis, we study in detail several leptogenesis scenarios in the
framework of the SUSY.  There have been proposed, in fact, various
leptogenesis scenarios depending on the production mechanisms of the
right-handed neutrinos:
%%%
\begin{itemize}
 \item In the simplest and the most conventional leptogenesis mechanism,
       the right-handed neutrino is produced by thermal
       scatterings~\cite{FY,LGthermal}.  The delayed decay of the
       right-handed neutrino satisfies the out-of-equilibrium condition
       \^^ a la the original GUT baryogenesis. This mechanism requires a
       relatively high reheating temperature to produce the heavy
       right-handed neutrino, and the aforementioned gravitino problem
       makes it somewhat difficult, depending on the gravitino mass.
 \item Another mechanism is given when the right-handed neutrinos are
       produced non-thermally in inflaton decay~\cite{LGinf}. In this
       scenario the reheating temperature can be lower than the case of
       thermal production, and the gravitino problem is avoided in a
       wider range of gravitino mass.
 \item The third mechanism is inherent in the SUSY. The lepton asymmetry is
       produced by the decay of coherent oscillation of the right-handed
       ``s''neutrino, which is the supersymmetric scalar partner of the
       right-handed neutrino~\cite{MSYY-1,MY}. If the right-handed
       sneutrino's oscillation dominates the energy density of the
       universe, the gravitino problem is drastically
       ameliorated~\cite{HMY}.
\end{itemize}
%%%
In the framework of SUSY, there is yet another, completely different
mechanism. The lepton asymmetry is produced not by the right-handed
neutrino decay, but by a coherent oscillation (actually, a rotation) of
a flat direction field including the lepton doublet $L$:
\begin{itemize}
 \item The leptogenesis via $L H_u$ flat
       direction~\cite{MY}, which is based on the Affleck-Dine
       mechanism~\cite{AD}.
\end{itemize}

\clearpage
%%%%%%%%%%%%%%%%%%%%%%%%%%%%%%%%%%%%%%%%%%%%%%%%%%%%%%%%%%%%%%%%%%%
\section{Outline of this thesis}%%%%%%%%%%%%%%%%%%%%%%%%%%%%%%%%%%%
%%%%%%%%%%%%%%%%%%%%%%%%%%%%%%%%%%%%%%%%%%%%%%%%%%%%%%%%%%%%%%%%%%%

%%%%%%%%%%%%%%%%%%%%%%%%%%%%%%%%%%%%%%%%%%%%%%%%%%%%%%%%%%%%
\begin{figure}[t]%%%%%%%%%%%%%%%%%%%%%%%%%%%%%%%%%%%%%%%%%%%
%%%%%%%%%%%%%%%%%%%%%%%%%%%%%%%%%%%%%%%%%%%%%%%%%%%%%%%%%%%%
 \vspace{6.5cm}
 %%%%%%%%%%%%%%%%%%%%%%%%%
 \begin{picture}(0,0)%%%%%
 %%%%%%%%%%%%%%%%%%%%%%%%%
  \put(10,100)
  {$\Frac{n_B}{s}\ne 0\quad\longleftarrow\quad \Frac{n_L}{s}\ne 0$}
  \put(50,70){sphaleron}
  \put(50,55){(\SEC{SEC-sphaleron})}
  \put(200,45){$L H_u$ flat direction}
  \put(220,20){(Chapter~\ref{Chap-LHu})}
  \put(200,30){\vector(-1,1){55}}
  \put(200,190){right-handed}
  \put(210,175){(s)neutrino}
  \put(200,180){\vector(-3,-4){50}}
  \put(190,150){decay}
  \put(180,135){(\SEC{SEC-LAfromN})}
  \put(210,230){(Chapter~\ref{Chap-Ndecay})}
  \put(330,230){\vector(-2,-1){50}}
  \put(340,230){thermal production}
  \put(350,215){(\SEC{SEC-thermal})}
  \put(330,160){\vector(-2,1){50}}
  \put(340,160){inflaton decay}
  \put(350,145){(\SEC{SEC-Inf})}
  \put(330,90){\vector(-2,3){50}}
  \put(340,90){coherent oscillation of}
  \put(340,75){right-handed sneutrino}
  \put(350,60){(\SEC{SEC-Ntilde})}
 %%%%%%%%%%%%%%%%%%%%%%%%%
 \end{picture}%%%%%%%%%%%%
 %%%%%%%%%%%%%%%%%%%%%%%%%
 \vspace{-0.5cm}
%%%%%%%%%%%%%%%%%%%%%%%%%%%%%%%%%%%%%%%%%%%%%%%%%%%%%%%%%%%%
\end{figure}%%%%%%%%%%%%%%%%%%%%%%%%%%%%%%%%%%%%%%%%%%%%%%%%
%%%%%%%%%%%%%%%%%%%%%%%%%%%%%%%%%%%%%%%%%%%%%%%%%%%%%%%%%%%%

The outline of this thesis is as follows.  The rest of this chapter is
devoted to some reviews.  We briefly mention the observed baryon
asymmetry in \SEC{SEC-BAU}. A review of the sphaleron effect, which is a
crucial ingredient of the leptogenesis, is given in \SEC{SEC-sphaleron}.
We also briefly review the results of the cosmological gravitino
problems in \SEC{SEC-grav}.

In Chapter~\ref{Chap-Ndecay}, we discuss leptogenesis scenarios by the
decay of the right-handed (s)neutrino. First, we study the asymmetric
decay of the right-handed neutrino into leptons and anti-leptons in
\SEC{SEC-LAfromN}.  The conventional leptogenesis mechanism where the
right-handed neutrinos are thermally produced is briefly discussed in
\SEC{SEC-thermal}. Then we perform a comprehensive study of the
leptogenesis mechanism in inflaton decay in \SEC{SEC-Inf}, adopting
various SUSY inflation models. In \SEC{SEC-Ntilde}, we investigate the
leptogenesis from coherent right-handed sneutrino. 

In the latter half of this thesis, in Chapter~\ref{Chap-LHu}, we
perform a detailed analysis on the leptogenesis via $L H_u$ flat
direction. This scenario may require another overview, which will be
given in \SEC{SEC-LHu-overview}. Here we mention one point, that the
most important parameter in this scenario which determines the baryon
asymmetry is the mass of the {\it lightest} neutrino, $\mnu{1}$.
(Notice that the data from neutrino-oscillation
experiments~\cite{SK-Atm,Solar,KamLAND} suggest the difference of the
neutrino mass squared, indicating the masses of the heavier neutrinos,
$\mnu{2}$ and $\mnu{3}$.) It is amazing that the observed baryon
asymmetry, which was generated in the very early universe, is directly
related to such a low-energy physics, the neutrino mass.

Chapter~\ref{SEC-Conc} is devoted to conclusions and discussion. Some of
our notations are summarized in Appendix~\ref{SEC-Notation}. Some
ingredients of the standard cosmology are reviewed very briefly in
Appendix.~\ref{SEC-App-SCInf}.

\clearpage
%%%%%%%%%%%%%%%%%%%%%%%%%%%%%%%%%%%%%%%%%%%%%%%%%%%%%%%%%%%%%%%%%%%
\section{Baryon asymmetry}%%%%%%%%%%%%%%%%%%%%%%%%%%%%%%%%%%%%%%%%%
%%%%%%%%%%%%%%%%%%%%%%%%%%%%%%%%%%%%%%%%%%%%%%%%%%%%%%%%%%%%%%%%%%%
\label{SEC-BAU}

We will use the following value for the baryon asymmetry in the present
universe~\cite{KT,PDB}:
%%%
\begin{eqnarray}
 \label{EQ-nBs-obs}
  \frac{n_B}{s}
  \equiv
  \frac{n_b - n_{\bar{b}}}{s}
  = 
  (0.4 - 1)\times 10^{-10}
  \,,
\end{eqnarray}
%%%
where $n_b$ and $n_{\bar{b}}$ are baryon and anti-baryon number density,
respectively, and $s$ is the entropy density of the universe. This ratio
takes a constant value, as long as the baryon number is conserved and no
entropy production takes place. Notice that $n_{\bar{b}}\ll n_b$ and
hence $n_B\simeq n_b$ in the present universe. (We sometimes call the
$n_B$ just ``baryon number density,'' for simplicity.) This value is
determined from the big-bang nucleosynthesis (BBN). The BBN occurs at
the temperature of $T\simeq (1$--$0.1)\MEV$, or equivalently at the
cosmic time $t\sim (1$--$100)$~sec, and generates light nuclei, D,
$^3$He, $^4$He, and $^7$Li.  (For reviews, see, for example,
Refs.~\cite{KT,PDB}.)  The agreement between the predictions of the BBN
theory for the abundances of these light nuclei and the primordial
abundances of them (which are inferred from observational data) is one
of the most important successes of the standard big-bang
cosmology. Actually, the theory of the BBN has basically only one free
parameter,\footnote{We assume the number of neutrino generations to be
three. As well known, this fact itself is one of the most important
implications of the BBN.} i.e., the baryon asymmetry, and all the above
light-element abundances are well explained with $n_B/s$ in the range of
\EQ{EQ-nBs-obs}.

To demonstrate the smallness of the baryon asymmetry, let us calculate
the baryon-number and anti-baryon number density at high temperature.
For example, consider a temperature between the QCD phase transition and
electroweak phase transition, say, $T\sim 10\GEV$. At this temperature,
all the baryons and anti-baryons are expected to exist as quarks and
anti-quarks, and they are well in thermal equilibrium. The ratio of the
baryon-number and anti-baryon-number densities to the entropy density
are then given by
%%%
\begin{eqnarray}
 \frac{n_b}{s}\simeq \frac{n_{\bar{b}}}{s}\simeq 0.02
  \,,
\end{eqnarray}
%%%
where we have used \EQ{EQ-nXs-thermal}. Compared with \EQ{EQ-nBs-obs},
this means, $(n_b - n_{\bar{b}})/(n_b + n_{\bar{b}})\sim {\cal
O}(10^{-9})$, i.e., there were (1,000,000,000+1) baryons per
1,000,000,000 anti-baryons at this epoch.

\clearpage
%%%%%%%%%%%%%%%%%%%%%%%%%%%%%%%%%%%%%%%%%%%%%%%%%%%%%%%%%%%%%%%%%%%
\section{Sphaleron}%%%%%%%%%%%%%%%%%%%%%%%%%%%%%%%%%%%%%%%%%%%%%%%%
%%%%%%%%%%%%%%%%%%%%%%%%%%%%%%%%%%%%%%%%%%%%%%%%%%%%%%%%%%%%%%%%%%%
\label{SEC-sphaleron}

In this section, we give a brief review of the ``sphaleron'' effect,
which is a crucial ingredient of the leptogenesis. If the lepton
asymmetry is successfully generated, it is partially converted into the
baryon asymmetry at equilibrium thanks to the sphaleron-induced baryon
and lepton number violating process, which then explains the observed
baryon asymmetry in the present universe.

\begin{figure}[t]%%%%%%%%%%%%%%%%%%%%%%%%%%%%%%%%%%%%%%%%%%%
%%%%%%%%%%%%%%%%%%%%%%%%%%%%%%%%%%%%%%%%%%%%%%%%%%%%%%%%%%%%
 \centerline{ {\psfig{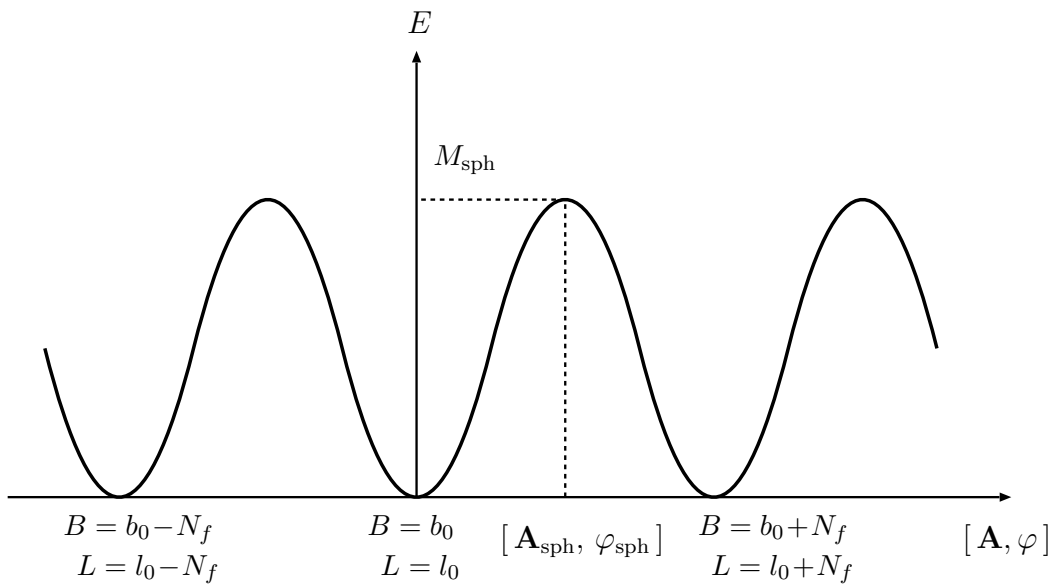}} }
 \vspace{-1cm}
 %%%%%%%%%%%%%%%%%%%%%%%%%
 \begin{picture}(0,0)%%%%%
 %%%%%%%%%%%%%%%%%%%%%%%%%
  \put(210,220){$E$}
  \put(220,170){\small $M_{\rm sph}$}
  \put(80,30){\small $B = b_0\!-\!N_f$}
  \put(85,15){\small $L = l_0\!-\!N_f$}
  \put(195,30){\small $B = b_0$}
  \put(200,15){\small $L = l_0$}
  \put(320,30){\small $B = b_0\!+\!N_f$}
  \put(325,15){\small $L = l_0\!+\!N_f$}
  \put(420,25){$[\,{\bf A}, \varphi\,]$}
  \put(245,25){$[\,{\bf A}_{\rm sph},\,\varphi_{\rm sph} \,]$}
 %%%%%%%%%%%%%%%%%%%%%%%%%
 \end{picture}%%%%%%%%%%%%
 %%%%%%%%%%%%%%%%%%%%%%%%%
 \caption{A Schematic behavior of the energy dependence on the
 configuration of the gauge and Higgs fields $[\,{\bf A}(x), \varphi
 (x)\,]$~\cite{sphaleron}. The minima correspond to topologically
 distinct vacua with different baryon ($B$) and lepton ($L$)
 numbers. The configuration $[\,{\bf A}_{\rm sph}(x),\,\varphi_{\rm sph}
 (x)\,]$ represents the saddle point of the energy functional, the
 sphaleron solution.}
 \vspace{1em}
 \label{FIG-sphaleron}
%%%%%%%%%%%%%%%%%%%%%%%%%%%%%%%%%%%%%%%%%%%%%%%%%%%%%%%%%%%%
\end{figure}

At the classical level, the Lagrangian of the standard ${\rm
SU}(3)_C\times {\rm SU}(2)_L\times {\rm U}(1)_Y$ gauge theory clearly
conserves the baryon ($B$) and lepton ($L$) numbers. However, as was
first pointed out by 't Hooft~\cite{tHooft}, both $B$ and $L$ are
violated by quantum effects:
%%%
\begin{eqnarray}
 \label{EQ-SU2-anomaly}
 \partial_\mu J_B^\mu
  =
  \partial_\mu J_L^\mu
  =
  N_f
  \frac{g_2^2}{32\pi^2}
  \epsilon_{\mu\nu\rho\sigma}
  {\rm Tr}
  F^{\mu\nu}
  F^{\rho\sigma}
  \,,
\end{eqnarray}
%%%
where $N_f$ is the number of fermion generations and $g_2$ and
$F_{\mu\nu}$ are the coupling constant and the field strength of the
${\rm SU}(2)_L$ gauge group, respectively. It was shown later that the
${\rm SU}(2)_L\times {\rm U}(1)_Y$ gauge theory has a non-contractible
loop (path) in the field configuration space, which connects
topologically distinct vacua with different baryon and lepton
numbers~\cite{Manton}, and it was found that the highest energy
configuration along this ``path'' corresponds to a spatially localized
and static, but unstable
solution~\cite{Kinkhamer-Manton,sph-others}. This solution was named
``sphaleron''. Since this sphaleron solution is a saddle point of the
field potential energy, its energy represents the height of the
barrier between the vacua with different baryon and lepton numbers
(see \FIG{FIG-sphaleron}).
%%%%%%%%%%%%%%%%%%%%%%%%%%%%%%%%%%%%%%%%%%%%%%%%%%%%%%%%%%%%
%%%%%%%%%%%%%%%%%%%%%%%%%%%%%%%%%%%%%%%%%%%%%%%%
%%%%%%%%%%%%%%%%%%%%%%%%%%%%%%%%%%%%%%%%%%%%%%%%%%%%%%%%%%%%

At zero temperature, the rate of baryon (and lepton) number violating
process via a tunneling between the topologically distinct vacua (for
example, proton decay) is extremely small~\cite{tHooft}, since it is
suppressed by the factor of $\exp (-16\pi^2/g_2^2)\sim 10^{-170}$.
However, as suggested by Kuzmin, Rubakov, and
Shaposhnikov~\cite{sphaleron,sph-early-comments}, such processes are not
suppressed and can even be efficient at temperatures close to (and
above) the electroweak phase transition. The transition of the fermion
number occurs simultaneously for each fermion doublets, so that the
total changes in the numbers of baryons and leptons are $\Delta B =
\Delta L = N_f$. Therefore, $B-L$ is conserved. (This is clear from the
absence of the anomaly in $B-L$ in \EQ{EQ-SU2-anomaly}.)

Let us roughly estimate the temperature $T_*$ above which this
$(B+L)$-violating process becomes in thermal equilibrium, according to
Ref.~\cite{sphaleron}. We assume that the electroweak phase transition
is the second order, and consider the temperature $T<T_C$, where $T_C$
is the critical temperature above which the vacuum expectation value
$v(T)$ of the Higgs field vanishes. In such a situation, the transition
from one vacuum (say, $B=b_0$ and $L=l_0$) to the next vacuum ($B=b_0\pm
N_f$ and $L = l_0\pm N_f$) occurs at the rate~\cite{sphaleron}
%%%
\begin{eqnarray}
 \label{EQ-sph-rate}
 \Gamma = C(T)\,\,T
  \exp
  \left(
   -\,\frac{M_{\rm sph}(T)}{T}
   \right)
   \,,
\end{eqnarray}
%%%
where dimensionless factor $C(T)$ depends on the ratio $v(T)/T$ and the
coupling constants.\footnote{See comments below.} $M_{\rm sph}(T)$
represents the free energy of the sphaleron configuration (at
temperature $T$), which is given by~\cite{Kinkhamer-Manton}
%%%
\begin{eqnarray}
 M_{\rm sph}(T)
  =
  4\pi
  B(T)
   \frac{v(T)}
   {g_2(T)}
   \,,
\end{eqnarray}
%%%
where $B(T)$ depends on the gauge coupling $g_2(T)$ and the 4-point
coupling constant of the Higgs potential $\lambda(T)$ as $B =
B(\lambda/g_2^2)$, varying from $1.5$ ($\lambda/g_2^2\to 0$) to $2.7$
($\lambda/g_2^2\to \infty$)~\cite{Kinkhamer-Manton}. The rate in
\EQ{EQ-sph-rate} should be compared with the Hubble expansion rate $H =
(\pi^2 g_*/90)^{1/2}\times T^2/M_G$. ($M_G = 2.4\times 10^{18}\GEV$ is
the reduced Planck scale and $g_*$ is defined in
Appendix~\ref{SEC-App-Thermo}.) Then, it is found that the sphaleron
rate in \EQ{EQ-sph-rate} indeed exceeds the Hubble expansion rate for
$T>T_*$, where $T_*$ is given by
%%%
\begin{eqnarray}
 T_* \simeq 4\pi B(T_*)
  \frac{v(T_*)}
   {g_2(T_*)}
   \times
   \left[
    \ln\left(\frac{M_G}{T_*}\right)
    \right]^{-1}
    \,.
\end{eqnarray}
%%%
Here, we have neglected the $\,\ln C(T)$ and $\,\ln (\pi^2 g_*/90)$
compared with the large number $\,\ln(M_G/T_*)$. Recalling that $v(T)$
approaches to zero for $T\to T_C$, this occurs below the critical
temperature, i.e., $T_* < T_C$.  Therefore, the sphaleron-induced
$(B+L)$-violating process becomes in equilibrium for $T_* < T \,(<
T_C)$. Precisely speaking, a more careful treatment including the
prefactor $C(T)$, which turns out to be proportional to
$\left(v(T)/T\right)^7$~\cite{sph-lowT}, is required. A numerical
estimation shows that $T_*$ is just below the critical temperature,
$(T_C - T_*)\lsim 10\GEV$, for $50\GEV \lsim T_C \lsim
200\GEV$~\cite{TCTstar}.

The evaluation of the ``sphaleron'' rate for $T>T_C$ is a complicated
problem, since the sphaleron configuration no longer exists in the
symmetric phase.\footnote{Although the saddle-point sphaleron solution
does not exist in the symmetric phase, we will call this anomalous
$(B+L)$-violating process ``sphaleron'' also for $T>T_C$.} Naively
thinking, there seems no reason for the $(B+L)$-violating processes to
be suppressed, since the potential barrier vanishes. Actually,
theoretical arguments as well as numerical calculations
suggest~\cite{sph-highT} that the sphaleron rate per unit time per unit
volume is $\Gamma/V \sim \alpha_2^5 T^4$ for $T > T_C$, where
$\alpha_2\equiv g_2^2/(4\pi)$. By using their result, it is found that
the sphaleron rate exceeds the Hubble expansion rate for $T\lsim
10^{12}\GEV$.

Therefore, the sphaleron-induced $(B+L)$-violating process is in thermal
equilibrium in the range of
%%%
\begin{eqnarray}
 \label{EQ-100-1012}
  100\GEV\sim T_* < T \lsim 10^{12}\GEV
  \,.
\end{eqnarray}
%%%

\subsubsection*{$\bullet$ relation between baryon and lepton asymmetry}

Let us calculate the relation between the baryon and lepton number in
the presence of the sphaleron process, by means of the analysis of the
chemical potentials~\cite{L-to-B,L-to-B-actual}. At first sight, it
seems that the relation is just given by $B+L=0$ since the sphaleron
effect violates $B+L$, preserving $B-L$. However, as we will see, a
nontrivial relation is derived if there exists a non-vanishing $B-L$
asymmetry.

We first consider the standard model without supersymmetric particles,
but with $N_{\varphi}$ Higgs doublets, $\varphi_i$ ($i = 1\cdots
N_{\varphi}$), and $N_f$ generations of fermions, i,e., left-handed
quark doublets $q_j$, right-handed up-type and down-type quarks
$\bar{u}_j$ and $\bar{d}_j$, left-handed lepton doublets $l_j$ and
right-handed charged leptons $\bar{e}_j$ ($j = 1\cdots N_f$). Hereafter,
we consider the symmetric phase $T>T_C$. Hence, the ${\rm SU}(2)_L\times
{\rm U}(1)_Y$ symmetry recovers. We denote the asymmetry of the number
density of particle $i$ by $\Delta n_i$, which is related to the
chemical potential of that particle as follows [see \EQ{EQ-chemical}]:
%%%
\begin{eqnarray}
  \label{EQ-deltan-mu}
  \Delta n_i 
  \equiv
  n_i^{(+)} - n_i^{(-)}
  =
  \left\{
   \begin{array}{ccc}
    \Frac{1}{6}g_i T^3
     \left(\Frac{\mu_i}{T}\right)
     \quad
     &{\rm for}&
     {\rm  fermion}
     \,,
     \\
    &&
     \\
    \Frac{1}{3}g_i T^3
     \left(\Frac{\mu_i}{T}\right)
     \quad
     &{\rm for}&
     {\rm  boson}
     \,.
   \end{array}
   \right.
\end{eqnarray}
%%%
Therefore, the baryon and lepton number asymmetries are given by
%%%
\begin{eqnarray}
 \label{EQ-nB-nL-mu}
 n_B
  &\equiv&
  \sum_j
  \left(
   \frac{1}{3}
   \Delta n_{q_j}
   -
   \frac{1}{3}
   \Delta n_{\bar{u}_j}
   -
   \frac{1}{3}
   \Delta n_{\bar{d}_j}
   \right)
   \nonumber \\
 &=&
  \frac{1}{6}T^2
  \times
  \sum_j
  \left(
   2\mu_{q_j}
   -
   \mu_{\bar{u}_j}
   -
   \mu_{\bar{d}_j}
   \right)
   \,,
   \nonumber\\
 n_{L_j}
  &\equiv&
  \Delta n_{l_j}
  -
  \Delta n_{\bar{e}_j}
  \nonumber\\
 &=&
  \frac{1}{6}T^2
  \times
  \left(
   2 \mu_{l_j}
   -
   \mu_{\bar{e}_j}
   \right)
   \,,
\end{eqnarray}
%%%%
where we have used the fact that the chemical potentials of particles in
the same gauge multiplet are the same, which is ensured by the gauge
interactions. (Note that the ${\rm SU}(2)_L$ is also recovered, since we
consider $T>T_C$.)

In the following, we derive the relations between the chemical
potentials $\mu_i$. First of all, we assume that all the Higgs doublets
have the same chemical potentials due to the mixings between themselves,
i.e., $\mu_{\varphi_i}\equiv \mu_{\varphi}$. (If there is a Higgs
doublet with a conjugate quantum number, like in the MSSM, we redefine
the chemical potential of that Higgs doublet with an additional minus
sign.) Then, the interactions via Yukawa couplings
%%%
\begin{eqnarray}
 y_{e,j}\,\varphi^*\, l_j\, \bar{e}_j
  \,,\qquad
  y_{u,jk}\,\varphi\,\, q_j\, \bar{u}_k
  \,,\qquad
  y_{d,jk}\,\varphi^*\, q_j\, \bar{d}_k
  \,,
\end{eqnarray}
%%%
lead to
%%%
\begin{eqnarray}
 \label{EQ-chemi-Yukawa}
 - \mu_{\varphi}
  + \mu_{l_j}
  + \mu_{\bar{e}_j}
  &=& 0
  \qquad
  j = 1\cdots N_f
  \,,
  \nonumber \\
 \mu_{\varphi}
  + \mu_{q_j}
  + \mu_{\bar{u}_k}
  &=& 0
  \qquad
  j,k = 1\cdots N_f
  \,,
  \nonumber \\
 - \mu_{\varphi}
  + \mu_{q_j}
  + \mu_{\bar{d}_k}
  &=& 0
  \qquad
  j,k = 1\cdots N_f
  \,.
\end{eqnarray}
%%%
Here, we have taken a basis of gauge eigenstates for the quarks. Thus,
because of the mixing in the Yukawa couplings, the chemical potentials
of the quarks become generation independent: $\mu_{q_j}\equiv \mu_q$,
$\mu_{\bar{u}_j}\equiv \mu_{\bar{u}}$ and $\mu_{d_j}\equiv
\mu_{\bar{d}}$ ($j=1\cdots N_f$). Notice that the relations in
\EQ{EQ-chemi-Yukawa} hold as long as these interactions are in thermal
equilibrium, and even the electron Yukawa coupling, which is the
smallest one, is in equilibrium for $T\lsim 10^4\GEV$.  Next, the charge
neutrality of the universe requires vanishing total ${\rm U}(1)_Y$
charge:
%%%
\begin{eqnarray}
 \sum_i 
  \left(
   \Delta n_i Y_i
   \right)
   = 0
   \,,
\end{eqnarray}
%%%
where $Y_i$ denotes the ${\rm U}(1)_Y$ charge of particle $i$. This
leads to
%%%
\begin{eqnarray}
 \label{EQ-U1Yneutral}
 N_f
  \left(
   \mu_q
   - 2\mu_{\bar{u}}
   + \mu_{\bar{d}}
   \right)
   +
   \sum_j
   \left(
    - \mu_{l_j}
    + \mu_{\bar{e}_j}
    \right)
    +
    2N_{\varphi}\,\mu_{\varphi}
    = 0
    \,.
\end{eqnarray}
%%%
Finally, the sphaleron interaction can be understood as
%%%
\begin{eqnarray}
 \sum_{\small \begin{array}{c}
  i = {\rm all\,\,{\rm SU}(2)_L}\\
	{\rm doublet\,\,fermions}
       \end{array}}
       \!\!\!\!\!\!\!\!\!\!\!\!
       \psi_i
       &\longleftrightarrow&
       {\rm vacuum}
       \,,
\end{eqnarray}
which leads to
\begin{eqnarray}
 \label{EQ-chemi-spha}
  3N_f \mu_q
  +
  \sum_j
  \mu_{l_j}
  =
  0
  \,.
\end{eqnarray}
%%%
{}From \EQS{EQ-chemi-Yukawa}, (\ref{EQ-U1Yneutral}) and
(\ref{EQ-chemi-spha}), all the chemical potentials can be written in
terms of $N_f$ independent ones, say, $\mu_{\bar{e}_j}$. Then, from
\EQ{EQ-nB-nL-mu}, we obtain
%%%
\begin{eqnarray}
 n_B
  &=&
  \frac{1}{6}T^2
  \times
  \frac{8 N_f + 4 N_{\varphi}}
  {2 N_f + 3 N_{\varphi}}
  \sum_j
  \mu_{\bar{e}_j}
  \,,
  \nonumber \\
 n_{L_j}
  &=&
  \frac{1}{6}T^2
  \times
  \left[
   \left(
    \frac{-8}
    {2 N_f + 3 N_{\varphi}}
    \sum_k
    \mu_{\bar{e}_k}
    \right)
    -
    3\mu_{\bar{e}_j}
   \right]
   \,,
   \nonumber \\
 n_L
  &\equiv&
  \sum_j
  n_{L_j}
  \nonumber \\
 &=&
  \frac{1}{6}T^2
  \times
  \frac{-14 N_f - 9 N_{\varphi}}
  {2 N_f + 3 N_{\varphi}}
  \sum_j
  \mu_{\bar{e}_j}
  \,.
\end{eqnarray}
%%%
Because the conserved quantity is the $B-L$ asymmetry in the present
analysis,\footnote{Actually, not only the total $B-L$ but each $(1/N_f)B
- L_j$ ($j = 1\cdots N_f$) is conserved in the present situation.} it is
suitable to rewrite the above relations in terms of $n_B - n_L$:
%%%
\begin{eqnarray}
 \label{EQ-nBnL}
 n_B = 
 C
 \left( n_B - n_L \right)
 \,,
\end{eqnarray}
where $C$ is given by~\cite{L-to-B}
\begin{eqnarray}
 \label{EQ-nBnLcoeff}
 C = \frac{8 N_f + 4 N_{\varphi}}{ 22 N_f + 13 N_{\varphi}}
  \,.
\end{eqnarray}
%%%
In the case of the standard model, $N_f = 3$ and $N_\varphi = 1$ yield
$C = 28/79$.  

%%%%%%%%%%%%%%%%%%%%%%%%%%%%%%%%%%%%%%%%%%%%%%%%%%%%%%%%%%%%
\begin{figure}[t]%%%%%%%%%%%%%%%%%%%%%%%%%%%%%%%%%%%%%%%%%%%
%%%%%%%%%%%%%%%%%%%%%%%%%%%%%%%%%%%%%%%%%%%%%%%%%%%%%%%%%%%%
 \centerline{ {\psfig{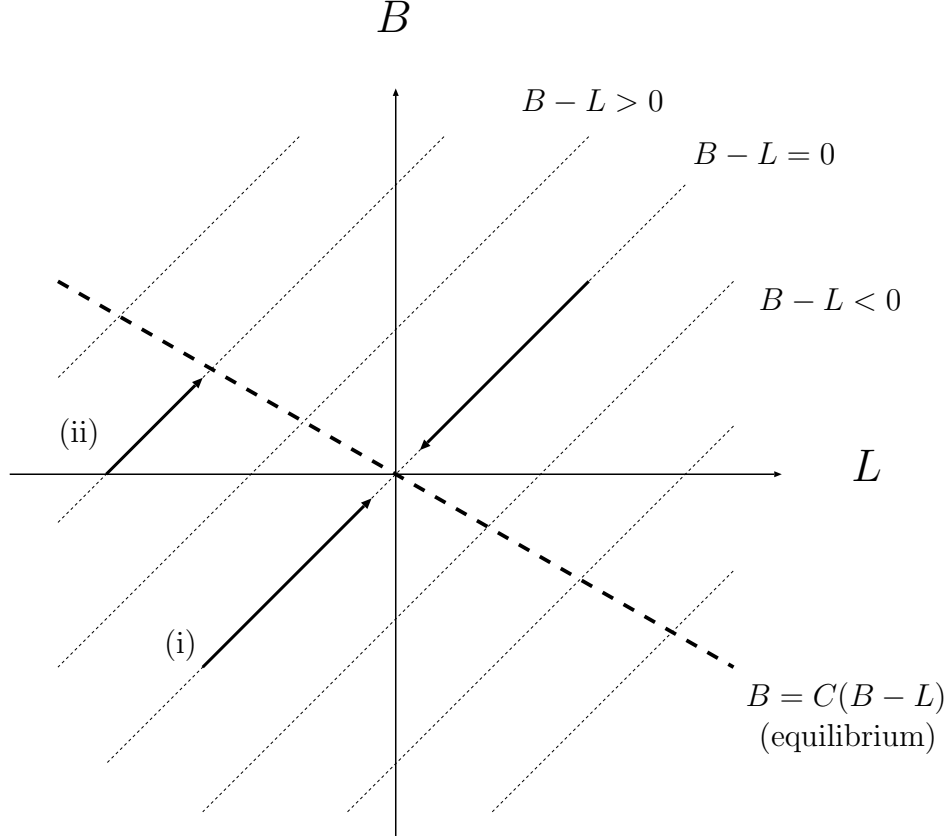}} }
 %%%%%%%%%%%%%%%%%%%%%%%%%
 \begin{picture}(0,0)%%%%%
 %%%%%%%%%%%%%%%%%%%%%%%%%
  \put(220,320){\Large $B$}
  \put(400,150){\Large $L$}
  \put(360,65){$B = C(B-L)$}
  \put(365,50){(equilibrium)}
  \put(340,270){$B-L = 0$}
  \put(365,215){$B-L < 0$}
  \put(275,290){$B-L > 0$}
  \put(140,85){(i)}
  \put(100,165){(ii)}
 %%%%%%%%%%%%%%%%%%%%%%%%%
 \end{picture}%%%%%%%%%%%%
 %%%%%%%%%%%%%%%%%%%%%%%%%
 \caption{The relation between baryon ($B$) and lepton ($L$)
 number. Thin dotted lines correspond to constant $B-L$, along which $B$
 and $L$ can move via the sphaleron process. At equilibrium, $B$ and $L$
 reach the thick dashed line, which represents $B = C(B-L)$.}
 \vspace{1em}
 \label{FIG-BandL}
%%%%%%%%%%%%%%%%%%%%%%%%%%%%%%%%%%%%%%%%%%%%%%%%%%%%%%%%%%%%
\end{figure}%%%%%%%%%%%%%%%%%%%%%%%%%%%%%%%%%%%%%%%%%%%%%%%%
%%%%%%%%%%%%%%%%%%%%%%%%%%%%%%%%%%%%%%%%%%%%%%%%%%%%%%%%%%%%

We can see from \EQ{EQ-nBnL} that, if $B-L$ asymmetry is absent, any
baryon asymmetry vanishes at equilibrium in the presence of the
sphaleron effect. (See the arrow (i) in \FIG{FIG-BandL}.) On the other
hand, if there occurs a successful leptogenesis, i,e., if nonzero lepton
asymmetry (and hence $B-L$ asymmetry) is generated in an
out-of-equilibrium way, it is partially converted into baryon
asymmetry~\cite{FY}. (See the arrow (ii) in \FIG{FIG-BandL}.) The amount
of the baryon asymmetry at equilibrium is obtained from \EQ{EQ-nBnL} as
%%%
\begin{eqnarray}
 \label{EQ-pre-L-to-B}
  \left.
   \frac{n_B}{s} 
   \right|_{\rm eq}
   =   
   C
   \left.
    \frac{n_B - n_L}{s}
    \right|_{\rm eq}
     =
     -C
     \left.
      \frac{n_L}{s}
      \right|_{\rm initial}
      \,,
\end{eqnarray}
%%%
where we have normalized the number densities by the entropy density, so
that they become constant against the expansion of the universe.

In the case of the MSSM, we have an additional Higgs doublet as well
as supersymmetric partners. Let us calculate the coefficient $C$ in
the presence of those particles.  First, the
(gaugino)-(fermion)-(sfermion)$^*$ interactions ensure that the
chemical potential of each sfermion\footnote{''sfermion'' denotes a
scalar partner of the fermion.} is the same as that of the
corresponding fermion. (Note that the chemical potentials of the
gauginos vanish since they are Majorana particles.) Next, the
supersymmetric mass term $W = \mu H_u H_d$ makes the chemical
potentials of the up-type and down-type Higgsinos have opposite signs,
and hence the relation induced by the sphaleron effect,
\EQ{EQ-chemi-spha}, does not change. However, the condition of the
vanishing ${\rm U}(1)_Y$ charge [\EQ{EQ-U1Yneutral}] changes. (Note
that the contributions from bosons and fermions are different. See
\EQ{EQ-deltan-mu}.)  Consequently, the coefficient $C$ becomes again
$C=28/79$, in spite of the presence of two Higgs doublets.  In the
actual thermal history, however, the Higgsino and sfermions likely
become massive and are essentially absent at the time of electroweak
phase transition. Hence, it is more appropriate to calculate $C$
without them, but with two Higgs doublets. This ($N_f = 3$, $N_\varphi
= 2$) leads to $C = 8/23$.

So far, we have considered the symmetric phase $T > T_C$. The actual
ratio of the baryon to lepton asymmetry in the present universe depends
on how the electroweak phase transition
occurs~\cite{L-to-B-actual}. More precisely, it depends on the value of
$v(T)/T$ just before the sphaleron decoupling $T\simeq T_*$.
Fortunately, however, numerically it does not change much from
\EQ{EQ-nBnLcoeff}. (The coefficient $C$ changes from $8/23$ $(v\ll T)$
to $10/31$ $(v\gg T)$ for $N_f = 3$ and $N_\varphi =
2$~\cite{L-to-B-actual}, so that the difference is at most a few
percent.) Thus, we will take $C = 8/23\simeq 0.35$ throughout this
thesis, for simplicity.

Finally, we should also note the sign of the baryon asymmetry. We know
that the sign of the present baryon asymmetry is positive, i.e., $n_B/s
> 0$.\footnote{This is not a matter of definition, since we can
distinguish the matter from antimatter by $CP$-violating processes in
the laboratory experiments (e.g., the asymmetry in the decay $K_L^0\to
\pi^{\pm}e^{\mp}\nu$), which is independent of the definition of
``matter'' we would name from the cosmological baryon asymmetry.} Thus,
from \EQ{EQ-pre-L-to-B}, it is found that the leptogenesis must generate
a lepton asymmetry with a negative sign, $n_L/s < 0$ (i.e., more
anti-leptons than leptons should be produced). In principle, the sign of
the generated lepton asymmetry could be determined if we know the sign
of the effective $CP$-violating phase in each leptogenesis mechanism. In
the case of leptogenesis by the decay of right-handed neutrino
(discussed in Chapter~\ref{Chap-Ndecay}), it depends on the phases of
the Yukawa couplings of the right-handed neutrino [see
\EQ{EQ-deltaeff}]. On the other hand, in the case of leptogenesis via $L
H_u$ flat direction (discussed in Chapter~\ref{Chap-LHu}), the effective
$CP$-violating phase is determined by the relative phase between the
phase of the SUSY-breaking term and the initial phase of the flat
direction field $\phi$, which depends on the coupling of $\phi$ to the
inflaton field.

In both of those cases, however, it is highly difficult to determine the
sign of the phases. Thus, we will simply assume that the leptogenesis
mechanisms we will discuss produce the lepton asymmetry with a correct
sign. Keeping in mind the discussion above, we will omit the relative
sign in \EQ{EQ-pre-L-to-B} for simplicity, and use the following
relation throughout this thesis:
%%%
\begin{eqnarray}
 \frac{n_B}{s} = 0.35\times 
  \left.\frac{n_L}{s}\right|_{\rm initial}
  \,.
\end{eqnarray}
%%%

\clearpage
%%%%%%%%%%%%%%%%%%%%%%%%%%%%%%%%%%%%%%%%%%%%%%%%%%%%%%%%%%%%%%%%%%%
\section{Cosmological gravitino problems}%%%%%%%%%%%%%%%%%%%%%%%%%%
%%%%%%%%%%%%%%%%%%%%%%%%%%%%%%%%%%%%%%%%%%%%%%%%%%%%%%%%%%%%%%%%%%%
\label{SEC-grav}

In this section, we briefly review the results of the cosmological
gravitino problems obtained in the literature, since they give a very
important and severe constraint on the baryogenesis, i.e., the upper
bound on the reheating temperature. For a review, see
Ref.~\cite{Moroi-D}.

\subsubsection*{$\bullet$ unstable gravitino}

\begin{figure}[t]%%%%%%%%%%%%%%%%%%%%%%%%%%%%%%%%%%%%%%%%%%%
%%%%%%%%%%%%%%%%%%%%%%%%%%%%%%%%%%%%%%%%%%%%%%%%%%%%%%%%%%%%
 \begin{center}
  \begin{tabular}{r|llcc}
   cosmic time & temperature &&&
   \\
   {\large $t$} & {\large $T$} &&&
   \\
   \hline
   && {\bf inflation} &&
   \\
   ?? & {\large $T_R$} & reheating
   \begin{picture}(0,0)
    \put(10,0){\vector(1,0){120}}
   \end{picture}
   && gravitino 
   \\
   &&& production & 
   \begin{picture}(0,0)
    \put(-3,0){\line(0,-1){80}}
    \put(-3,-80){\vector(-1,-4){15}}
    \put(-3,-80){\vector(1,-4){30}}
   \end{picture}
   \\
   &&&&
   \\
   && 
   $\left(
   \begin{array}{c}
    {\rm {\bf baryogenesis}}
     \\
    \Frac{n_B}{s}\simeq 10^{-10}
   \end{array}
   \right)$
   &&
   \\
   &&
   \begin{picture}(0,0)
    \put(30,5){\vector(-1,-3){10}}
   \end{picture}
   &&
   \\
   &&&&
   \\
   &&&&
   \\
   $\sim 1$~sec & $\sim 1\MEV$ & {\bf BBN} && 
   \hspace{-2em} (unstable)\qquad(stable)
   \\
   && $\to$ light elements &&
   \\
   &&
   \begin{picture}(0,0)
    \put(30,5){\vector(0,-1){60}}
   \end{picture}    
   &&
   \\
   &&&& \hspace{-2em}decay
   \begin{picture}(0,0)
    \put(-50,0){\vector(-1,0){130}}
   \end{picture}       
   \\
   &&\qquad\qquad\quad destroy ?? &&
   \\
   &&&&
   \\
   &&&&
   \\
   $\sim 10^{10}$ years & $2.73 K$ & \quad observed && \qquad\quad overclose ??
   \\
   &&&&
   \\
  \end{tabular}
 \end{center}
 \caption{A schematic figure of the thermal history of the universe
 relevant to the cosmological gravitino problems.}
 \vspace{1em}
%%%%%%%%%%%%%%%%%%%%%%%%%%%%%%%%%%%%%%%%%%%%%%%%%%%%%%%%%%%%
 \label{FIG-grav}%%%%%%%%%%%%%%%%%%%%%%%%%%%%%%%%%%%%%%%%%%%
\end{figure}
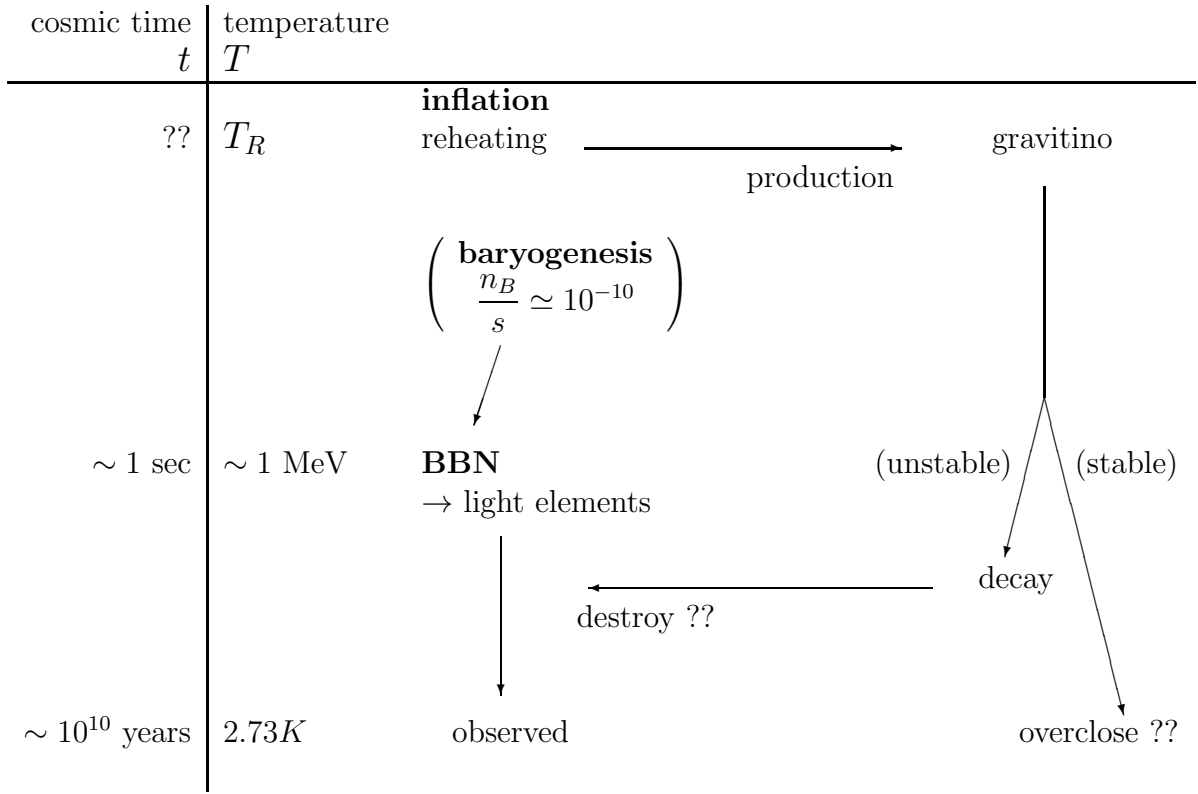

There are two cases; unstable and stable gravitino. (See
\FIG{FIG-grav}.) Let us first consider the unstable gravitino. Since the
couplings of the gravitino with ordinary matter are strongly suppressed
by the gravitational scale, it has a very long lifetime:
%%%
\begin{eqnarray}
 \label{EQ-grav-lifetime}
  \tau_{3/2}\simeq 4\times 10^5
  \times
  \left(
   \frac{m_{3/2}}{1\TEV}
   \right)^{-3}
   \,\,{\rm sec}
   \,,
\end{eqnarray}
%%%
where $m_{3/2}$ denotes the gravitino mass, and we have assumed that the
gravitino dominantly decays into a photon and a photino\footnote{If the
gravitino mainly decays into a neutrino and a sneutrino, the upper bound
on the reheating temperature becomes higher. See discussion in remarks
below.} and omitted the phase space suppression of the decay rate, for
simplicity. Therefore, it decays after the big-bang nucleosynthesis
(BBN) epoch ($t\sim 1$--$100$ sec), unless gravitino is heavier than
$\sim 10\TEV$~\cite{Gprob-W}. Then the energetic photon (or some charged
particle) produced in the gravitino decay induce electromagnetic cascade
process. This cascade might destroy the light elements and change their
abundances, and spoil the success of the BBN~\cite{Gprob}. Since the
abundance of the gravitinos produced from thermal environment at
reheating epoch is roughly proportional to the reheating temperature
$T_R$, usually there are upper bounds on the $T_R$ depending on the
gravitino mass. Recently, a detailed calculation of the gravitino
production rate in SUSY QCD at high temperature has been
done~\cite{BBB}.\footnote{Their calculation~\cite{BBB} shows a slightly
smaller abundance of the produced gravitinos than in earlier works.} A
recent analysis of the effect of the radiative decays of massive
particles on the BBN is found in Ref.~\cite{Gprob-recent}.  By using
their results, the upper bounds are given by $T_R\lsim 10^6$, $10^9$,
and $10^{12}\GEV$ for $m_{3/2} = 100\GEV$, $1\TEV$ and $3\TEV$,
respectively.\footnote{Here, we have taken the gluino mass to be
$m_{\tilde{g}}=1\TEV$. Precisely speaking, the abundance of the
gravitinos produced by thermal scatterings depends on the gluino mass as
$\propto \{1 + m_{\tilde{g}}^2(T) / (3
m_{3/2}^2)\}$~\cite{BBB}.\label{FN-grav-gluino}}

If the gravitino is even heavier, as in the anomaly mediated SUSY
breaking models~\cite{AMSB}, it decays during or near the BBN epoch
($t\sim 10^{-2}$--$10^2$ sec). In this case, there is another constraint
which comes from hadronic decay~\cite{Gprob-heavy}. If energetic hadrons
are emitted at this epoch, they interconvert the neutrons ($n$) and
protons ($p$) in the background even after the freeze-out time of the
$n/p$ ratio ($t\sim 1$~sec), which results in the change of the
abundances of the light elements. Thus, there are upper bounds on the
reheating temperature $T_R$ times the branching ratio of the gravitino
decay into hadrons $B_h$. By using the results of Ref.~\cite{BBB} and a
recent analysis of the effects of the hadronic decay on the BBN in
Ref.~\cite{Gprob-heavy-recent}, the upper bounds on the reheating
temperatures are given by $(B_h/0.1)\times T_R \lsim
10^9$--$10^{11}\GEV$ for $m_{3/2}\simeq$ (a few --
$100)\TEV$.\footnote{We have taken the gluino mass to be
$m_{\tilde{g}}=1\TEV$. See footnote~\ref{FN-grav-gluino}.}

%%%%%%%%%%%%%%%%%%%%%%%%%%%%%%%%%%%%%%%%%%%%%%%%%%%%%%%%%%%%
%%%%%%%%%%%%%%%%%%%%%%%%%%%%%%%%%%%%%%%%%%%%%%%%
%%%%%%%%%%%%%%%%%%%%%%%%%%%%%%%%%%%%%%%%%%%%%%%%%%%%%%%%%%%%

\subsubsection*{$\bullet$ stable gravitino}

Now let us turn to discuss the case of stable gravitino, which is the
case if the gravitino is the lightest SUSY particle (LSP). (We assume
the $R$-parity and hence the LSP is stable. A gravitino much lighter
than the weak scale is realized in a low-energy SUSY breaking scenario
such as gauge-mediated SUSY breaking models~\cite{GMSB}.\footnote{This
is not the case if the SUSY breaking is mediated by a bulk gauge field
in higher dimension spacetime~\cite{bulkU1}.})  In this case, the
upper bound on the reheating temperature $T_R$ comes from the
requirement that the energy density of the gravitino not
overclose\footnote{The word ``overclose'' might not be appropriate,
since the open or flat universe ($\Omega \le 1$) cannot change to a
closed ($\Omega > 1$) universe. The actual problem is that the Hubble
expansion would be too high compared with the observation at a
temperature of $T = T_0 = 2.73 K$, if the calculated value of $\Omega$
exceeds unity~\cite{KT}.} the present universe~\cite{Gprob-GMSB}. From
the result in Ref.~\cite{BBB}, the relic abundance of the gravitinos
which are produced by scattering processes of particles from the
thermal bath after the inflation is given by\footnote{Besides the
factor in \EQ{EQ-grav-abund}, the abundance of the gravitino depends
on the reheating temperature also through the ${\rm SU}(3)_C$ running
coupling, $\alpha_3(T_R)$. Precisely speaking, the final abundance is
proportional to $\alpha_3(T_R)^3\times \{$a logarithmic correction
(known)$\}$ for $m_{\tilde{g}}\gg m_{3/2}$~\cite{BBB}.}
\begin{eqnarray}
 \label{EQ-grav-abund}
  \Omega_{3/2}\,h^2
  \simeq
  0.3\times
  \left(
   \frac{m_{\tilde{g}}}
   {1\TEV}
   \right)^2
   \left(
   \frac{m_{3/2}}
   {10\MEV}
   \right)^{-1}
    \left(
     \frac{T_R}
     {10^6\GEV}
     \right)
     \,.
     \label{EQ-omega32}
\end{eqnarray}
Here, $m_{\tilde{g}}$ is the gluino mass, $h$ is the present Hubble
parameter in units of $100$ km sec$^{-1}$\,Mpc$^{-1}$ and $\Omega_{3/2}
= \rho_{3/2}/\rho_c$ .  ($\rho_{3/2}$ and $\rho_c$ are the present
energy density of the gravitino and the critical energy density of the
present universe, respectively.) It is found from \EQ{EQ-omega32} that
the overclosure limit $\Omega_{3/2} < 1$ puts a severe upper bound on
the reheating temperature $T_R$, depending on the gravitino mass
$m_{3/2}$. Here, we have omitted the contribution from the decays of
squarks and sleptons into gravitinos. This effect makes the upper bound
on the reheating temperature slightly severer for a smaller gravitino
mass region, $m_{3/2}\sim (1$--$100)\KEV$~\cite{Gprob-GMSB}. For a even
lighter gravitino $m_{3/2}\lsim 1\KEV$, there is no cosmological
gravitino problem~\cite{Gprob-1keV}, since in this case the gravitino
does not overclose the energy density of the universe even if it is
thermalized.

Meanwhile, we should also take care of the decays of the next-to-lightest
SUSY particle (NLSP)~\cite{Gprob-GMSB}. Since the NLSP can decay only to
the LSP gravitino through the suppressed interaction, it has a long
lifetime. Then its decay during or after the BBN might destroy the
success of the BBN, just like the case of unstable gravitino. Thus,
there are constraints on the abundance of the NLSP at the decay time
depending on the lifetime of the NLSP. (The lifetime of the NLSP depends
on the NLSP mass as well as the gravitino mass, and its abundance at the
time of the decay is determined by its annihilation cross section.) This
constraint leads to an upper bound on the gravitino mass
$m_{3/2}$~\cite{Gprob-GMSB}, depending on the mass and couplings of the
NLSP. Then combined with the constraint from the overclosure limit of
the gravitino explained above, we obtain an upper bound on the reheating
temperature $T_R$. Detailed analyses show that reheating temperature can
be as high as $T_R\sim 10^9$--$10^{10}\GEV$ for $m_{3/2}\sim
10$--$100\GEV$~\cite{LSPgrav}.

\subsubsection*{$\bullet$ remarks}

Several comments are in order. First, all of the above arguments have
assumed that no extra entropy production takes place after gravitinos
are produced. If there is a dilution of the gravitino by a late-time
entropy production, the bounds discussed in this section are relaxed.
Care has to be taken in this case, however, since baryon asymmetry is
also diluted if the baryogenesis occurs before that entropy production.

Next, in the case of unstable gravitino, if the gravitino decays mainly
into a neutrino and a sneutrino, the upper bound on the reheating
temperature becomes weaker since the neutrino have only weak
interactions. In this case, the high energy neutrinos emitted from the
gravitino decay scatter off the background neutrinos and produce charged
leptons, which cause electroweak cascade and produce many photons. The
requirement that those photons do not alter the abundances of light
elements gives an upper bound on the reheating temperature, $T_R\lsim
10^{10}$--$10^{12}\GEV$ for $m_{3/2}\simeq
100\GEV$--$1\TEV$~\cite{grav-neutrino}.

In some cases, considerably higher reheating temperatures are
allowed. For example, if the LSP is the axino (which is a fermionic
superpartner of the axion) and the gravitino is the NLSP, the
reheating temperature can be as high as $10^{15}\GEV$ for
$m_{3/2}\simeq 100\GEV$~\cite{Asa-Yana}. Another interesting case is
given when the leptogenesis takes place from the universe dominated by
the coherent oscillation of the right-handed sneutrino~\cite{HMY} (see
\SEC{SEC-Ntilde}). See also a model in \SEC{SEC-model-PQ}. In the
context of gauge-mediated SUSY breaking models, an attractive scenario
to solve the gravitino problem has been proposed~\cite{Fujii-Yana},
recently.

Finally, we mention the nonthermal production of the gravitino during
the preheating epoch after the inflation~\cite{grav-FKYY,NTgrav}. At
the preheating epoch, gravitinos can be produced either through the
scattering of particles which are created by the parametric resonance
of the oscillating inflaton~\cite{grav-FKYY} or directly from the
oscillating inflaton~\cite{NTgrav}. These nonthermal productions may
increase the primordial abundance of the gravitino and hence might
make the upper bound on the reheating temperature $T_R$ severer than
discussed in this section. However, in both cases the produced
gravitino abundance depends on the amplitude and coupling of the
oscillating inflaton. In particular, as for the second mechanism, it
was shown that the gravitino abundance produced directly from
oscillating inflaton is sufficiently small as long as the two sectors,
the one responsible for supersymmetry breaking at true vacuum and the
one for the inflation, are distinct and coupled only
gravitationally~\cite{NTgrav-OK}. (Notice that this is the case for
the SUSY inflation models which will be discussed in \SEC{SEC-Inf}.)
We will use, therefore, the constraint from the thermally produced
gravitinos as a conservative bound.

%%%%%%%%%%%%%%%%%%%%%%%%%%%%%%
%\include{Chap-Ndecay}%%%%%%%%%
%%%%%%%%%%%%%%%%%%%%%%%%%%%%%%%%%%%%%%%%%%%%%%%%%%%%%%%%%%%%%%%%%%%
%%%%%%%%%%%%%%%%%%%%%%%%%%%%%%%%%%%%%%%%%%%%%%%%%%%%%%%%%%%%%%%%%%%
%%%%%%%%%%%%%%%%%%%%%%%%%%%%%%%%%%%%%%%%%%%%%%%%%%%%%%%%%%%%%%%%%%%
\chapter{Leptogenesis by the decay of right-handed neutrino}%%%%%%% 
%%%%%%%%%%%%%%%%%%%%%%%%%%%%%%%%%%%%%%%%%%%%%%%%%%%%%%%%%%%%%%%%%%%
%%%%%%%%%%%%%%%%%%%%%%%%%%%%%%%%%%%%%%%%%%%%%%%%%%%%%%%%%%%%%%%%%%%
%%%%%%%%%%%%%%%%%%%%%%%%%%%%%%%%%%%%%%%%%%%%%%%%%%%%%%%%%%%%%%%%%%%
\label{Chap-Ndecay}

In this chapter, we discuss leptogenesis scenarios by the decays of
right-handed neutrino. The lepton asymmetry is produced by the
$CP$-violating decay of the right-handed neutrino $N$ into leptons $L$
and anti-leptons $\bar{L}$. As discussed in \SEC{SEC-sphaleron}, the
produced lepton asymmetry is partially converted to the baryon
asymmetry~\cite{FY} by the sphaleron process~\cite{sphaleron}.

In \SEC{SEC-LAfromN} we discuss the amount of the lepton asymmetry
produced in the decays of right-handed (s)neutrino. Then we turn to
discuss each scenario depending on the production mechanism of the
right-handed (s)neutrinos. The original, and the most extensively
studied mechanism is the thermal production. We briefly discuss it in
\SEC{SEC-thermal}. Next, we investigate the production of right-handed
(s)neutrinos in inflaton decay in \SEC{SEC-Inf}, adopting several SUSY
inflation models.  Here, we also study in detail the inflation dynamics
in each inflation model. In \SEC{SEC-Ntilde}, we discuss the
leptogenesis from coherent oscillation of the right-handed sneutrino. In
particular, we mainly discuss the most interesting case, the
leptogenesis from the universe dominated by the coherent oscillation of
the right-handed sneutrino.

\clearpage
%%%%%%%%%%%%%%%%%%%%%%%%%%%%%%%%%%%%%%%%%%%%%%%%%%%%%%%%%%%%%%%%%%%
\section{Asymmetric decay of the right-handed neutrino}%%%%%%%%%%%%
%%%%%%%%%%%%%%%%%%%%%%%%%%%%%%%%%%%%%%%%%%%%%%%%%%%%%%%%%%%%%%%%%%%
\label{SEC-LAfromN}

Let us start by introducing three generations of heavy right-handed
neutrinos to the minimal supersymmetric standard model (MSSM), which
have a superpotential;
%%%
\begin{eqnarray}
 \label{EQ-super}
  W = \frac{1}{2} M_i N_i N_i + h_{i\alpha} N_i L_{\alpha} H_u
  \,,
\end{eqnarray}
%%%
where $N_i$ ($i = 1, 2, 3$), $L_{\alpha}$ ($\alpha = e, \mu, \tau$) and
$H_u$ denote the supermultiplets of the heavy right-handed neutrinos,
lepton doublets and the Higgs doublet which couples to up-type quarks,
respectively. (Here and hereafter, we omit the SU$(2)_L$ indices for
simplicity.)  $M_i$ are the masses of the right-handed neutrinos.  Here,
we have taken a basis where the mass matrix for $N_i$ is diagonal and
real.

As can be seen in \EQ{EQ-super}, the masses of the right-handed
neutrinos violate the lepton number. This gives rise to the following
two important consequences. First, the tiny neutrino masses, which are
now strongly suggested by the neutrino-oscillation experiments, are
explained via the seesaw mechanism~\cite{seesaw}. {}From \EQ{EQ-super}
we obtain the mass matrix for the light neutrinos by integrating out the
heavy right-handed neutrinos:
%%%
\begin{eqnarray}
 \label{EQ-seesaw}
  (m_{\nu})_{\alpha\beta}
  = 
  - \sum_i h_{i\alpha} h_{i\beta}
  \frac{\vev{H_u}^2}{M_i}
  \,.
\end{eqnarray}
%%%

The second one is the production of the lepton asymmetry by the
right-handed neutrino decay~\cite{FY} which we discuss now. Because of
their Majorana masses, the right-handed neutrinos have two distinct
decay channels into leptons and anti-leptons.  At tree level, these two
kinds of decay channels have the same decay widths:
%%%
\begin{eqnarray}
 \label{EQ-all-channels}
  \begin{array}{lclclcl}
   \Frac{1}{4}
    \Gamma_{N_i}
    &\!\!\!=\!\!\!&
    \Gamma\,(N_i \to \widetilde{L} + \widetilde{h_u})
    &\!\!\!=\!\!\!&
    \Gamma\,(N_i \to \widetilde{L}^* + \overline{\widetilde{h_u}})
    &&
    \\
   &\!\!\!=\!\!\!&
    \Gamma\,(N_i \to l + H_u)
    &\!\!\!=\!\!\!&
    \Gamma\,(N_i \to \overline{l} + H_u^*)
    &\!\!\!=\!\!\!&
    \Frac{1}{16\pi}
    \sum_{\alpha}
    | h_{i\alpha} |^2
    M_i
    \,,
    \\
   &&&&&&
    \\
   \Frac{1}{2}
    \Gamma_{\widetilde{N_i}}
    &\!\!\!=\!\!\!&
    \Gamma\,(\widetilde{N_i}\to \widetilde{L} + H_u)
    &\!\!\!=\!\!\!&
    \Gamma\,(\widetilde{N_i}\to \overline{l} + \overline{\widetilde{h_u}})
    &\!\!\!=\!\!\!&
    \Frac{1}{8\pi}
    \sum_{\alpha}
    | h_{i\alpha} |^2
    M_i
    \,,
    \\
   &&&&&&
    \\
   \Frac{1}{2}
    \Gamma_{\widetilde{N_i}^*}
    &\!\!\!=\!\!\!&
    \Gamma\,(\widetilde{N_i}^*\to l + \widetilde{h_u})
    &\!\!\!=\!\!\!&
    \Gamma\,(\widetilde{N_i}^*\to \widetilde{L}^* + H_u^*)
    &\!\!\!=\!\!\!&
    \Frac{1}{8\pi}
    \sum_{\alpha}
    | h_{i\alpha} |^2
    M_i
    \,,
  \end{array}
\end{eqnarray}
%%%
where $N_i = \nu_{R i}^c + \overline{\nu_{R i}^c}^T$,
$\{\nu_R^c,\,l,\,\widetilde{h_u}\}$ and $\{
\widetilde{N},\,\widetilde{L},\, H_u\}$ denote fermionic and scalar
components of corresponding supermultiplets $\{N,\,L,\,H_u\}$, and
$\overline{y}$ and $Y^*$ represent antiparticles of fermion $y$ and
scalar $Y$, respectively. Here, we have summed the final states over
flavor ($\alpha = e, \mu, \tau$) and ${\rm SU}(2)_L$ indices. We can
symbolically write the above widths as
%%%
\begin{eqnarray}
 \frac{1}{2}\Gamma_{N_i}
  =
  \Gamma( N_i \to L + H_u)
  =  
  \Gamma( N_i \to \overline{L} + \overline{H_u})
  =  
  \frac{1}{8\pi}
  \sum_{\alpha}
  |h_{i\alpha}|^2
  M_i
  \,,
\end{eqnarray}
where $N$, $L$ and $H_u$ ($\overline{L}$ and $\overline{H_u}$) denote
fermionic or scalar components of corresponding supermultiplets (and
their anti-particles).

If $CP$ is violated in the Yukawa matrix $h_{i\alpha}$, the interference
between decay amplitudes of tree and one-loop diagrams results in
lepton-number violation~\cite{FY}. Hereafter, we concentrate on the
decay of lightest right-handed (s)neutrino $N_1$, since we will consider
only the $N_1$ decay in the following sections.\footnote{We will assume
$M_1\ll M_2,M_3$. In this case, even if the heavier right-handed
neutrinos $N_{2,3}$ produce lepton asymmetry, it is usually erased
before the decays of $N_1$.}  The lepton asymmetry produced in the $N_1$
decay is represented by the following parameter $\epsilon_1$:
%%%
\begin{eqnarray}
  \epsilon_1 
  &\equiv& 
  \frac{
  \Gamma (N_1 \to L + H_u) 
  -
  \Gamma (N_1 \to \overline{L} + \overline{H_u})
  }{  \Gamma_{N_1} }
  \,,
\end{eqnarray}
%%%
which means the lepton number asymmetry produced per one right-handed
neutrino decay. Summing up the one-loop vertex and self-energy
corrections~\cite{ep1-standard}, the $\epsilon_1$ has the following
form:
%%%
\begin{eqnarray}
 \epsilon_1 
  =
  -
  \frac{1}{8\pi}
  \frac{1}{\left(h h^{\dagger}\right)_{11}}
  \sum_{i = 2,3}
  {\rm Im} 
  \left[
   \{
   \left(
    h h^{\dagger}
    \right)_{1i}
    \}^2
   \right]
   \left[
    f^{V}
    \left(
     \frac{M_i^2}{M_1^2}
     \right)
     +
     f^{S}
     \left(
      \frac{M_i^2}{M_1^2}
      \right)
    \right]
    \,,
\end{eqnarray}
%%%
where $f^{V}(x)$ and $f^{S}(x)$ represent the contributions from vertex
and self-energy corrections, respectively. In the case of the
non-supersymmetric standard model with right-handed
neutrinos, they are given by~\cite{ep1-standard}
%%%
\begin{eqnarray}
 \label{EQ-ep1-function-nonSUSY}
  f^{V}_{\rm non-SUSY}(x)
  =
  \sqrt{x}
  \left[
   -1
   +
   (x+1)\ln\left(1+\frac{1}{x}\right)
   \right]
   \,,
   \qquad
   f^{S}_{\rm non-SUSY}(x)
   =
   \frac{\sqrt{x}}{x-1}
   \,,
\end{eqnarray}
%%%
while in the case of MSSM plus right-handed (s)neutrinos, they are given
by~\cite{ep1-covi},\footnote{See also
Refs.~\cite{camp-dav-oliv-2,MSYY-1}.}
%%%
\begin{eqnarray}
 \label{EQ-ep1-function}
  f^{V}_{\rm SUSY}(x)
  =
  \sqrt{x}\,
  \ln\left(1+\frac{1}{x}\right)
  \,,
  \qquad
  f^{S}_{\rm SUSY}(x)
  =
  \frac{2 \sqrt{x}}{x-1}
  \,.
\end{eqnarray}
%%%
Hereafter, we use \EQ{EQ-ep1-function} since we assume SUSY.  Notice
that all of the $N_1$, $\sneu$ and $\sneu^*$ in \EQ{EQ-all-channels}
produce the same amount of lepton asymmetry with the same
sign~\cite{camp-dav-oliv-2,ep1-covi,LG-Plu},\footnote{Here, we have
neglected the three body decay of the right-handed sneutrino, e.g.,
$\sneu \to \widetilde{L}^* + \widetilde{Q} + \widetilde{\bar{u}}$, which
gives only tiny corrections~\cite{LG-Plu}. We also neglect the effects
of the soft-SUSY breaking terms, since the mass scale of soft terms
($\sim 100\GEV$--$1\TEV$) are much smaller than right-handed neutrino
mass.}  given by the above $\epsilon_1$. Assuming a mass hierarchy
$M_1\ll M_2,M_3$ in the right-handed neutrino sector (i.e., $x\gg 1$),
the above formula is simplified to the following one:
%%%%%
\begin{eqnarray}
 \label{EQ-ep1-theirs}
  \epsilon_1 
  &\simeq&
  - \frac{3}{8\pi}
  \frac{1}{\left(h h^{\dagger}\right)_{11}}
  \sum_{i = 2,3}
  {\rm Im} 
  \left[
   \{
   \left(
    h h^{\dagger}
    \right)_{1i}
    \}^2
   \right]
   \frac{M_1}{M_i}
   \,.
\end{eqnarray}

Now let us rewrite this $\epsilon_1$ parameter in terms of the light
neutrino mass $m_\nu$, so that we can relate the lepton asymmetry (and
hence the baryon asymmetry in the present universe) to the neutrino
mass, which is observed by the neutrino-oscillation experiments. First
of all, by using ${\rm Im} \left[ \{\left( h h^{\dagger} \right)_{11}
\}^2 \right] = 0$, we can write the $\epsilon_1$ as follows:
%%%%%
\begin{eqnarray}
 \epsilon_1
  =
  -
  \frac{3}{8\pi}
  \frac{M_1}{\left(h h^{\dagger}\right)_{11}}
  {\rm Im} 
  \left[
   \left(
    h h^{\dagger}
    \frac{1}{M}
    h^* h^T
    \right)_{11}
   \right]
   \,,
\end{eqnarray}
%%%%%
where a matrix notation $M = {\rm diag}(M_1, M_2, M_3)$ is
adopted. Next, by using the seesaw formula in \EQ{EQ-seesaw}, it is
reduced to
%%%%%
\begin{eqnarray}
 \label{EQ-eps1}
  \epsilon_1
  &=&
  \frac{3}{8\pi}
  \frac{M_1}{\vev{H_u}^2}
  \frac{
  {\rm Im}\left[h (m_{\nu}^*) h^T\right]_{11}
  }
  {
  \left(h h^{\dagger}\right)_{11}
  }
  \,.
\end{eqnarray}
%%%%%
The neutrino mass matrix $(m_{\nu})_{\alpha\beta}$ can be diagonalized
by an unitary matrix $U_{\alpha i}$ (Maki-Nakagawa-Sakata
matrix~\cite{MNS}) as $(m_{\nu})_{\alpha\beta} = \sum_i U_{\alpha i}
U_{\beta i} (\widehat{m_\nu})_i$, where $\widehat{m_\nu}$ is a diagonal
mass matrix $\widehat{m_\nu} = {\rm diag}(\mnu{1}, \mnu{2}, \mnu{3})$
and $\mnu{1} < \mnu{2} < \mnu{3}$. Then, with rotated Yukawa couplings
\begin{eqnarray}
  \widehat{h}_{ik} &\equiv& \sum_{\alpha} h_{i\alpha} U_{\alpha k}^*
  \label{EQ-rtd-Ykw}
\end{eqnarray}
we obtain
%%%%%
\begin{eqnarray}
 \label{EQ-ep1-before-final}
 \epsilon_1
  &=&
  \frac{3}{8\pi}
  \frac{M_1}{\vev{H_u}^2}
  \frac{
  {\rm Im}
  \left[\widehat{h} 
   (\widehat{m_\nu}) 
   \widehat{h}^T\right]_{11}
  }
  {
  \left(\widehat{h}\,\widehat{h}^{\dagger}\right)_{11}
  }
  \nonumber\\
 &=&
  \frac{3}{8\pi}
  \frac{M_1}{\vev{H_u}^2}
  \mnu{3}
  \delta_{\rm eff}
  \,,
\end{eqnarray}
%%%%%
where the effective $CP$-violating phase $\delta_{\rm eff}$ is defined
by
%%%%%
\begin{eqnarray}
 \label{EQ-deltaeff}
 \delta_{\rm eff}
  \equiv
  \frac{
  {\rm Im}
  \left[
   \left(
    \widehat{h}_{13}
    \right)^2
    + \Frac{\mnu{2}}{\mnu{3}}
    \left(
     \widehat{h}_{12}
     \right)^2
     + \Frac{\mnu{1}}{\mnu{3}}
     \left(
      \widehat{h}_{11}
      \right)^2
   \right]
   }
   {
   \left|\widehat{h}_{13}\right|^2
   + \left|\widehat{h}_{12}\right|^2
   + \left|\widehat{h}_{11}\right|^2
   }
   \,.
\end{eqnarray}
%%%%%
As clearly seen from the above explicit expression, $\delta_{\rm eff}$
is always less than one~\cite{HMY}, but it is in general order one
unless the phase of the Yukawa coupling $\widehat{h}_{13}$ is
accidentally suppressed or the couplings have a inverted hierarchy,
i.e., $|\widehat{h}_{12}|^2 \gg (\mnu{2}/\mnu{1})
|\widehat{h}_{13}|^2$ or $|\widehat{h}_{11}|^2 \gg (\mnu{3}/\mnu{1})
|\widehat{h}_{13}|^2$. {}From \EQ{EQ-ep1-before-final}, the
$\epsilon_1$ parameter is given by
%%%%%
\begin{eqnarray}
 \label{EQ-ep1-final}
  \epsilon_1
  &\simeq&
  2.0 \times 10^{-10}
  \left(
   \frac{M_1}{10^6\GEV}
   \right)
   \left(
    \frac{\mnu{3}}{0.05\EV}
    \right)
    \delta_{\rm eff}
    \,.
\end{eqnarray}
%%%%%
This relation is consistent with the one obtained in
Ref.~\cite{BY}. Here, we have used $\vev{H_u} = 174\GEV \times
\sin\beta$, where $\tan\beta\equiv \vev{H_u}/\vev{H_d}$. ($H_d$ is the
Higgs field which couples to down-type quarks.) Here and hereafter, we
will take $\sin\beta \simeq 1$ for simplicity.\footnote{This is the case
as long as $\tan\beta > 1$. Even for $\tan\beta \simeq 1$, the final
lepton asymmetry changes (increases) by a factor of $2$.} As for the
heaviest neutrino mass, we take $\mnu{3}\simeq 0.05\EV$ as a typical
value throughout this chapter, suggested from the atmospheric neutrino
oscillation observed in the Super-Kamiokande experiments~\cite{SK-Atm}.

We should stress here that the $\epsilon_1$ parameter has an explicit
formula given in \EQ{EQ-ep1-final} with $\delta_{\rm eff}$ in
\EQ{EQ-deltaeff} (as long as $M_1\ll M_2$, $M_3$), although in the
literature it is sometimes treated as a free parameter.  In
particular, it is proportional to the right-handed neutrino mass $M_1$
for fixed values of $\mnu{3}$ and $\delta_{\rm eff}(\le 1)$.

Let us mention one last point, the possibility of an enhancement of
the asymmetry parameter $\epsilon_1$. For the self-energy contribution
$f^S(x)$ in \EQ{EQ-ep1-function} (and \EQ{EQ-ep1-function-nonSUSY}),
we have assumed the masses of the right-handed neutrinos are not so
degenerate, i.e., $|M_i-M_1| \gg \Gamma_{N_i}$. However, if the mass
difference becomes as small as the decay width, $|M_i-M_1|\sim
\Gamma_{N_i}$, one expects an enhancement of the self-energy
contribution $f^S(x)$~\cite{ep1-degenerate}. Actually, it was shown
that the asymmetry parameter $\epsilon_1$ can reach its maximum value
of $\epsilon_1\sim {\cal O}(1)$ for $|M_i-M_1|\simeq
\Gamma_{N_i}/2$~\cite{ep1-degenerate}. ($\epsilon_1$ cannot be
arbitrarily large. Notice that the lepton asymmetry vanishes in the
limit where the right-handed neutrinos become exactly mass degenerate,
since in this case $CP$-violating phases of the Yukawa couplings can
be eliminated by a change of basis.) Nevertheless, it requires an
extreme degeneracy of right-handed neutrino masses $|M_i -
M_1|/M_1\simeq h^2/(8\pi)$, and hence we do not consider this
possibility in the following discussion.

\clearpage
%%%%%%%%%%%%%%%%%%%%%%%%%%%%%%%%%%%%%%%%%%%%%%%%%%%%%%%%%%%%%%%%%%%
\section{Leptogenesis by thermally produced right-handed neutrino}%
%%%%%%%%%%%%%%%%%%%%%%%%%%%%%%%%%%%%%%%%%%%%%%%%%%%%%%%%%%%%%%%%%%%
\label{SEC-thermal}

In this section, we briefly discuss the leptogenesis by thermally
produced right-handed neutrinos. After first suggested by Fukugita and
Yanagida~\cite{FY}, this scenario has been extensively studied. (See,
for example,
Refs.~\cite{LG-Luty,camp-dav-oliv-2,ep1-standard,ep1-degenerate,BBP}. For
reviews and references, see Refs.~\cite{LGthermal,BBP}.) {}From the
viewpoint of the production mechanism of the right-handed neutrino,
compared with other production mechanisms, this scenario has a very
attractive point that it requires no extra assumption to create the
right-handed neutrinos, besides high enough temperature.

Suppose that the right-handed (s)neutrinos $N_1$ are produced thermally
and become as abundant as in thermal equilibrium at temperature $T >
M_1$. (Hereafter, we consider the lightest right-handed neutrino $N_1$,
since lepton asymmetry produced by the decays of the heavier two
right-handed neutrinos $N_{2(3)}$ is likely to be erased before the
$N_1$'s decay.) Then the ratio of the total number density of the
right-handed (s)neutrinos to the entropy density is simply given by the
following number [see \EQ{EQ-nXs-thermal}]:\footnote{Numerically the
value in \EQ{EQ-nLs-th-naive} is almost the same as $1/g_*$.}
%%%
\begin{eqnarray}
 \label{EQ-nLs-th-naive}
  \frac{n_{N_1}}{s} 
  &\simeq&
  \frac{n^{\rm eq}_{N_1}}{s}
  =
  \frac{45\zeta (3)}{2\pi^4}
  \frac{1}{g_*}
  \times 
  \left(
   \frac{3}{4}g_{N_1} + g_{\sneu}
   \right)
   \simeq
   \frac{1}{240}
   \,,
\end{eqnarray}
%%%
where $g_{N_1} = g_{\sneu} = 2$ are the numbers of degrees of freedom
for right-handed (s)neutrinos and $g_*= 232.5$ for the MSSM plus
right-handed neutrino multiplet [see \EQ{EQ-gstar-def}].

\paragraph{Boltzmann equations (a toy model)}
The evolution of the number density $n_{N_1}$ of the right-handed
neutrino and that of the lepton number density $n_L$ are described by
coupled Boltzmann equations. In order to understand qualitative
behaviors of the $n_{N_1}$ and $n_L$, in particular to understand the
``out-of-equilibrium condition,'' let us consider the following simple
set of Boltzmann equations:
%%%
\begin{eqnarray}
 \frac{d}{dt}n_{N_1} + 3 H n_{N_1}
  &=& 
  {}
  - \gamma(N_1 \to  l  \varphi) 
  - \gamma(N_1 \to \overline{l}  \varphi^*)
  \nonumber \\
 && 
  {}
  + \gamma(l  \varphi \to N_1) 
  + \gamma(\overline{l}  \varphi^* \to N_1)
  \,,
  \label{EQ-Boltz-1}
  \\
 \frac{d}{dt}n_{l} + 3 H n_{l}
  &=& 
  \gamma(N_1 \to  l  \varphi) 
  - \gamma(l  \varphi \to N_1)
  \nonumber \\
 && 
  {}
  - \gamma'(l  \varphi \to \overline{l}  \varphi^*)
  + \gamma'(\overline{l}  \varphi^* \to l  \varphi)
  \,,
  \label{EQ-Boltz-2}
  \\
 \frac{d}{dt}n_{\overline{l}} + 3 H n_{\overline{l}}
  &=& 
  \gamma(N_1 \to \overline{l}  \varphi^*)
  - \gamma(\overline{l}  \varphi^* \to N_1)
  \nonumber \\
 && 
  {}
  + \gamma'(l  \varphi \to \overline{l}  \varphi^*)
  - \gamma'(\overline{l}  \varphi^* \to l  \varphi)
  \,,
  \label{EQ-Boltz-3}
\end{eqnarray}
%%%
which describe the evolutions of the number densities of right-handed
neutrino $N_1$, lepton $l$, and anti-lepton $\overline{l}$. For a while,
we will consider a non-SUSY case, for simplicity. The terms proportional
to the Hubble parameter $H$ describe the effect of the expansion of the
universe. $\gamma(A\to B)$ denotes the reaction density, which means the
rate of the process $A\to B$ per unit time per unit volume. Thus, the
$\gamma$-terms in the right-hand sides of the equations describe the
change of the number densities due to the corresponding interactions. We
have included only the three kinds of processes ($N_1\leftrightarrow l
\varphi$, $N_1 \leftrightarrow \overline{l} \varphi^*$ and $l \varphi
\leftrightarrow \overline{l} \varphi^*$) which are necessary in the
following discussion. (In actual calculation, one must include many
other interactions. For other processes, which include the
super-partners $\sneu$, $\widetilde{L}$ and $\widetilde{h_u}$, see
Ref.~\cite{LG-Plu}.)  As for the primes in $\gamma'(l \varphi
\leftrightarrow \overline{l} \varphi^*)$, we will give an explanation
below.

The reaction densities of the decay and inverse decay are given by
%%%
\begin{eqnarray}
 \gamma(N_1 \to  l  \varphi)
  &=&
  \frac{1}{2}(1 + \epsilon_1)
  \vev{\Gamma_{N_1}}n_{N_1}
  \,,
  \nonumber \\
 \gamma(N_1 \to \overline{l}  \varphi^*)
  &=&
  \frac{1}{2}(1 - \epsilon_1) 
  \vev{\Gamma_{N_1}}n_{N_1}
  \,,
  \nonumber \\
 \gamma(l \varphi \to N_1) 
  &=&
  \frac{1}{2}(1 - \epsilon_1) 
  \vev{\Gamma_{N_1}}
  n_{N_1}^{\rm eq}
  \times \frac{n_l}{n_l^{\rm eq}}
  \,,
  \nonumber \\
 \gamma(\overline{l}  \varphi^* \to N_1)
  &=&
  \frac{1}{2}(1 + \epsilon_1) 
  \vev{\Gamma_{N_1}}
  n_{N_1}^{\rm eq}
  \times \frac{n_{\overline{l}}}{n_l^{\rm eq}}
  \,.
  \label{EQ-gammas}
\end{eqnarray}
%%%
Here, $\vev{\Gamma_{N_1}}$ denotes thermally averaged total decay rate
of $N_1$. For low temperature $T\ll M_1$, it is just given by the decay
rate at rest frame: $\vev{\Gamma_{N_1}}\simeq \Gamma_{N_1}$, while for
high temperature $T\gg M_1$, it is given by $\vev{\Gamma_{N_1}}\simeq
(M_1/2 T)\,\,\Gamma_{N_1}$ due to the time-dilatation effect.

{}From \EQS{EQ-Boltz-1}--(\ref{EQ-gammas}), the Boltzmann equations
of $Y_{N_1}\equiv n_{N_1}/s$ and $Y_L\equiv n_L/s = (n_l -
n_{\overline{l}})/s$ are given by
%%%
\begin{eqnarray}
 \frac{d}{dt}Y_{N_1}
  &=&
  {}
  - \vev{\Gamma_{N_1}}
  \left(
   Y_{N_1} - Y_{N_1}^{\rm eq}
   \right)
   -
   \frac{1}{2}\epsilon_1
   \vev{\Gamma_{N_1}}
   Y_{N_1}^{\rm eq}
   \times
   \frac{Y_L}{Y_l^{\rm eq}}
   \,,
   \label{Boltz-N1}
  \\
 \frac{d}{dt}Y_L
  &=&
  \epsilon_1
  \vev{\Gamma_{N_1}}
  \left(
   Y_{N_1} + Y_{N_1}^{\rm eq}
   \right)
   -
   \frac{1}{2}\vev{\Gamma_{N_1}}Y_{N_1}^{\rm eq}
   \times 
   \frac{Y_L}{Y_l^{\rm eq}}
   \nonumber \\
 &&
  {}
  - 2\gamma'(l  \varphi \to \overline{l}  \varphi^*)
  + 2\gamma'(\overline{l}  \varphi^* \to l  \varphi)
  \,.
  \label{EQ-Boltz-L}
\end{eqnarray}
%%%
Here, there is a subtle point one should take
care~\cite{Dolgov-Zeldovich,Kolb-Wolfram}.  As can be seen from
\EQ{EQ-gammas}, the inverse decay processes ($l \varphi \to N_1$ and
$\overline{l} \varphi^*\to N_1$) produce a net lepton asymmetry with
the {\it same} sign as the decay processes ($N_1\to l \varphi$ and
$N_1\to \overline{l} \varphi^*$) themselves. This can be seen by
applying the $CPT$ invariance to the matrix elements of those
processes: ${\cal M}(N_1\to l \varphi) = {\cal M}(\overline{l}
\varphi^*\to N_1)$ and ${\cal M}(N_1\to \overline{l} \varphi^*)$ =
${\cal M}(l \varphi \to N_1)$.  Therefore, if there would exist only
those decay and inverse decay processes, lepton asymmetry would not
vanish even in thermal equilibrium. In other words, if we ignore the
$\gamma'(l\varphi \leftrightarrow \overline{l}\varphi^*)$ terms, the
right-hand side of the \EQ{EQ-Boltz-L} would not vanish even for $Y_L
= 0$ and $Y_{N_1} = Y_{N_1}^{\rm eq}$.

On the other hand, there are lepton number violating two-body scattering
processes; $l \varphi \to \overline{l} \varphi^*$ and $\overline{l}
\varphi^* \to l \varphi$. By using unitarity, one can show that these
processes do not produce a net lepton asymmetry if we consider the $N_1$
only as a virtual
particle~\cite{Kolb-Wolfram,Buch-some-aspects}. However, the Boltzmann
equations already include $N_1$ as a real particle. Thus one should
subtract from the above two body scatterings the resonant $s$-channel
contribution mediated by $N_1$ (which is understood as an on-shell real
particle) to avoid a double counting of reactions. After subtracting the
resonant contribution, the reaction densities of the two-body
scatterings in \EQ{EQ-Boltz-L} leads
to~\cite{Kolb-Wolfram,Buch-some-aspects}
%%%
\begin{eqnarray}
 && {}- 2\gamma'(l  \varphi \to \overline{l}  \varphi^*)
  + 2\gamma'(\overline{l}  \varphi^* \to l  \varphi)
  \nonumber \\
 &=&
  {}
  -2\epsilon_1
  \vev{\Gamma_{N_1}}
  n_{N_1}^{\rm eq}
  - n_L
  n_{\varphi}^{\rm eq}
  \left[
   \vev{v \sigma'(l  \varphi \to \overline{l}  \varphi^*)}
   +
   \vev{v \sigma'(\overline{l}  \varphi^* \to l  \varphi)}
   \right]
   \,,
\end{eqnarray}
%%%
where the primes mean that the contribution from resonant $s$-channel
$N_1$ exchange has been subtracted. ($\sigma$ is the cross section, $v$
is the relative velocity between the initial particles and the bracket
denotes the thermal average.) Notice that there appears a term
$-2\epsilon_1 \vev{\Gamma_{N_1}} n_{N_1}^{\rm eq}$, which originates in
the subtraction of the resonant contribution~\cite{Kolb-Wolfram}. After
all, the Boltzmann equations are reduced to the following forms.
%%%
\begin{eqnarray}
 \frac{d}{dt}Y_{N_1}
  &=&
  {}
  - \vev{\Gamma_{N_1}}
  \left(
   Y_{N_1} - Y_{N_1}^{\rm eq}
   \right)
   -
   \frac{1}{2}
   \left(
   \epsilon_1
   \frac{Y_L}{Y_l^{\rm eq}}
   \right)
   \vev{\Gamma_{N_1}}
   Y_{N_1}^{\rm eq}
   \,,
   \label{Boltz-N1-final}
   \\
 \frac{d}{dt}Y_L
  &=&
  \epsilon_1
  \vev{\Gamma_{N_1}}
  \left(
   Y_{N_1} - Y_{N_1}^{\rm eq}
   \right)
   \nonumber \\
 {}
  &&
  -
  \left(
   \frac{1}{2}\vev{\Gamma_{N_1}}
   \frac{Y_{N_1}^{\rm eq}}{Y_l^{\rm eq}}
   +
   n_{\varphi}^{\rm eq}
   \left[
    \vev{v \sigma'(l  \varphi \to \overline{l}  \varphi^*)}
    +
    \vev{v \sigma'(\overline{l}  \varphi^* \to l  \varphi)}
    \right]
    \right)
    Y_L
    \,.
    \nonumber \\
    \label{EQ-Boltz-L-final}
\end{eqnarray}
%%%
We see that, for $Y_{N_1} = Y_{N_1}^{\rm eq}$ (and $Y_L = 0$), the
lepton asymmetry produced by [decay and inverse decay] processes is
canceled out by the lepton asymmetry from [two body scatterings minus
resonance] processes, which ensures vanishing lepton asymmetry at
equilibrium.

Now let us discuss the evolutions of $Y_{N_1}$ and $Y_L$. As can be seen
from \EQ{EQ-Boltz-L-final}, a deviation of the number density of the
$N_1$ from its equilibrium value ($Y_{N_1}\ne Y_{N_1}^{\rm eq}$) is
mandatory in order to produce a net lepton asymmetry $Y_L \ne 0$ from
$Y_L = 0$. This can be realized when the temperature of the universe $T$
cools down and becomes below the mass of the right-handed neutrino
$M_1$. To see the out-of-equilibrium condition for $N_1$, let us adopt a
dimensionless variable $z\equiv M_1/T$ instead of the cosmic time
$t$. Then \EQ{Boltz-N1-final} becomes
%%%
\begin{eqnarray}
 \frac{d Y_{N_1}}{d z}
  \simeq
  -
  \frac{\vev{\Gamma_{N_1}}}{z H}
  \left(
   Y_{N_1} - Y_{N_1}^{\rm eq}
   \right)
   \,.
\end{eqnarray}
Here, we have omitted the higher order term proportional to
$\epsilon_1 (Y_L/Y_l^{\rm eq})$, for simplicity. For $T < M_1$ ($z >
1$), the equilibrium value of the $N_1$'s abundance $Y_{N_1}^{\rm eq}$
decreases due to the Boltzmann suppression $\propto \exp (-M_1/T)$. At
this stage, if the dimensionless prefactor $\vev{\Gamma_{N_1}}/z H$ is
small enough, $Y_{N_1}$ can no longer catches up the decreasing
equilibrium value $Y_{N_1}^{\rm eq}$. Namely, $Y_{N_1}$ deviates from
$Y_{N_1}^{\rm eq}$, if $\vev{\Gamma_{N_1}}/z H \lsim 1$ for $z\gsim
1$. Thus, the out-of-equilibrium condition is roughly given by
%%%
\begin{eqnarray}
 \frac{\Gamma_{N_1}~}{\left.H \right|_{T = M_1}}
  &\lsim 1&
   \,.
   \label{EQ-OEC}
\end{eqnarray}
%%%
If this condition is satisfied, $Y_{N_1}>Y_{N_1}^{\rm eq}$ is realized,
and a net lepton asymmetry is produced [see \EQ{EQ-Boltz-L-final}].  

The ratio $\Gamma_{N_1}/H(T = M_1)$ is related to a mass parameter
$\widetilde{m_1}$, which is defined as~\cite{LGthermal}:
%%%
\begin{eqnarray}
 \label{EQ-m1parameter}
  \widetilde{m_1}
  &\equiv &
  \sum_\alpha |h_{1\alpha}|^2
  \frac{\vev{H_u}^2}{M_1}
  \nonumber\\
 &\simeq&
  0.8\times 10^{-3}\EV
  \times\left(
	 \frac{\Gamma_{N_1}~}{\left.H \right|_{T = M_1}}
	 \right)
   \,.
\end{eqnarray}
%%%
Notice that the above formula looks similar to the neutrino mass given
in \EQ{EQ-seesaw} but different from that. In terms of
$\widetilde{m_1}$, the out-of-equilibrium condition in \EQ{EQ-OEC} is
roughly equivalent to $\widetilde{m_1}\lsim 10^{-3}\EV$.

It is possible to show an important constraint on this parameter
$\widetilde{m_1}$~\cite{FHY-degenerate}:
\begin{eqnarray}
  \widetilde{m_1} &>& \mnu{1}\,.
\end{eqnarray}
To show this, let us rewrite the $\widetilde{m_1}$ parameter in terms
of the rotated Yukawa couplings $\widehat{h}_{ik}$ defined in
\EQ{EQ-rtd-Ykw};
\begin{eqnarray}
  \label{EQ-m1tilde-rewritten}
  \widetilde{m_1}
  &=&
  \sum_k | \widehat{h}_{1k}|^2
  \frac{\vev{H_u}^2}{M_1}
  \,.
\end{eqnarray}
On the other hand, from the seesaw formula \EQ{EQ-seesaw}, we obtain
\begin{eqnarray}
  \label{EQ-diagonal-seesaw}
  \mnu{j}
  \delta_{jk}
  &=&
  -
  \sum_i
  \widehat{h}_{ij}
  \widehat{h}_{ik}
  \frac{\vev{H_u}^2}{M_i}
  \,.
\end{eqnarray}
Let us define here a matrix $X_{ij}$ as follows:
\begin{eqnarray}
  X_{ij} \equiv i\,\widehat{h}_{ij}\frac{\vev{H_u}}{\sqrt{M_i \mnu{j}}}\,.
\end{eqnarray}
Then from \EQ{EQ-diagonal-seesaw} one can show that
\begin{eqnarray}
  \label{EQ-X-orth}
  \left(X^T X\right)_{ij}
  =
  \delta_{ij}
  =
  \left(X X^T\right)_{ij}
  \,,
\end{eqnarray}
while \EQ{EQ-m1tilde-rewritten} gives rise to
\begin{eqnarray}
  \widetilde{m_1}
  &=&
  \sum_k
  \mnu{k}
  \left|X_{1k}\right|^2
  \nonumber\\
  &>&
  \min_j\left\{\mnu{j}\right\}
  \sum_k
  \left|X_{1k}\right|^2
  \nonumber\\
  &\ge&
  \min_j\left\{\mnu{j}\right\}
  \left|
  \sum_k
  X_{1k}^2
  \right|
  =
  \min_j\left\{\mnu{j}\right\}
  \,.
\end{eqnarray}
In  the last  equation, we  have used  \EQ{EQ-X-orth}.  Therefore, the
$\widetilde{m_1}$ parameter is  bounded from below as $\widetilde{m_1}
> \mnu{1}$~\cite{FHY-degenerate}.

\paragraph{Lepton asymmetry}
Now let us discuss the amount of produced lepton asymmetry in the
present scenario. If the decaying right-handed neutrinos are as abundant
as in thermal equilibrium [see \EQ{EQ-nLs-th-naive}], and if there is no
wash-out process of the produced lepton asymmetry afterwards, the final
lepton asymmetry would be given by a very simple formula,
%%%
\begin{eqnarray}
 \frac{n_L}{s} = 
  \epsilon_1\,\frac{n_{N_1}}{s}
  \simeq
  \frac{\epsilon_1}{240}
  \,,
\end{eqnarray}
%%%
with the asymmetry parameter $\epsilon_1$ given in \EQ{EQ-ep1-final}. In
the actual case, however, a suppression factor $\kappa$ should be
multiplied:
%%%
\begin{eqnarray}
 \label{nLs-thermal}
  \frac{n_L}{s} =
  \kappa\,\epsilon_1\,\frac{n_{N_1}}{s}
  \simeq
  \kappa
  \,
  \frac{\epsilon_1}{240}
  \,.
\end{eqnarray}
%%%
The suppression factor $\kappa$ represents two effects. The first one is
the wash-out effect of the produced lepton asymmetry (and the strength
of the deviation from thermal equilibrium). If the interactions of
right-handed neutrinos $N_1$ are too strong, the produced lepton
asymmetry would be washed out by the interactions mediated by the $N_1$
itself. [See the second term in \EQ{EQ-Boltz-L-final}.]  Weakness of the
interaction is also required from the out-of-equilibrium condition for
$N_1$ in \EQ{EQ-OEC}.

The second effect represented by $\kappa$ is the efficiency of the
production of the right-handed (s)neutrinos $N_1$. Although we have
assumed that $N_1$ are produced as abundant as in thermal equilibrium
$Y_{N_1}\simeq Y_{N_1}^{\rm eq}$ for $T > M_1$, if the $N_1$'s
interactions are too weak, the thermal scatterings cannot produce enough
amount of $N_1$ and hence the number density $n_{N_1}$ cannot become as
abundant as that in thermal equilibrium, $n_{N_1}^{\rm
eq}$.\footnote{One can consider that enough amount of right-handed
neutrinos exist from the beginning and discuss their decay in thermal
background. We do not consider such a case in this section, however, and
assume that the right-handed neutrinos are produced by thermal
scatterings from $n_{N_1}\simeq 0$~\cite{LGthermal}.  Notice that as
long as an inflationary epoch is assumed, right-handed neutrinos should
be produced at some stage from $n_{N_1}\simeq 0$.}  Notice that those
two effects have double-edged behavior, i.e., the right-handed neutrino
$N_1$ should have couplings $h_{1\alpha}$ with intermediate strength.

In order to determine the precise value of the suppression factor
$\kappa$, one has to solve the Boltzmann equations including production,
decay and inverse decay of the right-handed (s)neutrinos, and all the
relevant lepton-number violating (and conserving) scatterings of lepton
fields~\cite{LG-Luty,LGthermal}. A detailed numerical calculation
solving the coupled Boltzmann equations in the case of SUSY has been
done in Ref.~\cite{LG-Plu}. (See also Ref.~\cite{LGthermal} and
references therein.) It was shown that both of the production rate of
the right-handed (s)neutrinos and the rate of the wash-out process of
the lepton asymmetry are proportional to the mass parameter
$\widetilde{m_1}$ defined in \EQ{EQ-m1parameter}. It turns out that the
suppression factor $\kappa$ becomes as large as $\kappa\simeq
0.05$--$0.3$ for~\cite{LG-Plu}
%%%
\begin{eqnarray}
 \label{EQ-m1favored}
 10^{-5}\EV\lsim
  \widetilde{m_1}\lsim 5\times 10^{-3}\EV
  \,.
\end{eqnarray}
%%%
For a smaller value of $\widetilde{m_1}$, the $N_1$ is not produced
enough and generated lepton asymmetry is suppressed.  On the other hand,
for a larger value of $\widetilde{m_1}$, enough amount of $N_1$ is
produced but wash-out of the lepton asymmetry is too strong, and hence
final lepton asymmetry is reduced.  

It is quite interesting to observe that the favored range of the mass
parameter $\widetilde{m_1}$ is just below the neutrino mass scale
observed by the atmospheric~\cite{SK-Atm} and solar~\cite{Solar}
($+$KamLAND~\cite{KamLAND}) neutrino-oscillation experiments,
$m_\nu\sim {\cal O}(10^{-3})$--${\cal O}(10^{-1})\EV$. We should note
that although $\widetilde{m_1}$ is not directly related to the
neutrino masses $\mnu{i}$, they are still indirectly related. (For
instance, if we adopt a Froggatt-Nielsen (FN) model~\cite{FN} which
will be discussed in \SEC{SEC-FN}, $\widetilde{m_1}$ is estimated as
$\widetilde{m_1}\sim {\cal O}(\mnu{3})\sim {\cal
O}(0.05\EV)$. Although this is a bit larger than the range in
\EQ{EQ-m1favored}, it can be consistent when we include ${\cal O}(1)$
ambiguities in the FN model. See also Ref.~\cite{BY-2}, where a more
general case was discussed for two generations of neutrinos.)

After being produced, a part of lepton asymmetry is immediately
converted to the baryon asymmetry~\cite{FY} via the sphaleron
effect~\cite{sphaleron} discussed in \SEC{SEC-sphaleron}. Then the
present baryon asymmetry is given by
%%%
\begin{eqnarray}
 \frac{n_B}{s}
  &=& 0.35\times \frac{n_L}{s}
  \nonumber\\
 &\simeq&
  0.3\times 10^{-10}
  \left(
   \frac{\kappa}{0.1}
   \right)
   \left(
    \frac{M_1}{10^9\GEV}
    \right)
    \times
    \left(
     \frac{\mnu{3}}{0.05\EV}
     \right)
     \delta_{\rm eff}
     \,,
\end{eqnarray}
where we have used \EQS{EQ-ep1-final} and
(\ref{nLs-thermal}). Therefore, the present baryon asymmetry
$n_B/s\simeq (0.4$--$1)\times 10^{-10}$ is naturally explained with the
mass of the lightest right-handed neutrino $M_1\sim
10^9$--$10^{10}\GEV$, for $\kappa \simeq 0.05$--$0.3$ and $\delta_{\rm
eff}\simeq {\cal O}(1)$.

In order to produce enough amount of right-handed neutrinos (i.e.,
$n_{N_1}\simeq n_{N_1}^{\rm eq}$), the temperature of the universe
should be higher than their mass $M_1$. This leads to a lower bound on
the reheating temperature as $T_R\gsim {\cal O}(10^{10})\GEV$. Thus the
overproduction of gravitinos might cause a difficulty depending on the
gravitino mass, as discussed in \SEC{SEC-grav}.  If the gravitino is
unstable, it must be relatively heavy, $m_{3/2}\gsim$ (a few)$\TEV$.
When the gravitino is stable, a consistent thermal history can be
obtained with gravitino mass $10$--$100\GEV$, avoiding the problem of
the decays of next-to-lightest SUSY particles after the big-bang
nucleosynthesis~\cite{LSPgrav}.\footnote{See also
Ref.~\cite{Buch-some-aspects}.}

~

Finally, we comment on the absolute upper bound on the neutrino masses
from thermal leptogenesis, which has been shown recently in
Ref.~\cite{BBP}. The crucial observation in Ref.~\cite{BBP} is that
the suppression factor $\kappa$ is determined only by three parameters
$\kappa = \kappa(\widetilde{m_1}$, $M_1$, $\overline{m})$, and hence
the final baryon asymmetry in thermal leptogenesis depends only on
four parameters: $\epsilon_1$, $\widetilde{m_1}$, $M_1$, and
$\overline{m}$. [See \EQ{nLs-thermal}.] Here, $\overline{m}\equiv
\sqrt{\mnu{1}^2 + \mnu{2}^2 + \mnu{3}^2}$. Furthermore, the maximal
value of $\epsilon_1$ is also determined by $M_1$ and $\overline{m}$,
$\epsilon_1 < \epsilon_1^{\rm max}(M_1, \overline{m})$, once the mass
squared differences of atmospheric and solar neutrino oscillations are
given [see \EQ{EQ-deltaeff} and
\EQ{EQ-ep1-final}]~\cite{HMY,Dav-Iba,BBP}. Thus, the maximal baryon
asymmetry is determined only by the set of three parameters
($\widetilde{m_1}$, $M_1$, $\overline{m}$). By solving the Boltzmann
equations for different points in this parameter space
($\widetilde{m_1}$, $M_1$, $\overline{m}$), and using the bound
$\widetilde{m_1} > \mnu{1}$, it has been shown in Ref.~\cite{BBP} that
the maximal baryon asymmetry can be larger than the empirical value
only if
\begin{eqnarray}
  \overline{m} < 0.30\EV\,.
\end{eqnarray}
Therefore, if the baryon asymmetry in the present universe was indeed
generated by thermal leptogenesis, we have stringent constraints on
the absolute neutrino masses,
\begin{eqnarray}
  \mnu{i} < 0.18\EV\,.
\end{eqnarray}

\clearpage
%%%%%%%%%%%%%%%%%%%%%%%%%%%%%%%%%%%%%%%%%%%%%%%%%%%%%%%%%%%%%%%%%%%
\section{Leptogenesis in inflaton decay}%%%%%%%%%%%%%%%%%%%%%%%%%%%
%%%%%%%%%%%%%%%%%%%%%%%%%%%%%%%%%%%%%%%%%%%%%%%%%%%%%%%%%%%%%%%%%%%
\label{SEC-Inf}

In this section,\footnote{This section is based on the works in a
collaboration with T.~Asaka, M.~Kawasaki, and T.~Yanagida~\cite{AHKY}.}
we discuss the leptogenesis scenario where the right-handed neutrino is
produced non-thermally in inflaton decays~\cite{LGinf}. We will find
that this scenario is fully consistent with existing various SUSY
inflation models such as hybrid, new, and topological inflation
models.\footnote{Leptogenesis in a ``natural chaotic inflation
model''~\cite{natural-chaotic} was also investigated in
Ref.~\cite{LG-chaotic}.}

The crucial difference between this scenario and the case of the
thermally produced right-handed neutrino discussed in the previous
section is that, the heavy right-handed neutrino $N$ can be produced
with relatively low reheating temperatures $T_R$ of the inflation. We
find that the required baryon asymmetry can be obtained even for $T_R
\simeq 10^6\GEV$ in some of the inflation models, and hence there is no
cosmological gravitino problem in the interesting wide region of the
gravitino mass $m_{3/2} \simeq 10\MEV$--$10\TEV$.\footnote{Here, we
consider both of the unstable and stable gravitino.} (See \SEC{SEC-grav}.)

On the other hand, the amount of the produced lepton asymmetry (and
hence baryon asymmetry) in the present scenario crucially depends on
the physics of the inflation, such as the mass of the inflaton $m_\chi$
and the reheating temperature $T_R$. Therefore, detailed analyses on the
inflation models are necessary.

In \SEC{SEC-nL-inf}, we calculate the amount of the resultant lepton
asymmetry in inflaton decay. Then we introduce a Froggatt-Nielsen
model in \SEC{SEC-FN} to estimate the mass (and decay rate) of the
right-handed neutrino. The subsequent subsections are devoted to each
SUSY inflation model. The leptogenesis in a hybrid inflationary
universe is discussed in \SEC{SEC-Hybrid}, where we consider two
different types of SUSY hybrid inflation models.  In \SEC{SEC-New} we
discuss the leptogenesis in a SUSY new inflation model.  The case of a
SUSY topological inflation is considered in \SEC{SEC-topological}.
Finally, we will briefly comment on the production of right-handed
neutrinos at preheating~\cite{LG-Pre} in \SEC{SEC-preheating}.

\subsection{Lepton asymmetry}
\label{SEC-nL-inf}

Let us first estimate the produced lepton asymmetry in the present
scenario. The result obtained here is a generic one, which can be
applied to all the inflation models discussed in subsequent subsections.

After the end of inflation, the inflaton decays into light particles and
the energy of the inflaton is transferred into the thermal bath. Then it
is very plausible that the right-handed neutrino is also produced in
inflaton decay, if its decay channel is kinematically allowed. We assume
that the inflaton decays into two right-handed neutrinos, which leads to
the following constraint:
%%%
\begin{eqnarray}
 \label{EQ-mass-bound}
 m_\chi > 2 M_1
  \,,
\end{eqnarray}
%%%
where $m_\chi$ is the inflaton mass. Here and hereafter, we consider
only the $N_1$ decay, provided that the mass $M_1$ is much smaller than
the others ($M_1 \ll M_2, M_3$).  As discussed in \SEC{SEC-LAfromN}, the
decay of $N_1$ into leptons and anti-leptons produces a lepton
asymmetry.  We will consider the case where the right-handed neutrino is
heavy enough compared with the reheating temperature, i.e., $M_1 \gsim
T_R$. In this case the produced right-handed neutrino $N_1$ is always
out of thermal equilibrium and it behaves like frozen-out, relativistic
particle with energy $E_{N_1}\simeq m_\chi/2$.  The ratio of the number
density of the right-handed (s)neutrinos $n_{N_1}$ to the entropy
density $s$ is then given by~\cite{LGinf}
%%%
\begin{eqnarray}
 \label{EQ-nN1s-inf}
  \frac{n_{N_1}}{s}
  &\simeq&
  \frac{\rho_{\rm rad}}{s}
  \times
  \frac{n_\chi}{\rho_\chi}
  \times
  \frac{n_{N_1}}{n_\chi}
  \nonumber \\
 &=&
  \frac{3}{4}T_R
  \times
  \frac{1}{m_\chi}
  \times
  2  B_r
  =
  \frac{3}{2}B_r \frac{T_R}{m_\chi}
  \,,
\end{eqnarray}
%%%
where $\rho_{\rm rad}$ is the energy density of the radiation just after
the reheating process completes, and $
n_\chi$ and $\rho_\chi = m_\chi
n_\chi \simeq \rho_{\rm rad}$ are the number and energy density of the
inflaton just before its decay, respectively. $B_r = B_r(\chi\to N_1
N_1)$ denotes the branching ratio of the inflaton decay into $N_1$
channel.  

As we will see, when we adopt the model which will be introduced in the
next subsection, the $N_1$ decays immediately after produced by the
inflaton decays. The lepton asymmetry is then given by
%%%
\begin{eqnarray}
 \label{EQ-nLs-inf}
  \frac{n_L}{s}
  &\simeq&
  \epsilon_1
  \frac{n_{N_1}}{s}  
  \nonumber \\
 &\simeq&
  3\times 10^{-10}\,
  B_r
  \left(
   \frac{T_R}{10^{6}\GEV}
   \right)
   \left(
    \frac{M_1}{m_\chi}
    \right)
    \times
    \left(
     \frac{\mnu{3}}{0.05\EV}
     \right)
     \delta_{\rm eff}
     \,,
\end{eqnarray}
%%%
where we have used the asymmetry parameter $\epsilon_1$ given in
\EQ{EQ-ep1-final}. Notice that there is no wash-out effect of the
produced lepton asymmetry,\footnote{Lepton-number violating 2-body
scatterings mediated by $N_i$ are out of thermal equilibrium as long as
$\sum_i \mnu{i}^2\lsim (10\KEV)^2 (10^9\GEV/T_R )$~\cite{FY-2}.} since
$N_1$ is out of thermal equilibrium.

Let us mention one point here. Although we will assume an explicit model
in the next subsection to make the subsequent discussion concrete, the
formula of the lepton asymmetry in \EQ{EQ-nLs-inf} is a generic one,
unless $M_1\lsim T_R$ or $N_1$ has a very long lifetime and dominates
the energy density of the universe before it decays. If $M_1\lsim T_R$,
the $N_1$ is produced more or less by thermal scatterings, and the
resultant lepton asymmetry is expected to be reduced to the case
discussed in \SEC{SEC-thermal}. As for the case where $N_1$ dominates
the energy density of the universe before its decay, we give a brief
discussion in Appendix~\ref{SEC-Inf-Dom}.

As discussed in \SEC{SEC-sphaleron}, the sphaleron process converts this
lepton asymmetry into baryon asymmetry as $n_B/s\simeq 0.35\times
n_L/s$. In order to explain the observed baryon asymmetry
$n_B/s\simeq (0.4$--$1)\times 10^{-10}$, we should have lepton asymmetry
%%%
\begin{eqnarray}
 \frac{n_L}{s}\simeq (1 - 3)\times 10^{-10}
  \,.
\end{eqnarray}
%%%
{}From \EQS{EQ-mass-bound} and (\ref{EQ-nLs-inf}), we can see that the
reheating temperature of the inflation $T_R$ is bounded from below as
$T_R \gsim 10^6\GEV$, since otherwise the produced lepton asymmetry is
too small as $n_L/s < 10^{-10}$.

\subsection{Froggatt-Nielsen model for neutrino}
\label{SEC-FN}

As we have just shown, the amount of generated lepton asymmetry depends
on the mass of the inflaton $m_\chi$ and the reheating temperature $T_R$
(and the branching ratio $B_r$), as well as the mass of the right-handed
neutrino $M_1$. Before discussing the inflaton sector, we here introduce
the Froggatt-Nielsen (FN) model~\cite{FN} in order to settle the mass
(and decay rate) of the right-handed neutrino and to make the discussion
in the subsequent subsections concrete.

The FN model is one of the most attractive framework for explaining the
observed hierarchies in quark and lepton mass matrices. This model is
based on an ${\rm U}(1)_{\rm FN}$ symmetry that is broken by the
vacuum-expectation value of $\Xi$, $\vev{\Xi} \neq 0$. Here $\Xi$ is a
gauge singlet field carrying the FN charge $Q_{\Xi} = -1$. Then, all
Yukawa couplings are realized as nonrenormalizable interactions
including $\Xi$, and are given by the following form;
%%%
\begin{eqnarray}
 W &=& \widetilde{y_{ij}}
  \left(
   \frac{\vev{\Xi}}{M_G}
   \right)^{Q_i + Q_j}
   \psi_i \psi_j H_{u(d)}
   \nonumber \\
 &=&
  \widetilde{y_{ij}}
  \,\,\varepsilon^{Q_i + Q_j}\,\,
  \psi_i \psi_j H_{u(d)} \,,
\end{eqnarray}
%%%
where $\widetilde{y_{ij}}$ are ${\cal O}(1)$ coupling constants, $Q_i$
are the FN charges of various chiral superfields $\psi_i$ and
$\varepsilon \equiv \vev{\Xi}/M_G$. Here, we have assumed that Higgs
doublets $H_u$ and $H_d$ (which couple to up-type and down-type quarks,
respectively) have zero FN charges. The observed mass hierarchies for
quarks and charged leptons are well explained by taking suitable FN
charges for them. For instance, we assign FN charges ($a+1$, $a$, $a$)
for lepton doublets $L_i$, while giving charges ($2$, $1$, $0$) to the
right-handed charged leptons $\overline{E}_i$, with $\varepsilon \simeq
0.05$--$0.1$~\cite{YanaRamo,SatoYana-paper}.  The charges are listed in
Table.~\ref{Table-FN}. We will take $a = 0$ or $1$ according to
Ref.~\cite{YanaRamo}. The Charges of the quarks are also determined in
the framework of the ${\rm SU}(5)$ grand unified theory~\cite{YanaRamo}.
%%%%%%%%%%%%%%%%%%%%%%%%%%%%%%%%%%%%%%%%%%%%%%%%
\begin{table}[t]%%%%%%%%%%%%%%%%%%%%%%%%%%%%%%%%
%%%%%%%%%%%%%%%%%%%%%%%%%%%%%%%%%%%%%%%%%%%%%%%%
 \begin{center}
   \begin{tabular}{|c||ccc|ccc|ccc|c|}
    \hline
    $\psi_i$ 
    & $L_1$ & $L_2$ & $L_3$ 
    & $\overline{E}_1$ & $\overline{E}_2$ & $\overline{E}_3$ 
    & $N_1$ & $N_2$ & $N_3$ 
    & $\Xi$
    \\ \hline
    $Q_i$ 
    & $a+1$ & $a$ & $a$ 
    & $2$ & $1$ & $0$
    & $d$ & $c$ & $b$
    & $-1$
    \\ \hline
   \end{tabular}
 \end{center}
 \caption{The Froggatt-Nielsen charges of various supermultiplets. $a =
 0$ or $1$. We assume $d > c \ge b \ge 0$, i.e., $M_1\ll M_2$, $M_3$.}
 \label{Table-FN}
%%%%%%%%%%%%%%%%%%%%%%%%%%%%%%%%%%%%%%%%%%%%%%%%
\end{table}%%%%%%%%%%%%%%%%%%%%%%%%%%%%%%%%%%%%%
%%%%%%%%%%%%%%%%%%%%%%%%%%%%%%%%%%%%%%%%%%%%%%%%

Let us apply the above mechanism to the neutrino sector as well. The
mass matrix for the heavy right-handed neutrinos $N_i$ is given by
%%%
\begin{eqnarray}
 M_{R\,ij} = \widetilde{\xi_{ij}}\,\,\varepsilon^{Q_i + Q_j}\,\,M_0 \,,
\end{eqnarray}
%%%
where $M_0$ represents some right-handed neutrino mass scale and
$\widetilde{\xi_{ij}}$ are coupling constants of order unity like
$\widetilde{y_{ij}}$. Charges for right-handed neutrinos are also found
in Table.~\ref{Table-FN}. We take $d > c \ge b \ge 0$, i.e., $M_1\ll
M_2$, $M_3$. Hereafter, we will take a basis where the mass matrix for
the charged leptons is diagonal.\footnote{Here, one might wonder if the
mixing matrix from the charged lepton sector would change the
discussion, since the mass matrix for the charged leptons has
off-diagonal elements in the above FN mechanism. However, the correction
from this effect yields higher order terms in $\varepsilon$, and hence
we can safely neglect it.}  Then, the neutrino Dirac mass matrix $m_D$
and the right-handed neutrino mass matrix $M_R$ are given by the
following forms;
%%%
\begin{eqnarray}
 m_D &=&
  \vev{H_u}
  \left(
   \begin{array}{ccc}
    \varepsilon^d & 0 & 0
     \\
    0 & \varepsilon^c & 0
     \\
    0 & 0 & \varepsilon^b
   \end{array}
   \right)
   \left(
    \begin{array}{ccc}
     \widetilde{y_{11}} & \widetilde{y_{12}} & \widetilde{y_{13}}
      \\
     \widetilde{y_{21}} & \widetilde{y_{22}} & \widetilde{y_{23}}
      \\
     \widetilde{y_{31}} & \widetilde{y_{32}} & \widetilde{y_{33}}
    \end{array}
    \right)
    \left(
     \begin{array}{ccc}
      \varepsilon^{a+1} & 0 & 0
       \\
      0 & \varepsilon^a & 0
       \\
      0 & 0 & \varepsilon^a
     \end{array}
     \right)
     \,,
     \nonumber \\
 M_R &=&
  M_0
  \left(
   \begin{array}{ccc}
    \varepsilon^d & 0 & 0
     \\
    0 & \varepsilon^c & 0
     \\
    0 & 0 & \varepsilon^b
   \end{array}
   \right)
   \left(
    \begin{array}{ccc}
     \widetilde{\xi_{11}} & \widetilde{\xi_{12}} & \widetilde{\xi_{13}}
      \\
     \widetilde{\xi_{12}} & \widetilde{\xi_{22}} & \widetilde{\xi_{23}}
      \\
     \widetilde{\xi_{13}} & \widetilde{\xi_{23}} & \widetilde{\xi_{33}}
    \end{array}
    \right)
    \left(
     \begin{array}{ccc}
      \varepsilon^d & 0 & 0
       \\
      0 & \varepsilon^c & 0
       \\
      0 & 0 & \varepsilon^b
     \end{array}
     \right)\,,
\end{eqnarray}
%%%
where $(m_D)_{i\alpha}$ is defined as $W = (m_D)_{i\alpha} N_i L_\alpha$
($\alpha = e, \mu, \tau$). The neutrino mass matrix is then given by
%%%
\begin{eqnarray}
 \label{EQ-mnu-FN}
  m_{\nu} &=&
  m_D^T \frac{1}{M_R} m_D
  \nonumber \\
 &=&
  \frac{\varepsilon^{2a} \vev{H_u}^2}{M_0}
  \left(
   \begin{array}{ccc}
    \varepsilon & 0 & 0
     \\
    0 & 1 & 0
     \\
    0 & 0 & 1
   \end{array}
   \right)
   \left(
    \,\{\widetilde{y_{ij}}\}\,
    \right)^T
    \left(
     \,\{\widetilde{\xi_{ij}}\}\,
     \right)^{-1}
     \left(
      \,\{\widetilde{y_{ij}}\}\,
      \right)
      \left(
       \begin{array}{ccc}
	\varepsilon & 0 & 0
	 \\
	0 & 1 & 0
	 \\
	0 & 0 & 1
       \end{array}
       \right)
       \nonumber \\
 &=&
  \frac{\varepsilon^{2a} \vev{H_u}^2}{M_0}
  \left(
   \begin{array}{ccc}
    {\cal O}(\varepsilon^2)
     & {\cal O}(\varepsilon)
     & {\cal O}(\varepsilon)
     \\
    {\cal O}(\varepsilon) 
     & {\cal O}(1) 
     & {\cal O}(1)
     \\
    {\cal O}(\varepsilon)
     & {\cal O}(1)
     & {\cal O}(1)
   \end{array}
   \right)\,.
\end{eqnarray}
%%%
As shown in Ref.~\cite{YanaRamo}, this mass matrix can naturally lead to
a large $\nu_{\mu}$--$\nu_{\tau}$ mixing angle, which is suggested from
the atmospheric neutrino oscillation~\cite{SK-Atm}. It is remarkable
that the FN charges of the right-handed neutrinos are completely
canceled out in the neutrino mass matrix in \EQ{EQ-mnu-FN} and hence the
hierarchy in the neutrino mass matrix is determined only by the charges
of the lepton doublets, ($a+1$, $a$, $a$).

In this model, the mass scale of the right-handed neutrino $M_0$ is
estimated as
%%%
\begin{eqnarray}
 \label{EQ-FN-M0}
  M_0
  &\sim&
  \varepsilon^{2a}
  \frac{\vev{H_u}^2}{\mnu{3}}
  \nonumber\\
 &\sim&
  \left\{
   \begin{array}{lcc}
    0.6\times 10^{15}\GEV & {\rm for} & a = 0\,,
     \\
    (1 - 6)\times 10^{12}\GEV & {\rm for} & a = 1\,.
   \end{array}
   \right.
\end{eqnarray}
%%%
Here, we have used $\mnu{3}\simeq 0.05\EV$ and $\varepsilon\simeq 0.05$ --
$0.1$. Then the mass of the lightest right-handed neutrino $M_1$ is given
by
%%%
\begin{eqnarray}
 \label{EQ-FN-M1}
 M_1
  &\sim&
  M_0 \varepsilon^{2d}
  \nonumber\\
 &\sim&
  \left\{
   \begin{array}{lcc}
    (1 - 6)\times 10^{12}\GEV & {\rm for} & a+d = 1\,,
     \\
    (0.3 - 6)\times 10^{10}\GEV & {\rm for} & a+d = 2\,.
   \end{array}
   \right.
\end{eqnarray}
%%%
In the following analysis, we only consider the case $a+d=1$ or $a+d=2$. 
Thus, our assumption of out-of-equilibrium condition $M_1\gsim T_R$ is
justified as far as $T_R\lsim 10^9\GEV$.  On the other hand, the total
decay width of the $N_1$, $\Gamma_{N_1}$, is given by
%%%
\begin{eqnarray}
 \Gamma_{N_1} 
  &\sim&
  \frac{ 1 }{ 4 \pi }
  \varepsilon^{ 2(a+d) } M_1
  \nonumber\\
 &\sim&
  \left\{
   \begin{array}{lcc}
    (0.3 - 5)\times 10^9\GEV & {\rm for} & a+d = 1\,,
     \\
    (0.02 - 5)\times 10^5\GEV & {\rm for} & a+d = 2\,.
   \end{array}
   \right.
\end{eqnarray}
In deriving the formula of the lepton asymmetry in \EQ{EQ-nLs-inf}, we
have assumed that the $N_1$ decays immediately after produced in
inflaton decay, which corresponds to $\Gamma_{N_1}\gg \Gamma_\chi$.
($\Gamma_\chi$ is the decay rate of the inflaton $\chi$.) In terms of
the reheating temperature $T_R \simeq (\pi^2 g_*/90)^{-1/4} \sqrt{M_G
\Gamma_\chi}$, this requirement corresponds to
%%%
\begin{eqnarray}
 T_R\ll 2\times
  10^{10}\GEV 
  \left(
   \frac{\Gamma_{N_1}}{10^3\GEV}
   \right)^{1/2}
  \,.
\end{eqnarray}
%%%
Therefore, our assumption is again guaranteed for $T_R\lsim 10^9\GEV$.

Finally, let us estimate the asymmetric parameter $\epsilon_1$ in the
present FN model. {}From \EQ{EQ-ep1-theirs}, it is given by~\cite{BY}
%%%
\begin{eqnarray}
 \epsilon_1
  &\sim&
  \frac{3}{8\pi}
  \frac{1}{\left(\varepsilon^{a+d}\right)^2}
  \left[
   \left(
    \varepsilon^{a+d}\cdot
    \varepsilon^{a+c}
    \right)^2
    \frac{\varepsilon^{2d}}{\varepsilon^{2c}}
    +
    \left(
     \varepsilon^{a+d}\cdot
     \varepsilon^{a+b}
     \right)^2
     \frac{\varepsilon^{2d}}{\varepsilon^{2b}}
   \right]
   \nonumber\\
 &\sim&
  \frac{3}{8\pi}
  \varepsilon^{2(a+d)}
  \nonumber\\
 &\sim&
  \frac{3}{8\pi}
  \frac{\mnu{1} M_1}{\vev{H_u}^2}
  \,.
\end{eqnarray}
%%%
We see that the above estimation agrees with the general result for
$\epsilon_1$ in \EQ{EQ-ep1-before-final}, with $\delta_{\rm eff}\sim
1$. Thus, we can use the formula of the lepton asymmetry in
\EQ{EQ-nLs-inf} consistently.

\subsection{Hybrid inflation}
\label{SEC-Hybrid}

In this subsection we perform a detailed analysis on hybrid inflation
models and examine whether they can produce sufficient lepton
asymmetry to account for the baryon asymmetry in the present universe,
avoiding the overproduction of the gravitinos.

\subsubsection{hybrid inflation with a $B-L$ symmetry}
\label{SEC-Hybrid-1}

Before discussing hybrid inflation models, let us first show a
particle-physics model for the heavy right-handed neutrinos $N_i$.  A
simple extension of the SUSY standard model is given by considering a
gauged ${\rm U}(1)_{B-L}$ symmetry, in which, as mentioned in
\SEC{SEC-1-1}, right-handed neutrinos $N_i$ are necessary to cancel
$B-L$ gauge anomaly.  We introduce standard-model gauge-singlet
supermultiplets $\Psi$ and $\overline{\Psi}$ carrying $B-L$ charges $+2$
and $-2$, respectively, and suppose that the $B-L$ symmetry is
spontaneously broken by the condensations $\vev{\Psi} =
\vev{\overline{\Psi}}$ at high energies. ($\vev{\Psi} =
\vev{\overline{\Psi}}$ is required from the $D$-term flatness condition
of the ${\rm U}(1)_{B-L}$.) Then the heavy neutrinos $N_i$, which carry
$B-L$ charge $-1$, acquire Majorana masses through the following
superpotential:\footnote{Leptogenesis in the hybrid inflation discussed
in Ref.~\cite{LGinf-Laza} assumes a nonrenormalizable superpotential $W
= (g'_i/2) N_i N_i \Psi \Psi$, where the $B-L$ charge of $\Psi$ is taken
to be $+1$.}
%%%
\begin{eqnarray}
 \label{EQ-W-PsiNN}
 W= \frac{1}{2} \xi_i \Psi N_i N_i
  \,.
\end{eqnarray}
%%%
Here, we have assumed that $\Psi$ and $\overline{\Psi}$ have zero FN
charges. Then the right-handed neutrino scale $M_0$ in \EQ{EQ-FN-M0} is
given by
%%%
\begin{eqnarray}
 M_0 = \vev{\Psi} = \vev{\overline{ \Psi }}
  \,.
\end{eqnarray}
%%%
Thus, the right-handed neutrino mass scale $M_0$ derived in the FN model
in \EQ{EQ-FN-M0} is explained in term of the $B-L$ breaking scale about
${\cal O}(10^{15})\GEV$ or ${\cal O}(10^{12})\GEV$.

A superpotential causing the $B-L$ breaking is given by
\begin{eqnarray}
 \label{EQ-W-HInf}
  W = \Phi
  \left( \lambda \Psi \overline{\Psi} - \mu^2
   \right)
   \,,
\end{eqnarray}
where $\Phi$ is a gauge-singlet supermultiplet, $\lambda$ a coupling
constant and $\mu$ a dimensionful mass parameter.  (We will take a basis
where $\mu^2$ and $\lambda$ are real and positive by using the phase
rotations of $\Phi$ and $\Psi \overline{\Psi}$.) Notice that this
superpotential possesses a ${\rm U}(1)$ $R$-symmetry where the $\Phi$
and $\Psi \overline{\Psi}$ have ${\rm U}(1)_R$ charges 2 and 0,
respectively.  The potential for scalar components of the
supermultiplets $\Phi$, $\Psi$ and $\overline{\Psi}$ is given by, in
supergravity,
\begin{eqnarray}
 \label{EQ-SUGRA}
  V =
  e^{K/M_G^2}
  \left\{
   \left(
    \frac{\partial^2 K}{\partial Z_i\partial Z_j^*}
    \right)^{-1}
    D_{Z_i}W D_{Z_j^*}W^*
    -
    3
    \frac{1}{M_G^2}
    |W|^2
    \right\}
     \,
     +
     \, D\,\,{\rm terms}
     \,,
\end{eqnarray}
where
\begin{eqnarray}
 D_{Z_i}W =
  \frac{\partial W}{\partial Z_i}
  +
  \frac{1}{M_G^2}
  \frac{\partial K}{\partial Z_i} W
  \,,
\end{eqnarray}
and $Z_i$ denote scalar components of supermultiplets. We assume the
$R$-invariant K\"ahler potential for $\Phi$, $\Psi$ and
$\overline{\Psi}$
\begin{eqnarray}
 \label{EQ-K-HInf}
    K = | \Phi |^2 +
    |\Psi|^2 + | \overline{\Psi} |^2
    + \frac{ \kappa_1}{4 M_G^2} | \Phi |^4 
    \cdots
    \,,
\end{eqnarray}
where the ellipsis denotes higher-order terms which we neglect in the
present analysis. {}From the above scalar potential, we have the
following SUSY-invariant vacuum (hereafter, we denote the scalar
components of the supermultiplets $\Phi$ and $\Psi$ by the same symbols
as the corresponding supermultiplets):  
\begin{eqnarray}
 \label{EQ-VEV-HInf}
  \vev{\Psi}= \vev{\overline{\Psi}}
  = \sqrt{ \frac{ \mu^2 }{ \lambda } }  
  \,,
  \qquad
  \vev{\Phi} = 0
  \,,
\end{eqnarray}
where we have chosen $\vev{\Psi}$ to be real and positive by using the
${\rm U}(1)_{B-L}$ rotation.

It is quite interesting to observe that the superpotential
(\ref{EQ-W-HInf}) is nothing but the one proposed in
Refs.~\cite{S-Hybrid-1,S-Hybrid-2} for a SUSY hybrid inflation model.
In this context, the real part of $\Phi$ is identified with the inflaton
field $\varphi / \sqrt{2}$.  The scalar potential is minimized at $\Psi
= \overline{\Psi}=0$ when $\varphi$ is larger than the following
critical value:
%%%
\begin{eqnarray}
 \varphi_C \equiv \sqrt{\frac{ 2 \mu^2 }{\lambda}}
  \,,
\end{eqnarray}
%%%
and hybrid inflation occurs for $\varphi_C < \varphi \lsim M_G$ and $k
\equiv - \kappa_1 \ge 0$.  Including one-loop corrections
\cite{S-Hybrid-1}, the potential for the inflaton $\varphi$ is given by,
for $\varphi > \varphi_C$,
%%%
\begin{eqnarray}
 \label{EQ-V-HInf}
  V 
  &\simeq& 
  \mu^4 
  \left[
   1
   + 
   \frac{ k }{ 2 } 
   \left(
    \frac{\varphi}{M_G}
    \right)^2
    + 
    \frac{1}{16} 
    \left( 4 k^2 + 7 k + 2  \right) 
    \left(
     \frac{\varphi}{M_G}
     \right)^4
     \right]
   \nonumber \\
 &+& \frac{ \lambda^4 }{ 128 \pi^2 }
  \left[ 
   2 \varphi_C^4\ln 
   \left( 
    \frac{ \lambda^2 \varphi^2 }{ 2 \Lambda^2 } 
    \right)
   +
   \left( \varphi^2 - \varphi_C^2 \right )^2
   \ln \left( 1 - \frac{\varphi_C^2}{\varphi^2} \right)
   + 
   \left( \varphi^2 + \varphi_C^2 \right)^2
   \ln \left( 1 + \frac{\varphi_C^2}{\varphi^2} \right)
    \right]
    \,,
    \nonumber\\
\end{eqnarray}
%%%
where $\Lambda$ denotes the renormalization scale.  Here, we have
included the higher order terms which come from the K\"ahler potential
(\ref{EQ-K-HInf}) up to $\mu^4 (\varphi/M_G)^4$ term, since the initial
value of the inflaton field is close to $M_G$.\footnote{However, we have
neglected the $\mu^4 (\varphi/M_G)^4$ terms coming from the higher order
interactions in the K\"ahler potential (e.g., $K = \kappa'
|\Phi|^6/M_G^4$), for simplicity.  }

Let us now discuss the inflation dynamics.  The slow-roll conditions for
inflation are given by~\cite{KT}
%%%
\begin{eqnarray}
 \frac{1}{2} 
  \left( M_G \frac{V'}{V} \right)^2 &<& 1
  \,,
  \nonumber \\
 M_G^2 \frac{ |V''| }{V} &<& 1
  \,,
  \label{EQ-SlowRoll}
\end{eqnarray}
%%%
where the prime denotes the derivative with the inflaton field
$\varphi$.  These conditions are satisfied when $\varphi > \varphi_C$,
$\lambda < 1$ and $0\le k < 1$.  Therefore, while the inflaton $\varphi$
rolls down along the potential in \EQ{EQ-V-HInf} from $\varphi_I$
($\varphi_C < \varphi_I \lsim M_G$) to $\varphi_C$, the vacuum energy
$\mu^4$ of the potential dominates the energy of the universe and hence
the hybrid inflation takes place \cite{Hybrid-Inf-Model}.

After the inflation ends, the vacuum energy is transferred into the
energies of the coherent oscillations of the following two fields: the
inflaton $\varphi$ and a scalar field $\Sigma = (\Psi +
\overline{\Psi})/\sqrt{2}$.  Notice that the inflaton $\varphi$ forms a
massive supermultiplet together with the $\Sigma$ field in the vacuum in
\EQ{EQ-VEV-HInf}, whose masses are given by
%%%
\begin{eqnarray}
 \label{EQ-MPHI-HInf}
  m_{\varphi} = m_\Sigma &=& \sqrt{2 \lambda}\,\mu 
  \nonumber\\
 &=& \sqrt{2}\,\,\lambda\vev{\Psi}
  \,.
\end{eqnarray}
%%%

The radiations of the universe are produced by the decays of the
$\varphi$ and/or $\Sigma$ field. In order to estimate the reheating
temperature $T_R$ we should know the total decay rates of these scalar
fields.  Through the interactions in the superpotentials in
\EQS{EQ-W-PsiNN} and (\ref{EQ-W-HInf}), the inflaton $\varphi$ decays
into scalar components $\widetilde{N_1}$ of the $N_1$ supermultiplet, if
kinematically allowed, with the rate
%%%
\begin{eqnarray}
 \Gamma_\varphi 
  &\simeq &
  \Gamma ( \varphi \to \widetilde{N_1}\widetilde{N_1} )
  \nonumber\\
 &=&
  \frac{ 1 }{ 64 \pi }
  \,
  m_\varphi
  \left(
   \frac{ M_1 }{ \vev{\Psi} }
   \right)^2
   \left( 1 - \frac{ 4 M_1^2 }{ m_\varphi^2 } \right)^{1/2}
   \,,
\end{eqnarray}
%%%
while the $\Sigma$ field decays into scalar ($\sneu$) and fermionic
($N_1$) components of the $N_1$ with the rates
%%%
\begin{eqnarray}
 \label{EQ-GammaSN1SN1}
  \Gamma (\Sigma \to {\sneu}{\sneu} ) 
  &=& 
  \frac{ 1 }{ 16 \pi }
  \,
  \frac{ 1 }{ m_\Sigma }
  \left(
   \frac{M_1^2}{\vev{\Psi}}
   \right)^2
   \left( 1 - \frac{ 4 M_1^2 }{ m_\Sigma^2 } \right)^{1/2}
   \,,\\
 \Gamma (\Sigma \to N_1 N_1) 
  &=& 
  \frac{ 1 }{ 64 \pi }
  \,
  m_\Sigma
  \left(
   \frac{ M_1 }{\vev{\Psi}}
   \right)^2
   \left( 1 - \frac{ 4 M_1^2 }{ m_\Sigma^2 } \right)^{3/2}
   \,.
   \label{EQ-GammaN1N1}
\end{eqnarray}
%%%
Here, we have used $M_1 = \xi_1 \vev{\Psi}$ and $m_{\varphi} = m_\Sigma
= \sqrt{2}\,\,\lambda\vev{\Psi}$.

Since the $\Sigma$ field has a non-zero vacuum expectation value, it
can decay also through nonrenormalizable interactions in the K\"ahler
potential,
%%%
\begin{eqnarray}
 \label{EQ-KTR-HInf}
  K = \sum_i \frac{c_i}{M_G^2} | \Sigma |^2  | \psi_i |^2
  \,,
\end{eqnarray}
%%%
where $\psi_i$ denote supermultiplets of the SUSY standard-model
particles including the right-handed neutrinos $N_i$, and $c_i$ are
coupling constants of order unity.  Then the partial decay rate through
these interactions is estimated as
%%%
\begin{eqnarray}
 \label{EQ-GAMK-HInf}
  \Gamma (\Sigma \to \psi_i \overline{\psi_i}) 
  \simeq \frac{1}{8 \pi } C
  \left(
   \frac{\vev{\Psi}}{M_G^2}
   \right)^2
   m_\Sigma^3
   \,,
\end{eqnarray}
%%%
where $C = \sum c_i^2$ is a parameter of order unity.  In the following
analysis we take $C\simeq1$, for simplicity.  Comparing this rate
(\ref{EQ-GAMK-HInf}) with the rates in \EQS{EQ-GammaSN1SN1} and
(\ref{EQ-GammaN1N1}), we see that the total decay rate of the $\Sigma$
field is determined by that in \EQ{EQ-GammaN1N1}, i.e.,
%%%
\begin{eqnarray}
 \Gamma_\Sigma\simeq \Gamma (\Sigma \to N_1 N_1)
  \,,
\end{eqnarray}
%%%
since $\Gamma (\Sigma \to N_1 N_1) > \Gamma (\Sigma \to \sneu \sneu) \gg
\Gamma (\Sigma \to \psi_i \overline{\psi_i})$ for $2 M_1 < m_\Sigma \ll
M_1(M_G/\vev{\Psi})^2$.

We assume $2 M_1 < m_\varphi = m_\Sigma$ as discussed in
\SEC{SEC-nL-inf}. Then the inflaton $\varphi$ and the $\Sigma$ field
decay at the almost same time because of $\Gamma_\varphi \simeq
\Gamma_\Sigma$.  Thus, the reheating temperature $T_R$ is estimated by
the total decay width of the $\Sigma$ field as
%%%
\begin{eqnarray}
 T_R \simeq
  \left(
   \frac{\pi^2}{90}\pi^2
   \right)^{-1/4}
   \sqrt{ \Gamma_\Sigma M_G}
   \,.
\end{eqnarray}
%%%
Here it should be noted that the branching ratio of the decay rate of
the $\Sigma$ field and that of the inflaton $\varphi$ into two $N_1$
($\sneu$) is almost unity, and hence we have $B_r\simeq 1$ in
\EQ{EQ-nLs-inf}.

The above hybrid inflation must explain the following two
observations:\footnote{The spectrum index in the hybrid inflation is
predicted to be scale-invariant, $n_s\simeq 1$.}
%%%
\begin{enumerate}
 \item the $e$-hold number $N_e$ of the present horizon, and
 \item the density fluctuations observed by the cosmic background
       explorer (COBE) satellite~\cite{COBE}.
\end{enumerate}
%%%
First, while the inflaton $\varphi$ rolls down along the potential
(\ref{EQ-V-HInf}) from $\varphi_{N_e}$ to $\varphi_C$, the scale factor
of the universe increases by $e^{N_e}$.  This $e$-fold number $N_e$ is
given by
%%%
\begin{eqnarray}
 \label{EQ-Ne-1}
  N_e \simeq \int_{\varphi_C}^{\varphi_{N_e}} d \varphi ~
  \frac{ V(\varphi) }{M_G^2 V'(\varphi)}
  \,.
\end{eqnarray}
In order to explain the present horizon scale, the $e$-fold number
should be 
\begin{eqnarray}
 \label{EQ-Ne-2}
  N_e = 67 + \frac{1}{3} \ln \left( \frac{H_{\rm inf} T_R}{M_G^2} \right)
  \,,
\end{eqnarray}
%%%
where $H_{\rm inf} = \sqrt{V}/(\sqrt{3} M_G) \simeq \mu^2 /(\sqrt{3}
M_G)$ denotes the Hubble parameter during the inflation.  Next, the
amplitude of the primordial density fluctuations $\delta \rho / \rho$
predicted by the hybrid inflation,
%%%
\begin{eqnarray}
 \frac{ \delta \rho }{ \rho }
  \simeq \frac{ 1 }{ 5 \sqrt{3} \pi }
  \frac{ V^{3/2} (\varphi_{N_e} ) }
  {M_G^3 \left| V'(\varphi_{N_e} )\right|}
  \,,
\end{eqnarray}
%%%
should be normalized by the data on anisotropies of the cosmic microwave
background radiation observed by the COBE satellite~\cite{COBE}, which
gives the following constraint:
\begin{eqnarray}
 \label{EQ-COBE-Norm}
  \frac{ V^{3/2} (\varphi_{N_e} ) }
  {M_G^3 \left| V'(\varphi_{N_e} )\right|}
  \simeq 5.3 \times 10^{-4}
  \,.
\end{eqnarray}
%%%

\begin{figure}[t]%%%%%  FIGURE Hybrid-mu  %%%%%%%%%%%%%%%%%%
%%%%%%%%%%%%%%%%%%%%%%%%%%%%%%%%%%%%%%%%%%%%%%%%%%%%%%%%%%%%
 \centerline{ {\psfig{figure=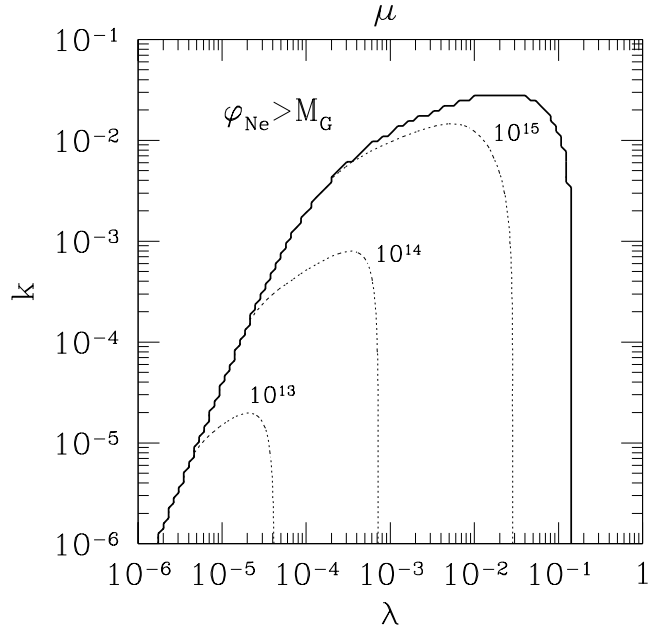,height=10cm}} }
 \vspace{-1cm}
 \caption{ The contour lines of the scale $\mu$ in the hybrid inflation
 model.  The contour lines are shown by the dotted lines and
 corresponding values of $\mu$ are represented in unit of GeV.  The
 upper bound on $k$ from the requirement $\varphi_{N_e} < M_G$ is shown
 by the thick solid line.} 
%%%%%%%%%%%%%%%%%%%%%%%%%%%%%%%%%%%%%%%%%%%%%%%%%%%%%%%%%%%%
 \label{FIG-Hybrid-mu}%%%%%%%%%%%%%%%%%%%%%%%%%%%%%%%%%%%%%%
\end{figure}

The parameters of the inflation are determined by the constraints in
\EQS{EQ-Ne-2} and (\ref{EQ-COBE-Norm}). We have performed numerical
calculations using the full scalar potential in \EQ{EQ-V-HInf}. The
scale $\mu$ of the hybrid inflation is determined for given couplings
$\lambda$ and $k$, and it is shown in
\FIG{FIG-Hybrid-mu}.\footnote{Although the scale $\mu$ slightly depends
on the mass of the right-handed neutrino $M_1$, we find no sizable
difference in the scale $\mu$ between the cases of $a+d=1$ and $a+d=2$,
which are discussed below. This is because $M_1$ enters in the
calculation only through the small correction of $\ln(T_R)$ in
\EQ{EQ-Ne-2}.}  Here and hereafter, we exclude the region where
$\varphi_{N_e} \gsim M_G$, because in that region the higher order terms
of $(\varphi/M_G)^n$ become large and our effective treatment of the
inflaton potential in \EQ{EQ-V-HInf} would be invalid.

%%%%%%%%%%%%%%%%%%%%%%%%%%%%%%%%%%%%%%%%%%%%%%%%%%%%%%%%%%%%
%%%%%%%%%%%%%%%%%%%%%%%%%%%%%%%%%%%%%%%%%%%%%%%%
%%%%%%%%%%%%%%%%%%%%%%%%%%%%%%%%%%%%%%%%%%%%%%%%%%%%%%%%%%%%

%%%%%%%%%%%%%%%%%%%%%%%%%%%%%%%%%%%%%%%%%%%%%%%%%%%%%%%%%%%%
\begin{figure}%%%%%  FIGURE Hybrid-Psi  %%%%%%%%%%%%%%%%%%%%
%%%%%%%%%%%%%%%%%%%%%%%%%%%%%%%%%%%%%%%%%%%%%%%%%%%%%%%%%%%%
 \centerline{ {\psfig{figure=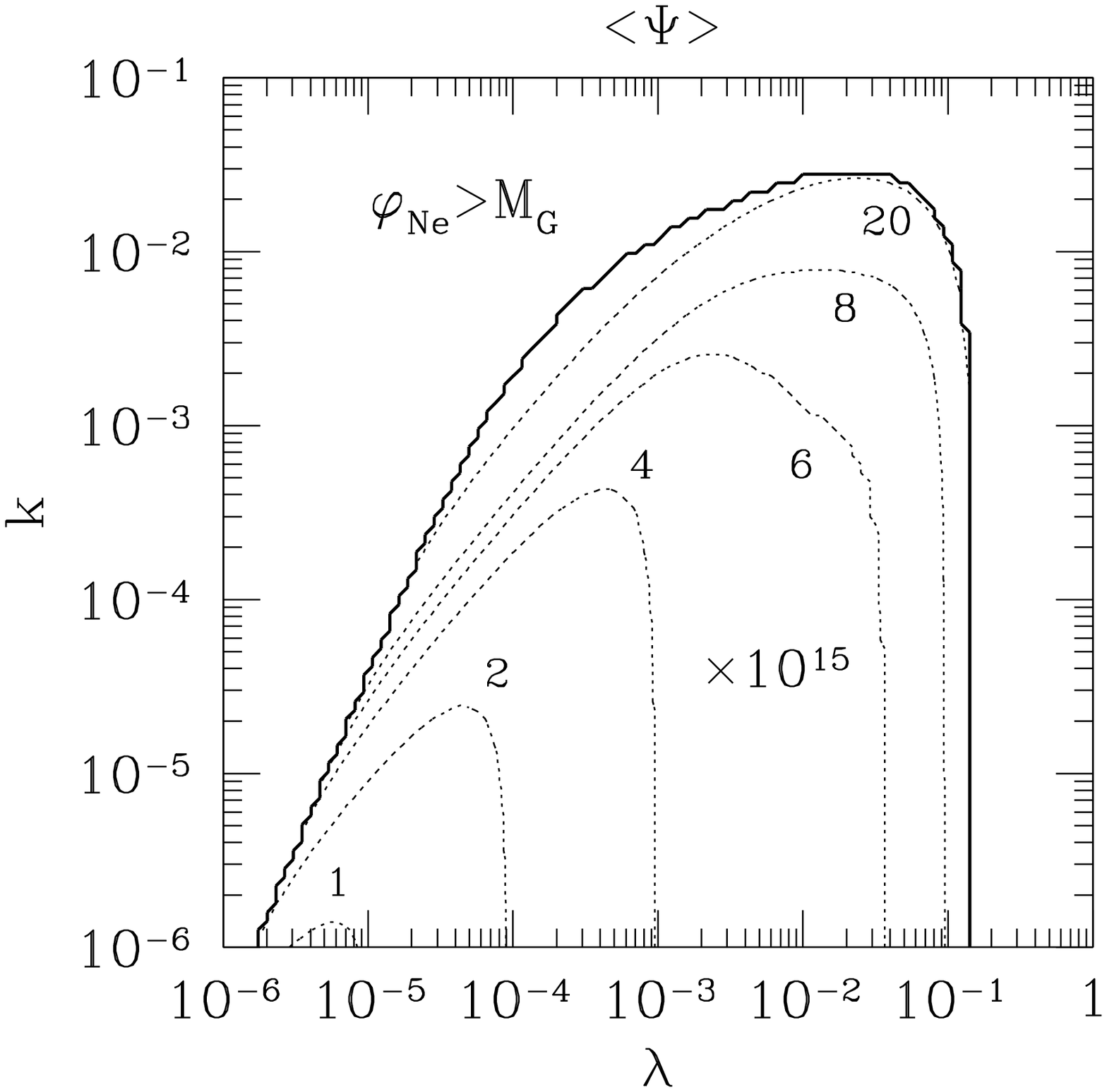,height=10cm}} }
 \vspace{-1cm}
 \caption{ The contour lines of the $\vev{\Psi}$ in the hybrid inflation
 model.  The contour lines are shown by the dotted lines and
 corresponding values of $\vev{\Psi}$ are represented in unit of
 $10^{15}$ GeV.}  
%%%%%%%%%%%%%%%%%%%%%%%%%%%%%%%%%%%%%%%%%%%%%%%%%%%%%%%%%%%%
 \label{FIG-Hybrid-Psi}%%%%%%%%%%%%%%%%%%%%%%%%%%%%%%%%%%%%%
%%%%%%%%%%%%%%%%%%%%%%%%%%%%%%%%%%%%%%%%%%%%%%%%%%%%%%%%%%%%
%%%%%%%%%%%%%%%%%%%  FIGURE Hybrid-Psi  %%%%%%%%%%%%%%%%%%%%
%%%%%%%%%%%%%%%%%%%%%%%%%%%%%%%%%%%%%%%%%%%%%%%%%%%%%%%%%%%%
 \centerline{ {\psfig{figure=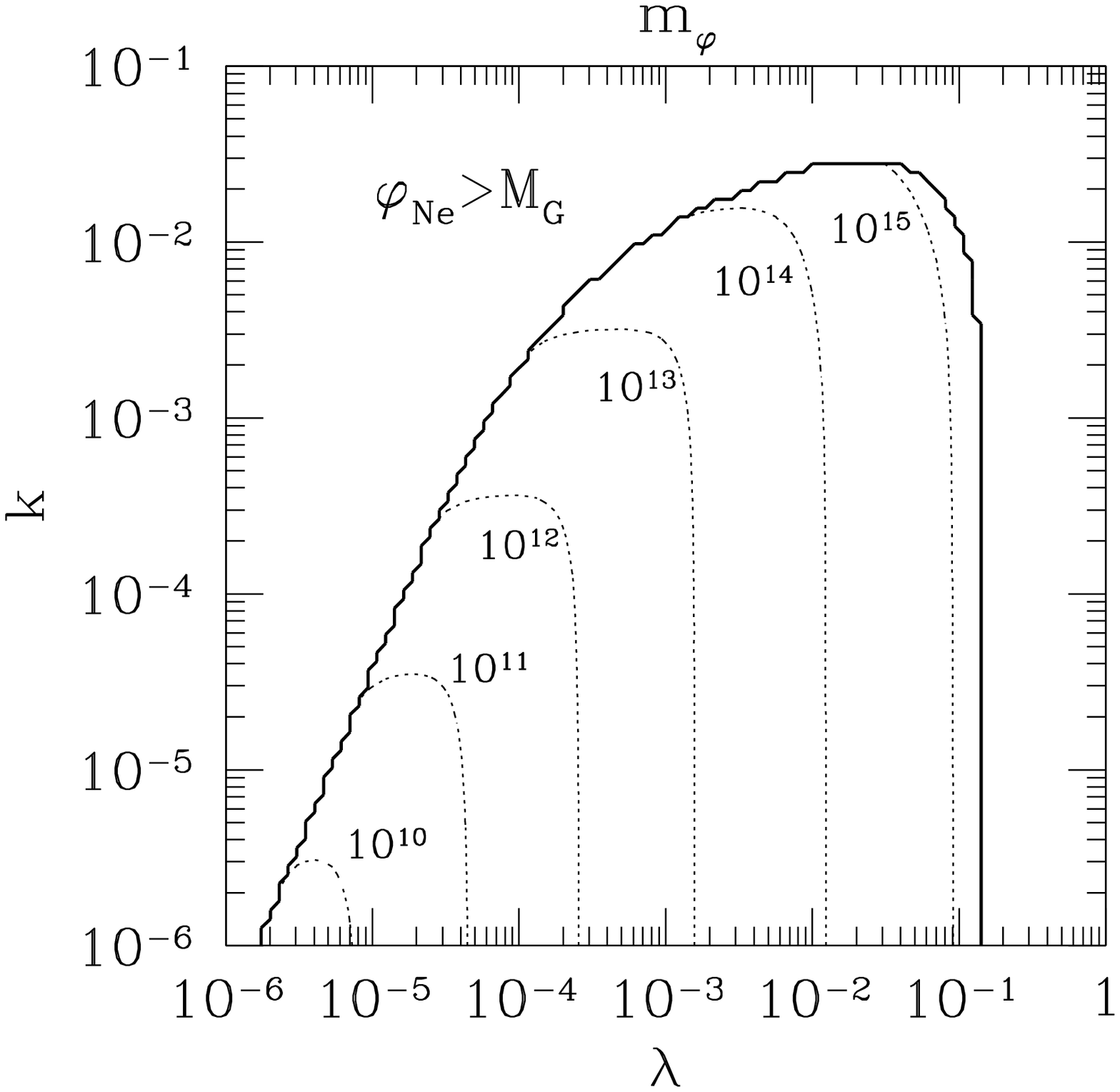,height=10cm}} }
 \vspace{-1cm}
 \caption{ The contour lines of the inflaton mass $m_\varphi$ in the
 hybrid inflation model.  The contour lines are shown by the dotted
 lines and corresponding values of $m_\varphi$ are represented in unit
 of GeV.}  
%%%%%%%%%%%%%%%%%%%%%%%%%%%%%%%%%%%%%%%%%%%%%%%%%%%%%%%%%%%%
 \label{FIG-Hybrid-mphi}%%%%%%%%%%%%%%%%%%%%%%%%%%%%%%%%%%%%
\end{figure}%%%%%%%%%%%%%%%%%%%%%%%%%%%%%%%%%%%%%%%%%%%%%%%%
%%%%%%%%%%%%%%%%%%%%%%%%%%%%%%%%%%%%%%%%%%%%%%%%%%%%%%%%%%%%
In \FIG{FIG-Hybrid-Psi} and \FIG{FIG-Hybrid-mphi}, we also show the
$B-L$ breaking scale $\vev{\Psi}$ and the inflaton mass $m_\varphi$.  It
is interesting that, as shown in \FIG{FIG-Hybrid-Psi}, the scale of the
$B-L$ breaking is predicted as $\vev{\Psi} \simeq
(1$--$5)\times10^{15}\GEV$ in a wide parameter region, $10^{-6} \lsim
\lambda \lsim 10^{-2}$ and $k \lsim 10^{-3}$, which is consistent with
$M_0 \sim 10^{15}$ GeV (i.e., $a=0$) derived from the observed neutrino
mass [see \EQ{EQ-FN-M0}].  On the other hand, the lower value of the
$B-L$ breaking scale of $M_0 \sim 10^{12}$ GeV ($a=1$) cannot be
obtained in the present hybrid inflation model.

\begin{figure}%%%%%%%%  FIGURE Hybrid-TR1 and -TR2  %%%%%%%%
%%%%%%%%%%%%%%%%%%%%%%%%%%%%%%%%%%%%%%%%%%%%%%%%%%%%%%%%%%%%
 \centerline{ {\psfig{figure=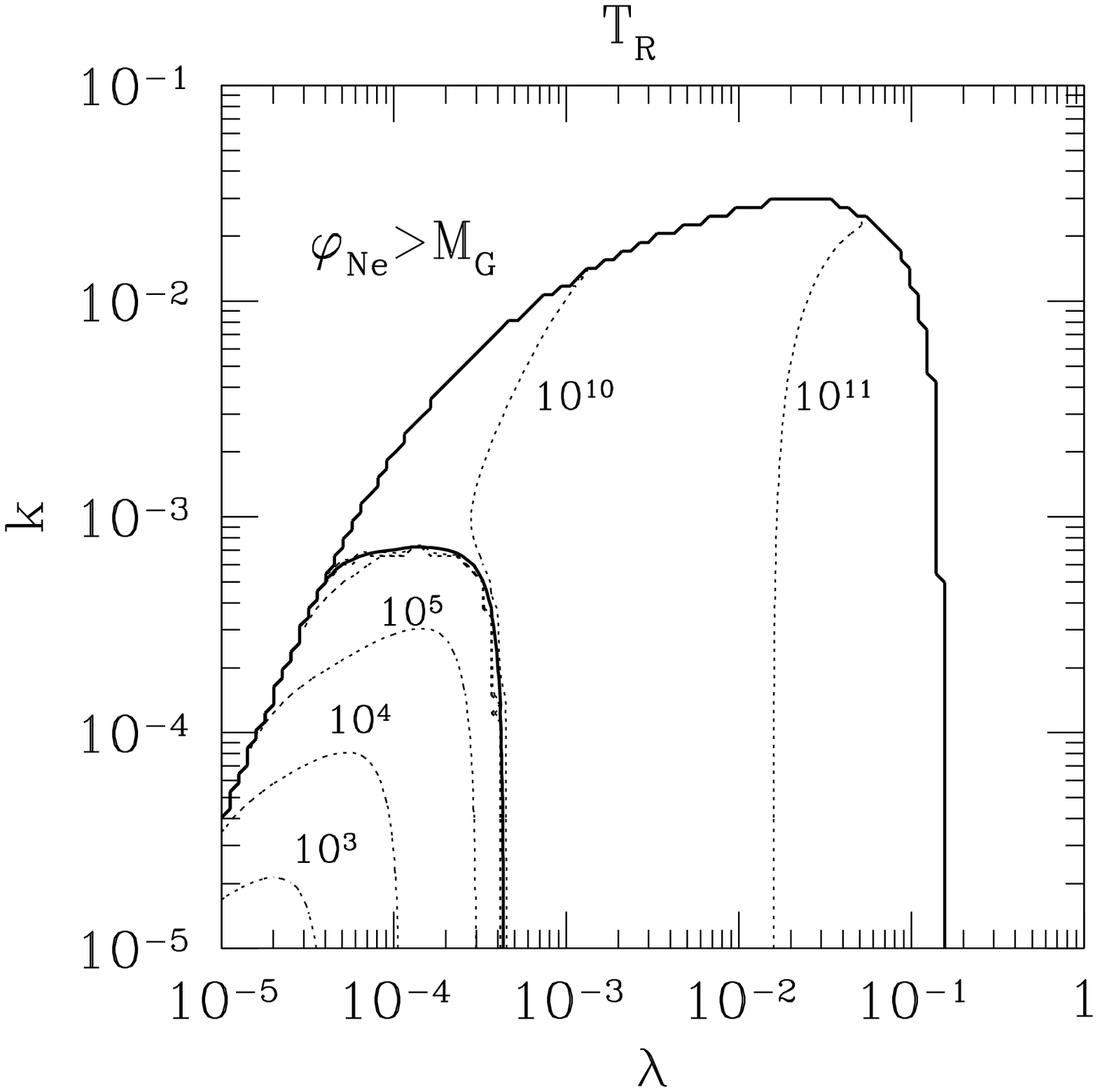,height=10cm}} }
 \vspace{-1cm}
 \caption{The contour lines of the reheating temperature $T_R$ in the
 hybrid inflation model for the case $M_1 \simeq 10^{12}$ GeV ($a+d=1$).
 The contour lines are shown by the dotted lines and corresponding
 values of $T_R$ are represented in unit of GeV.  The upper bound on $k$
 from the requirement $\varphi_{N_e} < M_G$ is shown by the thick solid
 line.  The $k$ yielding $m_\varphi = m_\Sigma = 2M_1$ is also shown by
 the thick solid line.}
%%%%%%%%%%%%%%%%%%%%%%%%%%%%%%%%%%%%%%%%%%%%%%%%%%%%%%%%%%%%
 \label{FIG-Hybrid-TR1}%%%%%%%%%%%%%%%%%%%%%%%%%%%%%%%%%%%%%
%%%%%%%%%%%%%%%%%%%%%%%%%%%%%%%%%%%%%%%%%%%%%%%%%%%%%%%%%%%%
 \centerline{ {\psfig{figure=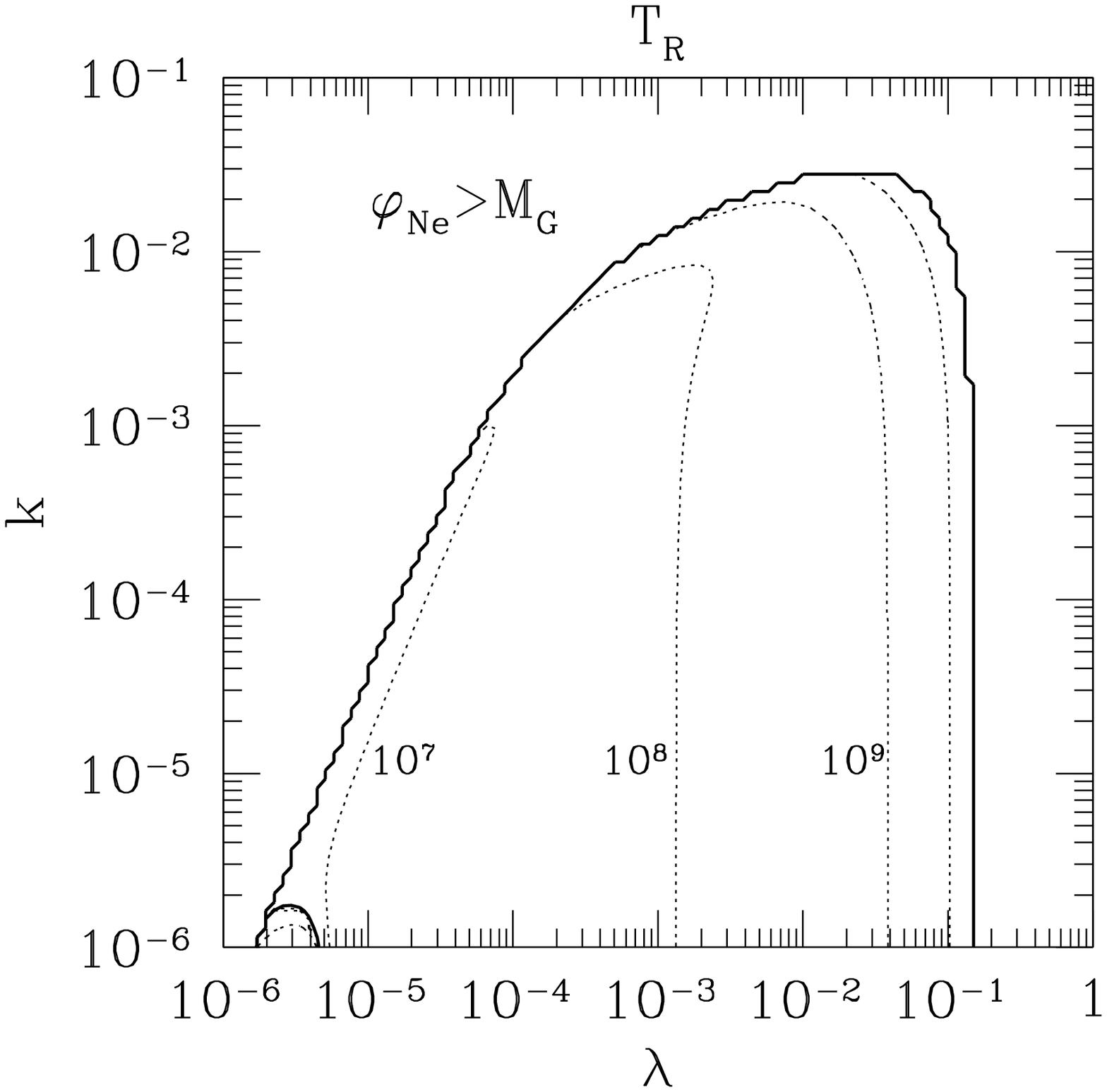,height=10cm}} }
 \vspace{-1cm}
 \caption{ The same as \FIG{FIG-Hybrid-TR1}, but for the case $M_1
 \simeq 3\times 10^9$ GeV ($a+d=2$).}
%%%%%%%%%%%%%%%%%%%%%%%%%%%%%%%%%%%%%%%%%%%%%%%%%%%%%%%%%%%%
 \label{FIG-Hybrid-TR2}%%%%%%%%%%%%%%%%%%%%%%%%%%%%%%%%%%%%%
\end{figure}

The reheating temperature $T_R$ depends on the mass of the right-handed
neutrino, $M_1$, since the decay rate of the $\Sigma$ (and $\varphi$)
depends on $M_1$. We take $M_1 \simeq 10^{12}\GEV$ ($a+d=1$) and
$M_1\simeq 3 \times 10^9\GEV$ ($a+d=2$) for representation. The obtained
reheating temperatures $T_R$ are shown in \FIG{FIG-Hybrid-TR1} and
\FIG{FIG-Hybrid-TR2}.
%%%%%%%%%%%%%%%%%%%%%%%%%%%%%%%%%%%%%%%%%%%%%%%%%%%%%%%%%%%%
%%%%%%%%%%%%%%%%%%%%%%%%%%%%%%%%%%%%%%%%%%%%%%%%
%%%%%%%%%%%%%%%%%%%%%%%%%%%%%%%%%%%%%%%%%%%%%%%%%%%%%%%%%%%%

Here, we require the reheating temperatures $T_R$ to be lower than
$10^8$ GeV to avoid overproduction of the gravitinos in a relatively
wide range of gravitino mass (see \SEC{SEC-grav}). It is found that, for
the region of the inflaton mass $m_\varphi = m_\Sigma \gg 2 M_1$, the
reheating temperature $T_R \lsim 10^8\GEV$ is obtained only for the case
$M_1 \simeq 3 \times 10^9$ GeV ($a+d=2$). In the case of $M_1 \simeq
10^{12}$ GeV ($a+d=1$), the desired low reheating temperature is
obtained for the region $m_\varphi \le 2 M_1$ because the decay into
$N_1$ is kinematically forbidden and the decay rate is determined by the
suppressed decay width in \EQ{EQ-GAMK-HInf}. However, such cases are not
interesting since the $N_1$ are not produced in the $\varphi$ and
$\Sigma$ decays and leptogenesis does not take place.\footnote{See,
however, also \SEC{SEC-preheating}.}

\begin{figure}%%%%%  FIGURE  Hybrid-nLs1 and -nLs2  %%%%%%%%
%%%%%%%%%%%%%%%%%%%%%%%%%%%%%%%%%%%%%%%%%%%%%%%%%%%%%%%%%%%%
 \centerline{ {\psfig{figure=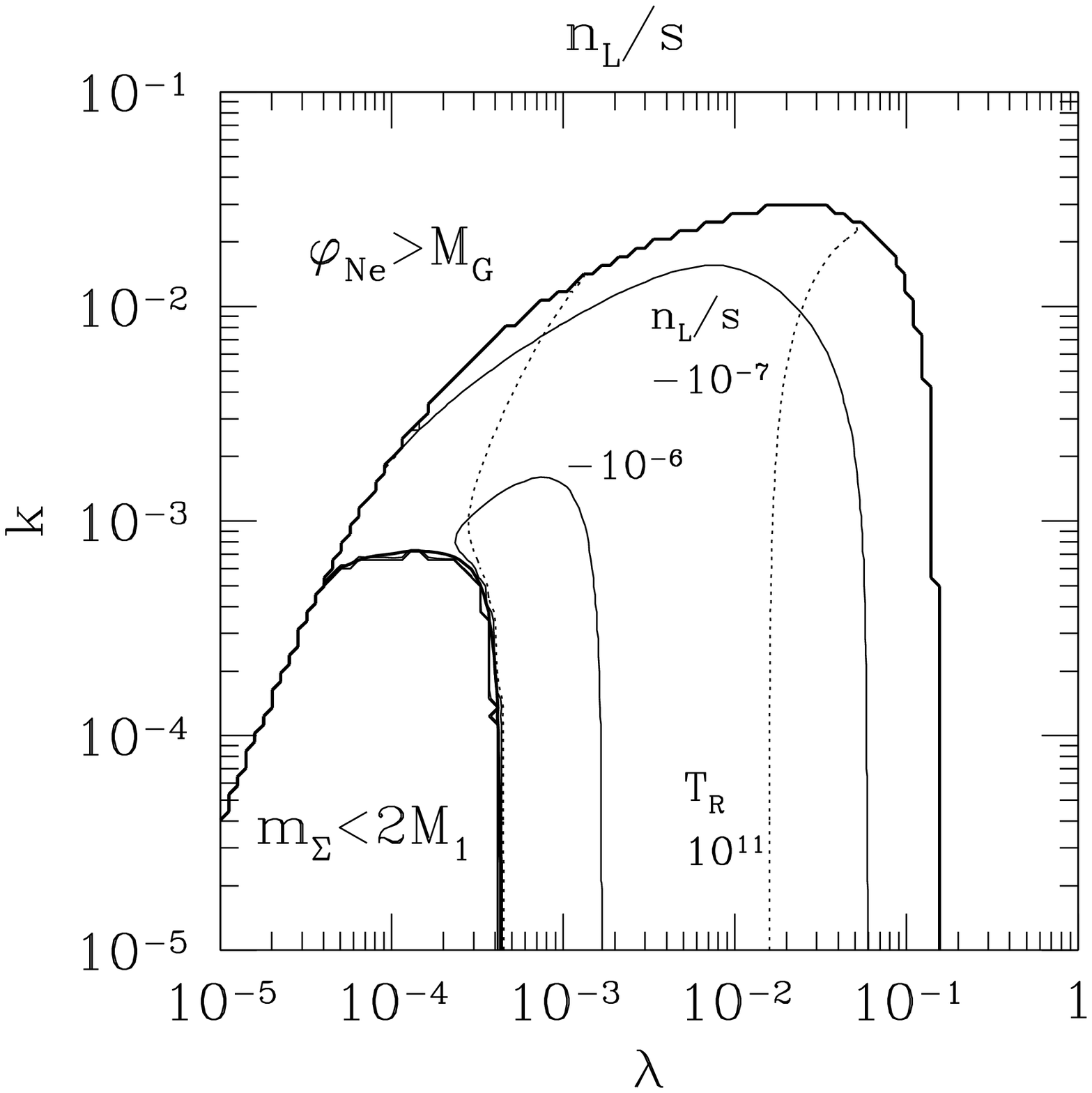,height=10cm}} }
 \vspace{-1cm}
 \caption{The contour lines of the lepton asymmetry $n_L/s$ in the
 hybrid inflation model for the case $M_1 \simeq 10^{12}$ GeV ($a+d=1$).
 The contour lines are shown by the thin solid lines and corresponding
 values of $n_L/s$ are represented.  We also show the contour lines of
 the reheating temperature by the dotted lines and corresponding values
 of $T_R$ are represented in unit of GeV.  The upper bound on $k$ from
 the requirement $\varphi_{N_e} < M_G$ and the lower bound on $k$ from
 $m_\varphi = m_\Sigma > 2 M_1$ are both shown by the thick solid
 lines.}  
%%%%%%%%%%%%%%%%%%%%%%%%%%%%%%%%%%%%%%%%%%%%%%%%%%%%%%%%%%%%
 \label{FIG-Hybrid-nLs1}%%%%%%%%%%%%%%%%%%%%%%%%%%%%%%%%%%%%
%%%%%%%%%%%%%%%%%%%%%%%%%%%%%%%%%%%%%%%%%%%%%%%%%%%%%%%%%%%%
 \centerline{ {\psfig{figure=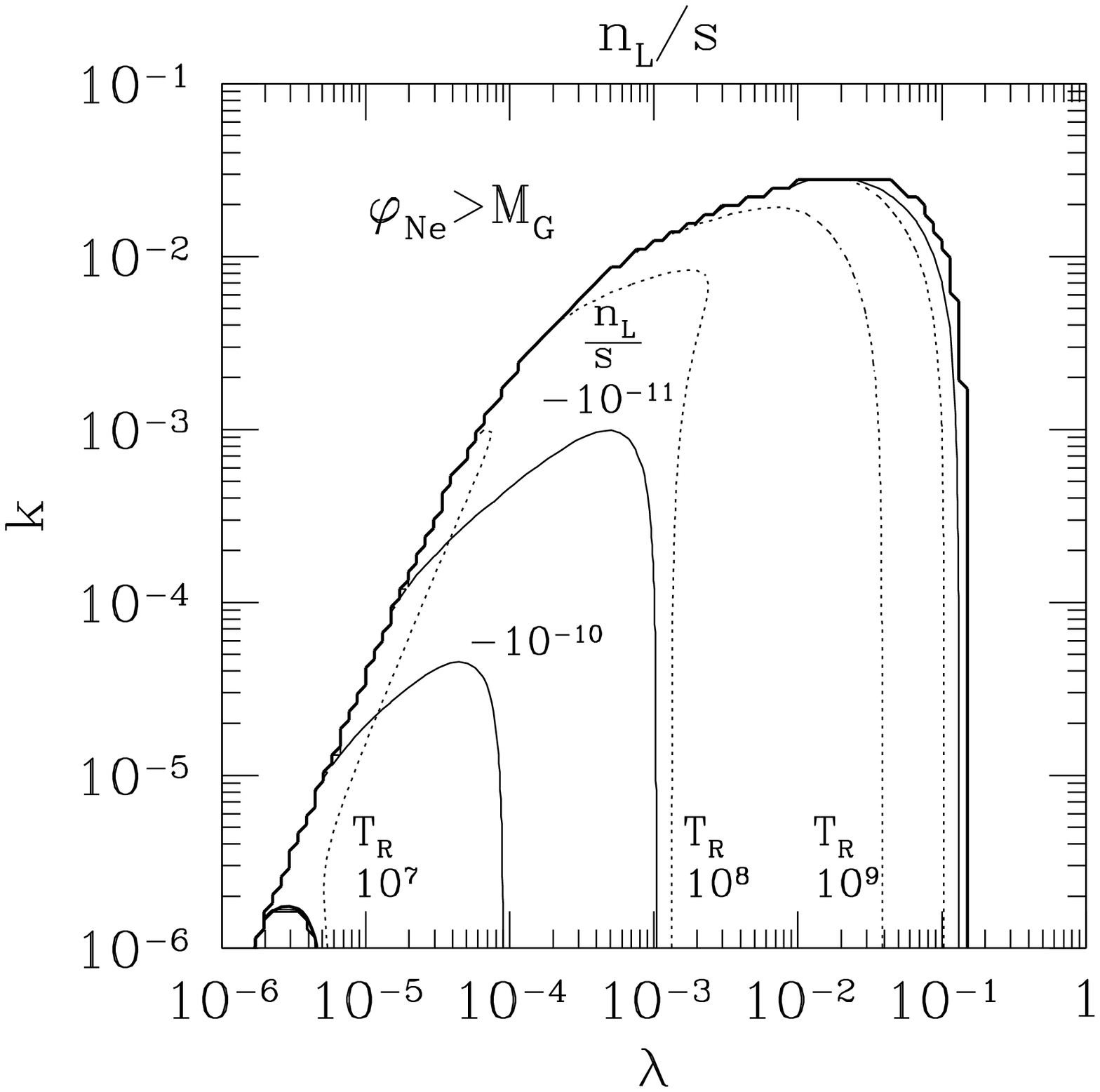,height=10cm}} }
 \vspace{-1cm}
 \caption{ The same as \FIG{FIG-Hybrid-nLs1}, but for the case $M_1 \simeq
 3\times 10^9$ GeV ($a+d=2$).} 
%%%%%%%%%%%%%%%%%%%%%%%%%%%%%%%%%%%%%%%%%%%%%%%%%%%%%%%%%%%%
 \label{FIG-Hybrid-nLs2}%%%%%%%%%%%%%%%%%%%%%%%%%%%%%%%%%%%%
\end{figure}

Now let us examine whether the leptogenesis works well or not in the
above hybrid inflation model.  Since the heavy Majorana neutrinos $N_1$
are produced in the decays of the inflaton $\varphi$ and the $\Sigma$
field, the mass of the inflaton should satisfy $m_\varphi = m_\Sigma > 2
M_1$. As derived in \SEC{SEC-nL-inf}, the ratio of the produced lepton
number to the entropy $n_L/s$ is given by \EQ{EQ-nLs-inf}. Here, we
stress again that the branching ratio is automatically given by $B_r
\simeq 1$ in the present hybrid inflation. We show in
\FIG{FIG-Hybrid-nLs1} and \FIG{FIG-Hybrid-nLs2} the $n_L/s$ for the
cases $a+d=1$ and $a+d=2$, respectively.
%%%%%%%%%%%%%%%%%%%%%%%%%%%%%%%%%%%%%%%%%%%%%%%%%%%%%%%%%%%%
%%%%%%%%%%%%%%%%%%%%%%%%%%%%%%%%%%%%%%%%%%%%%%%%
%%%%%%%%%%%%%%%%%%%%%%%%%%%%%%%%%%%%%%%%%%%%%%%%%%%%%%%%%%%%

First, we consider the case of $M_1 \simeq 10^{12}$ GeV ($a+d=1$).  We
find from \FIG{FIG-Hybrid-nLs1} that the lepton asymmetry enough to
explain the present baryon asymmetry can be generated in a wide
parameter region. However, we have a too high reheating temperature of
$T_R \simeq 10^{9}$--$10^{12}$ GeV as mentioned before. Therefore, only
the small region of $m_\Sigma \simeq 2 M_1$, where the reheating
temperature becomes as small as $T_R \simeq 10^6$--$10^8\GEV$ due to the
phase space suppression,\footnote{In this region, we should include the
decay rates in \EQS{EQ-GammaSN1SN1} and (\ref{EQ-GAMK-HInf}) in
estimating $T_R$.}  is free from the cosmological gravitino problem for
a wide range of gravitino mass. We can obtain in this very narrow region
the required lepton asymmetry of $n_L/s \simeq (1$--$3) \times 10^{-10}$
to account for the present baryon asymmetry.

Next, we consider the case of $M_1 \simeq 3 \times 10^9$ GeV ($a+d=2$).
It is found from \FIG{FIG-Hybrid-nLs2} that the required lepton
asymmetry of $n_L/s \simeq 10^{-10}$ as well as the low enough reheating
temperature of $T_R \simeq 10^7$--$10^8\GEV$ are naturally offered in
the region of $k \lsim 10^{-3}$ and $\lambda \simeq 10^{-6}$--$10^{-3}$.
Therefore, we conclude that the hybrid inflation with $M_1 \simeq 3
\times 10^9\GEV$ can produce a sufficient baryon asymmetry, giving a
reheating temperature low enough to solve the cosmological gravitino
problem.  However, even lower reheating temperature of $T_R \simeq
10^6\GEV$, which is required to avoid the cosmological difficulty for a
very wide range of gravitino mass $m_{3/2} \simeq 10\MEV$--$10\TEV$, is
achieved only in the narrow parameter region of $m_\varphi = m_\Sigma
\simeq 2 M_1$, where $T_R$ is reduced due to the phase suppression as in
the previous case ($a+d=1$).

\subsubsection{hybrid inflation without a $B-L$ symmetry}

We have, so far, identified the ${\rm U}(1)$ gauge symmetry in the
hybrid inflation model with the $B-L$ symmetry. We now consider the case
where the ${\rm U}(1)$ symmetry is not related to the $B-L$ symmetry and
even completely decoupled from the SUSY standard-model sector.  The role
of the ${\rm U}(1)$ gauge symmetry here is only to eliminate an unwanted
flat direction in the superpotential in \EQ{EQ-W-HInf}.

\begin{figure}%%%%%%%%  FIGURE Hybrid-TR3 and -nLs3  %%%%%%%
%%%%%%%%%%%%%%%%%%%%%%%%%%%%%%%%%%%%%%%%%%%%%%%%%%%%%%%%%%%%
 \centerline{ {\psfig{figure=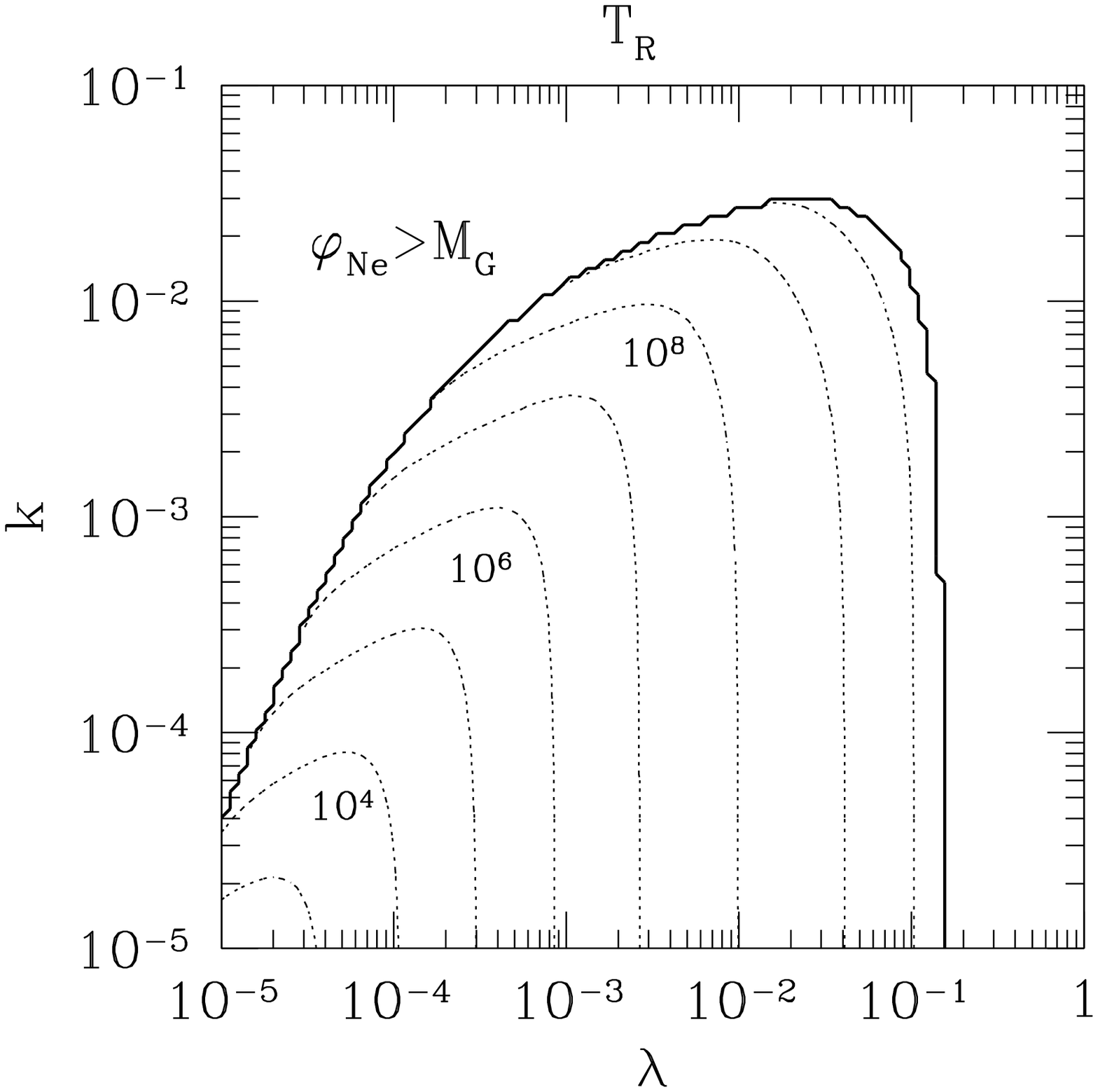,height=10cm}} }
 \vspace{-1.3cm}
 \caption{The contour lines of the reheating temperature $T_R$ in the
 hybrid inflation model without the $B-L$ symmetry.  The contour lines
 are shown by the dotted lines and corresponding values of $T_R$ are
 represented in unit of GeV.}
%%%%%%%%%%%%%%%%%%%%%%%%%%%%%%%%%%%%%%%%%%%%%%%%%%%%%%%%%%%%
 \label{FIG-Hybrid-TR3}%%%%%%%%%%%%%%%%%%%%%%%%%%%%%%%%%%%%%
%%%%%%%%%%%%%%%%%%%%%%%%%%%%%%%%%%%%%%%%%%%%%%%%%%%%%%%%%%%%
%%%%%%%%%%%%%%%%%%%%%%%%%%%%%%%%%%%%%%%%%%%%%%%%%%%%%%%%%%%%
 \centerline{ {\psfig{figure=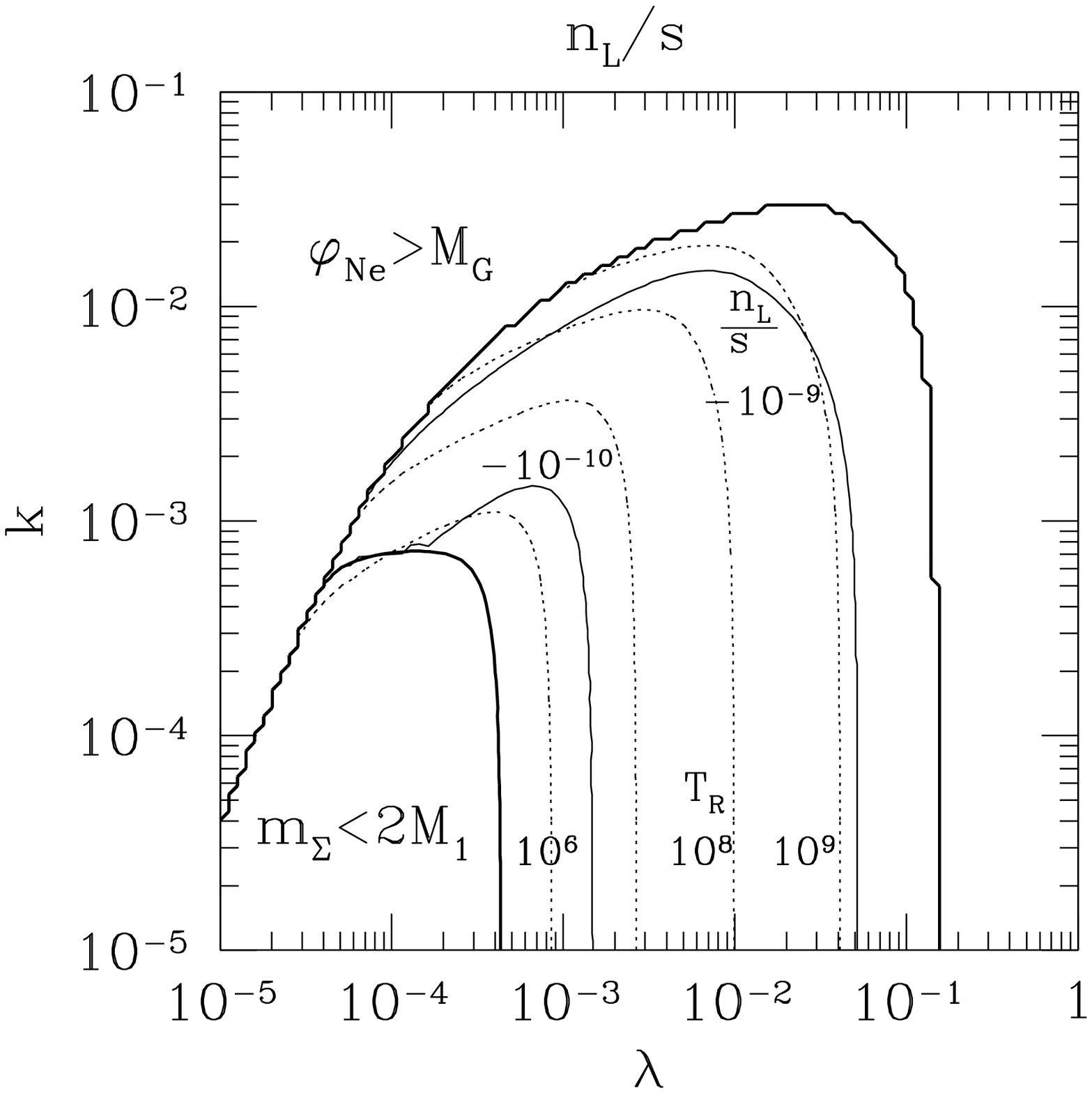,height=10cm}} }
 \vspace{-1.3cm}
 \caption{The contour lines of the lepton asymmetry $n_L/s$ in the
 hybrid inflation model without the $B-L$ symmetry for the case $M_1
 \simeq 10^{12}$ GeV ($a+d$=1).  The contour lines are shown by the thin
 solid lines and corresponding values of $n_L/s$ are represented.  Here
 we have assumed the branching ratio $B_r(\Sigma \to N_1 N_1)$ is $B_r =
 1$ for the estimation of $n_L/s$.  We also show the contour lines of
 the reheating temperature by the dotted lines and corresponding values
 of $T_R$ in units of GeV.  The upper bound on $k$ from the requirement
 $\varphi_{N_e} < M_G$ and the lower bound on $k$ from $m_\varphi =
 m_\Sigma > 2 M_1$ are both shown by the thick solid lines.}
%%%%%%%%%%%%%%%%%%%%%%%%%%%%%%%%%%%%%%%%%%%%%%%%%%%%%%%%%%%%
 \label{FIG-Hybrid-nLs3}%%%%%%%%%%%%%%%%%%%%%%%%%%%%%%%%%%%%
\end{figure}

We reanalyze the leptogenesis in hybrid inflation in the absence of the
superpotential in \EQ{EQ-W-PsiNN}. In this case the $\Sigma$ field
decays through the nonrenormalizable interactions in \EQ{EQ-KTR-HInf}.
On the other hand, the decay of the inflaton $\varphi$ is much
suppressed due to the absence of the interaction in
\EQ{EQ-W-PsiNN}. Thus we introduce a new interaction in the K\"ahler
potential as
%%%
\begin{eqnarray}
 \label{EQ-KH-HInf} 
  K = \frac{h}{M_G} \Phi^\ast H_u H_d + {\rm H.c.}
  \,.
\end{eqnarray}
%%%
Through this interaction the inflaton $\varphi$ can decay faster than
the $\Sigma$ field for the coupling $h \gsim C\,(\vev{\Psi}/M_G)^2
\simeq 10^{-6}$, and the reheating temperature $T_R$ is given by the
decay of the $\Sigma$ field. (Notice that the reheating temperature is
determined by the slower decay rate.) Since the decay rate of the
$\Sigma$ field in \EQ{EQ-GAMK-HInf} is very small compared with
\EQ{EQ-GammaN1N1}, the reheating temperature $T_R$ becomes much lower
than in the previous model. We show the obtained reheating temperature
$T_R$ in \FIG{FIG-Hybrid-TR3}.

The inflation dynamics is almost the same as in the previous hybrid
inflation model. The differences in the results for parameters, $\mu$,
$\vev{\Psi}$ and $m_\varphi$, only come from the the small correction of
$\ln(T_R)$ in \EQ{EQ-Ne-2}, and we find no sizable difference in these
values between the present and the previous models.

%%%%%%%%%%%%%%%%%%%%%%%%%%%%%%%%%%%%%%%%%%%%%%%%%%%%%%%%%%%%
%%%%%%%%%%%%%%%%%%%%%%%%%%%%%%%%%%%%%%%%%%%%%%%%
%%%%%%%%%%%%%%%%%%%%%%%%%%%%%%%%%%%%%%%%%%%%%%%%%%%%%%%%%%%%

Now let us turn to estimate the lepton asymmetry in this hybrid
inflation model. First, we find that too small lepton asymmetry is
obtained for $M_1 \simeq 3 \times 10^9\GEV$ ($a+d=2$). This is because
the $T_R$ and $M_1$ are too low to produce enough lepton asymmetry [see
\EQ{EQ-nLs-inf}]. Hence, we concentrate on the case of $M_1 \simeq
10^{12}$ GeV ($a+d=1$) in the following analysis.

We show the obtained lepton asymmetry $n_L/s$ in
\FIG{FIG-Hybrid-nLs3}. Here, we have assumed $B_r =1$. Notice that the
$\Sigma$ field decays not only into the heavy neutrinos $N_1$ but also
into the SUSY standard-model particles through the nonrenormalizable
interactions in \EQ{EQ-KTR-HInf} and hence $B_r \simeq 1$ is not
automatic in this model. Therefore, the lepton asymmetry shown in
\FIG{FIG-Hybrid-nLs3} is understood as a maximal value.  It is found
that the required lepton asymmetry to account for the empirical baryon
asymmetry can be generated in a wide parameter region $k \lsim 10^{-2}$
and $\lambda \simeq 10^{-3}$-- $10^{-2}$ with the reheating temperature
of $T_R \simeq 10^6$--$10^8\GEV$. It is remarkable that we can obtain
$n_L/s \simeq 10^{-10}$ and $T_R \simeq 10^6\GEV$ simultaneously for
$\lambda \simeq 10^{-3}$ and $k \lsim 10^{-3}$. The overproduction of
gravitinos can be avoided in the full gravitino mass region of $m_{3/2}
\simeq 10\MEV$--$10\TEV$ with such a low reheating temperature
$T_R\simeq 10^6$ GeV.\footnote{Although gravitinos are also produced in
the reheating process of the inflaton $\varphi$ decay, they are diluted
by the subsequent entropy production of the $\Sigma$ decays and become
negligible.}

\subsection{New inflation}
\label{SEC-New}

In the previous subsection, we have seen that the SUSY hybrid inflation
can successfully produce the lepton asymmetry to explain the baryon
number in the present universe, even with low reheating temperatures of
$T_R \simeq 10^6$--$10^8\GEV$. In was found, however, that relatively
small couplings of $k \lsim 10^{-2}$ and $\lambda \lsim 10^{-2}$ are
necessary to realize such low reheating temperatures. Although this is
not so problematic, it is important to consider also other inflation
models which naturally realize a low reheating temperature. The new
inflation~\cite{New-Inflation} is a well-known candidate for such an
inflation.  In this subsection, therefore, we investigate the
leptogenesis in the new inflation.

In order to make the discussion concrete, we propose a SUSY new
inflation model which has the following superpotential $W$ and K\"ahler
potential $K$:
%%%
\begin{eqnarray}
 W 
  &=&
  X ( v^2 - g \frac{\Phi^n}{M_G^{n-2}} )
  \,,
  \label{EQ-W-NewInf}
  \\
 K 
  &=&
  |\Phi|^2 
  + |X|^2 
  + \frac{\kappa_1}{4 M_G^2} |\Phi|^4 
  + \frac{\kappa_2}{M_G^2} |\Phi|^2|X|^2 
  + \frac{\kappa_3}{4 M_G^2}|X|^4 
  + \cdots
  \,,
  \label{EQ-K-NewInf}
\end{eqnarray} 
%%%
where $\Phi$ and $X$ denote supermultiplets, $v$ is the energy scale of
the inflation, $g$, $\kappa_1$, $\kappa_2$ and $\kappa_3$ are constants
of order unity, and the ellipsis denotes higher order terms.  The
superpotential in \EQ{EQ-W-NewInf} is naturally obtained, for example,
by imposing ${\rm U}(1)_R\times Z_n$ symmetry. Hereafter, we will take
$g$ and $v^2$ real and positive by using the phase rotations of $X$ and
$\Phi$.

The SUSY vacuum of the scalar potential is obtained from \EQS{EQ-SUGRA},
(\ref{EQ-W-NewInf}) and (\ref{EQ-K-NewInf}), which is given by
\begin{eqnarray}
 \label{EQ-VEV-of-phi}
  \vev{\Phi} =  M_G \left( \frac{v^2}{g M_G^2} \right)^{1/n}
  \,,
  \qquad
  \vev{X} = 0
  \,.
\end{eqnarray}
Here and hereafter, we use the same symbols $\Phi$ and $X$ for the
scalar components of corresponding supermultiplets. The scalar potential
for $|\Phi|$, $|X| \ll M_G$ is given by
\begin{eqnarray}
 V 
  &\simeq&
  v^4 
  + (1-\kappa_2)
  \,v^4 
  \left(
   \frac{|\Phi|}{M_G}
   \right)^2 
   - g
   \,v^2
   \,\frac{\Phi^n + {\Phi^*}^n}{M_G^{n-2}}
   + g^2 \frac{|\Phi|^{2n}}{M_G^{2n-4}}
   \nonumber\\
 &&-\, \kappa_3 v^4 
  \left(
   \frac{|X|}{M_G}
   \right)^2
   \,.
\end{eqnarray}
Here, we have neglected irrelevant higher order terms. As for the $X$
field, it is found that if $\kappa_3 < - 1/3$, $X$ receives a positive
mass squared which is larger than $H_{\rm inf}^2$, where $H_{\rm inf} =
v^2 /(\sqrt{3}M_G)$ is the Hubble parameter during the new inflation.
Thus $X$ settles down at $X = 0$ for $\kappa_3 < -1/3$.  Hereafter, we
assume $\kappa_3 < -1/3$ and set $X=0$.

For $g>0$ and $k\equiv\kappa_2-1>0$, we can identify the inflaton field
$\varphi$ with the real part of the scalar field $\Phi$.  The inflaton
potential near the origin ($\varphi \simeq 0$) is obtained
as,\footnote{The new inflation model proposed in
Ref.~\cite{Izawa-Yanagida-New} offers a scalar potential for the
inflaton field similar to \EQ{EQ-V-NewInf}. Thus, our discussion in this
subsection can also be applied to that model.}
\begin{eqnarray}
 \label{EQ-V-NewInf}
  V(\varphi) \simeq v^4 - \frac{k}{2}v^4
  \left(
   \frac{\varphi}{M_G}
   \right)^2
   - \frac{g}{ 2^{n/2-1} }v^2
   \frac{\varphi^n}{M_G^{n-2}}
   \,.
\end{eqnarray}
The slow-roll conditions for the inflation in \EQ{EQ-SlowRoll} are
satisfied when
\begin{eqnarray}
 &&
  0 < k   \lsim   1
  \,,
  \\
 &&
  0 < \varphi   \lsim  \varphi_f
  \equiv
  \sqrt{2}
  M_G
  \left[
   \frac{ 1-k }{ g n (n-1) }
   \,
   \frac{v^2}{M_G^2}
   \right]^{ 1/(n-2) }
   \,,
   \label{EQ-New-phif}
\end{eqnarray}
and the new inflation takes place when the inflaton $\varphi$ rolls down
along the potential in \EQ{EQ-V-NewInf} from $\varphi \simeq 0$ to
$\varphi_f$.\footnote{For a successful new inflation, the initial value
of the inflaton field $\varphi$ should be taken near the local maximum
of the potential $\varphi \simeq 0$~\cite{Linde}. Although this seems
unnatural, there have been found some dynamical
mechanisms~\cite{New-Inf-Initial} to solve this initial value problem.}

Now let us discuss the decay of the inflaton. We assume that the
inflaton decays through nonrenormalizable interactions in the K\"ahler
potential such as
\begin{eqnarray}
 \label{EQ-Kdecay}
  K = \sum_i c'_i |\Phi|^2|\psi_i|^2
  \,,
\end{eqnarray}
where $\psi_i$ denote supermultiplets for SUSY standard-model particles
including the right-handed neutrinos, and $c'_i$ are coupling constants
of order unity. (Here, a K\"ahler potential $K = (1/M_G)\Phi^* H_u H_d$
is forbidden by suitable charge assignment of the $Z_n$ symmetry for the
Higgs supermultiplets $H_u$ and $H_d$.) With the interactions in the
above K\"ahler potential, the decay rate $\Gamma_{\varphi}$ is estimated
as
\begin{eqnarray}
 \label{EQ-Gamma-K}
 \Gamma_{\varphi}
  \simeq 
  \frac{1}{8\pi}
  C'
  \left(
   \frac{\vev{\Phi}}{M_G^2}
   \right)^2
   m_{\varphi}^3
   \,,
\end{eqnarray}
where $C'\equiv \sum_i {c'_i}^2$ and we take $C'=1$ in the following
discussion. Here, the inflaton mass $m_\varphi$ in the true vacuum
\EQ{EQ-VEV-of-phi} is given by
%%%
\begin{eqnarray}
 m_{\varphi} \simeq 
  n 
  \,g^{1/n}
  M_G
  \left(
  \frac{v}{M_G}
  \right)^{2-2/n}
   \,.
\end{eqnarray}
%%%
Then the reheating temperature $T_R$ is given by $T_R\simeq (\pi^2
g_*/90)^{-1/4} \sqrt{\Gamma_\varphi M_G}$ with the decay rate in
\EQ{EQ-Gamma-K}.

\begin{figure}[t]%%%%%%%%%%%%%%%%%%%%%%%%%%%%%%%%%%%%%%%%%%%
%%%%%%%%%%%%%%%%%%%%%%%%%%%%%%%%%%%%%%%%%%%%%%%%%%%%%%%%%%%%
 \centerline{ {\psfig{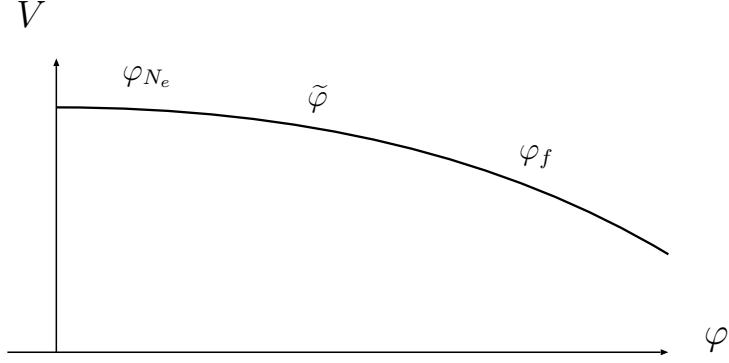}} }
 %%%%%%%%%%%%%%%%%%%%%%%%%
 \begin{picture}(0,0)%%%%%
 %%%%%%%%%%%%%%%%%%%%%%%%%
  \put(120,140){\large $V$}
  \put(160,120){$\varphi_{N_e}$}
  \put(230,110){$\widetilde{\varphi}$}
  \put(310,90){$\varphi_f$}
  \put(380,20){\large $\varphi$}
 %%%%%%%%%%%%%%%%%%%%%%%%%
 \end{picture}%%%%%%%%%%%%
 %%%%%%%%%%%%%%%%%%%%%%%%%
 \caption{A Schematic behavior of the new inflation potential for
 $\varphi \ll M_G$.}
 \label{FIG-New-Pot}
 \vspace{1em}
%%%%%%%%%%%%%%%%%%%%%%%%%%%%%%%%%%%%%%%%%%%%%%%%%%%%%%%%%%%%
\end{figure}

The scale $v$ of the new inflation is determined from (i) the $e$-fold
number $N_e$ of the present horizon, (ii) the amplitude and (iii) the
spectrum index $n_s$ of the primordial density fluctuations
$\delta\rho/\rho$. First, the number of $e$-foldings $N_e$ should
satisfy \EQ{EQ-Ne-2} to explain the present horizon scale.  In the
present new inflation model, $N_e$ is given by
\begin{eqnarray}
 \label{EQ-New-horizon}
  N_e
  &=&
  \int_{\varphi_{N_e}}^{\varphi_f}
  d\varphi ~
  \frac{V(\varphi)}{M_G^2 |V'(\varphi)|} 
  \nonumber
  \\
 &\simeq&
  \int_{\varphi_{N_e}}^{\widetilde{\varphi}}
  d\varphi  ~
  \frac{v^4}{k v^4\Frac{\varphi}{M_G}}
  +
  \int_{\widetilde{\varphi}}^{\varphi_f}
  d\varphi ~
  \frac{v^4}{\Frac{ng}{2^{n/2-1}} v^2 \Frac{\varphi^{n-1}}{M_G^n}  }
  \,.
\end{eqnarray}
where
\begin{eqnarray}
 \label{EQ-New-phitilde}
  \widetilde{\varphi}
  =
  \sqrt{2}M_G
  \left(
   \frac{k}{ng}
   \,
   \frac{v^2}{M_G^2}
   \right)^{1/(n-2)}
   \,,
\end{eqnarray}
and $\varphi_{N_e}$ is the value of the $\varphi$ field when the present
universe crossed the horizon.  Here we have assumed that $\varphi_{N_e}
< \widetilde{\varphi}$ (see \FIG{FIG-New-Pot}). This condition
corresponds to 
%%%
\begin{eqnarray}
 \label{EQ-New-k}
  k \gsim 1/[N_e(n-2)]
  \,.
\end{eqnarray}
%%%%%%%%%%%%%%%%%%%%%%%%%%%%%%%%%%%%%%%%%%%%%%%%%%%%%%%%%%%%
%%%%%%%%%%%%%%%%%%%%%%%%%%%%%%%%%%%%%%%%%%%%%%%%
%%%%%%%%%%%%%%%%%%%%%%%%%%%%%%%%%%%%%%%%%%%%%%%%%%%%%%%%%%%%

{}From \EQS{EQ-New-phif}, (\ref{EQ-New-horizon}), and
(\ref{EQ-New-phitilde}) we obtain the field value of $\varphi_{N_e}$:
\begin{eqnarray}
 \label{EQ-New-phine}
  \varphi_{N_e} \simeq
  \sqrt{2}M_G
  \left(
   \frac{k}{ng}
   \,
   \frac{v^2}{M_G^2}
   \right)^{1/(n-2)}
  \exp
  \left[
   - k 
   \left( N_e + 
    \frac{ n k - 1 }{ (n-2) k (1-k) }
    \right)
   \right]
   \,.
\end{eqnarray}
This $\varphi_{N_e}$ should satisfy the COBE normalization in
\EQ{EQ-COBE-Norm}:
\begin{eqnarray}
 \frac{v^2}{ k M_G \varphi_{N_e} }
  \simeq
  5.3 \times 10^{-4}
  \,.
\end{eqnarray}

On the other hand, the COBE observations show the spectrum index $n_s$
as $n_s = 1.0 \pm 0.2$~\cite{COBE}, while in the present new inflation
model it is given by~\cite{Izawa-Yanagida-New}
\begin{eqnarray}
 n_s \simeq 1 - 2 k
  \,.
\end{eqnarray}
Therefore, we take $0.01 \lsim k \lsim 0.1$ in the following analysis
assuming the coupling $k$ not be extremely small. (Notice that this
justifies the assumption $\varphi_{N_e} < \widetilde{\varphi}$ [see
\EQ{EQ-New-k}].)

\begin{figure}%%%%%  FIGURE  New-v and -Phi  %%%%%%%%%%%%%%%
%%%%%%%%%%%%%%%%%%%%%%%%%%%%%%%%%%%%%%%%%%%%%%%%%%%%%%%%%%%%
 \centerline{ {\psfig{figure=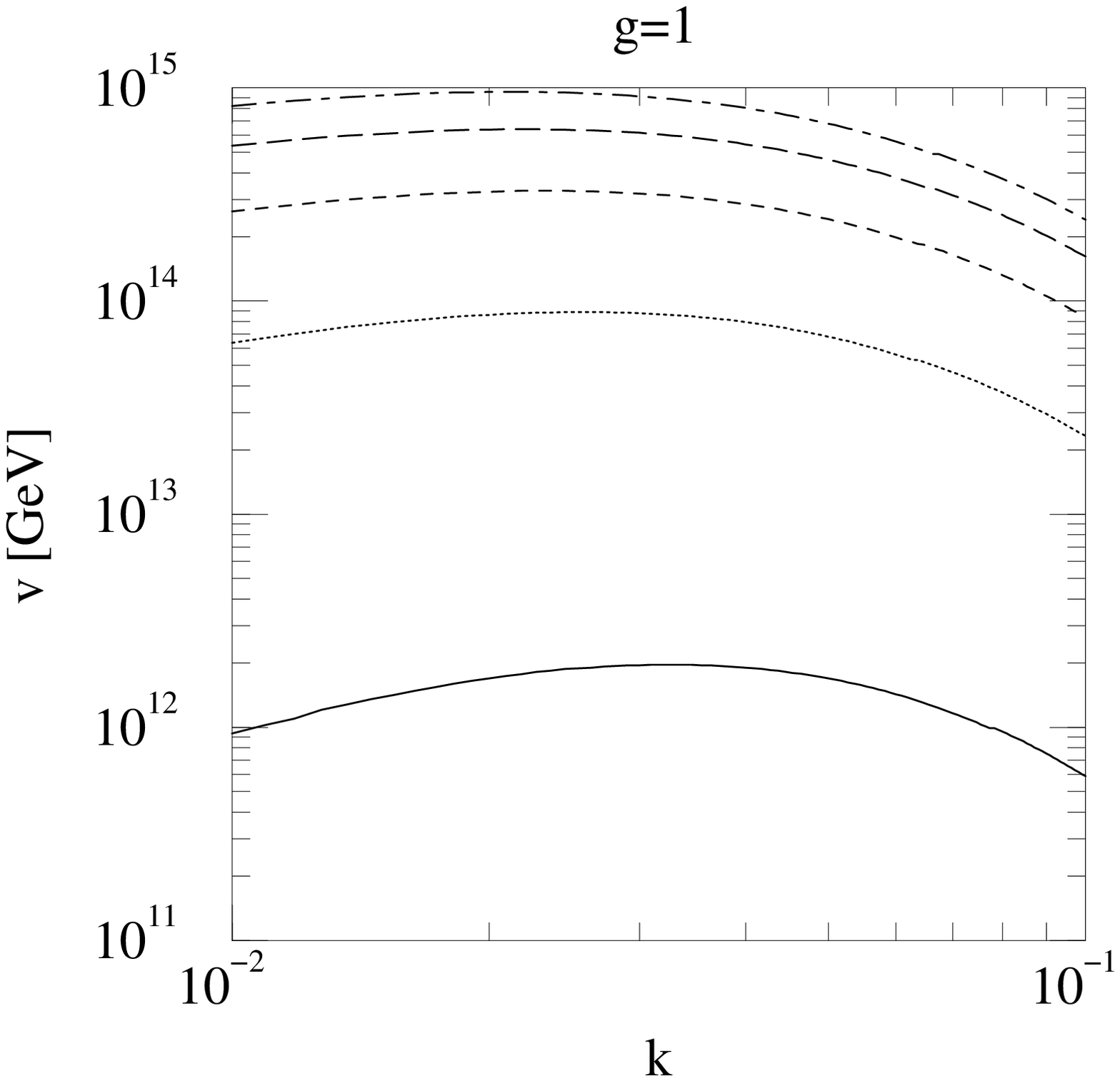,height=9cm}} }
 \caption{ The scale $v$ of the new inflation for $g=1$.  We take the
 index $n$ as $n=4$, 5, 6, 7 and 8 from the bottom to the top.}
%%%%%%%%%%%%%%%%%%%%%%%%%%%%%%%%%%%%%%%%%%%%%%%%%%%%%%%%%%%%
 \label{FIG-New-v}%%%%%%%%%%%%%%%%%%%%%%%%%%%%%%%%%%%%%%%%%%
 \vspace{1cm}%%%%%%%%%%%%%%%%%%%%%%%%%%%%%%%%%%%%%%%%%%%%%%%
%%%%%%%%%%%%%%%%%%%%%%%%%%%%%%%%%%%%%%%%%%%%%%%%%%%%%%%%%%%%
 \centerline{ {\psfig{figure=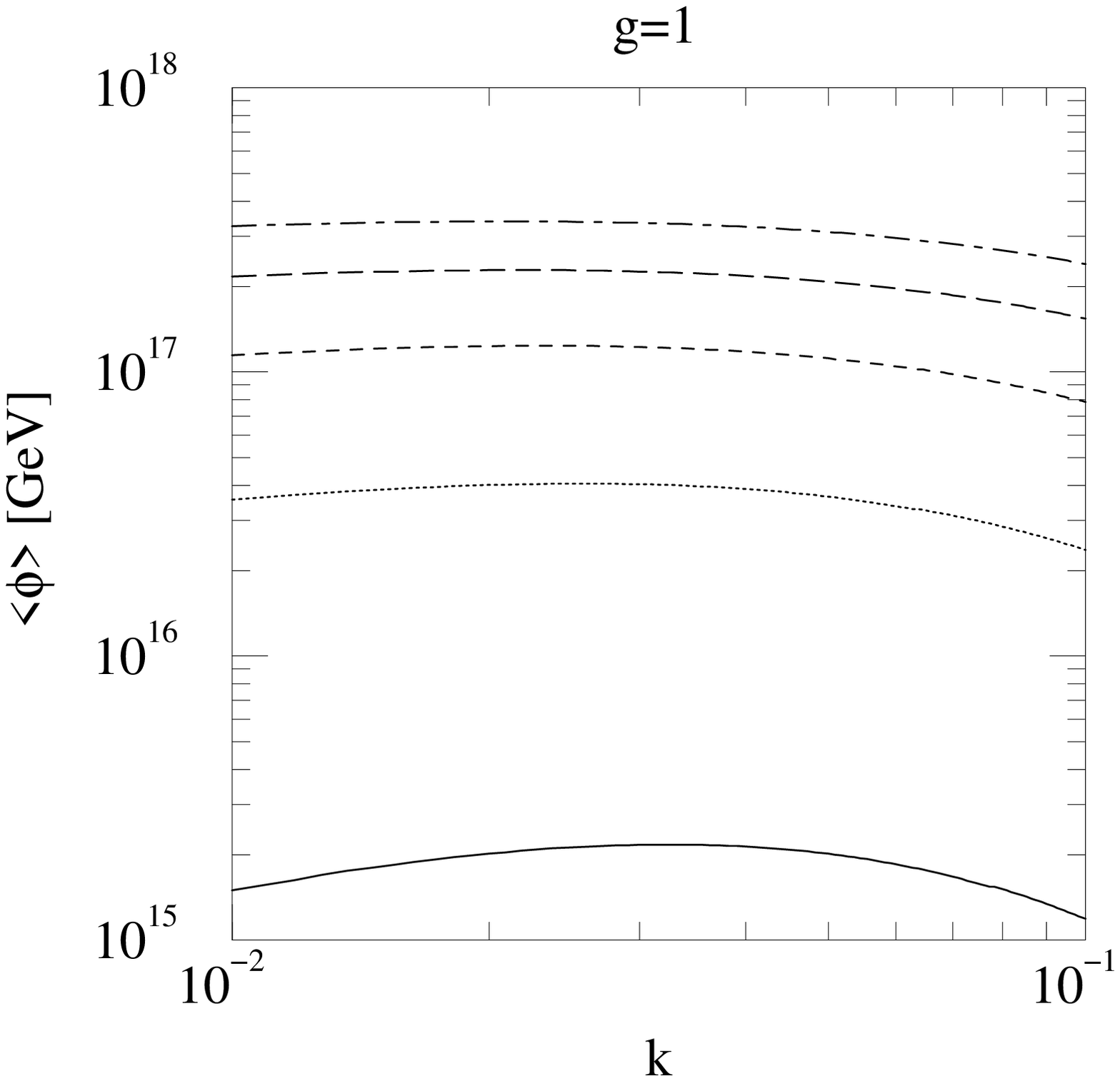,height=9cm}} } \caption{
 The vacuum expectation value $\vev{\Phi}$ of the new inflation for
 $g=1$.  We take the index $n$ as $n=4$, 5, 6, 7 and 8 from the bottom
 to the top.  }
%%%%%%%%%%%%%%%%%%%%%%%%%%%%%%%%%%%%%%%%%%%%%%%%%%%%%%%%%%%%
 \label{FIG-New-Phi}%%%%%%%%%%%%%%%%%%%%%%%%%%%%%%%%%%%%%%%%
\end{figure}

\begin{figure}%%%%%  FIGURE  New-mphi and -TR  %%%%%%%%%%%%%
%%%%%%%%%%%%%%%%%%%%%%%%%%%%%%%%%%%%%%%%%%%%%%%%%%%%%%%%%%%%
 \centerline{ {\psfig{figure=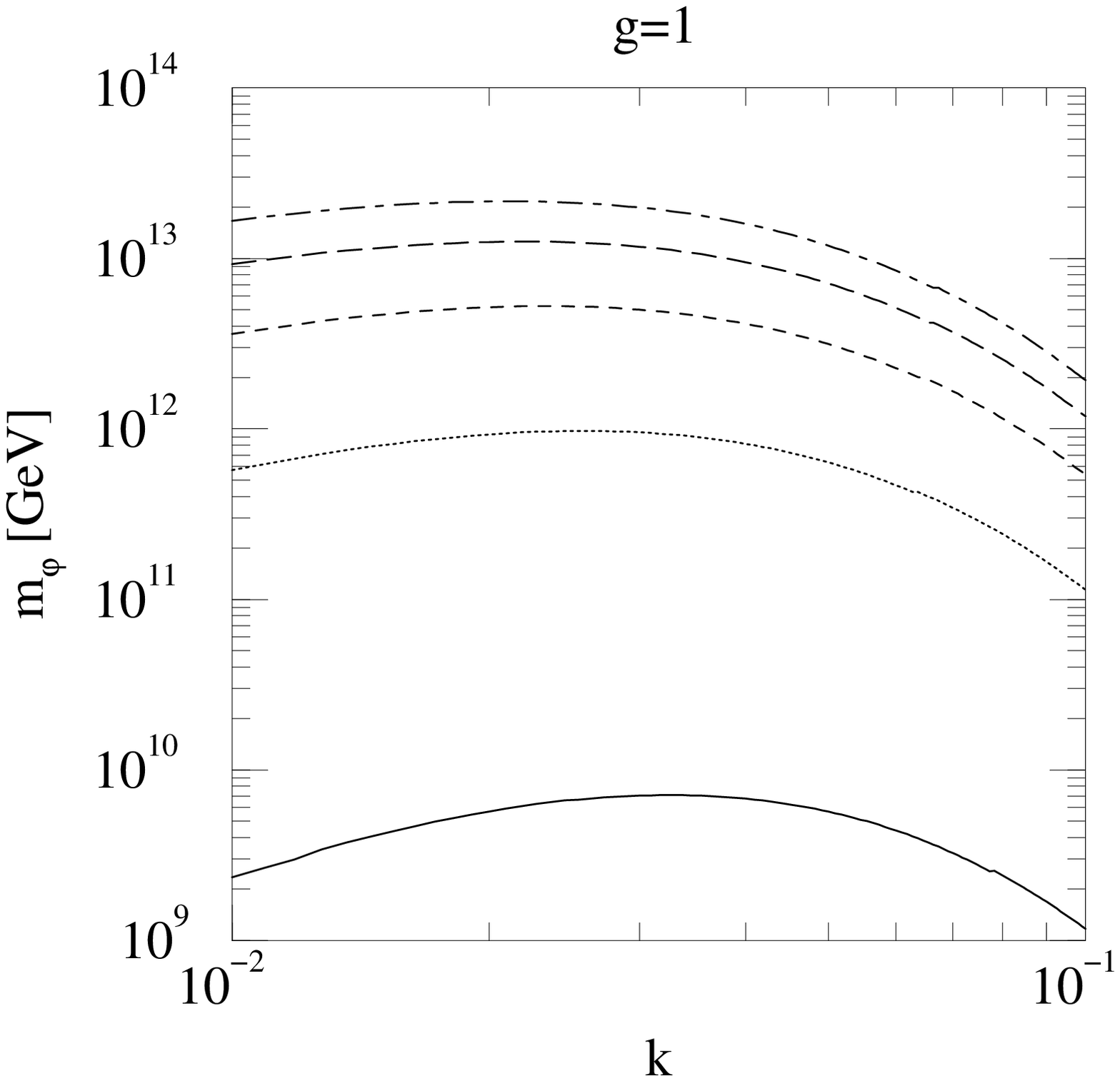,height=9cm}} }
 \caption{ The inflation mass $m_\Phi$ of the new inflation for $g=1$.
 We take the index $n$ as $n=4$, 5, 6, 7 and 8 from the bottom to the
 top.}
%%%%%%%%%%%%%%%%%%%%%%%%%%%%%%%%%%%%%%%%%%%%%%%%%%%%%%%%%%%%
 \label{FIG-New-mphi}%%%%%%%%%%%%%%%%%%%%%%%%%%%%%%%%%%%%%%%
 \vspace{1cm}%%%%%%%%%%%%%%%%%%%%%%%%%%%%%%%%%%%%%%%%%%%%%%%
%%%%%%%%%%%%%%%%%%%%%%%%%%%%%%%%%%%%%%%%%%%%%%%%%%%%%%%%%%%%
 \centerline{ {\psfig{figure=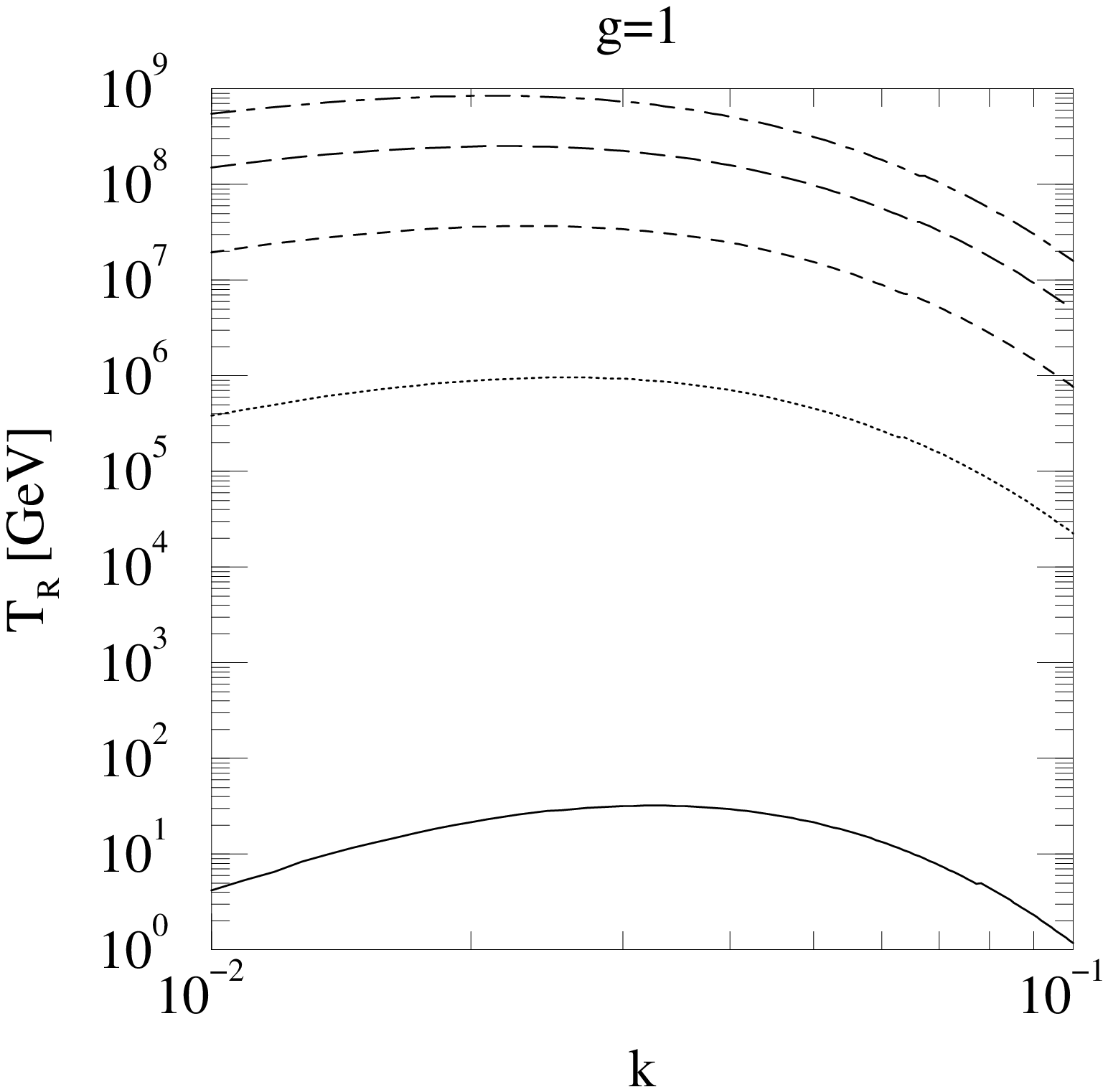,height=9cm}} }
 \caption{ The reheating temperature $T_R$ of the new inflation for
 $g=1$.  We take the index $n$ as $n = $ 4, 5, 6, 7 and 8 from the
 bottom to the top.
 } 
%%%%%%%%%%%%%%%%%%%%%%%%%%%%%%%%%%%%%%%%%%%%%%%%%%%%%%%%%%%%
 \label{FIG-New-TR}%%%%%%%%%%%%%%%%%%%%%%%%%%%%%%%%%%%%%%%%%
\end{figure}

{}From the above relations, we calculate the inflation scale $v$ for
given $n$, $g$ and $k$, which are shown in \FIG{FIG-New-v}.  The vacuum
expectation value $\vev{\Phi}$, the inflaton mass $m_{\varphi}$ and the
reheating temperature $T_R$ are found in \FIG{FIG-New-Phi},
\FIG{FIG-New-mphi}, and \FIG{FIG-New-TR}, respectively.  {}From
\FIG{FIG-New-TR}, we find that the new inflation model with the power
indices $n = 4,5$ and $6$ naturally offer the low reheating temperature
$T_R \lsim 10^8\GEV$.  For the cases $n=7,8$, we obtain $T_R \lsim 10^8$
GeV for the region $k \simeq 0.05$--$0.1$ and $0.07$--$0.1$,
respectively.

%%%%%%%%%%%%%%%%%%%%%%%%%%%%%%%%%%%%%%%%%%%%%%%%%%%%%%%%%%%%
%%%%%%%%%%%%%%%%%%%%%%%%%%%%%%%%%%%%%%%%%%%%%%%%
%%%%%%%%%%%%%%%%%%%%%%%%%%%%%%%%%%%%%%%%%%%%%%%%%%%%%%%%%%%%

%%%%%%%%%%%%%%%%%%%%%%%%%%%%%%%%%%%%%%%%%%%%%%%%%%%%%%%%%%%%
%%%%%%%%%%%%%%%%%%%%%%%%%%%%%%%%%%%%%%%%%%%%%%%%
%%%%%%%%%%%%%%%%%%%%%%%%%%%%%%%%%%%%%%%%%%%%%%%%%%%%%%%%%%%%

Now we turn to discuss the leptogenesis in this new inflation model.
The ratio of the lepton-number density $n_L$ to the entropy density $s$
is again given by \EQ{EQ-nLs-inf}. Notice that the new inflation model
with $n=4$ gives such a low reheating temperature as $T_R\simeq
1\GEV$--$10\GEV$ (see \FIG{FIG-New-TR}) that the required amount of
lepton asymmetry cannot be generated, and hence we discard this case.

\begin{figure}%%%%%  FIGURE  New-nLs1 and -nLs2  %%%%%%%%%%%
%%%%%%%%%%%%%%%%%%%%%%%%%%%%%%%%%%%%%%%%%%%%%%%%%%%%%%%%%%%%
 \centerline{ {\psfig{figure=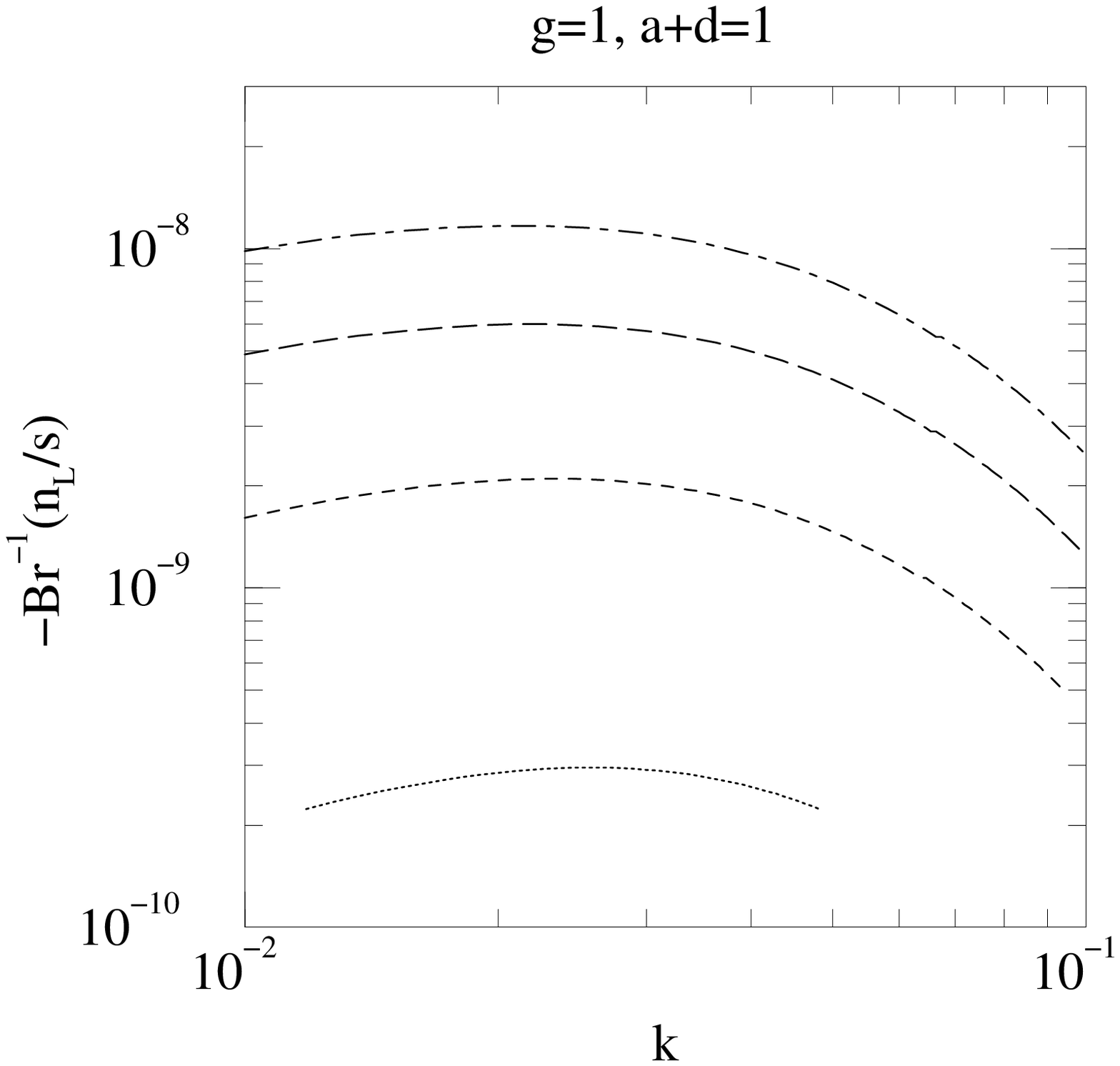,height=9cm}} }
 \caption{The lepton asymmetry produced via the decays of $N_1$ in the
 new inflation.  The index $n$ is taken as $n=5,6,7$ and $8$ from the
 bottom to the top.  We take $M_1 \simeq 3\times 10^{11}\GEV$ ($a+d =
 1$). For $n=5$, the regions $k < 1.2 \times 10^{-2}$ and $k > 4.8\times
 10^{-2}$ are excluded since $m_{\varphi}\le 2M_1$.}
%%%%%%%%%%%%%%%%%%%%%%%%%%%%%%%%%%%%%%%%%%%%%%%%%%%%%%%%%%%%
 \label{FIG-New-nLs1}%%%%%%%%%%%%%%%%%%%%%%%%%%%%%%%%%%%%%%%
 \vspace{1cm}%%%%%%%%%%%%%%%%%%%%%%%%%%%%%%%%%%%%%%%%%%%%%%%
%%%%%%%%%%%%%%%%%%%%%%%%%%%%%%%%%%%%%%%%%%%%%%%%%%%%%%%%%%%%
 \centerline{ {\psfig{figure=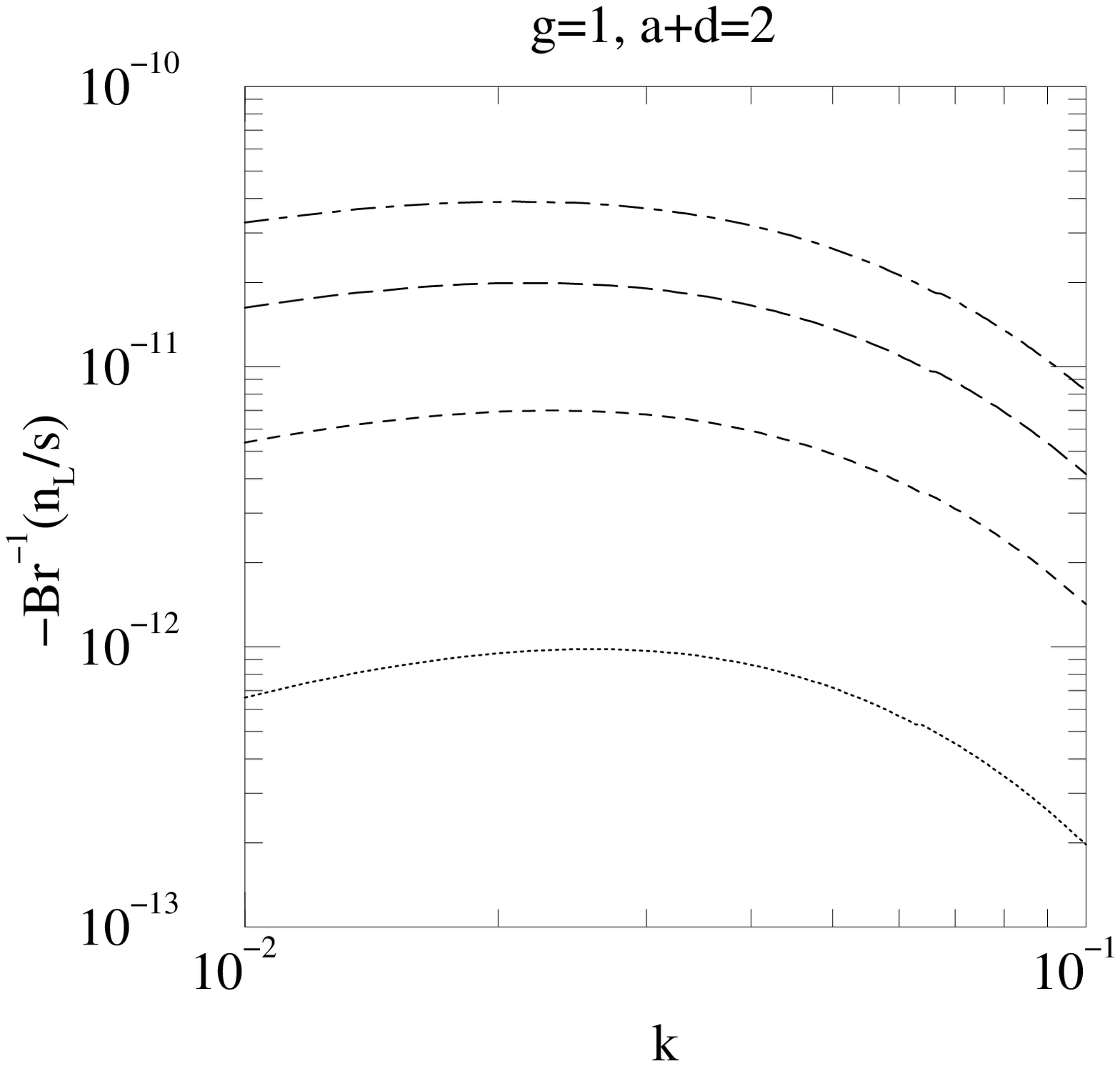,height=9cm}} }
 \caption{The same as \FIG{FIG-New-nLs1}, but for $M_1 \simeq 10^9\GEV$
 ($a+d = 2$).}
%%%%%%%%%%%%%%%%%%%%%%%%%%%%%%%%%%%%%%%%%%%%%%%%%%%%%%%%%%%%
 \label{FIG-New-nLs2}%%%%%%%%%%%%%%%%%%%%%%%%%%%%%%%%%%%%%%%
\end{figure}

The obtained lepton asymmetry for the cases $n=5,6,7$ and $8$ are shown
in \FIG{FIG-New-nLs1} and \FIG{FIG-New-nLs2} by taking $M_1 \simeq 3
\times 10^{11}\GEV$ ($a+d=1$) and $M_1 \simeq 10^9\GEV$ ($a+d=2$),
respectively.  Here we have taken relatively smaller values of $M_1$
[see \EQ{EQ-FN-M1}] to obtain a wider allowed region, taking account of
${\cal O}(1)$ ambiguities in the FN model.

Interesting results are given for $M_1 \simeq 3 \times
10^{11}\GEV$. (Here, we find that the condition $m_\varphi > 2 M_1$
excludes the regions $k < 1.2 \times 10^{-2}$ and $k > 4.8\times
10^{-2}$ for the case $n=5$.)  {}From \FIG{FIG-New-TR} and
\FIG{FIG-New-nLs1}, we see that the sufficient lepton asymmetry can be
generated for $n=5,6,7$ and $8$ with the low enough reheating
temperature of $T_R\simeq 10^6$--$10^8\GEV$. In particular, for $n=5$,
the required lepton asymmetry $n_L/s\simeq - 10^{-10}$ is obtained for
$B_r \simeq 1$ with low reheating temperature $T_R\simeq
10^6\GEV$. Thus, we can avoid the overproduction of gravitinos for a
wide range of gravitino mass $m_{3/2}\simeq 10\MEV$--$10\TEV$ in this
case.

On the other hand, when $M_1 \simeq 10^9\GEV$ ($a+d=2$), we find from
\FIG{FIG-New-nLs2} that the leptogenesis does not work well.

%%%%%%%%%%%%%%%%%%%%%%%%%%%%%%%%%%%%%%%%%%%%%%%%%%%%%%%%%%%%
%%%%%%%%%%%%%%%%%%%%%%%%%%%%%%%%%%%%%%%%%%%%%%%%
%%%%%%%%%%%%%%%%%%%%%%%%%%%%%%%%%%%%%%%%%%%%%%%%%%%%%%%%%%%%

\clearpage
\subsection{Topological inflation}
\label{SEC-topological} 

In this subsection, we discuss the leptogenesis in a topological
inflation. If the vacuum expectation value of the inflaton exceeds the
gravitational scale, the inflaton potential of the new inflation in
\EQ{EQ-V-NewInf} becomes nothing but the inflaton potential for the
topological inflation \cite{Top-Inf}.

We adopt the SUSY topological inflation model proposed in
Ref.~\cite{IKY-TopInf}.  The superpotential and K\"ahler potential in
that model are given by
\begin{eqnarray}
 \label{EQ-W-TopInf}
  W & = & v^2 X ( 1 - g \frac{\Phi^2}{M_G^2})
  \,, \\
 \label{EQ-K-TopInf}
  K & = & |X|^2 + |\Phi|^2
  + \frac{ \kappa_1 }{ 4 M_G^2 } |\Phi|^4
  + \frac{ \kappa_2 }{ M_G^2 }|X|^2|\Phi|^2 
  + \frac{ \kappa_3 }{ 4 M_G^2 }  |X|^4 + \cdots
  \,,
\end{eqnarray}
where $v$ is the energy scale of the inflation and $g$, $\kappa_1$,
$\kappa_2$ and $\kappa_3$ are coupling constants of order unity.
Hereafter, we take $g$ and $v$ to be real and positive by using the
redefinition of the field $X$ and $\Phi$. These potentials possess ${\rm
U}(1)_R \times Z_2$ symmetry; the ${\rm U}(1)_R$ charge of $\Phi$ and
$X$ are $0$ and $2$, respectively, and $\Phi$ ($X$) is odd (even) under
the $Z_2$.  This discrete symmetry is crucial to allow a topological
defect (in this case, a domain wall) and hence to realize the
topological inflation~\cite{Top-Inf}.

{}From \EQS{EQ-SUGRA}, (\ref{EQ-W-TopInf}) and (\ref{EQ-K-TopInf}) we
find a SUSY-invariant vacuum:
\begin{eqnarray}
 \vev{X} = 0
  \,,
  \quad
  \vev{\Phi} = \frac{1}{\sqrt{g}} M_G
  \,.
  \label{EQ-VEVs-TopInf}
\end{eqnarray}
where the scalar components of the supermultiplets are denoted by the
same letters as the corresponding supermultiplets. The potential for the
region $|X|$, $|\Phi| \ll M_G$ is written approximately as
\begin{equation}
 V \simeq v^4
  \left\{
   \left|
    1 - g \frac{\Phi^2}{M_G^2}
    \right|^2
    + 
    (1 - \kappa_2)
    \frac{|\Phi|^2}{M_G^2}
    - \kappa_3
    \frac{|X|^2}{M_G^2}
	\right\}
	 \,.
\end{equation}
Hereafter, we set $X=0$ assuming $\kappa_3 < - 1/3$ as we have done in
the new inflation.

For $g > 0$ and $\kappa_2 < 1$, the inflaton field $\varphi$ is
identified with the real part of $\Phi$ and the potential around the
origin is given by
\begin{equation}
 V(\varphi) \simeq v^4
  \left(
   1 - \frac{k}{2} \frac{\varphi^2}{M_G^2}
   \right)
   \,,
\end{equation}
where $k \equiv 2g + \kappa_{2} - 1$.

A topological inflation takes place if the vacuum expectation value of
$\Phi$ is of order of the gravitational scale $M_G$~\cite{Top-Inf}. It
was shown in Ref.~\cite{Top-Sakai} that, in order for a topological
defect (in our case, a domain wall) to inflate, its vacuum expectation
value should be larger than the following critical value.
\begin{eqnarray}
 \vev{\Phi} \ge \vev{\Phi}_C \simeq 1.7 M_G
  \,.
\end{eqnarray}
The slow-roll conditions in \EQ{EQ-SlowRoll} are satisfied when $0 < k <
1$ and $\varphi \lsim \varphi_f$, where $\varphi_f$ is expected to be of
order of the gravitational scale. Hereafter, we take $\varphi_f = M_G$.

As for the decay of the inflaton $\varphi$, we assume that $\varphi$
decays through the interactions in the K\"ahler potential as shown in
\EQ{EQ-Kdecay}, and hence the reheating temperature $T_R$ of the
topological inflation is again given by $T_R\simeq (\pi^2
g_*/90)^{-1/4}\sqrt{\Gamma_\varphi M_G}$ with the decay rate in
\EQ{EQ-Gamma-K}. Here, the inflaton mass in the true vacuum in
\EQ{EQ-VEVs-TopInf} is estimated as
\begin{equation}
 m_{\varphi} \simeq 2 \sqrt{g} \frac{v^2}{M_G}
  = \frac{ 2 v^2 }{ \vev{\Phi} }
  \,.
\end{equation}

As in the other inflation models, the scale of this topological
inflation is determined by (i) the $e$-fold number $N_e$, (ii) the
density fluctuations $\delta \rho/\rho$ and (iii) the spectrum index
$n_s$.  The $e$-fold number in the present model is estimated as
\begin{eqnarray}
 N_e 
  &\simeq&
  \int_{\varphi_f}^{\varphi_{N_e}} d\varphi ~
  \frac{V}{M_G^2 V'}
  \simeq
  \frac{ 1 }{ k } \ln 
  \left( \frac{ \varphi_f }{ \varphi_{N_e}} \right)
  \,,
\end{eqnarray}
%%%
which should satisfy the present Hubble horizon scale in \EQ{EQ-Ne-2},
while the COBE normalization in \EQ{EQ-COBE-Norm} requires
\begin{eqnarray}
 \frac{v^2}{ k M_G \varphi_{N_e} }
  \simeq
  5.3 \times 10^{-4}
  \,.
\end{eqnarray}
The spectrum index $n_s$ is again given by $n_s \simeq 1 - 2 k$, and
hence we take $0.01 \lsim k \lsim 0.1$ as in the new inflation model.

%%%%%%%%%%%%%%%%%%%%%%%%%%%%%%%%%%%%%%%%%%%%%%%%%%%%%%%%%%%%
\begin{figure}%%     FIGURE  Top-mTR and -nLs    %%%%%%%%%%%
%%%%%%%%%%%%%%%%%%%%%%%%%%%%%%%%%%%%%%%%%%%%%%%%%%%%%%%%%%%%
 \centerline{ {\psfig{figure=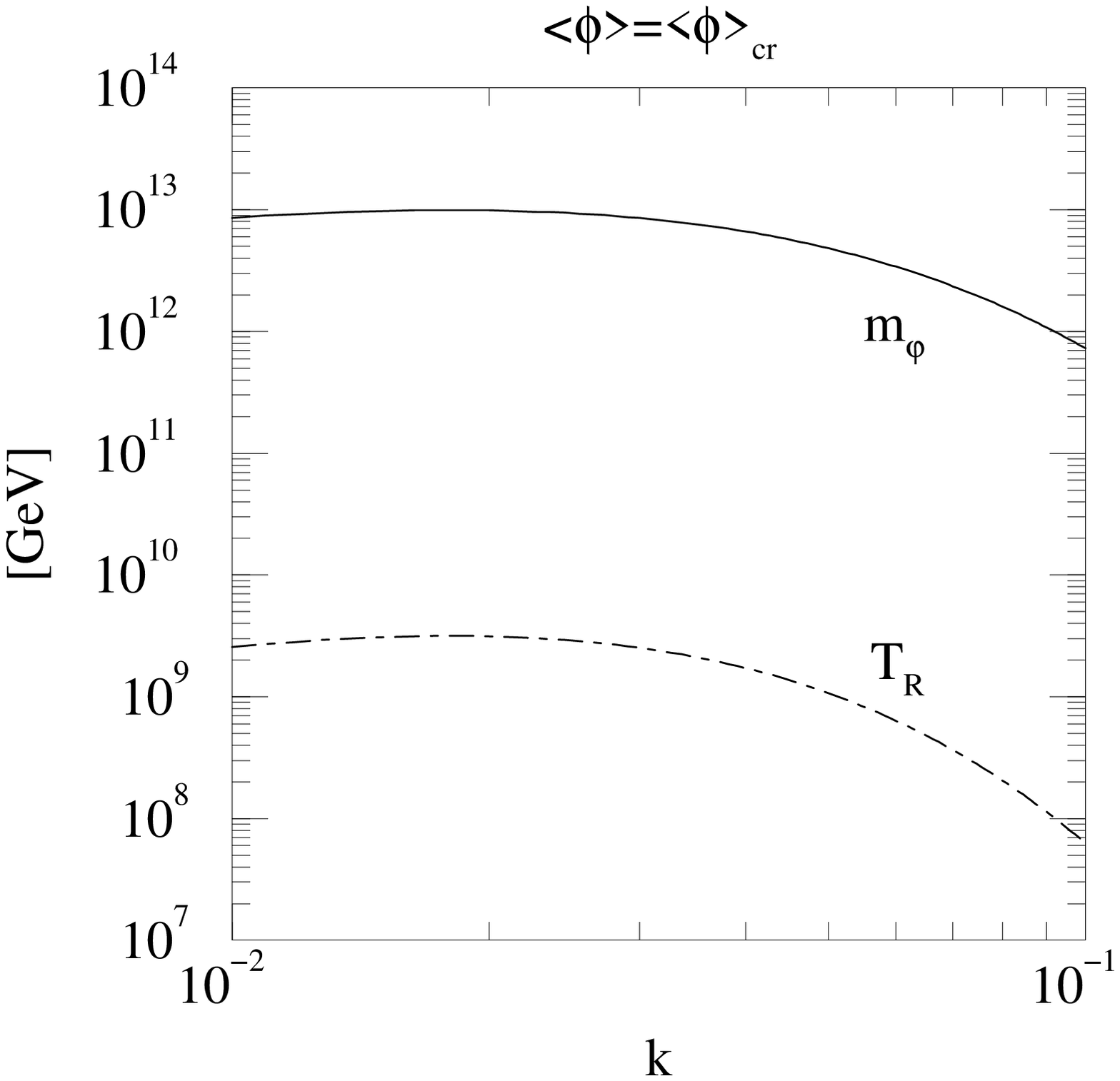,height=9cm}} }
 \caption{The inflaton mass $m_\varphi$ and the reheating temperature
 $T_R$ of the topological inflation model.  We take $\vev{\Phi} =
 \vev{\Phi}_C \simeq 1.7 M_G$.  The solid and dot-dashed lines represent
 $m_\varphi$ and $T_R$, respectively. }
%%%%%%%%%%%%%%%%%%%%%%%%%%%%%%%%%%%%%%%%%%%%%%%%%%%%%%%%%%%%
 \label{FIG-Top-mTR}%%%%%%%%%%%%%%%%%%%%%%%%%%%%%%%%%%%%%%%%
 \vspace{1cm}%%%%%%%%%%%%%%%%%%%%%%%%%%%%%%%%%%%%%%%%%%%%%%%
%%%%%%%%%%%%%%%%%%%%%%%%%%%%%%%%%%%%%%%%%%%%%%%%%%%%%%%%%%%%
 \centerline{ {\psfig{figure=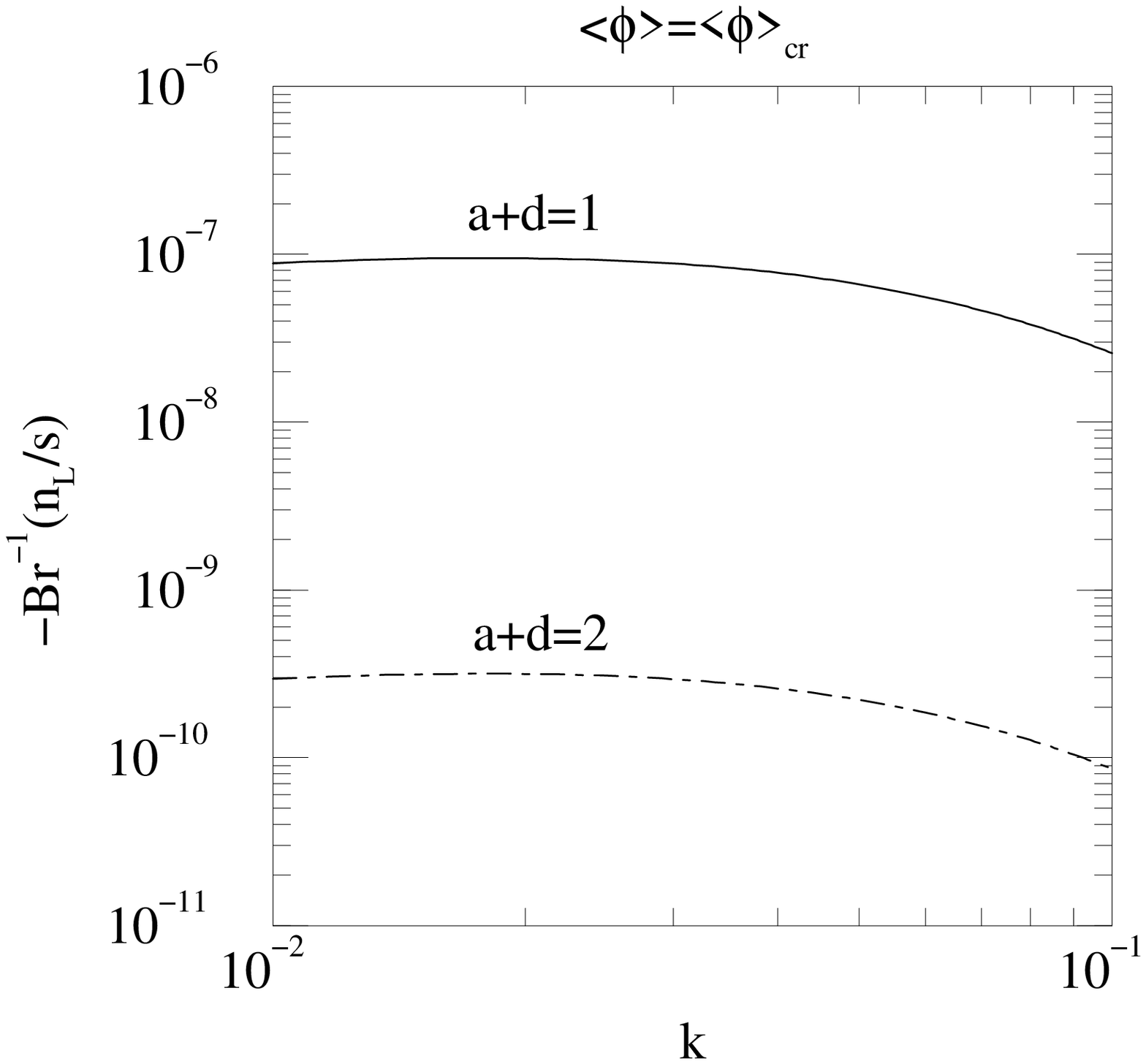,height=9cm}} }
 \caption{The lepton asymmetries, $n_L/s$, in the topological inflation
 model for the cases of $M_1\simeq 3 \times 10^{11}\GEV$ ($a+d=1$, the
 solid line) and $M_1\simeq 10^9\GEV$ ($a+d=2$, the dot-dashed line).}
%%%%%%%%%%%%%%%%%%%%%%%%%%%%%%%%%%%%%%%%%%%%%%%%%%%%%%%%%%%%
 \label{FIG-Top-nLs}%%%%%%%%%%%%%%%%%%%%%%%%%%%%%%%%%%%%%%%%
\end{figure}%%%%%%%%%%%%%%%%%%%%%%%%%%%%%%%%%%%%%%%%%%%%%%%%
%%%%%%%%%%%%%%%%%%%%%%%%%%%%%%%%%%%%%%%%%%%%%%%%%%%%%%%%%%%%

{}From the above relations, we can calculate the inflation scale $v$ and
the mass $m_\varphi$, and hence the reheating temperature $T_R$. We show
the results in \FIG{FIG-Top-mTR}. Here, we have taken $\vev{\Phi} =
\vev{\Phi}_C \simeq 1.7 M_G$. A crucial deference between this model and
the previous new inflation model is the reheating temperature. In this
topological model, as shown in \FIG{FIG-Top-mTR}, relatively high
reheating temperatures are obtained compared with those in the new
inflation. This is because the large value of the vacuum expectation
value $\vev{\Phi}$. (See \EQ{EQ-Gamma-K}.)

Finally, we show the obtained lepton asymmetry $n_L/s$ in
\FIG{FIG-Top-nLs}.  Here, we have taken $M_1 \simeq 3 \times
10^{11}\GEV$ ($a+d=1$) and $M_1 \simeq 10^9\GEV$ ($a+d=2$) as in the
new inflation model discussed in the previous subsection. We can see
that for the both cases enough lepton asymmetry $n_L/s > 10^{-10}$ is
obtained for $B_r \simeq 1$. On the other hand, the requirement of a
low enough reheating temperature ($T_R \lsim 10^8\GEV$) is satisfied
only in a small range of the coupling $0.092\lsim k \lsim 0.1$.  (The
lowest reheating temperature $T_R = 6.3 \times 10^7$ GeV is obtained
for $k=0.1$.) For $k\simeq 0.1$, the spectrum index $n_s$ deviates
from the scale invariant one as $n_s \simeq 0.8$, and it will be
testable in current and future satellite experiments on anisotropies
of the cosmic microwave background radiation~\cite{MAP-Planck}.

One final comment: in the case of $M_1 \simeq 3 \times 10^{11}\GEV$, we
may accommodate a late-time entropy production of order $10^2$--$10^3$
since the produced lepton asymmetry is as large as $n_L/s \simeq
10^{-7}$ as shown in \FIG{FIG-Top-nLs}. If this is the case, the energy
density of the gravitinos is diluted by the factor $10^2$--$10^3$, which
allows the region of $T_R \simeq 10^9\GEV$ in \FIG{FIG-Top-mTR}.

\subsection{Production of right-handed neutrinos at preheating}%%%%
\label{SEC-preheating}

Here, we briefly mention the production of right-handed neutrinos at
preheating epoch, which was investigated in Ref.~\cite{LG-Pre}.

So far, we have assumed that the reheating process after the inflation
occurs in such a way that energy density of the zero mode of the
inflaton decays perturbatively into light particles. However, the heavy
right-handed neutrinos might be predominantly produced
nonperturbatively, through the parametric
resonance~\cite{Para-Res-boson,Para-Res-fermion}.

In Ref.~\cite{LG-Pre}, production of heavy right-handed neutrinos via
preheating of fermion~\cite{Para-Res-fermion} was discussed, adopting a
chaotic inflation with the inflaton potential $V(\varphi) =
(1/2)m_\varphi \varphi^2$. ($m_\varphi$ is the inflaton mass.) The point
is that, if there is a Yukawa coupling between the inflaton $\varphi$
and heavy fermion $X$
%%%
\begin{eqnarray}
 {\cal L} = g\, \varphi \overline{X} X
  \,,
\end{eqnarray}
%%%
then the effective mass $m_{X {\rm eff}}$ of the fermion $X$ depends on
time $t$ through the oscillating inflaton $\varphi(t)$ during the
inflaton-oscillating period:
%%%
\begin{eqnarray}
 \label{EQ-mXeff}
  m_{X {\rm eff}}(t) = m_X + g\,\varphi(t)
  \,.
\end{eqnarray}
%%%
Therefore, the effective mass $m_{X {\rm eff}}(t)$ oscillates with time, 
and can even be zero if the amplitude of the oscillating inflaton is
large enough, which is very crucial to produce very massive fermions
heavier than the inflaton itself.

The production of the $X$ fermion crucially depends on the parameter
%%%
\begin{eqnarray}
 q \equiv g^2\frac{\varphi(0)^2}{4 m_\varphi^2}
  \,.
\end{eqnarray}
%%%
where $\varphi(0)$ denotes the initial amplitude of the inflaton
oscillation. An efficient production of very massive fermions requires a
large value of $q$~\cite{LG-Pre}, and hence large enough initial
amplitude $\varphi(0)$. In the case of chaotic inflation,
$\varphi(0)\simeq M_G$ and $m_\varphi\simeq 10^{13}\GEV$ can lead to
$q\simeq g\times {\cal O}(10^{10})\gg 1$.

The maximum mass of the produced fermion is given by~\cite{LG-Pre}
%%%
\begin{eqnarray}
 \left(m_X\right)_{\rm max} \simeq 0.25 g\,\varphi(0)
  \,.
\end{eqnarray}
%%%
If $m_X \lsim \left(m_X\right)_{\rm max}$, the effective mass $m_{X {\rm
eff}}(t)$ can vanish more than once [see \EQ{EQ-mXeff}]\footnote{Their
numerical calculation shows~\cite{LG-Pre} that the minimum of the
inflaton amplitude $\varphi^{(-)}_0$ during the first oscillation is
given by $\varphi^{(-)}_0\simeq -0.25 \varphi(0)$.} and the production
of $X$ through parametric resonance occurs efficiently, whereas if
otherwise the production is much less effective. The production
continues during vanishing effective mass can be achieved. Namely, the
production stops when the amplitude of the oscillating inflaton
$\varphi_0$ becomes below the following critical value:
%%%
\begin{eqnarray}
 \left(\varphi_0\right)_{\rm crit}
  =
  \frac{1}{g}m_X
  \,.
\end{eqnarray}
%%%
The energy density of the inflaton at this time is thus given by
$\rho_\varphi = (1/2)m_\varphi^2 \left(\varphi_0\right)_{\rm crit}^2$.
On the other hand, the energy density of the produced fermion $X$ at
this frozen-out time is given by~\cite{LG-Pre}
%%%
\begin{eqnarray}
 \rho_X \simeq \frac{1}{6\pi^2} m_X^2 m_\varphi^2
  \,.
\end{eqnarray}
%%%
After that, i.e., for $\varphi_0 < \left(\varphi_0\right)_{\rm crit}$,
the ratio of the energy density of the $X$ to the inflaton $\varphi$,
$\rho_X/\rho_\varphi$, is fixed. {}From the above equations, this ratio
is simply given by
%%%
\begin{eqnarray}
 \frac{\rho_X}{\rho_\varphi}
  \simeq
  \frac{g^2}{3\pi^2}
  \,.
\end{eqnarray}
%%%

Now let us identify the $X$ field to the right-handed neutrino
$N_1$. {}From above equation, the resultant lepton asymmetry after the
reheating process of the inflation completes is given by
%%%
\begin{eqnarray}
 \frac{n_L}{s}
  \simeq
  \frac{3}{4}
  \epsilon_1
  \left(\frac{T_R}{M_1}\right)
  \frac{g^2}{3\pi^2}
  \,.
\end{eqnarray}
%%%
Thus, the baryon asymmetry $n_B/s\simeq 0.35\times n_L/s$ is given by
%%%
\begin{eqnarray}
 \frac{n_B}{s}
  \simeq
  2\,g^2\times 10^{-10}
  \left(
   \frac{T_R}{10^8\GEV}
   \right)
   \times
   \left(
    \frac{\mnu{3}}{0.05\EV}
    \right)
    \delta_{\rm eff}
    \,,  
\end{eqnarray}
%%%
where we have used \EQ{EQ-ep1-final}.

Before closing this section, let us see what happens if we apply the
above mechanism to the first one of the SUSY hybrid inflation models
discussed in \SEC{SEC-Hybrid-1}, where the inflaton has a Yukawa
coupling to the right-handed neutrino. In this case, the parameter $q$
is estimated as
%%%
\begin{eqnarray}
 q &\simeq& \xi_1^2\frac{\vev{\Psi}^2}{4 m_\varphi^2}
  \nonumber
  \\
 &\simeq&
  \frac{1}{16}
  \left(
   \frac{2 M_1}{m_\varphi}
   \right)^2
   \,.
\end{eqnarray}
%%%
Therefore, in the region we have studied, where the perturbative decay
is kinematically allowed ($m_\varphi > 2 M_1$), preheating takes place
only through a narrow resonance, and hence only small fraction of the
inflaton oscillating energy can be converted to heavy right-handed
neutrinos via preheating. {}For the case of $M_1\simeq 10^{12}\GEV$
($a+d = 1$), in the small coupling region $k\lsim 10^{-4}$ and
$\lambda\lsim 10^{-4}$ where $m_\varphi\lsim (1/2)M_1$ [see
\FIG{FIG-Hybrid-mphi}], preheating mechanism might work effectively,
although it requires another detailed analysis.

\clearpage
%%%%%%%%%%%%%%%%%%%%%%%%%%%%%%%%%%%%%%%%%%%%%%%%%%%%%%%%%%%%%%%%%%%
\section{Leptogenesis from $\widetilde{N}$-dominant universe}%%%%%%
%%%%%%%%%%%%%%%%%%%%%%%%%%%%%%%%%%%%%%%%%%%%%%%%%%%%%%%%%%%%%%%%%%%
\label{SEC-Ntilde}

In the previous sections, we have discussed the leptogenesis
mechanisms by the decay of right-handed (s)neutrino, which is produced
as a {\it particle}.  However, there is a new possibility once we
introduce the SUSY, that is, the condensation of the scalar component
of the right-handed neutrino, $\widetilde{N}$~\cite{MY}.\footnote{See
also Refs.~\cite{MSYY-1,MSYY-2}, where the right-handed sneutrino is
considered as an inflaton of the chaotic inflation.}  The lepton
asymmetry is produced in the same way as the previous mechanisms, by
the asymmetric decay of the right-handed ``s''neutrino into leptons
and anti-leptons.  However, the thermal history of the universe in
this scenario is quite different from the others. In this
section,\footnote{This section is based on the work in a collaboration
with H.~Murayama and T.~Yanagida~\cite{HMY}.} we investigate this
scenario in detail. In particular, we mainly discuss the case in which
the coherent oscillation of the right-handed sneutrino $\widetilde{N}$
dominates the energy density of the universe before it decays.
Remarkably, the amount of produced baryon asymmetry is determined
almost only by the decay rate of the right-handed neutrino, whatever
happened before the coherent oscillation dominates the
universe. (Thus, the baryon asymmetry is independent of the details of
the inflation~!)

Furthermore, as a big bonus, the abundance of the gravitinos which are
created in the reheating phase after inflation are diluted by the
entropy production due to the decay of the coherent right-handed
sneutrino, so that the cosmological gravitino problems can be avoided
even when the reheating temperature $T_R$ of the inflation is higher
than $10^{11}\GEV$, in a wide range of the gravitino mass $m_{3/2}\sim
10\MEV$--$10\TEV$. In particular, this dilution of the gravitinos has
great advantages in the gauge-mediated SUSY breaking (GMSB)
models~\cite{GMSB}.

We will also comment on the cosmic density perturbation caused by the
quantum fluctuation of the right-handed sneutrino, which may be
detected in the future.

\subsection{Lepton asymmetry from $\widetilde{N}$-dominant universe}

The main character in the present scenario is the right-handed
sneutrino, $\widetilde{N}$.  During inflation, it can acquire a large
amplitude~\cite{MY,MSYY-1,MSYY-2} if the Hubble expansion rate of the
inflation $H_{\rm inf}$ is larger than the mass of the
$\widetilde{N}$.  (The mechanism which makes a scalar field develop a
large initial amplitude during inflation will be explained in
\SEC{SEC-Initial}, in the context of the Affleck-Dime condensation.)
Let us assume that there exists (at least) one right-handed neutrino
with a mass lighter than $H_{\rm inf}$, and that it develops a large
expectation value during the inflation. Hereafter, we focus on the
lightest right-handed sneutrino $\sneu$ for simplicity.  (Possible
contributions from the heavier right-handed sneutrinos
$\widetilde{N_2}$ and $\widetilde{N_3}$ will be discussed in
\SEC{SEC-Remarks-N2N3}.)  It is assumed here that the potential for
the right-handed neutrino is given simply by the mass term
\begin{equation}
 V = M_1^2 |\sneu|^2
  \,.
\end{equation}
We do not settle the initial amplitude of the $\sneu$ at this stage,
which will be discussed later. Because of the large amplitude of the
$\sneu$, the lepton $L$ and Higgs $H_u$ doublets receive large masses,
and the expectation values of their scalar components
$\vev{\widetilde{L}}$ and $\vev{H_u}$ vanish.

The evolution of the $\sneu$ is as follows.  After the end of the
inflation, the Hubble parameter $H$ decreases with cosmic time $t$ as
$H\propto t^{-1}$, and $\sneu$ begins to oscillate around the origin
when $H$ becomes smaller than the mass of the right-handed sneutrino
$M_1$. ($H \simeq M_1$.) Then, the coherent oscillation eventually
decays when the Hubble expansion rate $H$ becomes comparable to the
decay rate $\Gamma_{\sneu}$ of $\sneu$. ($H\simeq \Gamma_{N_1}$.)  As
explained in \SEC{SEC-LAfromN}, the asymmetric decay of $\sneu$ into
leptons and anti-leptons produces a net lepton asymmetry. The generated
lepton number density is given by
%%%%%
\begin{eqnarray}
\label{EQ-nL}
n_L = \epsilon_1 M_1 |\sneu_d|^2
\,,
\end{eqnarray}
%%%%%
where $|\sneu_d|$ is the amplitude of the oscillation when it decays,
and $\epsilon_1$ is given by \EQ{EQ-ep1-final}.

The fate of the generated lepton asymmetry depends on whether or not
the coherent oscillation of $\sneu$ dominates the energy density of
the universe before it decays~\cite{MY}.  Here, we focus on the
leptogenesis scenario from the universe dominated by $\sneu$. (We will
mention the case where $\sneu$ does not dominate the universe in
\SEC{SEC-Remarks-general}.)  As we shall show soon, once the $\sneu$
dominant universe is realized, the present baryon asymmetry is
determined only by the properties of the right-handed neutrino,
whatever happened before the $\sneu$ dominates the universe.  We first
derive the amount of the generated lepton asymmetry just assuming that
the $\sneu$ dominates the universe, and after that we will discuss the
necessary conditions of the present scenario.

Once $\sneu$ dominates the universe before it decays, the universe is
{\it reheated again} at $H \simeq \Gamma_{N_1}$ by the decay of $\sneu$.
The energy density of the resulting radiation, with a temperature
$T_{N_1}$, is given by the following relation;
%%%%%
\begin{eqnarray}
\label{EQ-TN}
\frac{\pi^2}{30} g_* T_{N_1}^4
&=&
M_1^2 |\sneu_d|^2
\nonumber \\
&=&
3 M_G^2 \Gamma_{N_1}^2
\,,
\end{eqnarray}
%%%%%
while the entropy density is given by [see \EQ{EQ-entropy}]
%%%%%
\begin{eqnarray}
\label{EQ-s}
s = \frac{2\pi^2}{45} g_* T_{N_1}^3
\,.
\end{eqnarray}
%%%%%
{}From the \EQS{EQ-ep1-final}, (\ref{EQ-nL}), (\ref{EQ-TN}) and
(\ref{EQ-s}), the ratio of the lepton number density to the entropy
density is given by the following very simple form;
%%%%%
\begin{eqnarray}
 \label{EQ-nL-s}
  \frac{n_L}{s} 
  &=& 
  \frac{3}{4}\epsilon_1 \frac{T_{N_1}}{M_1}
  \nonumber \\
 &\simeq& 1.5 \times 10^{-10}
  \left(
   \frac{T_{N_1}}{10^6\GEV}
   \right)
   \times
   \left(
    \frac{\mnu{3}}{0.05\EV}
    \right)
    \delta_{\rm eff}
    \,.
\end{eqnarray}
%%%%%
Here, it is assumed that the decay of the $\sneu$ occurs in an
out-of-equilibrium way, namely, $T_{N_1} < M_1$, so that the produced
lepton-number asymmetry not be washed out by lepton-number violating
interactions mediated by $N_1$. As explained in \SEC{SEC-sphaleron},
this lepton asymmetry is partially converted into the baryon asymmetry
through the sphaleron effect:
%%%%%
\begin{eqnarray}
 \label{EQ-nBs-Ntilde}
  \frac{n_B}{s} &\simeq& 0.35\times \frac{n_L}{s}
  \nonumber\\
 &\simeq&
  0.5\times 
  10^{-10}
  \left(
   \frac{T_{N_1}}{10^6\GEV}
   \right)
   \times
   \left(
    \frac{\mnu{3}}{0.05\EV}
    \right)
    \delta_{\rm eff}
    \,.
\end{eqnarray}
%%%%%
Therefore, as stressed previously, the baryon asymmetry in the present
universe is indeed determined only by the decay temperature of the
right-handed sneutrino $T_{N_1}$ (and the effective $CP$ violating phase
$\delta_{\rm eff}$). Thus it is independent of unknown parameters of the
inflation such as the reheating temperature $T_R$. Assuming the
effective $CP$ violating phase $\delta_{\rm eff} \,(\le 1)$ to be not
too small, the empirical baryon asymmetry $n_B/s \simeq (0.4$--$1)\times
10^{-10}$ is obtained by taking
%%%%%
\begin{eqnarray}
\label{EQ-TN-required}
T_{N_1} \simeq 10^6 - 10^7\GEV
\,.
\end{eqnarray}
%%%%%

\subsection{Necessary conditions}
\label{SEC-Ntilde-conditions}

So far, we have required the following two conditions;
\begin{enumerate}
 \item $\sneu$ dominates the universe before it decays.
 \item $\sneu$ decays in an out-of-equilibrium way.
\end{enumerate}
Let us first consider the easier one, the second condition (ii).  By
taking the $T_{N_1}$ in \EQ{EQ-TN-required}, the condition of the
out-of-equilibrium decay is simply given by\footnote{ The condition $M_1
> T_{N_1}$ is formally equivalent to
\begin{eqnarray}
 \left. H \right|_{T = M_1} > \Gamma_{N_1}
  \nonumber
  \,,
\end{eqnarray}
if we take the usual relation $3 M_G H^2 = (\pi^2/30)g_* T^4$. However,
this interpretation is not appropriate in the present case, since the
universe is not radiation-dominant for $H > \Gamma_{N_1}$. In any case,
the condition \EQ{EQ-mass} is necessary to avoid the thermal production
of $N_1$ after the decay of the coherent $\sneu$.  }
%%%%%
\begin{eqnarray}
 \label{EQ-mass}
  M_1 > T_{N_1} \simeq 10^6 - 10^7\GEV
  \,.
\end{eqnarray}
%%%%%
Notice that this condition gives rise to a constraint on the Yukawa
couplings $h_{1\alpha}$, since the temperature $T_{N_1}$ is determined
by the decay rate of the $\sneu$, and hence is related to the mass and
couplings of $\sneu$. Actually, from \EQ{EQ-TN} and $\Gamma_{N_1} =
(1/4\pi) \sum_{\alpha} |h_{1\alpha}|^2 M_1$, the following relation is
derived:
%%%%%
\begin{eqnarray}
 \label{EQ-coupling}
  \sqrt{\sum_{\alpha} |h_{1\alpha}|^2} 
  \simeq 5\times 10^{-6}
  \left( \frac{T_{N_1}}{10^6\GEV} \right)^{1/2} 
  \left( \frac{T_{N_1}}{M_1}\right)^{1/2} 
  \,.
\end{eqnarray}
%%%%%
Hence, it is found from \EQ{EQ-mass} that Yukawa couplings
$h_{1\alpha}$ of $N_1$ should be smaller than ${\cal
O}(10^{-5})$. (Possible explanations for the smallness of
$h_{1\alpha}$ will be discussed in \SEC{SEC-Remarks-mnus}.)

Now let us turn to discuss the first condition (i). In order to discuss
whether or not $\sneu$ dominates the universe, it is necessary to
consider the history of the universe before it decays. Here, we assume
that the potential of the $\sneu$ is ``flat'' up to the Planck scale,
namely, the potential is just given by the mass term $M_1^2 |\sneu|^2$
up to the Planck scale. (This may not be the case when the masses of the
right-handed neutrinos are induced by a breaking of an additional gauge
symmetry. We will discuss such a case in \SEC{SEC-Remarks-general}.)

Assuming the flatness of the $\sneu$'s potential up to the Planck scale
(i.e., only the mass term), the initial amplitude of the oscillation is
naturally given by $|\sneu_i| \simeq M_G$, since above the Planck scale
the scalar potential is expected to be exponentially lifted by the
supergravity effects [see \EQ{EQ-SUGRA}].\footnote{Even though it is
possible that $\sneu$ has a larger initial amplitude $|\sneu_i| > M_G$
(see, e.g., Ref~\cite{MSYY-1,MSYY-2}), it depends on the scalar
potential beyond the Planck scale, so that we do not discuss this
possibility here.} Then the energy density of $\sneu$ when it starts the
coherent oscillation is given by $\rho_{N_1} = M_1^2 |\sneu_i|^2\simeq
M_1^2 M_G^2$.

\begin{figure}[h!]%%%%%%%%%%%%%%%%%%%%%%%%%%%%%%%%%%%%%%%%%%%
%%%%%%%%%%%%%%%%%%%%%%%%%%%%%%%%%%%%%%%%%%%%%%%%%%%%%%%%%%%%
  \psfrag{I}{\hspace{-3em}\bf\huge (I)}
  \psfrag{II}{\hspace{-4em}\bf\huge (II)}
  \psfrag{rho}{\huge $\rho$}
  \psfrag{R}{\qquad\huge $R$}
  \psfrag{HGC}{\large $H = \Gamma_\chi$}
  \psfrag{HM1}{\large $H = M_1$}
  \psfrag{GC}{\large $\Gamma_\chi$}
  \psfrag{GN1}{\large $\Gamma_{N_1}$}
  \psfrag{M1}{\large $M_1$}
  \psfrag{T}{\large $(T =$}
  \psfrag{TR}{\large $T_R$}
  \psfrag{TTR}{\large $(T = T_R$}
  \psfrag{TN1}{\large $T_{N_1})$}
  \psfrag{TM1}{\large $T_{M_1}$}
  \psfrag{rhoc}{\large $\rho_\chi\propto R^{-3}$}
  \psfrag{rhoN1}{\large $\rho_{N_1}\propto R^{-3}$}
  \psfrag{rhorad}{\large $\rho_{\rm rad}\propto R^{-4}$}
  \psfrag{MM}{\hspace{-3.5em}\large $M_1^2 M_G^2$}
  \centerline{ \scalebox{0.7}{\includegraphics{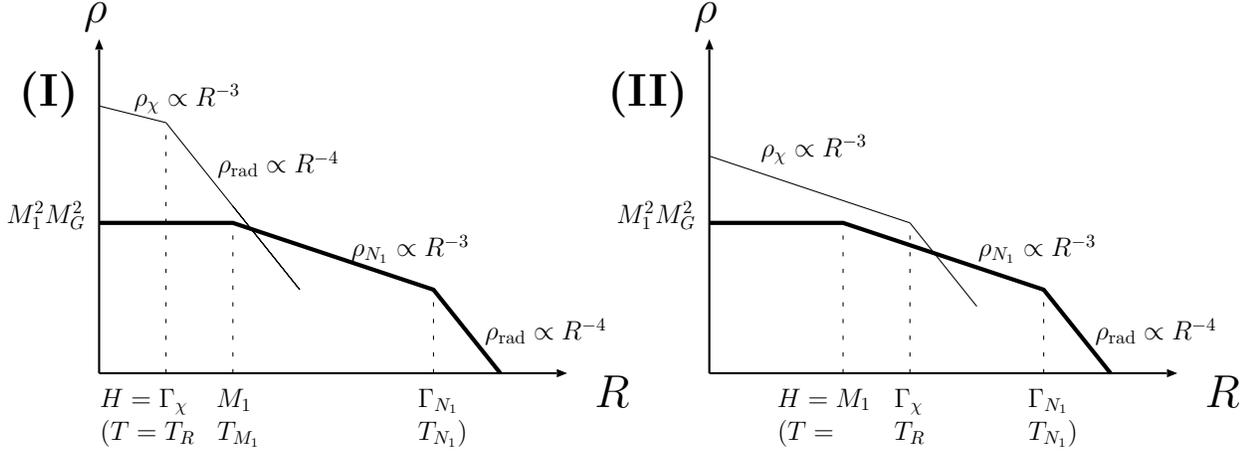}} }
 %%%%%%%%%%%%%%%%%%%%%%%%%
 \vspace{0.5cm}
 \caption{Schematic behaviors of the energy densities. $R$ denotes the
 scale factor of the expanding universe. It is found that the coherent
 oscillation $\sneu$ can dominate the energy density of the universe if
 the condition $\Gamma_{\chi} \gg \Gamma_{N_1}$ ($T_R \gg T_{N_1}$) is
 satisfied.}  \label{FIG-N-dom}
%%%%%%%%%%%%%%%%%%%%%%%%%%%%%%%%%%%%%%%%%%%%%%%%%%%%%%%%%%%%
\end{figure}

The rest of the total energy density of the universe at $H = M_1$ is
dominated by 
(I) the radiation or
(II) the oscillating inflaton $\chi$, 
depending on the $M_1$ and the decay rate of the inflaton
$\Gamma_{\chi}$.  (See \FIG{FIG-N-dom}.)

If $\Gamma_{\chi} > M_1$ [case (I)], the inflaton decay has already
completed before $H = M_1$, and the energy density of the radiation at
$H = M_1$ is given by $\rho_{\rm rad} = (\rho_{\rm total} -
\rho_{N_1}) \simeq 2 M_1^2 M_G^2$. In this case, the oscillating
$\sneu$ dominates the universe soon after it starts the oscillation
and hence before its decay.

On the other hand, if $M_1 > \Gamma_{\chi}$ [case (II)], the reheating
process of the inflation has not completed yet at $H = M_1$, and the
inflaton $\chi$ is still oscillating around its minimum, whose energy
density is given by $\rho_{\chi} \simeq 2 M_1^2 M_G^2$.  The ratio of
the energy density of $\sneu$ to that of the inflaton, $\rho_{N_1} /
\rho_{\chi} \simeq 1 / 2$, takes a constant value until either of
these oscillations decays. Because the energy density of the radiation
$\rho_{\rm rad}$ resulting from the inflaton decay is diluted faster
than $\rho_{N_1}$, the oscillating $\sneu$ can dominate the universe
if its decay rate $\Gamma_{N_1}$ is slow enough compared with that of
the inflaton $\Gamma_{\chi}$; $\Gamma_{\chi} \gg
\Gamma_{N_1}$.\footnote{More precisely, $\rho_{N_1} \ge \rho_{\rm
rad}$ occurs for $\Gamma_{N_1} \le (1/2\sqrt{3}) \Gamma_{\chi}$ in
case (II).}
%%%%%%%%%%%%%%%%%%%%%%%%%%%%%%%%%%%%%%%%%%%%%%%%%%%%%%%%%%%%
%%%%%%%%%%%%%%%%%%%%%%%%%%%%%%%%%%%%%%%%%%%%%%%%
%%%%%%%%%%%%%%%%%%%%%%%%%%%%%%%%%%%%%%%%%%%%%%%%%%%%%%%%%%%

Therefore, the condition for $\sneu$ to dominate the universe is just
given by $ \Gamma_{\chi} \gg \Gamma_{N_1}$. In terms of the reheating
temperature $T_R$, it is
%%%%%
\begin{eqnarray}
\label{EQ-dom-cond}
T_R \gg  T_{N_1}
\simeq 10^6 - 10^7\GEV
\,,
\end{eqnarray}
%%%%%
which is easily satisfied in various SUSY inflation models. (Recall that
the SUSY inflation models in \SEC{SEC-Inf} naturally lead to $T_R \gg
10^6$--$10^7\GEV$.) Thus, the present leptogenesis scenario from $\sneu$
dominated early universe is almost automatic as long as the right-handed
neutrino has suitable mass and couplings given in \EQS{EQ-mass} and
(\ref{EQ-coupling}).

\subsection{Dilution of the gravitinos}
\label{SEC-Nlep-grav}

A very attractive feature in the present ``$\sneu$-dominant universe
scenario'' is the dilution of the gravitino due to the entropy
production caused by the $\sneu$ decay. We discuss this point in this
subsection. 

Let us first calculate the dilution factor of the entropy production
caused by the $\sneu$'s decay. It depends on the two cases discussed
in the previous subsection, $M_1 < \Gamma_\chi$ [case~(I)] and $M_1 >
\Gamma_\chi$ [case~(II)]. We first consider the latter case, $M_1 >
\Gamma_\chi$ [see \FIG{FIG-N-dom}~(II)]. At the time of inflaton decay
($H = \Gamma_\chi$), the ratio of the energy density of $\sneu$ to the
entropy density is given by
%%%
\begin{eqnarray}
 \label{EQ-N1s-1}
  \left.
  \frac{\rho_{N_1}}{s}
  \right|_{{\rm before}\,\sneu\,{\rm decay}}
  =
  \frac{\rho_{N_1}}{\rho_{\rm rad}}
  \times
  \frac{\rho_{\rm rad}}{s}
  &\simeq&
  \frac{3}{8}T_R
  \,,
\end{eqnarray}
where we have used $\rho_{N_1}/\rho_{\rm rad}\simeq 1/2$ and $\rho_{\rm
rad}/s = (3/4)T_R$ at $H = \Gamma_\chi$. This ratio takes a constant
value until the $\sneu$ decays, since $\rho_{N_1}$ and $s$ are diluted
at the same rate $R^{-3}$ as the universe expands. Then at the time of
the $\sneu$'s decay, the energy density of the $\sneu$ is converted to
the radiation. The ratio of the energy density to the entropy density in
the {\it new} radiation is given by
%%%
\begin{eqnarray}
 \label{EQ-rad-s-1}
 \left.
  \frac{\rho_{\rm rad}}{s}
  \right|_{{\rm after}\,\sneu\,{\rm decay}}
  =
  \frac{3}{4}T_{N_1}
  \,.
\end{eqnarray}
%%%
By equating the $\rho_{N_1}$ (just before the $\sneu$'s decay) in
\EQ{EQ-N1s-1} with the $\rho_{\rm rad}$ (just after the $\sneu$'s decay)
in \EQ{EQ-rad-s-1}, we obtain the dilution factor of the entropy
production:
%%%
\begin{eqnarray}
 \label{EQ-Delta-1}
  \Delta 
  \equiv
  \frac{s_{\rm after}}{s_{\rm before}}
  =
  \frac{T_R}{2\,T_{N_1}}
  \,.
\end{eqnarray}
%%%

Next, we consider the case $\Gamma_\chi > M_1$ [case (I)]. In this
case, the ratio $\rho_{N_1}/s$ takes a constant value after the
$\sneu$ starts its oscillation, i.e., for $H \lsim M_1$. This ratio is
evaluated in the same way as \EQ{EQ-N1s-1}:
%%%
\begin{eqnarray}
 \left.
  \frac{\rho_{N_1}}{s}
  \right|_{{\rm before}\,\sneu\,{\rm decay}}
  =
  \frac{3}{8}T_{M_1}
\end{eqnarray}
%%%
where $T_{M_1}$ is the temperature at $H = M_1$ in the case (I), which
is given by
%%%%%
\begin{eqnarray}
 \label{EQ-TRC}
  T_{M_1} \equiv 7\times 10^{11}
  \left(\frac{M_1}{10^6\GEV}\right)^{1/2} \GEV
  \,. 
\end{eqnarray}
%%%%%
Then the dilution factor at $H = \Gamma_{N_1}$ is obtained in the same
way as in \EQ{EQ-Delta-1}, which results in $\Delta =
T_{M_1}/(2\,T_{N_1})$.

To summarize, the dilution factor $\Delta$ is given by
%%%%%
\begin{eqnarray}
 \label{EQ-Delta}
\Delta 
\simeq 
\left\{
\begin{array}{lc}
\Frac{T_R}{2\,T_{N_1}} 
&
({\rm for}\quad T_R < T_{M_1})
\\
\Frac{T_{M_1}}{2\,T_{N_1}} 
&
({\rm for}\quad T_R > T_{M_1})
\end{array}
\right.
\,,
\end{eqnarray}
%%%%%
where we have used the fact that $\Gamma_{\chi}<M_1$
($\Gamma_{\chi}>M_1$) corresponds to $T_R<T_{M_1}$ ($T_R>T_{M_1}$).

As discussed in \SEC{SEC-grav}, the abundance of the gravitino
produced at the reheating epoch after the inflation is proportional to
the reheating temperature of the inflation $T_R$. Therefore, there are
usually severe upper bounds on $T_R$ depending on the gravitino
mass. However, in the present scenario, the number density of the
gravitino is diluted thanks to the entropy production caused by the
$\sneu$'s decay, which was discussed just above. Namely, the
constraint coming from the gravitino problems should apply, not to the
reheating temperature $T_R$, but to an effective temperature given by
%%%%%
\begin{eqnarray}
\label{EQ-Tgrav}
 T_{R\,{\rm eff}} \equiv \frac{1}{\Delta} T_R 
 &\simeq& 
 \left\{
  \begin{array}{l}
   2\, T_{N_1}
    \\
   2\, T_{N_1}
    \left(
     \Frac{T_R}{T_{M_1}}
     \right)
  \end{array}
   \right.
\nonumber
\\
&\simeq& 
\left\{
\begin{array}{lc}
2 \times 10^6 - 2\times 10^7\GEV
&
({\rm for}\quad T_R < T_{M_1})
\\
2\times 10^6 - 2\times 10^7\GEV 
\times
\left(
\Frac{T_R}{T_{M_1}}
\right)
&
({\rm for}\quad T_R > T_{M_1})
\end{array}
\right.
\,,
\end{eqnarray}
%%%%%
which is much below the original reheating temperature $T_R$. Actually,
for $T_R < T_{M_1}$, the gravitino abundance which was originally
produced at the reheating after primordial inflation is diluted and
becomes comparable to the abundance which is produced at the
``reheating'' after $\sneu$'s decay.

In the case of unstable gravitino, \EQ{EQ-Tgrav} means that the
cosmological gravitino problem coming from the big-bang
nucleosynthesis bounds can be avoided in a wide range of the gravitino
mass $m_{3/2}\simeq 100\GEV$--$10\TEV$, even if the reheating
temperature $T_R$ of the inflation is much higher than
$10^{11}\GEV$. (Recall that the $T_{M_1}$ is given by \EQ{EQ-TRC}.) We
should stress that the fact that such high reheating temperature is
allowed makes it very easy to construct SUSY inflation models (see
\SEC{SEC-Inf}).

In the case of stable gravitino as well, we see from \EQ{EQ-Tgrav} that
the overclosure problem can be avoided in a wide range of the gravitino
mass. Actually, in our scenario, the relic abundance of the gravitino is
obtained by dividing the original abundance in \EQ{EQ-omega32} by the
dilution factor $\Delta$ in \EQ{EQ-Delta}:
\begin{eqnarray}
 \left.
  \Omega_{3/2}\,h^2
  \right|_{{\rm with}\,\,\sneu\,\,{\rm decay}}
  &\simeq&
  \frac{1}{\Delta}
  \left.
   \Omega_{3/2}\,h^2
   \right|_{{\rm without}\,\,\sneu\,\,{\rm decay}}
   \nonumber \\
 &\simeq&
  0.3\times
  \left(
   \frac{m_{\tilde{g}}}
   {1\TEV}
   \right)^2
   \left(
   \frac{m_{3/2}}
   {10\MEV}
   \right)^{-1}
    \left(
     \frac{T_{R\,{\rm eff}}}
     {10^6\GEV}
     \right)
     \,.
\end{eqnarray}
%%%
Therefore, the overclosure problem can be avoided almost independently
of the reheating temperature $T_R$, and a reheating temperature even
higher than $10^{11}\GEV$ is possible for $m_{3/2} \gsim
10\MEV$. Moreover, it is found from this equation that the present
energy density of the gravitino is independent of the reheating
temperature $T_R$, in a very wide range of $T_{N_1} \ll T_R <
T_{M_1}$. Thus, we can {\it predict} the gravitino mass by requiring
that the gravitino is the dominant component of the dark
matter:\footnote{Notice that with such a small gravitino mass as in
\EQ{EQ-m32-predicted}, the problem of the decay of the
next-to-lightest SUSY particle into LSP gravitino after the big-bang
nucleosynthesis is also avoided~\cite{MMY,LSPgrav}. See
\SEC{SEC-grav}.}
\begin{eqnarray}
 \label{EQ-m32-predicted}
 m_{3/2}
  &\simeq&
  20\MEV
  \times
  \left(
   \frac{m_{\tilde{g}}}
   {1\TEV}
   \right)^2
   \left(
    \frac{\Omega_{\rm matter}\,h^2}
    {0.15}
    \right)^{-1}
    \left(
     \frac{T_{R\,{\rm eff}}}
     {10^6\GEV}
     \right)
     \nonumber \\
 &\simeq&
  (40 - 400)\MEV
  \times
  \left(
   \frac{m_{\tilde{g}}}
   {1\TEV}
   \right)^2
   \left(
    \frac{\Omega_{\rm matter}\,h^2}
    {0.15}
    \right)^{-1}
    \,,
\end{eqnarray}
for $T_{N_1} \ll T_R < T_{M_1}$. Here, we have taken the present matter
density $\Omega_{\rm matter}\simeq 0.3$ and $h \simeq 0.7$~\cite{PDB}.

Such a light (stable) gravitino as in \EQ{EQ-m32-predicted} naturally
exists in gauge-mediated SUSY breaking (GMSB) models~\cite{GMSB}. The
GMSB mechanism has been regarded as a very attractive candidate for the
SUSY breaking, since it suppresses quite naturally the flavor changing
processes, which are inherent problems in the SUSY standard model. It is
remarkable that the gravitino with mass $m_{3/2}\simeq 40$--$400\MEV$
becomes naturally the dominant component of the dark matter in the
present scenario, independently of the reheating temperature $T_R$ for
$T_{N_1}\ll T_R < T_{M_1}$. This prediction comes from the fact that the
present energy density of the gravitino is determined by the effective
temperature $T_{R\,{\rm eff}} = 2\,T_{N_1}$ (for $T_R < T_{M_1}$), while
the decay temperature of the right-handed neutrino $T_{N_1}$ is fixed by
the baryon asymmetry in the present universe [see \EQ{EQ-TN-required}].

\subsection{Remarks}
\label{SEC-Remarks}

We give several comments on the present scenario.

\subsubsection{neutrino masses $\mnu{i}$}
\label{SEC-Remarks-mnus}

The first one is about the neutrino mass $m_{\nu}$. Let us first rewrite
the neutrino masses in terms of the three contributions from $N_1$,
$N_2$ and $N_3$:
%%%
\begin{eqnarray}
 \mnu{j} = \mnu{j}^{\rm from\,\,N_1}
  + \mnu{j}^{\rm from\,\,N_2}
  + \mnu{j}^{\rm from\,\,N_3}
  \,,
\end{eqnarray}
where
\begin{eqnarray}
 \mnu{j}^{\rm from\,\,N_i}
  \equiv
  -\widehat{h}_{ij}\widehat{h}_{ij}
  \frac{\vev{H_u}^2}{M_i}
  \,,
\end{eqnarray}
and the $\widehat{h}_{ij}$ denote the Yukawa couplings of the
neutrinos in the basis where mass matrices of heavy and light
neutrinos are diagonal, which are defined in \EQ{EQ-rtd-Ykw}. Then we
see that the contribution from $N_1$ satisfies the following
relation:\footnote{The last expression in \EQ{EQ-mnu-from1} is nothing
but the $\widetilde{m_1}$ parameter in \EQ{EQ-m1parameter}.}
\begin{eqnarray}
 \label{EQ-mnu-from1}
  \left|
   \mnu{j}^{\rm from\,\,N_1}  
   \right|
   &=&
   \left|\widehat{h}_{1j}\widehat{h}_{1j}\right|
   \frac{\vev{H_u}^2}{M_1}
   \nonumber \\
 &\le&
  \sum_k|\widehat{h}_{1k}|^2
   \frac{\vev{H_u}^2}{M_1}
   \nonumber \\
 &=&
   \sum_{\alpha}|h_{1\alpha}|^2
   \frac{\vev{H_u}^2}{M_1}
   \simeq
   8\times 10^{-4} \EV
   \left(\frac{T_{N_1}}{M_1}\right)^2
   \,,
\end{eqnarray}
Here, we have used the relation in \EQ{EQ-coupling}. Therefore, it is
understood that the mass scale of the neutrinos suggested from the
atmospheric~\cite{SK-Atm} and solar~\cite{Solar}
($+$KamLAND~\cite{KamLAND}) neutrino oscillations, $m_\nu\sim {\cal
O}(10^{-3})$--${\cal O}(10^{-1})\EV$, should be induced from the
heavier right-handed neutrinos, $N_2$ and $N_3$.

The relative hierarchy between the mass and couplings of $N_1$ and
those of the $N_2$ and $N_3$ might be naturally explained by a broken
flavor symmetry. For example, a broken discrete $Z_6$
symmetry~\cite{FHY-ADL-1}\footnote{The model here looks similar to the
one in \SEC{SEC-model-FN}, but different from that.} with a breaking
parameter $\varepsilon \simeq 1/17$ and charges $Q(L_e,L_{\mu},L_\tau)
= (a+1$, $a$, $a)$ and $Q(N_1,N_2,N_3) = (3+d$, $c$, $b)$ gives rise
to the following superpotential;
\begin{eqnarray}
 W = 
  \frac{1}{2} \widetilde{\xi_{11}}\,
  M_0 \,\varepsilon^{2d} N_1 N_1
  + 
  \frac{1}{2} \sum_{(i,j)\ne (1,1)} 
  \widetilde{\xi_{ij}}\,
  M_0 \,\varepsilon^{Q_i + Q_j} N_i N_j
  + 
  \widetilde{h_{i\alpha}} \,
  \varepsilon^{Q_i + Q_{\alpha}} 
  N_i L_{\alpha} H_u
  \,,
  \nonumber
\end{eqnarray}
where $\widetilde{\xi_{ij}}$ and $\widetilde{h_{i\alpha}}$ are ${\cal
O}(1)$ couplings. The above charge assignments for lepton doublets
naturally lead to the realistic neutrino mass matrix including the
maximal mixing for the atmospheric neutrino oscillation~\cite{YanaRamo}. 
(See \SEC{SEC-FN}.)  The overall mass scale of the right-handed neutrino
$M_0$ is determined by $\mnu{3}\sim \varepsilon^{2a}\vev{H_u}^2/M_0$. By
taking $a+d=2$, this model gives $M_1\sim \varepsilon^{2d}M_0\sim
7\times 10^9\GEV$, $\sqrt{\sum_{\alpha} |h_{1\alpha}|^2}\sim
\varepsilon^5\sim 7\times 10^{-7}$, and hence $T_{N_1}\sim 1\times
10^7\GEV$.\footnote{In terms of $\widetilde{m_1}$ parameter, this
corresponds to $\widetilde{m_1}\sim \varepsilon^6 \mnu{3}\sim {\cal
O}(10^{-9})\EV$.}

Recently, it has been shown~\cite{Heb-Mar-Y} that the small mass and
the couplings of the $N_1$, required in the present scenario,
naturally arise if the right-handed neutrino is a bulk field in higher
dimensional theory.

\subsubsection{contributions from $\widetilde{N_2}$ and $\widetilde{N_3}$}
\label{SEC-Remarks-N2N3}

In the previous subsections, we have only considered the lightest
right-handed sneutrino, $\sneu$.  The heavier right-handed sneutrino
$\widetilde{N_2}_{(3)}$ can also develop a large amplitude during the
inflation (if $M_{2(3)} < H_{\rm inf}$) and it may produce lepton
asymmetry in a similar way to the $\sneu$. However, it is found that the
decay temperatures of the $\widetilde{N_2}$ and $\widetilde{N_3}$ cannot
satisfy the out-of-equilibrium condition $T_{2(3)} < M_1$, since $N_2$
and $N_3$ must explain the mass scales of the neutrino oscillations. (In
other words, if $T_{N_{2(3)}}$ would also satisfy the condition
$T_{N_{2(3)}} < M_1 < M_{2(3)}$, the mass scale for the neutrino
oscillations could no longer be explained by the contributions from
$N_2$ and $N_3$ [see \EQ{EQ-mnu-from1}].) Therefore, even if the
$\widetilde{N_2}_{(3)}$'s decay produces additional lepton asymmetry, it
is washed out and hence it cannot contribute to the resultant total
lepton asymmetry.

\subsubsection{thermal effects}

We should also comment on the effects of the thermal plasma, which might
cause an early oscillation of the right-handed sneutrino $\sneu$ before
$H = M_1$. (Notice that there is a dilute plasma with a temperature
$T\simeq (T_R^2 M_G H)^{1/4}$ even before the reheating process of the
inflation completes~\cite{KT}.) There are basically two possible thermal
effects~\cite{ACE,SomeIssues}. (Those thermal effects will be also
explained in detail in \SEC{SEC-LHu-evol}.)

First, when the temperature $T$ is higher than the effective mass for
$L$ and $H_u$, i.e., $T > m_{\rm eff} = \sqrt{\sum_{\alpha}
|h_{1\alpha}|^2} |\sneu|$, the $\sneu$ receives an additional thermal
mass $\delta M_{\rm th}^2 = (1/4)\sum_{\alpha} |h_{1\alpha}|^2 T^2$ from
the Yukawa coupling to $L$ and $H_u$~\cite{ACE}. Thus, the $\sneu$ field
would start an early oscillation if the additional thermal mass becomes
larger than the Hubble expansion rate before $H = M_1$. However, even if
$\sneu$ receives the thermal mass, the ratio of the thermal mass to the
Hubble expansion rate is given by
%%%%%
\begin{eqnarray}
\frac{\delta M_{\rm th}^2}{H^2}
\simeq 
\left\{
\begin{array}{lc}
0.07\times
\left(
\Frac{10 T_{N_1}}{M_1}
\right)^2
\left(
\Frac{M_1}{H}
\right)^{3/2}
\left(
\Frac{T_R}{T_{M_1}}
\right) 
&
{\rm for}\quad T_R < T_{M_1}
\,,
\\
0.03\times
\left(
\Frac{10 T_{N_1}}{M_1}
\right)^2
\left(
\Frac{M_1}{H}
\right)
&
{\rm for}\quad T_R > T_{M_1}
\,,
\end{array}
\right.
\end{eqnarray}
%%%%%
where we have used the relation given in \EQ{EQ-coupling}. Therefore, we
can safely neglect the above thermal effect as long as $M_1$ is a bit
larger than $T_{N_1}$.

There is also another thermal effect~\cite{SomeIssues}. If the
temperature is lower than the effective mass for $L$ and $H_u$, i.e., $T
< m_{\rm eff} = \sqrt{\sum_{\alpha} |h_{1\alpha}|^2} |\sneu|$, the
evolution of the running gauge and/or Yukawa coupling constants $f(\mu)$
which couple to them are modified below the scale $\mu = m_{\rm
eff}$. Thus, these running coupling constants depend on $|\sneu|$, and
there appears an additional thermal potential for $\sneu$;
\begin{eqnarray}
 \delta V(\sneu) = a T^4
  \log
  \left(
   \frac{|\sneu|^2}{T^2}
   \right)
   \,,
\end{eqnarray}
where $a$ is a constant of order ${\cal O}(f^4)$. However, again, it
turns out that the effective thermal mass for $\sneu$ is less than the
Hubble expansion rate;
\begin{eqnarray}
\label{EQ-T4effect}
\frac{\delta M_{\rm th}^{'\, 2}}{H^2}
&=&
\frac{a T^4}{H^2 |\sneu|^2}
\nonumber\\
&\simeq& 
\left\{
\begin{array}{lc}
0.2\times
a
\left(
\Frac{M_G}{|\sneu|}
\right)^2
\left(
\Frac{M_1}{H}
\right)
\left(
\Frac{T_R}{T_{M_1}}
\right)^2 
&
{\rm for}\quad T_R < T_{M_1}
\,,
\\
0.05\times
a
\left(
\Frac{M_G}{|\sneu|}
\right)^2
&
{\rm for}\quad T_R > T_{M_1}
\,,
\end{array}
\right.
\end{eqnarray}
and hence this thermal effect is also irrelevant to the present scenario.

\subsubsection{more general cases}
\label{SEC-Remarks-general}

So far, we have assumed that the initial amplitude of the $\sneu$'s
oscillation is $|\sneu_i| \simeq M_G$. This can be realized when the
right-handed neutrino has only the mass term up to the Planck scale.
We have also assumed that the $\sneu$ dominates the energy density of
the universe before its decay. Here, we consider more general cases
for completeness.

Let us assume that the masses of the right-handed neutrinos are
dynamically induced by a spontaneously broken gauge symmetry, for
example, by the ${\rm U}(1)_{B-L}$ symmetry, which is an attractive
candidate for such a gauge symmetry as discussed in \SEC{SEC-1-1}.  We
denote the chiral superfields whose vacuum expectation values break
the ${\rm U}(1)_{B-L}$ by $\Psi$ and $\overline{\Psi}$.  (We need two
fields with opposite charges in order to cancel ${\rm U}(1)_{B-L}$
gauge anomalies.) Due to the $D$-term and the $F$-term coming from the
superpotential which gives the right-handed neutrino masses, the
scalar potential of the right-handed sneutrino $\sneu$ is lifted above
the ${\rm U}(1)_{B-L}$ breaking scale
$\vev{\Psi}$~\cite{MY}. Therefore, the initial amplitude of the
$\sneu$'s oscillation at $H\simeq M_1$ is given by $|\sneu_i|\sim
\vev{\Psi}$.

The breaking scale of the ${\rm U}(1)_{B-L}$ gauge symmetry is model
dependent. If it is broken at the Planck scale, $\vev{\Psi}\simeq
M_G$, the discussion so far does not change at all.\footnote{In this
case, we need small couplings in order to explain the intermediate
right-handed neutrino mass scale. For example, a superpotential $W =
(1/2)g_i \Psi N_i N_i$ with $\vev{\Psi}\simeq M_G$ and $g_3 \sim
10^{-4}$ gives the mass $M_3 \sim 10^{14}\GEV$ to the heaviest
right-handed neutrino.  Such a small Yukawa coupling could well be a
consequence of broken flavor symmetries.}  On the other hand, if
$\vev{\Psi}$ is below the Planck scale, the initial amplitude of the
$\sneu$'s oscillation is reduced and its energy density when it starts
the oscillation is given by
%%%
\begin{eqnarray}
 \rho_{N_1} = M_1^2 |\sneu_i|^2 \ll \rho_{\rm total} = 3 M_1^2 M_G^2 
  \qquad
  {\rm for}
  \quad
  H = M_1
  \,.
\end{eqnarray}
%%%
Compared with the previous case (see \FIG{FIG-N-dom}), we can see that
the initial energy density is much suppressed.  

%%%%%%%%%%%%%%%%%%%%%%%%%%%%%%%%%%%%%%%%%%%%%%%%%%%%%%%%%%%%
\begin{figure}[h!]%%%%%%%%%%%%%%%%%%%%%%%%%%%%%%%%%%%%%%%%%%%
%%%%%%%%%%%%%%%%%%%%%%%%%%%%%%%%%%%%%%%%%%%%%%%%%%%%%%%%%%%%
  \psfrag{Ia}{\hspace{-4em}\bf\Huge (Ia)}
  \psfrag{Ib}{\hspace{-4em}\bf\Huge (Ib)}
  \psfrag{IIa}{\hspace{-4.7em}\bf\Huge (IIa)}
  \psfrag{IIb}{\hspace{-4.7em}\bf\Huge (IIb)}
  \psfrag{III}{\hspace{-4em}\bf\Huge (III)}
  \psfrag{rho}{\huge $\rho$}
  \psfrag{R}{\qquad\huge $R$}
  \psfrag{HGC}{\large$H = \Gamma_\chi$}
  \psfrag{HM1}{\large$H = M_1$}
  \psfrag{GC}{\large$\Gamma_\chi$}
  \psfrag{GN1}{\large$\Gamma_{N_1}$}
  \psfrag{M1}{\large$M_1$}
  \psfrag{T}{\large$T = T_R$}
  \psfrag{TN}{\large$T = T_{N_1}$}
  \psfrag{TM}{\large$T = T_{M_1}$}
  \psfrag{c}{\Large $\chi$}
  \psfrag{r}{\Large rad.}
  \psfrag{N}{\Large $\sneu$}
  \psfrag{NN}{\huge $\sneu$}
  \psfrag{I}{\hspace{-4.5em}\large$M_1^2|\sneu_i|^2$}
  \centerline{\scalebox{0.7}{\includegraphics{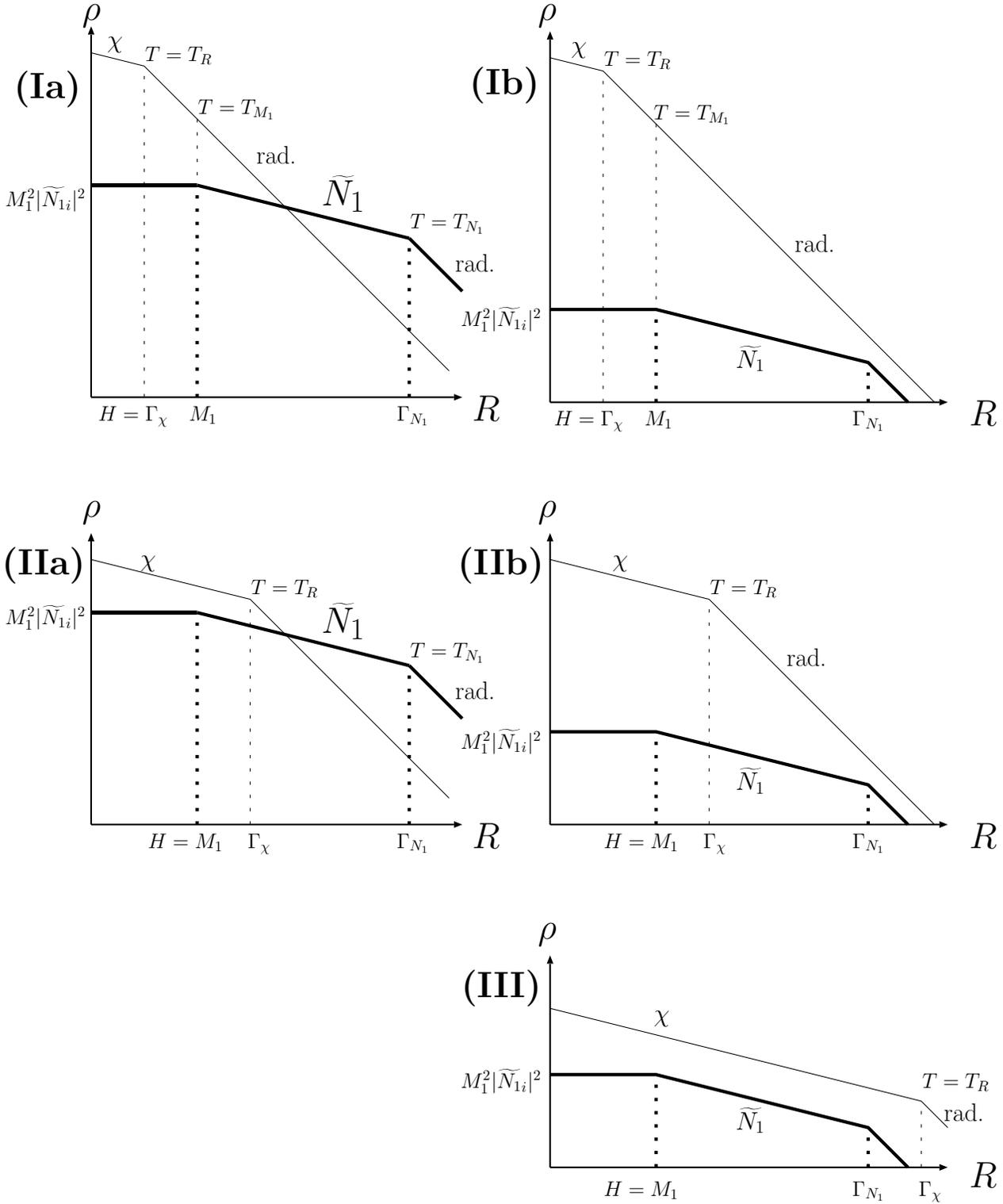}} }
 %%%%%%%%%%%%%%%%%%%%%%%%%
 \caption{Schematic behaviors of the energy densities of the inflaton
 $\chi$, right-handed sneutrino $\sneu$, and radiation for general
 cases.}
 \label{FIG-N-dom-generic}
%%%%%%%%%%%%%%%%%%%%%%%%%%%%%%%%%%%%%%%%%%%%%%%%%%%%%%%%%%%%
\end{figure}%%%%%%%%%%%%%%%%%%%%%%%%%%%%%%%%%%%%%%%%%%%%%%%%
%%%%%%%%%%%%%%%%%%%%%%%%%%%%%%%%%%%%%%%%%%%%%%%%%%%%%%%%%%%

In \FIG{FIG-N-dom-generic}, We show the behavior of the energy
densities of the inflaton $\chi$, the right-handed sneutrino $\sneu$
and the radiation, for generic cases.\footnote{For the reduced initial
amplitude, the thermal effect from the $aT^4\log(|\sneu|^2)$ potential
becomes larger than the case of $|\sneu_i|\simeq M_G$. However, it is
still irrelevant as long as the coupling $a\sim {\cal O}(f^4)$ is
small enough.  [See \EQ{EQ-T4effect}.]} Figure (Ia) and (Ib) represent
the case of $\Gamma_\chi > M_1$ ($T_R > T_{M_1}$), (IIa) and (IIb)
represent the case of $M_1 > \Gamma_\chi > \Gamma_{N_1}$ ($T_{M_1} >
T_R > T_{N_1}$) and (III) represents that of $\Gamma_{N_1} >
\Gamma_\chi$ ($T_{N_1} > T_R$). Among them, (Ia) and (IIa) show the
case of $\sneu$ dominant universe, corresponding to (I) and (II) in
\FIG{FIG-N-dom}, respectively. Figures (Ib), (IIb), and (III)
represent the cases in which the coherent oscillation of the
right-handed sneutrino $\widetilde{N}$ does not dominate the energy
density of the universe.

\begin{figure}[t!]%%%%%%%%%%%%%%%%%%%%%%%%%%%%%%%%%%%%%%%%%%%
%%%%%%%%%%%%%%%%%%%%%%%%%%%%%%%%%%%%%%%%%%%%%%%%%%%%%%%%%%%%
  \psfrag{Ia}{\bf\Large (Ia)}
  \psfrag{Ib}{\bf\Large (Ib)}
  \psfrag{IIa}{\bf\Large (IIa)}
  \psfrag{IIb}{\bf\Large (IIb)}
  \psfrag{III}{\bf\Large (III)}
  \psfrag{GC}{\Large$\Gamma_\chi$}
  \psfrag{TR}{\large$(T_R)$}
  \psfrag{GN1}{\large$\Gamma_{N_1}$}
  \psfrag{M1}{\large$M_1$}
  \psfrag{TN1}{($T_{N_1}$)}
  \psfrag{TM1}{($T_{M_1}$)}
  \psfrag{Ni}{\Large $|\sneu_i|$}
  \psfrag{3M}{\hspace{-1em}\large $\sqrt{3}M_G$}
  \psfrag{notNdom}{\large\bf \underline{not\,\,$\sneu$ dominant}}
  \psfrag{Ndom}{\Large\bf \underline{$\sneu$ \,\,dominant}}
  \psfrag{Ni3MGN1GC}{$|\sneu_i | >
     \sqrt{3}M_G
     \left(\Frac{\Gamma_{N_1}}{\Gamma_\chi}\right)^{1/4}$}
  \psfrag{Ni3MGN1M1}{$|\sneu_i | > 
    \sqrt{3}M_G
    \left(\Frac{\Gamma_{N_1}}{M_1}\right)^{1/4}$}
  \centerline{\scalebox{1.2}{\includegraphics{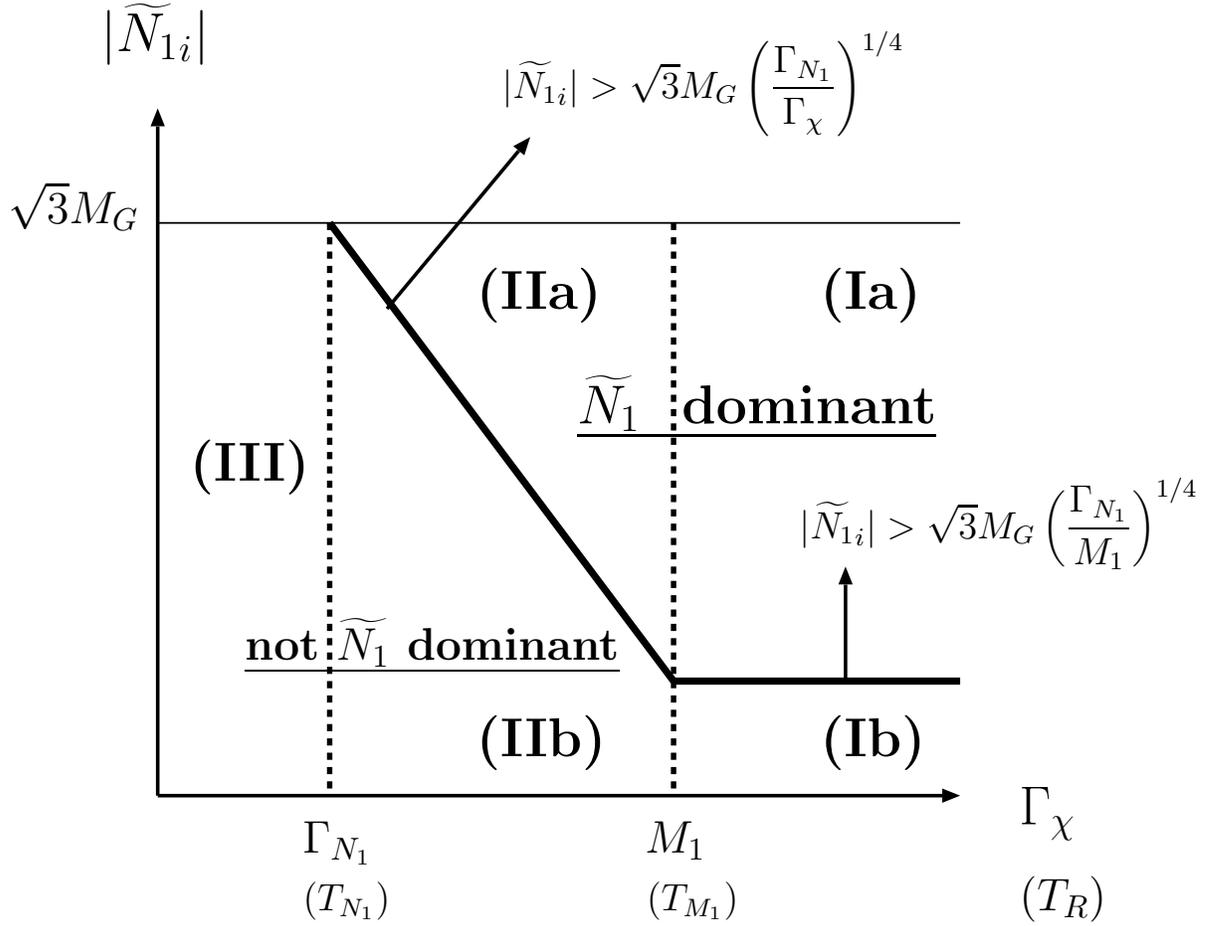}} }
 %%%%%%%%%%%%%%%%%%%%%%%%%
  \caption{The condition of the $\sneu$ dominant universe is
  represented in ($\Gamma_\chi$\,,\,$|\sneu_i|$) plane. The behaviors
  of the energy densities for the regions of (Ia), (Ib), (IIa), (IIb)
  and (III) are shown in the corresponding figures in
  \FIG{FIG-N-dom-generic}. }
  \label{FIG-N-dom-cond}
%%%%%%%%%%%%%%%%%%%%%%%%%%%%%%%%%%%%%%%%%%%%%%%%%%%%%%%%%%%%
\end{figure}

Let us first discuss the condition of the $\sneu$ dominant universe,
which was given by [\EQ{EQ-dom-cond}] for $|\sneu|\simeq M_G$.
Because of the smaller initial amplitude, the condition now becomes
\begin{eqnarray}
  |\sneu_i | &>&
  \sqrt{3}M_G
  \left(\Frac{\Gamma_{N_1}}{M_1}\right)^{1/4}
  \qquad{\rm for}\quad
  \Gamma_\chi > M_1
  \qquad{\rm \bf (Ia)}
  \,,
  \\
  |\sneu_i | &>& 
  \sqrt{3}M_G
  \left(\Frac{\Gamma_{N_1}}{\Gamma_\chi}\right)^{1/4}
  \qquad{\rm for}\quad
  \Gamma_{N_1} < \Gamma_\chi < M_1
  \qquad{\rm \bf (IIa)}
  \,.
\end{eqnarray}
These conditions are summarized in \FIG{FIG-N-dom-cond}.  Notice that
the $\sneu$ cannot dominate the universe for $\Gamma_\chi <
\Gamma_{N_1}$ [the case (III)].

%%%%%%%%%%%%%%%%%%%%%%%%%%%%%%%%%%%%%%%%%%%%%%%%%%%%%%%%%%%%
%%%%%%%%%%%%%%%%%%%%%%%%%%%%%%%%%%%%%%%%%%%%%%%%
%%%%%%%%%%%%%%%%%%%%%%%%%%%%%%%%%%%%%%%%%%%%%%%%%%%%%%%%%%%

Next we estimate the baryon asymmetry in the above general cases. It
can be calculated in the same manner as before, and is given by, for
each case,
\begin{eqnarray}
  \begin{array}{rll}
    {\rm\bf (Ia)(IIa)}
    &
    \Frac{n_B}{s} = 0.35\times \Frac{3}{4}\epsilon_1
    \Frac{T_{N_1}}{M_1}
    &
    =
    0.5\times 10^{-10}
    \left(\Frac{T_{N_1}}{10^6\GEV}\right)
    \eta\,,
    \\
      {\rm\bf (Ib)}
      &
      \Frac{n_B}{s} = 0.35\times \Frac{3}{4}\epsilon_1
      \left(\Frac{|\sneu_i|}{\sqrt{3}M_G}\right)^2
      \Frac{T_{M_1}}{M_1}
      &
      =
      0.5\times 10^{-10}
      \left(\Frac{T_{M_1}}{10^6\GEV}\right)
      \left(\Frac{|\sneu_i|}{\sqrt{3}M_G}\right)^2
      \eta\,,
      \\
	{\rm\bf (IIb)(III)}
	&
	\Frac{n_B}{s} = 0.35\times \Frac{3}{4}\epsilon_1
	\left(\Frac{|\sneu_i|}{\sqrt{3}M_G}\right)^2
	\Frac{T_R}{M_1}
	&
	=
	0.5\times 10^{-10}
	\left(\Frac{T_R}{10^6\GEV}\right)
	\left(\Frac{|\sneu_i|}{\sqrt{3}M_G}\right)^2
	\eta\,,
  \end{array}
  \nonumber\\
\end{eqnarray}
where $\eta \equiv (m_{\nu 3}/0.05\EV)\delta_{\rm eff}$. Notice that
  the baryon asymmetry which was given in \EQ{EQ-nBs-Ntilde} does not
  change and independent of the initial amplitude $|\sneu_i|$ as long
  as the $\sneu$ dominant universe is realized [(Ia) and (IIa)].

Finally, we comment on the gravitino problem. The effective
temperature of the cosmological gravitino problem, which was given by
[\EQ{EQ-Tgrav}], is now given by
\begin{eqnarray}
  \Omega_{3/2}\,h^2
  &\propto&
  T_{R\,{\rm eff}}
  =
  \left\{
      \begin{array}{ll}
        T_{N_1}\left(\Frac{\sqrt{3}M_G}{|\sneu_i|}\right)^2
        \left(\Frac{T_R}{T_{M_1}}\right)
        & {\rm\bf (Ia)}
	\,,
	\\
        T_{N_1}\left(\Frac{\sqrt{3}M_G}{|\sneu_i|}\right)^2
        & {\rm\bf (IIa)}
	\,,
        \\
        T_R & {\rm\bf (Ib)(IIb)(III)}
	\,.
      \end{array}
      \right.
\end{eqnarray}

\subsubsection{cosmic microwave background (CMB) anisotropy from $\sneu$}

Let us mention the cosmic density perturbation coming from the
fluctuation of the initial amplitude of the right-handed sneutrino,
$|\sneu_i|$, according to the recent study in Ref.~\cite{MM-Ntilde}.

In the present scenario, the CMB radiation we observe today has two
origins; the inflaton $\chi$ and the right-handed sneutrino
$\sneu$. Furthermore, if the Hubble expansion rate of the inflation is
larger than the mass of the right-handed sneutrino $M_1$, as we have
assumed, the initial amplitude of the $\sneu$ has a non-vanishing
fluctuation during the inflation. Therefore, the CMB anisotropy also
has two origins, the primordial fluctuations of $\chi$ and that of
$\sneu$.  If the $\sneu$ completely dominates the energy density of
the universe, the perturbation becomes adiabatic~\cite{curvaton}. On
the other hand, if the radiation coming from $\sneu$ is negligible
compared to that from $\chi$, purely baryonic isocurvature
perturbation is produced from $\sneu$. Particularly interesting is
between these two cases (i.e., the region near the border line between
the two regions (Ia)-(IIa) and (Ib)-(IIb)-(III) in
\FIG{FIG-N-dom-cond}), in which correlated mixture of the adiabatic
and isocurvature fluctuation is generated, that can be observed in the
current and future satellite experiments on the CMB
anisotropies~\cite{MAP-Planck}.

\clearpage
%%%%%%%%%%%%%%%%%%%%%%%%%%%%%%%%%%%%%%%%%%%%%%%%%%%%%%%%%%%%%%%%%%%
% Change the label of section to *.A.  %%%%%%%%%%%%%%%%%%%%%%%%%%%%
%%%%%%%%%%%%%%%%%%%%%%%%%%%%%%%%%%%%%%%%%%%%%%%%%%%%%%%%%%%%%%%%%%%
\renewcommand{\thesection}{\thechapter.A}%%%%%%%%%%%%%%%%%%%%%%%%%%
%%%%%%%%%%%%%%%%%%%%%%%%%%%%%%%%%%%%%%%%%%%%%%%%%%%%%%%%%%%%%%%%%%%
\section{$N$-dominant universe from inflaton decay}%%%%%%%%%%%%%%%%
%%%%%%%%%%%%%%%%%%%%%%%%%%%%%%%%%%%%%%%%%%%%%%%%%%%%%%%%%%%%%%%%%%%
\label{SEC-Inf-Dom}

In this appendix, we briefly discuss the case where the $N_1$ which are
produced in inflaton decay dominates the universe before it
decays. (This case was also discussed in Ref.~\cite{LG-Pre}.)

The baryon asymmetry produced by the $N_1$ decay in this case is the
same as that of the $\widetilde{N}$-dominant universe
(\SEC{SEC-Ntilde}), and is given by \EQ{EQ-nBs-Ntilde}:
%%%
\begin{eqnarray}
 \label{EQ-nBs-app}
  \frac{n_B}{s}
  &\simeq&
  0.5\times 
  10^{-10}
  \left(
   \frac{T_{N_1}}{10^6\GEV}
   \right)
   \left(
    \frac{\mnu{3}}{0.05\EV}
    \right)
    \delta_{\rm eff}
    \,,
\end{eqnarray}
which requires the decay temperature of the $N_1$ to be $T_{N_1}\simeq
10^6$--$10^7\GEV$ for $\delta_{\rm eff}\simeq 0.1$--$1$.\footnote{This
point was not discussed in Ref.~\cite{LG-Pre}.} 

On the other hand, the condition for the $N_1$ to dominate the energy
density of the universe in this case is derived as follows. The point is
that the ratio $n_{N_1}/s$, which is given by \EQ{EQ-nN1s-inf}, takes a
constant value until the time of $N_1$'s decay. Thus, the ratio of the
energy density of the $N_1$ to that of the radiation at $N_1$'s decay is
given by
%%%
\begin{eqnarray}
 \frac{\rho_{N_1}}{\rho_{\rm rad}}
  &=&
  M_1 \frac{n_{N_1}}{s}
  \times
  \frac{s}{\rho_{\rm rad}}
  \nonumber\\
 &=&
  M_1
  \frac{3}{2}
  B_r
  \frac{T_R}{m_\chi}
  \times
  \frac{4}{3 T_{N_1}}
  \nonumber\\
 &=&
  2 B_r
  \frac{M_1}{m_\chi}
  \frac{T_R}{T_{N_1}}
  \,.
\end{eqnarray}
%%%
Thus, the dominant condition $\rho_{N_1}/\rho_{\rm rad} > 1$ gives
a lower bound on the $M_1$ as follows:
%%%
\begin{eqnarray}
 M_1 > \frac{m_\chi}{2 B_r T_R}
  T_{N_1}
  \,,
\end{eqnarray}
which is stronger than the out-of-equilibrium condition ($M_1 >
T_{N_1}$) for $m_\chi > 2 B_r T_R$. Then using the relation in
\EQ{EQ-coupling}, we obtain a constraint on the Yukawa couplings:
%%%
\begin{eqnarray}
 \sum_{\alpha}|h_{1\alpha}|^2
  \lsim
  5\times 10^{-15}
  B_r
  \left(
   \frac{T_R}{10^9\GEV}
   \right)
   \left(
    \frac{10^{13}\GEV}{m_\chi}
    \right)
    \times
    \left(
     \frac{T_{N_1}}{10^6\GEV}
     \right)
     \,.
\end{eqnarray}
If this is satisfied, $N_1$ produced from inflaton decay can dominate
the energy density of the universe, and the produced baryon asymmetry is 
given by \EQ{EQ-nBs-app}.

%%%

%%%%%%%%%%%%%%%%%%%%%%%%%%%%%%%%%%%%%%%%%%%%%%%%%%%%%%%%%%%%%%%%%%%
% Change again the label of the section to *.number. %%%%%%%%%%%%%%
%%%%%%%%%%%%%%%%%%%%%%%%%%%%%%%%%%%%%%%%%%%%%%%%%%%%%%%%%%%%%%%%%%%
\renewcommand{\thesection}{\thechapter.\arabic{section}}%%%%%%%%%%%
%%%%%%%%%%%%%%%%%%%%%%%%%%%%%%%%%%%%%%%%%%%%%%%%%%%%%%%%%%%%%%%%%%%

%%%%%%%%%%%%%%%%%%%%%%%%%%%%%%
%\include{Chap-LHu}%%%%%%%%%%%%
%%%%%%%%%%%%%%%%%%%%%%%%%%%%%%%%%%%%%%%%%%%%%%%%%%%%%%%%%%%%%%%%%%%
%%%%%%%%%%%%%%%%%%%%%%%%%%%%%%%%%%%%%%%%%%%%%%%%%%%%%%%%%%%%%%%%%%%
%%%%%%%%%%%%%%%%%%%%%%%%%%%%%%%%%%%%%%%%%%%%%%%%%%%%%%%%%%%%%%%%%%%
\chapter{Leptogenesis via $L H_u$ flat direction}%%%%%%%%%%%%%%%%%%
%%%%%%%%%%%%%%%%%%%%%%%%%%%%%%%%%%%%%%%%%%%%%%%%%%%%%%%%%%%%%%%%%%%
%%%%%%%%%%%%%%%%%%%%%%%%%%%%%%%%%%%%%%%%%%%%%%%%%%%%%%%%%%%%%%%%%%%
%%%%%%%%%%%%%%%%%%%%%%%%%%%%%%%%%%%%%%%%%%%%%%%%%%%%%%%%%%%%%%%%%%%
\label{Chap-LHu}

In the previous chapter, we have investigated various leptogenesis
mechanisms in which lepton asymmetry is produced by the
$CP$-violating, asymmetric decay of the heavy right-handed (s)neutrino
into leptons and anti-leptons. In this chapter,\footnote{This chapter
is based on the works with T.~Asaka, M.~Fujii and
T.~Yanagida~\cite{AFHY,FHY-ADL-1,FHY-ADL-BL,FHY-ADL-2}.}  we will
discuss a completely different leptogenesis scenario.  The main
character in this chapter is not a heavy right-handed (s)neutrino, but
a flat direction scalar field including the charged lepton doublet
$L$. Actually, as we will see, most of the necessary ingredients for
this scenario exist in the minimal supersymmetric standard model
(MSSM), and basically it does not rely on any new physics beyond
supersymmetry, the existence of small Majorana neutrino masses and
inflation.

\clearpage
%%%%%%%%%%%%%%%%%%%%%%%%%%%%%%%%%%%%%%%%%%%%%%%%%%%%%%%%%%%%%%%%%%%
\section{Overview}%%%%%%%%%%%%%%%%%%%%%%%%%%%%%%%%%%%%%%%%%%%%%%%%%
%%%%%%%%%%%%%%%%%%%%%%%%%%%%%%%%%%%%%%%%%%%%%%%%%%%%%%%%%%%%%%%%%%%
\label{SEC-LHu-overview}

Affleck and Dine~\cite{AD} proposed a completely new mechanism to
produce cosmological baryon asymmetry, which is inherent in the SUSY
framework. In the SUSY theories, there are many flat directions in the
scalar potential which contain scalar quark (squark) and scalar lepton
(slepton) fields, carrying baryon and/or lepton charges. During the
inflation, such a scalar field (which will be denoted by $\phi$) can
acquire a large vacuum expectation value along the flat direction,
since the potential is flat. After the inflationary stage ends, the
flat direction field $\phi$ starts a coherent oscillation. At this
stage, if there exists baryon (or lepton) number violating operator in
the scalar potential, $\phi$ field can have a nonzero phase rotational
motion: $\Frac{d}{dt}\left[\arg(\phi)\right]\ne 0$, which means, a
nonzero $\phi$-number
\begin{eqnarray}
 \label{EQ-3-1}
  n_\phi\equiv i
  \left(
   \dot{\phi}^* 
   \phi 
   - 
   \phi^* 
   \dot{\phi}
   \right)
   \ne 0
   \,,
\end{eqnarray}
is produced. (The dot denotes a derivative with time $t$.) Since the
$\phi$ field carries baryon (or lepton) number, this represents nothing
but baryon (or lepton) number production.

Among various flat directions, the $L H_u$ flat direction proposed by
Murayama and Yanagida~\cite{MY} is especially attractive and had been
investigated in detail later as
well~\cite{DRT,MM,AFHY,FHY-ADL-1,FHY-ADL-BL,FHY-ADL-2}. The most
interesting point is that it is closely related to the neutrino
physics. Actually, as we will see, the most important parameter which
determines the present baryon asymmetry is the mass of the lightest
neutrino, $\mnu{1}$.  Furthermore, the $L H_u$ flat direction is very
special among various flat directions, since it is free from a serious
problem of the Q-ball formation.\footnote{We do not discuss the Q-ball
problem in this thesis. For details and references, see, for example,
Refs.~\cite{FHY-ADBL,FHY-ADL-BL,FH}.}

Meantime, it has been pointed out~\cite{DRT,ACE,SomeIssues} that the
dynamics of a flat direction field is affected by thermal
effects. Although production of baryon and/or lepton number usually
takes place before the reheating process of the inflation completes,
there is a dilute plasma even in this epoch. This dilute plasma can
modify the scalar potential~\cite{ACE,SomeIssues}, affect the dynamics
of $\phi$, and change (actually, suppress) the resultant baryon
asymmetry.

In \SEC{SEC-LHu}, therefore, we will perform a detailed analysis on the
dynamics of the $L H_u$ flat direction field $\phi$, including all the
relevant thermal effects, in order to calculate the amount of produced
lepton (and hence baryon) asymmetry. We see that evolution of the $\phi$
field is indeed affected by the thermal effects, and that the produced
lepton asymmetry is suppressed by those thermal effects. On the other
hand, we will find a very interesting result: the amount of produced
baryon asymmetry is almost independent of the reheating temperature
$T_R$, and is determined mainly by the mass of the lightest neutrino
$\mnu{1}$. We show that the empirical baryon asymmetry $n_B/s\simeq
(0.4$--$1)\times 10^{-10}$ predicts an ultralight neutrino
$\mnu{1}\simeq (0.3$--$1)\times 10^{-9}\EV$ for a wide range of
reheating temperature $T_R\gsim 10^5\GEV$. In \SEC{SEC-somemodels} we
show some explicit models to explain such a small mass of the lightest
neutrino.

Then we study in \SEC{SEC-withB-L} what happens in this leptogenesis
scenario via $L H_u$ flat direction when we introduce a gauged ${\rm
U}(1)_{B-L}$ symmetry. We will see that if the initial amplitude of the
flat direction field exceeds the $B-L$ breaking scale, we can avoid the
suppression of the lepton-asymmetry production due to the thermal
effects.

Finally, in \SEC{SEC-double-beta}, we show that if we assume a very
small mass of the lightest neutrino as suggested by the present
leptogenesis mechanism, the rate of the neutrino-less double-beta decay
can be predicted in a high accuracy.

A comment: we do not discuss other lepton-number carrying flat
directions (see, e.g., early works in
Refs.~\cite{camp-dav-oliv-1,camp-dav-oliv-2}, other general directions
in Ref.~\cite{DRT}, and a recent one in Ref.~\cite{AD-triplet}) nor
the case of multi-scalar manifolds~\cite{AD-multi}. Notice that the $L
H_u$ direction is special not only because it is a leptonic direction,
but also because it is directly related to observable physics: the
neutrino mass.

\clearpage
%%%%%%%%%%%%%%%%%%%%%%%%%%%%%%%%%%%%%%%%%%%%%%%%%%%%%%%%%%%%%%%%%%%
\section{Lepton asymmetry from $L H_u$ flat direction}%%%%%%%%%%%%%
%%%%%%%%%%%%%%%%%%%%%%%%%%%%%%%%%%%%%%%%%%%%%%%%%%%%%%%%%%%%%%%%%%%
\label{SEC-LHu}

In this section, we will discuss the dynamics of the $L H_u$ flat
direction field in detail and calculate the amount of generated lepton
asymmetry.

\subsection{The $L H_u$ flat direction}

Let us start by writing down the effective dimension-five operator in
the superpotential
%%%
\begin{eqnarray}
 \label{EQ-LHu-start}
 W = \frac{\mnu{i}}{2\vev{H_u}^2}
  \left(L_i H_u \right)
  \left(L_i H_u \right)
  =
  \frac{1}{2 M_{{\rm eff},i}}
  \left(L_i H_u \right)
  \left(L_i H_u \right)
  \,,
\end{eqnarray}
which induces neutrino masses $\mnu{i}$ after the neutral component of
the Higgs field $H_u$ obtains its vacuum expectation value $\vev{H_u} =
174\GEV\times \sin\beta$. Here, we have taken a basis where the neutrino
mass matrix is diagonal. As we will see, the effective mass scales
$M_{{\rm eff},i}$ play crucial roles in the present scenario. Notice
that they are directly related to the neutrino masses $\mnu{i}$ by the
following equation:
%%%
\begin{eqnarray}
 \label{EQ-Meff-mnu}
 M_{{\rm eff},i}\equiv \frac{\vev{H_u}^2}{\mnu{i}}
 =
 3\times 10^{13}\GEV
 \left(
  \frac{1\EV}{\mnu{i}}
  \right)
  \,.
\end{eqnarray}
Here and hereafter, we take $\sin\beta\simeq 1$, for
simplicity.\footnote{This is the case for $\tan\beta > 1$. Even for
$\tan\beta\simeq 1$, the final lepton (and hence baryon) asymmetry
changes (decreases) by a factor of $2$.}

We adopt the following scalar field $\phi_i$ as the Affleck-Dine flat
direction~\cite{AD} for leptogenesis, which was originally proposed in
Ref.~\cite{MY}:
%%%
\begin{eqnarray}
 \widetilde{L_i} = 
  \frac{1}{\sqrt{2}}
  \left(
   \begin{array}{c}
    \phi_i\\
    0
   \end{array}
   \right)
   \,,
   \quad
   H_u = 
   \frac{1}{\sqrt{2}}
   \left(
    \begin{array}{c}
     0\\
     \phi_i
    \end{array}
    \right)
    \,,
\end{eqnarray}
%%%
where the factor $\sqrt{2}$ is introduced to ensure the canonical
kinetic term for $\phi_i$. Here, we have chosen the neutral components
of the $\widetilde{L_i}$ and $H_u$ for illustration, which is always
possible by using the ${\rm SU}(2)_L$ gauge transformation. We also used
an ${\rm U}(1)_Y$ gauge rotation to make their phases to be the same.

There are three flat directions ($i = 1,2,3$).  (See
Appendix~\ref{SEC-App-Dflat} for more details.) Hereafter, we will
suppress the family index $i$ for simplicity. As we will see later, it
turns out that the relevant flat direction $\phi_i$ for the most
effective leptogenesis corresponds to the first family $i = 1$. Thus,
our flat direction field is parameterized by one complex scalar field,
$\phi\equiv \phi_1$. We will use the same symbol $\phi$ to denote the
supermultiplet corresponding to the scalar field $\phi$, hereafter. Its
superpotential is then given by
%%%
\begin{eqnarray}
 \label{EQ-super-for-phi}
 W = \frac{1}{8 M_{\rm eff}}\phi^4
  \,,
\end{eqnarray}
%%%
while the scalar potential is given by
%%%
\begin{eqnarray}
 \label{EQ-V0}
 V_0 = 
  m_{\phi}^2
  |\phi|^2
  +
  \frac{m_{3/2}}{8 M_{\rm eff}}
  \left(
   a_m\phi^4
   +
   {\rm H.c.}
   \right)
   +
   \frac{1}{4 M_{\rm eff}}
   |\phi|^6
   \,.
\end{eqnarray}
%%%
Here, $m_{\phi}$ and $m_{3/2}a_m$ are SUSY-breaking mass
parameters,\footnote{Here, we have included the possible contribution
from the $\mu$-term, $W = \mu H_u H_d$, which gives a mass term
$(1/2)\mu^2 |\phi|^2$, in the mass $m_\phi^2$.} and the last term
directly comes from the superpotential \EQ{EQ-super-for-phi}. Here and
hereafter, we assume that the SUSY breaking is mediated by the
gravity. Therefore, it is expected that $m_{\phi}\sim m_{3/2}\sim 1\TEV$
and $|a_m|\sim 1$.

\subsection{Initial amplitude}
\label{SEC-Initial}

The potential for $\phi$ is $D$-flat, and it is also $F$-flat in the
limit of $m_{\nu}\to 0$. Thus, it is expected that the $\phi$ field can
develop a large amplitude during the inflaton. It depends on the
coupling between $\phi$ and the inflaton $\chi$ whether or not the
$\phi$ field can have a large initial amplitude. Suppose that there are
general nonminimal couplings of the $\phi$ field to the inflaton $\chi$
in the K\"ahler potential:
%%%
\begin{eqnarray}
 K &=& \phi^{\dagger}\phi + \chi^{\dagger}\chi
  \nonumber\\
 &+&
  \left(
   \frac{c_{\phi}}{M_G}
   \chi
   \phi^{\dagger}\phi
   + {\rm H.c.}
   \right)
   +
   \frac{b_{\phi}}{M_G^2}
   \chi^{\dagger}\chi
   \phi^{\dagger}\phi
   +
   \cdots
   \,,
\end{eqnarray}
%%%
where $c_{\phi}$ and $b_{\phi}$ are complex and real coupling constants,
respectively, and the ellipsis denotes higher order terms which are
irrelevant to the following discussion. During inflation, the energy
density of the universe is dominated by the inflaton $\chi$. Thus, the
total energy density of the universe is given by $3 H_{\rm inf}^2
M_G^2\simeq |F_{\chi}|^2$, where $H_{\rm inf}$ is the Hubble parameter
during the inflation and $F_{\chi}$ is the $F$-component of the inflaton
supermultiplet $\chi$. Because of this nonzero energy density of $\chi$,
there appears additional SUSY-breaking effects during the
inflation~\cite{DRT}. Using the scalar potential in the supergravity in
\EQ{EQ-SUGRA}, we obtain
%%%
\begin{eqnarray}
 \delta V =
  3 
  \left( 1 - b_{\phi} + |c_{\phi}|^2 \right) 
  H_{\rm inf}^2 |\phi|^2
  -\frac{\sqrt{3}}{2}
  \frac{H_{\rm inf}}{M_{\rm eff}}
  \left(
   c_{\phi}\phi^4 + {\rm H.c.} 
   \right)
   +
   \cdots
   \,,
\end{eqnarray}
%%%
where we redefined the phase of $c_{\phi}$ to include the phase of
$F_\chi$. A crucial observation here is that the $\phi$ field gets a
large mass term of the order of the Hubble parameter. (We will call it a
Hubble mass term.) In general, the coefficient of this Hubble mass term,
$3( 1 - b_{\phi} + |c_{\phi}|^2 )$, is expected to be order unity. If it
is positive, therefore, the $\phi$ field is driven exponentially towards
the origin during the inflation, and the present mechanism cannot
work. (Notice that this would happen even if the K\"ahler potential is
minimal, i.e., $c_{\phi} = b_{\phi} = 0$.) Hence, we will assume that it
is negative, and hereafter take
%%%
\begin{eqnarray}
 \label{EQ-Vinf}
 \delta V =
  - c_H
  H_{\rm inf}^2 |\phi|^2
  +
  \frac{H_{\rm inf}}{8 M_{\rm eff}}
  \left(
   a_H\,\phi^4 + {\rm H.c.}
   \right)
   \,,
\end{eqnarray}
%%%
with $c_H\simeq |a_H|\simeq 1$.

The negative Hubble mass term causes an instability of $\phi$ at the
origin. From \EQS{EQ-V0} and (\ref{EQ-Vinf}), the minima of the
potential are given by
%%%
\begin{eqnarray} 
 |\phi| &\simeq& \sqrt{M_{\rm eff} H_{\rm inf}}
  \,,
  \label{EQ-minimum-inf1}
  \\
 \arg(\phi) &\simeq& \frac{- \arg(a_H) + ( 2 n + 1 )\pi}{4}
  \,,
  \qquad n = 0\cdots 3
  \,,
  \label{EQ-minimum-inf2}
\end{eqnarray}
%%%
where we have used $m_{\phi}$, $m_{3/2}|a_m| \ll H_{\rm inf}$. Since
curvatures of the potential around the minimum along both the radius and
phase directions are positive and of the order of the Hubble parameter
$H_{\rm inf}$, the flat direction $\phi$ runs to one of the four minima
from any given initial value and is settled down it.

Here, we comment on another possibility that the flat direction field
does not receive a mass term of the order of the Hubble parameter, which
occurs with a K\"ahler potential of the no-scale type
supergravity~\cite{GMO}. The evolution of the $\phi$ field in this case
was discussed in Ref.~\cite{MM}, and it turns out that the initial
amplitude is likely to be given by \EQ{EQ-minimum-inf1} in this case as
well.

\subsection{Evolution of the flat direction}
\label{SEC-LHu-evol}

Before discussing the detailed dynamics of the $\phi$ field, we first
roughly describe the evolution of $\phi$ and note the relevant epoch for
the leptogenesis. During the inflation, the $\phi$ field takes a large
value $\sim \sqrt{M_{\rm eff}H_{\rm inf}}$ as discussed in the previous
subsection.  After the end of inflation, the value of $\phi$ gradually
decreases as the Hubble parameter $H$ decreases and then, at some time,
the $\phi$ starts its coherent oscillation.  As we will see, the net
lepton number is fixed when the flat direction field $\phi$ starts the
coherent oscillation.

Hereafter, we will assume that the net lepton number is fixed before the
reheating process of the inflation completes, namely, during the epoch
when the energy density of the universe is dominated by the oscillating
inflaton $\chi$. (This assumption will be justified after we calculate
the amount of the generated lepton asymmetry in \SEC{SEC-resultinLHu}.)

\subsubsection{potential for $\phi$}

Let us first show the total effective potential for the flat direction
field $\phi$ relevant to the leptogenesis:
%%%
\begin{eqnarray}
 \label{EQ-Vtotal}
  V_{\rm total} &=&
  \left( m_{\phi}^2 - H^2 + \sum_{f_k|\phi| < T} c_k f_k^2 T^2 \right)
  |\phi|^2
  \nonumber\\ &&
  + \,\,\frac{m_{3/2}}{8 M_{\rm eff}} \left(a_m \phi^4 + H.c.\right)
  + \frac{H}{8 M_{\rm eff}} \left(a_H \phi^4 + H.c.\right)
  \nonumber\\ &&
  + a_g \alpha_S(T)^2 \,T^4\ln\left(\frac{|\phi|^2}{T^2}\right)
  \nonumber\\ &&
  + \frac{1}{4 M_{\rm eff}^2}|\phi|^6
  \,.
\end{eqnarray}
%%%
We explain each term in turn.

First of all, the $F$-term potential and the SUSY-breaking terms give
rise to the scalar potential at vacuum $V_0$, which is given by
\EQ{EQ-V0}.

Next, because the energy density of the universe is still dominated by
the inflaton $\chi$ during the inflaton-oscillation epoch, the $\phi$
field receives additional SUSY-breaking terms caused by the finite
energy density of $\chi$:
%%%
\begin{eqnarray}
 \label{EQ-VH}
 \delta V_H =
  - c_H
  H^2 |\phi|^2
  +
  \frac{H}{8 M_{\rm eff}}
  \left(
   a_H\,\phi^4 + {\rm H.c.}
   \right)
   \,.
\end{eqnarray}
%%%
This can be derived in the same way as \EQ{EQ-Vinf}. Recall that we have
taken $c_H\simeq |a_H|\simeq 1$. We will call the second term Hubble
$A$-term.

Finally, the rests in \EQ{EQ-Vtotal} correspond to the thermal effects
which we discuss now. The crucial point here is that the decay of the
inflaton $\chi$ occurs during the coherent oscillation of $\chi$, while
it completes much later than the beginning of the oscillation.  Thus,
even in the $\chi$ oscillation period, there is a dilute plasma
consisting of the decay products of the inflaton $\chi$, although most
of the energy density of the universe is carried by the coherent
oscillation of $\chi$.  The temperature of this dilute plasma is given
by~\cite{KT}
%%%
\begin{eqnarray}
 T \simeq \left( T_R^2 M_G H \right)^{1/4}
  \,.
\end{eqnarray}
%%%
Notice that in this dilute plasma the temperature decreases as $T\propto
H^{1/4}$, rather than as $T\propto H^{1/2}$ like in the usual radiation
dominated universe.

This dilute plasma has crucial effects on the dynamics of the $\phi$
field~\cite{DRT,ACE,SomeIssues}. There are basically two possible
thermal effects. First, the fields $\psi_k$ which couple to $\phi$ are
produced by the inflaton decay and/or by thermal scatterings if their
effective masses are less than the temperature,
%%%
\begin{eqnarray}
 \label{EQ-thermass-c}
 m_{\rm eff,k} = f_k |\phi| < T 
  \,,
\end{eqnarray}
%%%
and hence they must be included in the plasma.  Here, $f_k$ denote the
Yukawa or gauge coupling constants between the flat direction $\phi$ and
$\psi_k$ (we take $f_k$ real and positive). Therefore, the flat
direction $\phi$ receives a thermal mass of order $\sim f_k T$, if the
condition \EQ{EQ-thermass-c} is satisfied. The induced mass term is
given by~\cite{ACE}
%%%
\begin{eqnarray}
 \delta V^{\rm th}_1 = \sum_{f_k|\phi| < T} c_k f_k^2 T^2 |\phi|^2 
  \,,
\end{eqnarray}
%%%
where $c_k$ are real positive constants of order unity (see below). The
summation means that only the fields in thermal plasma [i.e., $\psi_k$
whose couplings $f_k$ satisfy the condition \EQ{EQ-thermass-c}] can
induce the thermal mass for $\phi$.

We include all the couplings relevant for the flat direction $H_u =
\widetilde{L_1} = \phi / \sqrt{2}$ in the SUSY standard model, i.e.,
the gauge couplings for the ${\rm SU}(2)_L$ and ${\rm U}(1)_Y$ gauge
groups, and Yukawa couplings for up-type quarks and charged leptons.
The couplings $f_k$, which are redefined so that the effective masses
for the fields $\psi_k$ become $m_{\rm eff,k} = f_k|\phi|$, are listed
in Table.~\ref{Table-fk} with the coefficients $c_k$ of the thermal
mass for $\phi$. Among the gauge supermultiplets, the flat direction
$\phi$ does not couple to one ${\rm U}(1)$ group which remains
unbroken by the condensation of $\phi$.  Thus, there are one $Z$-like
and two $W$-like massive gauge multiplets as in the SUSY standard
model.  For the couplings of the charged leptons, we should be
careful.  The flat direction $\phi$ receives its thermal mass from
only one linear combination of charged leptons.  The effective Yukawa
coupling $y_{L1}$ is determined by the Yukawa couplings for the
charged leptons and the mixing matrix among gauge eigenstates of the
neutrinos.  (Note that the relevant $\phi$ is the flattest direction
which corresponds to the first family $\nu_1$ of mass eigenstates.)
Taking large mixing angles both for atmospheric and solar neutrino
oscillations suggested by the recent data~\cite{SK-Atm,Solar,KamLAND},
we find $y_{L 1} \simeq {\cal O}(0.1$--$1)\times y_{\tau}$. We will
see in \SEC{SEC-resultinLHu} that the Yukawa coupling for the up quark
induces a relevant thermal effect, while other couplings turn out to
be too large to satisfy the condition \EQ{EQ-thermass-c} in most of
the parameter space.
%%%%
\begin{table}[t]
 \begin{center}
  \begin{tabular}{|c||c|c|c|c|}
   \hline 
   &&&&\\
   $f_k$ (coupling for $\phi$) 
   & $\sqrt{\Frac{g_1^2 + g_2^2}{2}}$
   & $\Frac{g_2}{\sqrt{2}}$ 
   & $\Frac{y_{u,a}}{\sqrt{2}}$ ( $a = {\rm up, charm, top}$ )
   & $\Frac{y_{L1}}{\sqrt{2}}$ 
   \\
   &&&&
   \\
   $c_k$ 
   & $\Frac{1}{4}$ 
   & $\Frac{1}{2}$ 
   & $\Frac{3}{4}$ 
   & $\Frac{1}{4}$ 
   \\ 
   &&&&
   \\ \hline
  \end{tabular}
 \end{center}
 \caption{The couplings of the flat direction field $\phi$ to other
 fields and the coefficients $c_k$ of the thermal mass of $\phi$ induced
 by these fields. $g_1$ and $g_2$ denote the gauge couplings for the
 ${\rm U}(1)_Y$ and ${\rm SU}(2)_L$ gauge groups, respectively, and
 $y_{u,a}$ ($a =$ up, charm, top) and $y_{L1}$ are the Yukawa couplings
 for up-type quarks and charged lepton, respectively.}  \label{Table-fk}
 \vspace{2em}
\end{table}
%%%%

Next, there is another thermal effect which was pointed out in
Ref.~\cite{SomeIssues}. Along the $L H_u$ flat direction, the ${\rm
SU}(3)_C$ gauge symmetry remains unbroken and hence gluons and gluinos
are massless. Furthermore, the down type (s)quarks also remain massless
since they have no direct coupling to the $\phi$ field.  These massless
fields produce the free energy which depends on the ${\rm SU}(3)_C$
gauge coupling constant $g_S$, and we obtain an effective potential
$\delta V\propto g_S(T)^2 T^4$ at two loop level.\footnote{For the
coefficient of this free energy, see, for example, Ref.~\cite{T4term}.}
At first sight, there seems no dependence on the $\phi$ field in this
free energy. However, we can see this is not the case by means of the
following arguments.

\begin{figure}[t]%%%%%%%%%%%%%%%%%%%%%%%%%%%%%%%%%%%%%%%%%%%
%%%%%%%%%%%%%%%%%%%%%%%%%%%%%%%%%%%%%%%%%%%%%%%%%%%%%%%%%%%%
 \centerline{\psfig{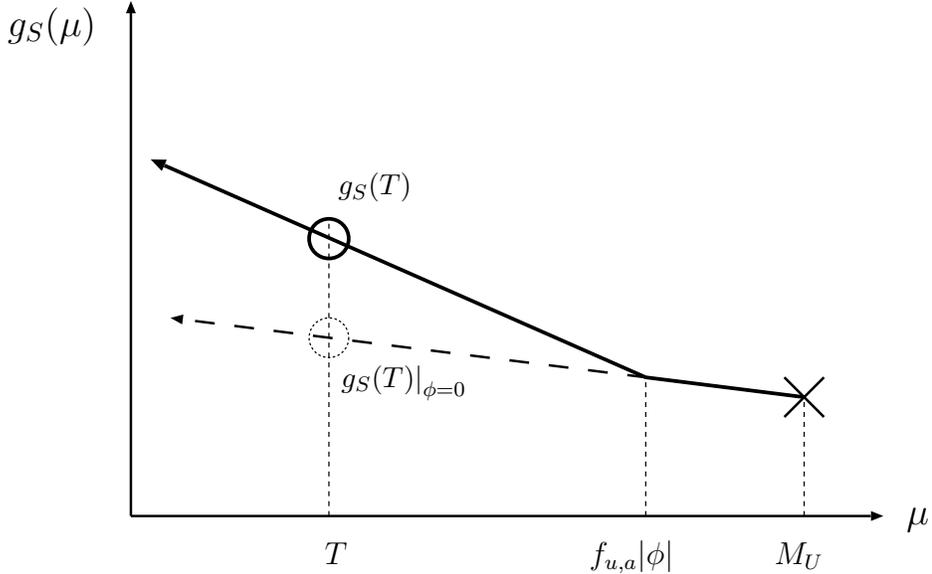}}
 %%%%%%%%%%%%%%%%%%%%%%%%%
 \begin{picture}(0,0)%%%%%
 %%%%%%%%%%%%%%%%%%%%%%%%%
  \put(40,200){\large $g_S(\mu)$}
  \put(160,0){$T$}
  \put(165,67){\small $\left.g_S(T)\right|_{\phi = 0}$}
  \put(165,140){\small $g_S(T)$}
  \put(260,0){$f_{u,a}|\phi|$}
  \put(330,0){$M_U$}
  \put(380,15){\large $\mu$}
 %%%%%%%%%%%%%%%%%%%%%%%%%
 \end{picture}%%%%%%%%%%%%
 %%%%%%%%%%%%%%%%%%%%%%%%%
 \caption{A schematic behavior of the SU$(3)_C$ gauge coupling
 $g_S$. The dashed line represents the running coupling when the $\phi$
 field does not have a vacuum-expectation value.}  \label{FIG-running}
%%%%%%%%%%%%%%%%%%%%%%%%%%%%%%%%%%%%%%%%%%%%%%%%%%%%%%%%%%%%
\end{figure}

The evolution of the running coupling $g_S$ is given by
%%%
\begin{eqnarray}
 \frac{d}{d\,\ln \mu}
  g_S(\mu)
  =
  \frac{g_S^3}{16 \pi^2}
  \left(
   - 3 C_2 
   + \sum_{m_i < \mu}
   C(R_i)
   \right)
   \,,
\end{eqnarray}
where $C_2 = 3$ and $C(R_i) = 1/2$ for fundamental representations
$R_i$. Notice that the evolution changes due to the decoupling when the
scale $\mu$ passes through a mass of a field $m_i$. Since the up type
(s)quarks get large masses from the couplings to the $\phi$ field
$f_{u,a} = y_{u,a}/\sqrt{2}$ ($a =$ up, charm, top), they change the
trajectory of the running at $\mu = m_{u,a} = f_{u,a}|\phi|$, as shown
in \FIG{FIG-running}.
%%%%%%%%%%%%%%%%%%%%%%%%%%%%%%%%%%%%%%%%%%%%%%%%%%%%%%%%%%%%
%%%%%%%%%%%%%%%%%%%%%%%%%%%%%%%%%%%%%%%%%%%%%%%%
%%%%%%%%%%%%%%%%%%%%%%%%%%%%%%%%%%%%%%%%%%%%%%%%%%%%%%%%%%%%
Hence, if $f_{u,a}|\phi|>T$, the running coupling $g_S(T)$ depends on
the value of the $\phi$ field:
%%%
\begin{eqnarray}
 \label{EQ-su3running}
  g_S(T)
  =
  \left.
   g_S(T)
   \right|_{\phi = 0}
   +
   \frac{g_S(M_U)^3}{32\pi^2}
   \sum_{f_{u,a} |\phi| > T}
   C(R_{u,a})
   \ln
   \left(
    \frac{f_{u,a}^2 |\phi|^2}{T^2}
    \right) 
    \,,
\end{eqnarray}
%%%
where $M_U$ is the ultraviolet scale where $g_S$ is fixed.  Then, there
appears an additional potential through the modification of the gauge
coupling constant in \EQ{EQ-su3running}
%%%
\begin{eqnarray}
 \delta V^{\rm th}_2
  = a_g \alpha_S(T)^2\,T^4
  \left(
   \frac{2}{3}
   \sum_{f_{u,a} |\phi| > T}
   \,
   C(R_{u,a})
   \right)
   \ln\left(\frac{|\phi|^2}{T^2}\right) 
   \,,
\end{eqnarray}
%%%
where $a_g$ is a constant which is a bit larger than unity\footnote{By
using the free energy at two-loop order given in Ref.~\cite{T4term},
$a_g$ is given by $a_g = 1.125$ in the case of the $L H_u$ flat
direction. The author thanks Masaaki Fujii for helpful discussion.} and
$\alpha_S(T) \equiv g_S(T)_{\phi=0}^2/(4\pi)$. (We will omit the
subscript $\phi=0$ for simplicity.) Hereafter, we take the factor
$(2/3)\sum C(R)$ to be unity since it does not change the result
much.\footnote{At least the top Yukawa coupling $f_{u,{\rm top}}$ always
satisfies $f_{u,{\rm top}}|\phi| > T$ before the oscillation of
$\phi$. Thus, the resultant baryon asymmetry changes by only a factor of
$\sqrt{3}$ at most. [See \EQS{EQ-Hosc-final} and (\ref{EQ-nBs-final}).]}

\subsubsection{evolution of $\phi$}

Now we eventually obtain the total effective potential \EQ{EQ-Vtotal}:
%%%
\begin{eqnarray}
 V_{\rm total} = V_0 + \delta V_H + \delta V^{\rm th}_1 +
 \delta V^{\rm th}_2
 \,.
\end{eqnarray}
%%%
The evolution of $\phi$ is described by the equation of motion with this
$V_{\rm total}$ as
%%%
\begin{eqnarray}
 \label{EQ-EOM}
 \ddot{\phi} + 3 H \dot{\phi}
  + \frac{\partial V_{\rm total}}{\partial \phi^*} = 0 
  \,,
\end{eqnarray}
%%%
where the dot denotes a derivative with time. During the inflation,
there is no thermal plasma and hence no thermal potentials. The vacuum
expectation value of the $\phi$ field at this stage is given by
\EQS{EQ-minimum-inf1} and (\ref{EQ-minimum-inf2}). Let us call the
valley to which the $\phi$ field rolls down $n = n_0$.

After the inflation ends, the inflaton $\chi$ starts to oscillate and
its decay produces a dilute plasma. However, the potential is still
dominated by Hubble-induced terms [in \EQ{EQ-VH}] and $|\phi|^6$ term
[in \EQ{EQ-V0}] at the first stage of the oscillation. Thus, the flat
direction field $\phi$ is trapped in the following minimum:
%%%
\begin{eqnarray} 
 |\phi| &\simeq& \sqrt{M_{\rm eff} H}
  \,,
  \\
 \arg(\phi) &\simeq& \frac{- \arg(a_H) + ( 2 n_0 + 1 )\pi}{4}
  \label{EQ-minim-valley}
  \,,
\end{eqnarray}
%%%
This is because the curvatures around the minimum along both radius and
phase directions are of the order of $H$ also in this period, and hence
the $\phi$ field always catches up the instantaneous minimum~\cite{DRT}.

Then, as the Hubble parameter decreases, the negative Hubble-induced
mass term is eventually exceeded by one of the following three terms in
the potential:\footnote{Here, we have compared $(1/\phi)(\partial
V/\partial \phi^*)$ instead of the potential terms themselves, since the
evolution of $\phi$ is governed by the equation of motion \EQ{EQ-EOM}.}
%%%
\begin{eqnarray}
 \label{EQ-preHO}
 H^2 
  &\lsim& 
  m_{\phi}^2
  \,\,
  +
  \sum_{f_k|\phi| < T} c_k f_k^2 T^2
  \,\,
  +
  a_g \alpha_S^2(T) \frac{T^4}{|\phi|^2} 
  \,.
\end{eqnarray}
%%%
As we show now, it is this time when the oscillation of $\phi$
starts. Let us denote the Hubble parameter at this time by $H_{\rm
osc}$. The evolution of the $\phi$ after $H \simeq H_{\rm osc}$ depends
on which term in \EQ{EQ-Vtotal} dominates the effective potential. There
are basically three cases; the potential is dominated by (i) $m_{\phi}^2
|\phi|^2$ term, (ii) $T^2 |\phi|^2$ term, or (iii) $T^4 \ln(|\phi|^2)$
term. 

{\bf (i)} First, if the potential is dominated by the
$m_{\phi}^2|\phi|^2$ term, the equation of motion \EQ{EQ-EOM} is given
by
%%%
\begin{eqnarray}
 \label{EQ-EOMmass}
  \ddot{\phi} + 3 H \dot{\phi}
  + m_{\phi}^2 \phi \simeq 0 
  \,.
\end{eqnarray}
%%%
It is clear in this case that the field $\phi$ oscillates around the
origin ($\phi = 0$) and the amplitude of the oscillation damps as
$|\phi|\propto H\propto t^{-1}$.  

{\bf (ii)} Second, when the potential is dominated by the thermal mass
term $c_k f_k^2 T^2 |\phi|^2$, the equation of motion becomes
%%%
\begin{eqnarray}
 \ddot{\phi} + 3 H \dot{\phi} + c_k f^2_k T^2 \phi  
  \simeq 0
  \,.
\end{eqnarray}
This equation can be solved analytically and the solution is given by
\begin{eqnarray}
 \label{EQ-ev-thmass}
 \phi(t)
  &=&
  \phi_1
  \left[ \frac{1}{z^{2/3}} J_{2/3}(z) \right]
  +
  \phi_2
  \left[ \frac{1}{z^{2/3}} J_{-2/3}(z) \right]
  \nonumber
  \\
 z &=& \frac{4}{3}
  \left(
   \frac{2}{3} c_k^2 f^4_k M_G T_R^2 \,\,t^3
   \right)^{1/4},
\end{eqnarray}
where $\phi_1$ and $\phi_2$ are constants and $J_{\nu}$ is the Bessel
function.  We see that the $\phi$ oscillates around the origin in this
case as well.  The time scale of the oscillation is $\sim \left(f^4_k
M_G T_R^2 \right)^{-1/3}$ and the amplitude of the oscillation is damped
as $|\phi| \sim t^{-7/8} \sim H^{7/8}$. 

{\bf (iii)} The third case is given when the $a_g \alpha_S^2 T^4 \ln
(|\phi|^2/T^2)$ term dominates the potential. If we neglect the time
dependence of $T^4$, the damping rate of the oscillation amplitude due
to such a flat potential, $V\sim \ln (|\phi|^2)$, can be estimated by
using the virial theorem, and it is given by $|\phi|\propto H^2\propto
t^{-2}$~\cite{GoMoMu}. In the actual case, however, the potential itself
gradually decreases with time as $T^4\propto t^{-1}$. We have
numerically checked that the amplitude damps as $|\phi|\propto
H^{\alpha} \propto t^{-\alpha}$ with $\alpha\simeq 1.5$.

Thus, in any case, the $\phi$ field starts a coherent
oscillation. Notice that, in all the above cases, the damping rate is
faster than the rate before the beginning of the $\phi$'s oscillation
($|\phi|\propto H^{1/2} \propto t^{-1/2}$).

Now let us estimate the cosmic time $t_{\rm osc}$ when the $\phi$ field
starts its oscillation, or equivalently the Hubble parameter at that
time, $H_{\rm osc}=(2/3)t_{\rm osc}^{-1}$. As we will see in
\SEC{SEC-resultinLHu}, this oscillation time plays an important role to
determine the final lepton asymmetry. $H_{\rm osc}$ is determined by the
term which exceeds the Hubble mass term at first. (Therefore, again,
there are three cases.)

{\bf (i)} If the soft mass term dominates the potential at first,
$H_{\rm osc}$ is just given by $H_{\rm osc} = m_{\phi}$.

{\bf (ii)} The case when the thermal mass term dominates at first is a
bit complicated. Let us first consider a simple case in which there is
only one field $\psi_1$ with a coupling $f_1$. {}From
\EQS{EQ-thermass-c} and (\ref{EQ-preHO}), it is found that the following
two conditions must be satisfied in order to cause the $\phi$'s
oscillation by the thermal mass term:
%%%
\begin{eqnarray}
 H 
  &<& 
  \frac{1}{f_1^4}
  \frac{M_G T_R^2}{M_{\rm eff}^2}
  \,, 
  \label{EQ-thosc-cond1} \\ 
 H
  &<& 
  \left(c_1^2 f_1^4 M_G T_R^2 \right)^{1/3}
  \,, 
  \label{EQ-thosc-cond2}
\end{eqnarray}
%%%
where we have used the relations $|\phi|\simeq \sqrt{M_{\rm eff}H}$ and
$T = (T_R^2 M_G H)^{1/4}$.  The meanings of these conditions are as
follows. If the coupling $f_1$ is small enough, the $\psi_1$ can easily
enter in the thermal plasma ($f_1 |\phi| < T$), but the induced thermal
mass for $\phi$ is smaller than the Hubble-induced mass $H$ at the
beginning ($c_1 f_1^2 T^2 < H^2$).  However, since the temperature
decreases more slowly than the Hubble parameter $H$ ($T\propto H^{1/4}$)
as the universe expands, the thermal mass eventually exceeds $H$ ($c_1
f_1^2 T^2 > H^2$). Then the $\phi$'s oscillation starts at the time when
the condition \EQ{EQ-thosc-cond2} is satisfied.  On the other hand, if
$f_1$ is very large, the would-be thermal mass can be large enough to
exceed the Hubble mass ($c_1 f_1^2 T^2 > H^2$).  However, the thermal
mass does not appear until the $|\phi|$ becomes small enough to satisfy
$f_1 |\phi| < T$. In this case the $\phi$ starts to oscillate at the
time when the condition \EQ{EQ-thosc-cond1} is satisfied.  Therefore, we
find the Hubble parameter $H_{\rm osc}$ to be
%%%
\begin{eqnarray}
 H_{\rm osc} 
  &=& 
  {\rm min}
  \left[ 
   \frac{1}{f_1^4}
   \frac{M_G T_R^2}{M_{\rm eff}^2}
   \,
   \,,
   \,\,
   \left(c_1^2 f_1^4
    M_G T_R^2 \right)^{1/3}
   \right]
   \,.
\end{eqnarray}
%%%
It is easy to apply the above discussion to the case of more than one
couplings.  If there is another coupling $f_i$ which can satisfy the
both conditions \EQS{EQ-thosc-cond1} and (\ref{EQ-thosc-cond2}) earlier
than $f_1$, the flat direction $\phi$ starts its oscillation earlier.
Therefore, the Hubble parameter $H_{\rm osc}$ is given by\footnote{We
assume here hierarchical couplings and neglect effects of the
summation.}
%%%
\begin{eqnarray}
 H_{\rm osc} = \max_i\left( H_i \right)
  \,,
\end{eqnarray}
where
\begin{eqnarray}
 \label{EQ-Hi}
 H_i
  &=& 
  {\rm min}
  \left[ 
   \frac{1}{f_i^4}
   \frac{M_G T_R^2}{M_{\rm eff}^2}
   \,
   \,,
   \,\,
   \left(c_i^2 f_i^4
    M_G T_R^2 \right)^{1/3}
   \right]
   \,.
\end{eqnarray}
%%%

{\bf (iii)} The third case is again given when the $a_g \alpha_s^2 T^4
\ln (|\phi|^2/T^2)$ term dominates the potential at first. In this case,
it is found from \EQ{EQ-preHO} that the oscillation time of the $\phi$
field is given by\footnote{The running coupling $\alpha_S(T)$ here is
obtained by solving the equation $H^2 = a_g \alpha_S(T)T^4/|\phi|^2$
iteratively.}
%%%
\begin{eqnarray}
 H_{\rm osc}
  &=&
  \alpha_S T_R \left(\frac{a_g M_G}{M_{\rm eff}}\right)^{1/2}
  \,.
\end{eqnarray}
%%%

After all, the Hubble parameter at the time when the $\phi$ field starts 
its coherent oscillation is given by
%%%
\begin{eqnarray}
 \label{EQ-Hosc-final}
  H_{\rm osc} \simeq 
  \max
  \left[
   m_{\phi}
   \, ,\,
   H_i
   \, ,\,
   \alpha_S T_R \left(\frac{a_g M_G}{M_{\rm eff}}\right)^{1/2}
   \right] 
   \,,
\end{eqnarray}
%%%
where $H_i$ is given by \EQ{EQ-Hi}. We stress here that the thermal
effects always make the oscillation time earlier. (Notice that if there
is no thermal effect, the oscillation time is just given by $H_{\rm osc}
= m_{\phi}$.) As we shall show soon, this ``early oscillation''
drastically affect the amount of the produced lepton asymmetry.

\subsection{Lepton asymmetry}
\label{SEC-resultinLHu}

Now we are at the point to calculate the lepton asymmetry produced by
the $\phi$ field.  Since the $\phi$ field carries lepton charge, its
number density is related to the lepton number density $n_L$ as
\begin{eqnarray}
 \label{EQ-nLandphi}
 n_L = \frac{1}{2} i 
  \left(
   \dot{\phi}^* 
   \phi 
   - 
   \phi^* 
   \dot{\phi}
   \right)
   \,.
\end{eqnarray}
{}From \EQS{EQ-Vtotal}, (\ref{EQ-EOM}) and (\ref{EQ-nLandphi}), the
evolution of $n_L$ is described by the following equation:
\begin{eqnarray}
 \dot{n}_L + 3 H n_L 
  &=&
  \frac{m_{3/2}}{2 M_{\rm eff}}
  {\rm Im}
  \left(
   a_m \phi^4
   \right)
  + 
  \frac{H}{2 M_{\rm eff}}
  {\rm Im}
  \left(
   a_H \phi^4
   \right)
   \,.
   \label{EQ-EOM-nL}
\end{eqnarray}
The production of the lepton asymmetry occurs as follows. Suppose that
the original $A$-term would vanish, i.e., $m_{3/2}\,a_m = 0$.  If this
is the case, the flat direction $\phi$ is always trapped in one of the
valleys induced by the Hubble $A$-term [see \EQ{EQ-minim-valley}] during
the both periods $t<t_{\rm osc}$ and $t>t_{\rm osc}$, and the direction
of the valley does not change with time. Therefore, there is no force
which causes the motion of $\phi$ along the phase direction and no
lepton-number asymmetry is produced (i.e., the right-hand side of
\EQ{EQ-EOM-nL} vanishes.) However, we have the original $A$-term, and
the phase of $\phi$ is kicked by the relative phase difference between
$a_m$ and $a_H$ in \EQ{EQ-Vtotal}.  The phase of $\phi$ changes during
its rolling towards the origin, because the Hubble parameter $H$
decreases and the direction of the true valleys changes with time.
Therefore, the original $A$-term, which corresponds to the first term in
\EQ{EQ-EOM-nL}, plays a role of the source of the lepton asymmetry.

One might wonder if the Hubble $A$-term, which corresponds to the second
term in \EQ{EQ-EOM-nL}, gives a larger contribution to the lepton
asymmetry since $H \gg m_{3/2}$. However, since the $\phi$ almost traces
one of the valleys determined mainly by the Hubble $A$-term, ${\rm
Im}(a_H\phi^4)$ in \EQ{EQ-EOM-nL} is highly suppressed compared with
${\rm Im}(a_m\phi^4)$.  In fact, we have found numerically that the
contribution of the second term in \EQ{EQ-EOM-nL} is always comparable
or less than that from the first term.  Thus, we neglect the lepton
asymmetry produced from the Hubble $A$-term in our analytic calculation,
for simplicity. By integrating \EQ{EQ-EOM-nL}, we obtain the resultant
lepton number at the time $t$,
%%%
\begin{eqnarray}
 \label{EQ-nL-integral}
  \left[ R^3 n_L \right] (t)
  &\simeq& 
  \frac{m_{3/2}}{2 M_{\rm eff}} 
  \int^t dt' 
  R^3 
  \,
  {\rm Im}
  \left(
   a_m \phi^4 
   \right) 
   \,,
\end{eqnarray}
%%%
where $R$ denotes the scale factor of the expanding universe, which
scales as $R^3\propto H^{-2} \propto t^2$ in the universe dominated by
the oscillation energy of the inflaton $\chi$. We can see that the total
lepton number increases with time as $R^3 n_L \propto t$ until the
oscillation of $\phi$ starts ($t < t_{\rm osc}$), since $\phi^4 \propto
H^2$ and hence $R^3 \phi^4 \sim const$ in this stage.

On the other hand, after the $\phi$ starts its oscillation, the
production of lepton number is strongly suppressed. This is because
${\rm Im}\left(a_m \phi^4 \right)$ changes its sign rapidly due to the
oscillation of $\phi$, and also because the amplitude of $\phi$'s
oscillation is damped as $R^3 \phi^4\sim t^{-n}$ with $n > 1$. [See
discussion around \EQS{EQ-EOMmass}--(\ref{EQ-ev-thmass}).]  Therefore,
the net lepton asymmetry is fixed when the oscillation of $\phi$ starts. 
The generated lepton number at this epoch ($t = t_{\rm osc}$) is given
approximately by
%%%
\begin{eqnarray}
 n_L (t_{\rm osc})  
  &=&
  \frac{m_{3/2}}{2 M_{\rm eff}}
  {\rm Im}
  \left[
   a_m \phi^4 (t_{\rm osc})
   \right]
   t_{\rm osc}
   \nonumber\\
 &\simeq&
  \frac{1}{3}
  \left(
   m_{3/2}
   |a_m|
   \right)
   \,
   M_{\rm eff} 
   H_{\rm osc} 
   \,
   \delta_{\rm ph} 
   \,,
\end{eqnarray}
%%%
where $\delta_{\rm ph}\simeq \sin(4\arg\phi + \arg a_m)$ represents an
effective $CP$ violating phase. Here, we have used $\phi(t_{\rm osc})
\simeq \sqrt{M_{\rm eff} H_{\rm osc}}$ and $t_{\rm osc} = (2/3)H_{\rm
osc}^{-1}$. Recall that we have assumed that the production of net
lepton asymmetry occurs before the reheating process of the inflation
completes, i.e., $H_{\rm osc} > \Gamma_{\chi}$ where $\Gamma_{\chi}$ is
the decay rate of the inflaton $\chi$. Thus, the lepton number when the
reheating process completes ($t = t_R$, $H\simeq \Gamma_{\chi}$) is
given by
%%%
\begin{eqnarray}
 \label{EQ-nLtR-nLtosc}
  n_L(t_R)
  &=&
  n_L(t_{\rm osc})
  \left(
   \frac{ R(t_{\rm osc}) }
   { R(t_R)}
   \right)^3
   \nonumber \\
 &=&
  n_L(t_{\rm osc})
  \left(
   \frac{ \Gamma_{\chi} }
   { H_{\rm osc} }
   \right)^2
   \,.
\end{eqnarray}
Then, the lepton-to-entropy ratio is estimated as
%%%
\begin{eqnarray}
 \label{EQ-nLs-final}
  \frac{n_L}{s} &=& \frac{M_{\rm eff} T_R}{12 M_G^2}
  \left(
   \frac{m_{3/2}|a_m|}{H_{\rm osc}}
   \right)
   \delta_{\rm ph} 
   \,,
\end{eqnarray}
%%%
when the reheating process of inflation completes. Here, we have used $3 
M_G^2 \Gamma_{\chi}^2 = \rho_{\rm rad}(t_R) = (3/4)T_R\times s(t_R)$.

As explained in \SEC{SEC-sphaleron}, this lepton asymmetry is partially
converted into the baryon asymmetry through the sphaleron effect. Thus,
after all, the present baryon asymmetry is given by
%%%
\begin{eqnarray}
 \label{EQ-nBs-final}
  \frac{n_B}{s} &\simeq&
  0.35\times
  \frac{n_L}{s}
  \nonumber \\
 &=&
  0.029\times
  \frac{M_{\rm eff} T_R}{M_G^2}
  \left(
   \frac{m_{3/2}|a_m|}{H_{\rm osc}}
   \right)
   \delta_{\rm ph} 
   \,.
\end{eqnarray}
%%%

{}From \EQ{EQ-nBs-final}, we see that the baryon asymmetry is a
monotonically increasing function of the scale $M_{\rm eff}$. [Notice
that the $H_{\rm osc}$ depends on $M_{\rm eff}$ as shown in
\EQ{EQ-Hosc-final}.] Since $M_{\rm eff}$ is directly related to the
neutrino mass $m_{\nu}$ as $M_{\rm eff} = \vev{H_u}^2/m_{\nu}$, this
means that the baryon asymmetry becomes larger as the neutrino mass
$m_{\nu}$ decreases. Therefore, as mentioned at the beginning of this
chapter, the most effective flat direction corresponds to the lightest
neutrino $\nu_1$, i.e., the first family field $\phi/\sqrt{2} = L_1 =
H_u$.

The factor $H_{\rm osc}^{-1}$ in \EQ{EQ-nBs-final} represents the
suppression of the lepton asymmetry due to the thermal effects. When the
$\phi$'s oscillation is caused by the thermal effects, i.e., $H_{\rm
osc} > m_{\phi}$, the production of lepton asymmetry stops earlier than
the case without thermal effects, and hence the resultant baryon
asymmetry becomes suppressed.

We should also note that the amount of the produced baryon asymmetry is
proportional to the gravitino mass, $m_{3/2}$. This is because the force
which rotates the $\phi$ field along the phase direction comes from the
SUSY-breaking $A$-term $(m_{3/2}/8 M_{\rm eff})(a_m \phi^4 + {\rm
H.c.})$. Thus, the resultant baryon asymmetry would be very much
suppressed if the gravitino is light, like in gauge-mediated SUSY
breaking models~\cite{GMSB}.\footnote{If we introduce a gauged ${\rm
U}(1)_{B-L}$, however, the $L H_u$ flat direction can produce enough
lepton asymmetry even with a light gravitino. See \SEC{SEC-withB-L}.}
This is the reason why we have assumed gravity-mediated SUSY-breaking
and $m_{3/2}\simeq 1\TEV$.

\begin{figure}[t]%%%%%%%%%%%%%%%%%%%%%%%%%%%%%%%%%%%%%%%%%%%
%%%%%%%%%%%%%%%%%%%%%%%%%%%%%%%%%%%%%%%%%%%%%%%%%%%%%%%%%%%%
 \centerline{\psfig{figure=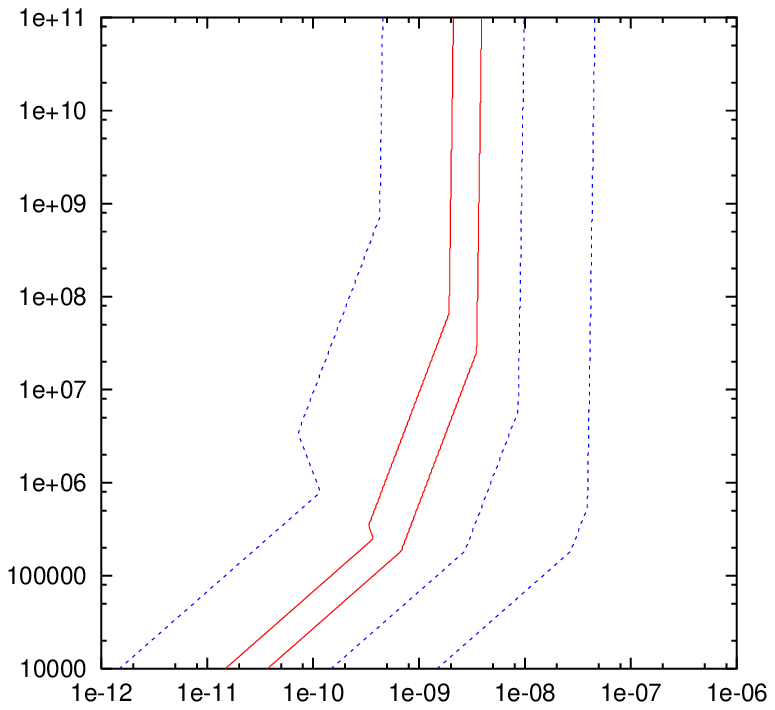,height=10cm}}
 %%%%%%%%%%%%%%%%%%%%%%%%%
 \begin{picture}(0,0)%%%%%
 %%%%%%%%%%%%%%%%%%%%%%%%%
  \put(60,180){$T_R$}  
  \put(53,163){$[$GeV$]$}  
  \put(230,1){$\mnu{1}\,\,[$eV$]$}
 %%%%%%%%%%%%%%%%%%%%%%%%%
 \end{picture}%%%%%%%%%%%%
 %%%%%%%%%%%%%%%%%%%%%%%%%
 \caption{The contour plot of the baryon asymmetries $n_B/s$ in the
 $\mnu{1}$--$T_R$ plane.  The lines represent the contour plots for
 $n_B/s = 10^{-9}$, $10^{-10}$, $0.4\times 10^{-10}$, $10^{-11}$, and
 $10^{-12}$ from the left to the right.}  \label{FIG-BA-LHu}
%%%%%%%%%%%%%%%%%%%%%%%%%%%%%%%%%%%%%%%%%%%%%%%%%%%%%%%%%%%%
\end{figure}

\FIG{FIG-BA-LHu} shows the contour plot of the produced baryon asymmetry
in the $\mnu{1}$--$T_R$ plane, given by the analytic formula
\EQ{EQ-nBs-final}.
%%%%%%%%%%%%%%%%%%%%%%%%%%%%%%%%%%%%%%%%%%%%%%%%%%%%%%%%%%%%
%%%%%%%%%%%%%%%%%%%%%%%%%%%%%%%%%%%%%%%%%%%%%%%%
%%%%%%%%%%%%%%%%%%%%%%%%%%%%%%%%%%%%%%%%%%%%%%%%%%%%%%%%%%%%
Here, we have taken $m_{\phi} = m_{3/2} = 1\TEV$, $|a_m| = 1$ and
$\delta_{\rm ph} = 1$, and we have used the relation $\mnu{1} =
\vev{H_u}^2 / M_{\rm eff}$. (Although the big-bang nucleosynthesis
constraint on the gravitino abundance requires $T_R\lsim 10^9\GEV$ for
$m_{3/2} = 1\TEV$, we have plotted the contour in \FIG{FIG-BA-LHu} up to
$T_R \le 10^{11}\GEV$, keeping in mind that even $T_R\simeq 10^{11}\GEV$
can be allowed for a slightly heavier gravitino, say, $m_{3/2}\sim
3\TEV$. See \SEC{SEC-grav}.)

A remarkable observation here is that the present baryon asymmetry
$n_B/s$ is determined almost independently of the reheating temperature
for a wide range of $T_R \gsim 10^5\GEV$. In particular, for a
relatively high reheating temperature $T_R \gsim 10^8\GEV$, the baryon
asymmetry derived from the \EQS{EQ-Hosc-final} and (\ref{EQ-nBs-final})
is given by the following simple from;
\begin{eqnarray}
 \frac{n_B}{s}
  \simeq
  10^{-11}
  \delta_{\rm eff}
  \times
  \left(
   \frac{\mnu{1}}{10^{-8}\EV}
   \right)^{-3/2}
   \left(
    \frac{m_{3/2} |a_m|}{1\TEV}
    \right)
    \,.
\end{eqnarray}
The reason why it is independent of the reheating temperature $T_R$ is
that the oscillation time $H_{\rm osc}$ is determined by the thermal
potential $T^4\ln(|\phi|^2)$ in the higher temperature regime and
becomes proportional to the reheating temperature $T_R$. [See
\EQ{EQ-Hosc-final}.] Thus, the $T_R$ dependence is canceled out in
\EQ{EQ-nBs-final}.\footnote{Precisely speaking, $H_{\rm osc}$ depends on
$T_R$ through the running coupling $\alpha_S(T)$ even in this
case. However, as can be seen in \FIG{FIG-BA-LHu}, this dependence is
negligibly mild.} Even in the lower reheating temperature region $10^5
\lsim T_R \lsim 10^8\GEV$, where $H_{\rm osc}$ is determined by the
thermal-mass term potential $T^2 |\phi|^2$, $T_R$ dependence is still
mild, i.e., $n_B/s \propto T_R^{1/3}$. The ``reheating-temperature
independence'' discussed here is a very attractive feature of the
present mechanism since the produced baryon asymmetry crucially depends
on $T_R$ in many other baryogenesis scenarios.

For the analytic calculation in \FIG{FIG-BA-LHu}, we have taken
$\delta_{\rm ph}= 1$. It is expected that $\delta_{\rm ph}\sim {\cal
O}(1)$, say, $\delta_{\rm ph}\simeq 0.1$--$1$, unless there is an
unnatural cancellation between $\arg(a_m)$ and $\arg(a_H)$. Thus, it is
found from \FIG{FIG-BA-LHu} that the present baryon asymmetry in our
universe $n_B/s\simeq (0.4$--$1)\times 10^{-10}$ suggests an ultralight
neutrino of a mass,
%%%
\begin{eqnarray}
 \label{EQ-mnu1-predicted}
  \mnu{1}
  &\simeq&
  (0.1-3)\times 10^{-9}\EV
  \,,
\end{eqnarray}
for $T_R\gsim 10^5\GEV$ and $\delta_{\rm ph}\simeq 0.1$--$1$.

\subsubsection*{$\bullet$ remarks}

Before closing this section, we give several comments.  First, let us
justify our assumption that the production of the lepton asymmetry stops
before the reheating process of the inflation completes, namely, $H_{\rm
osc}>\Gamma_{\chi}$. We can see this condition is satisfied in all the
parameter region in \FIG{FIG-BA-LHu} by using the fact that $H_{\rm
osc}$ is bounded from below, as shown in \EQ{EQ-Hosc-final}. For
example, it is enough to show $\alpha_S T_R (a_g M_G/M_{\rm eff})^{1/2}
> \Gamma_{\chi}$. By using $\Gamma_{\chi} = (\pi^2
g_*/90)^{1/2}T_R^2/M_G$, it is found that this condition is indeed
satisfied for $T_R\lsim 10^{14}\GEV\times (\mnu{1}/10^{-9}\EV)^{1/2}$.

\begin{figure}[t]%%%%%%%%%%%%%%%%%%%%%%%%%%%%%%%%%%%%%%%%%%%
%%%%%%%%%%%%%%%%%%%%%%%%%%%%%%%%%%%%%%%%%%%%%%%%%%%%%%%%%%%%
 \centerline{\psfig{figure=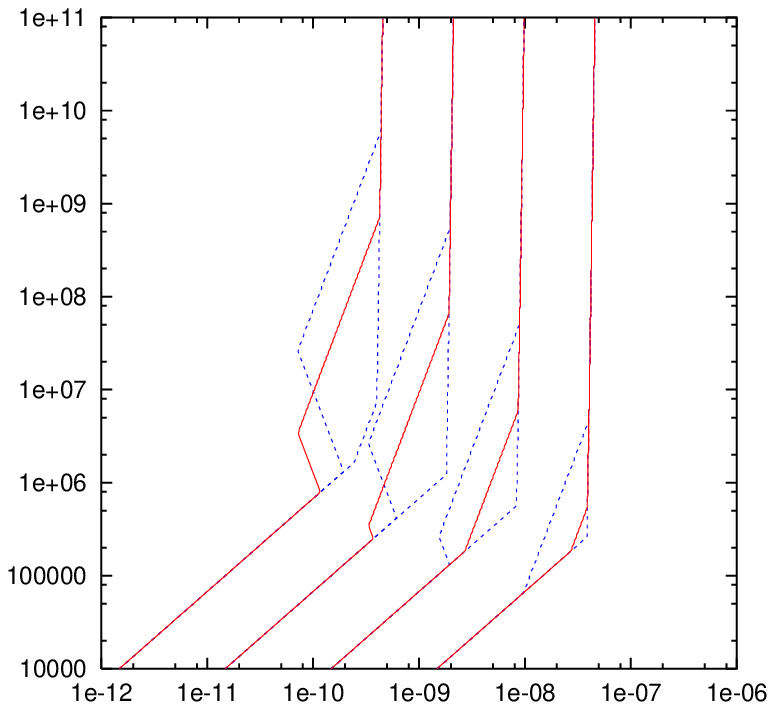,height=10cm}}
 %%%%%%%%%%%%%%%%%%%%%%%%%
 \begin{picture}(0,0)%%%%%
 %%%%%%%%%%%%%%%%%%%%%%%%%
  \put(60,180){$T_R$}  
  \put(53,163){$[$GeV$]$}  
  \put(230,1){$\mnu{1}\,\,[$eV$]$}
 %%%%%%%%%%%%%%%%%%%%%%%%%
 \end{picture}%%%%%%%%%%%%
 %%%%%%%%%%%%%%%%%%%%%%%%%
 \caption{The contour plot of the baryon asymmetries $n_B/s$ in the
 $\mnu{1}$--$T_R$ plane.  The lines represent the contour plots for
 $n_B/s = 10^{-9}$, $10^{-10}$, $10^{-11}$, and $10^{-12}$ from the left
 to the right. The solid lines are the same as those in \FIG{FIG-BA-LHu}
 ($m_{\rm up} = 3 \MEV$). The left (right) dashed lines correspond to
 $m_{\rm up} = 5 (1) \MEV$.}  \label{FIG-change-up}
%%%%%%%%%%%%%%%%%%%%%%%%%%%%%%%%%%%%%%%%%%%%%%%%%%%%%%%%%%%%
\end{figure}

Next, let us investigate the behavior of \FIG{FIG-BA-LHu} in more
detail. \FIG{FIG-change-up} shows the dependence of the baryon asymmetry
on the Yukawa coupling of the up quark. Notice that the up-quark Yukawa
coupling is given by $y_{u,{\rm up}} = m_{\rm up}/\vev{H_u} = m_{\rm
up}/(174\GEV\times \sin\beta)$. In the previous figure \ref{FIG-BA-LHu},
we have taken $m_{\rm up} = 3\MEV$ (and $\tan\beta = 10$). In
\FIG{FIG-change-up}, we have taken $m_{\rm up} = 1\MEV$ and $m_{\rm up}
= 5\MEV$ for comparison.
%%%%%%%%%%%%%%%%%%%%%%%%%%%%%%%%%%%%%%%%%%%%%%%%%%%%%%%%%%%%
%%%%%%%%%%%%%%%%%%%%%%%%%%%%%%%%%%%%%%%%%%%%%%%%
%%%%%%%%%%%%%%%%%%%%%%%%%%%%%%%%%%%%%%%%%%%%%%%%%%%%%%%%%%%%
It is found that the final baryon asymmetry changes at most one order of
magnitude. However, the dependence on the reheating temperature is still
mild, and hence our conclusion does not change much. 

Meantime, we can also see from \FIG{FIG-change-up} that the up-quark
Yukawa coupling is the only coupling which can give a dominant
contribution to the early oscillation of $\phi$, since below this
``up-quark Yukawa'' region the oscillation is caused by the soft mass
term $m_{\phi}$ (where $n_B/s\propto T_R \mnu{1}^{-1}$), while in the
higher $T_R$ region the oscillation is caused by the thermal logarithmic
term $T^4\ln(|\phi|^2)$ (where $n_B/s\propto \mnu{1}^{-3/2}$). This is
because other couplings $f_k$ are too large and the fields with those
couplings cannot satisfy the condition $f_k|\phi|< T$.

\begin{figure}[t!]%%%%%%%%%%%%%%%%%%%%%%%%%%%%%%%%%%%%%%%%%%%
%%%%%%%%%%%%%%%%%%%%%%%%%%%%%%%%%%%%%%%%%%%%%%%%%%%%%%%%%%%%
 \centerline{\psfig{figure=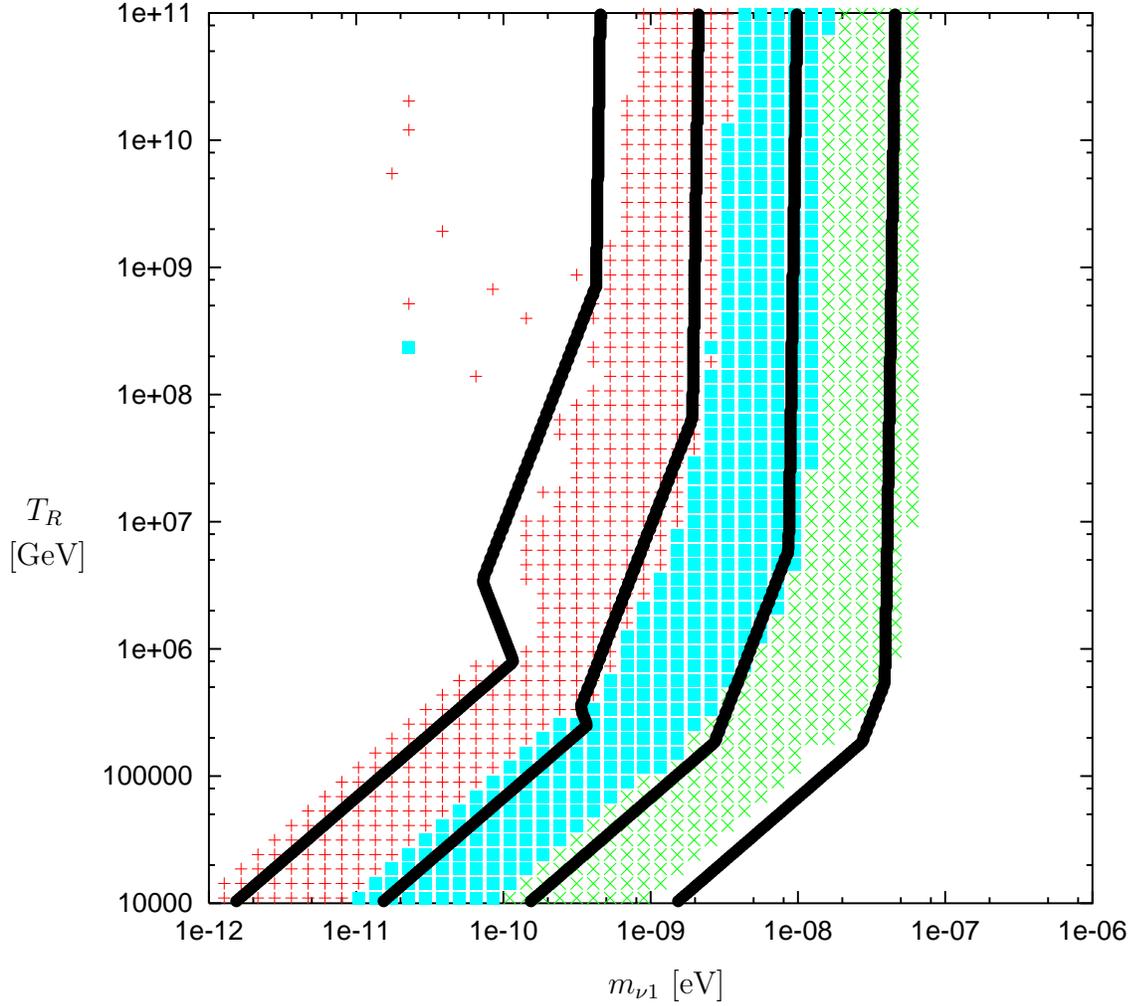,height=13cm}}
 %%%%%%%%%%%%%%%%%%%%%%%%%
 \begin{picture}(0,0)%%%%%
 %%%%%%%%%%%%%%%%%%%%%%%%%
  \put(10,180){$T_R$}  
  \put(3,163){$[$GeV$]$}  
  \put(230,1){$\mnu{1}\,\,[$eV$]$}
 %%%%%%%%%%%%%%%%%%%%%%%%%
 \end{picture}%%%%%%%%%%%%
 %%%%%%%%%%%%%%%%%%%%%%%%%
 \caption{The plots of the baryon asymmetries $n_B/s$ in the
 $\mnu{1}$--$T_R$ plane, which are obtained by numerically solving the
 equation of motion \EQ{EQ-EOM}. The regions with points show the result
 of the numerical calculation. They represent $10^{-9} > n_B/s >
 10^{-10}$, $10^{-10} > n_B/s > 10^{-11}$ and $10^{-11} > n_B/s >
 10^{-12}$ from the left to the right. The solid lines are the same as
 those in \FIG{FIG-BA-LHu}, which represent the contour plots for $n_B/s
 = 10^{-9}$, $10^{-10}$, $10^{-11}$, and $10^{-12}$ from the left to the
 right, given by the analytic formula \EQ{EQ-nBs-final}.} 
 \label{FIG-LHu-numerical}
%%%%%%%%%%%%%%%%%%%%%%%%%%%%%%%%%%%%%%%%%%%%%%%%%%%%%%%%%%%%
\end{figure}

We have also solved the equation of motion (\ref{EQ-EOM}) numerically
with the full scalar potential \EQ{EQ-Vtotal}. The result is shown in
\FIG{FIG-LHu-numerical}, where the final baryon asymmetry $n_B/s
\simeq 0.35\times n_L/s$ is plotted. ($n_L$ is given by
\EQ{EQ-nLandphi}, $n_L= (i/2)(\dot{\phi}^* \phi - \phi^*
\dot{\phi})$.) Here, we have taken $m_{\phi} = m_{3/2}|a_m| = 1\TEV$,
$\arg(a_m) = \pi/3$ and $\arg(a_H) = 0$. As for the initial condition,
we have taken $H_{\rm inf} = 10^{12}\GEV$, and put the $\phi$ field at
the minimum of the potential during the inflation. {}From
\FIG{FIG-LHu-numerical} we confirm that the analytic estimation
discussed above reproduces very well the result obtained by the
numerical calculation.

Finally, we comment on the possibility to avoid the early oscillation
caused by the thermal effects discussed in this section, which has been
recently pointed out in Ref.~\cite{Asaka}. The point is that, if the
scale of the inflation is very low, the Hubble parameter during the
inflation $H_{\rm inf}$ can be below the $H_{\rm osc}$ which is given by
\EQ{EQ-Hosc-final}. If $H_{\rm inf} < H_{\rm osc}$, our (implicit)
assumption that the Hubble parameter of the universe gradually decreases
from $H_{\rm inf}$ to $H_{\rm osc}$ breaks down. What happens in this
case is that the flat direction field $\phi$ starts its oscillation just
after the inflation ends, i.e., at $H\simeq H_{\rm inf}$. Thus lepton
asymmetry is also fixed at this time. The final baryon asymmetry is then
given by replacing $H_{\rm osc}$ in \EQ{EQ-nBs-final} with $H_{\rm
inf}$:
%%%
\begin{eqnarray}
 \left.
 \frac{n_B}{s}
 \right|_{H_{\rm inf} < H_{\rm osc}}
  &\simeq&
  0.029\times
  \frac{M_{\rm eff} T_R}{M_G^2}
  \left(
   \frac{m_{3/2}|a_m|}{H_{\rm inf}}
   \right)
   \delta_{\rm ph} 
   \,.
\end{eqnarray}
%%%
It was shown~\cite{Asaka} that $H_{\rm inf} \lsim H_{\rm osc}$ is
indeed satisfied for very low $H_{\rm inf}$ and high enough reheating
temperature $T_R$, e.g, $H_{\rm inf}\simeq 10^5\GEV$ and $T_R\gsim
10^7$--$10^8\GEV$. For example, in an extreme case where $H_{\rm
inf}\simeq m_{\phi}$ and $T_R\simeq (\pi^2 g_*/90)^{-1/4}\sqrt{M_G
H_{\rm inf}}$ (sudden reheating), baryon asymmetry is given
by~\cite{Asaka}
%%%
\begin{eqnarray}
 \frac{n_B}{s}\simeq 0.3\times 10^{-10}
  \left(
   \frac{10^{-4}\EV}{\mnu{1}}
   \right)
   \left(
    \frac{m_{3/2}}{H_{\rm inf}}
    \right)^{1/2}
    \left(
     \frac{m_{3/2}}{1\TEV}
     \right)^{1/2}
     |a_m|\delta_{\rm ph}
     \,.
\end{eqnarray}
%%%
Thus the empirical baryon asymmetry can be explained even with
$\mnu{1}\simeq 10^{-4}\EV$ in this extreme case.

%%%%%%%%%%%%%%%%%%%%%%%%%%%%%%%%%%%%%%%%%%%%%%%%%%%%%%%%%%%%
%%%%%%%%%%%%%%%%%%%%%%%%%%%%%%%%%%%%%%%%%%%%%%%%
%%%%%%%%%%%%%%%%%%%%%%%%%%%%%%%%%%%%%%%%%%%%%%%%%%%%%%%%%%%%

\clearpage
%%%%%%%%%%%%%%%%%%%%%%%%%%%%%%%%%%%%%%%%%%%%%%%%%%%%%%%%%%%%%%%%%%%
\section{Some models}%%%%%%%%%%%%%%%%%%%%%%%%%%%%%%%%%%%%%%%%%%%%%%
%%%%%%%%%%%%%%%%%%%%%%%%%%%%%%%%%%%%%%%%%%%%%%%%%%%%%%%%%%%%%%%%%%%
\label{SEC-somemodels}

In \SEC{SEC-resultinLHu} we have shown that the baryon asymmetry in
the present universe predicts the mass for the lightest neutrino in a
narrow region, $\mnu{1}\simeq (0.1$--$3)\times 10^{-9}\EV$. Together
with the neutrino masses required to solve the solar~\cite{Solar} and
atmospheric~\cite{SK-Atm} neutrino anomalies, this suggests a very
large mass hierarchy between the lightest and the heavier two
neutrinos.

Although the present leptogenesis mechanism via $L H_u$ direction does
not rely on the origin of the effective operator \EQ{EQ-LHu-start}, it
is natural to consider the seesaw mechanism~\cite{seesaw} as the origin
of that operator. In this case, the effective mass scale $M_{\rm eff}$
is related to the masses of the right-handed neutrinos $M_i$. Then, one
might wonder if the discussion so far is reliable, since successful
leptogenesis requires the effective mass scale $M_{\rm eff}$ to be far
beyond the Planck scale $M_G$, i.e., $M_{\rm eff} =
\vev{H_u}^2/\mnu{1}\simeq 3\times 10^{22}\GEV \times
(\mnu{1}/10^{-9}\EV)^{-1}$. Here, we first comment on this point. In the
framework of the seesaw mechanism, the neutrino masses are given by
\EQ{EQ-seesaw}:
%%%
\begin{eqnarray}
 \label{EQ-seesaw-again}
 \mnu{i} = -\sum_k
  \widehat{h}_{ki}
  \widehat{h}_{ki}
  \frac{\vev{H_u}^2}{M_k}
  \,,
\end{eqnarray}
where we have rotated the Yukawa coupling so that the neutrino mass
matrix becomes diagonal [see \SEC{SEC-LAfromN} and
\EQ{EQ-rtd-Ykw}]. Then, from \EQS{EQ-Meff-mnu} and
(\ref{EQ-seesaw-again}) we obtain
%%%
\begin{eqnarray}
 M_{\rm eff}
  \equiv
  M_{{\rm eff},1} = -
  \left(
   \sum_i
   \frac{
   \widehat{h}_{i1}
   \widehat{h}_{i1}
   }{M_i}
   \right)^{-1}
   \,.
\end{eqnarray}
%%%
Therefore, the large mass scale $M_{\rm eff}$ (or equivalently small
neutrino mass $\mnu{1}$) can be obtained by taking the Yukawa
couplings $\widehat{h}_{i1}$ to be small, keeping the masses of the
right-handed neutrinos $M_i$ to be below the Planck scale.

Nonetheless, the reader might still consider that the mass scale
$M_{\rm eff}$ is too large. In other words, the mass of the lightest
neutrino $\mnu{1}\simeq (0.1$--$3)\times 10^{-9}\EV$ might seem to be
too small, compared with the masses of the two heavier neutrinos,
$m_\nu\sim {\cal O}(10^{-3})$--${\cal O}(10^{-1})\EV$ suggested from
the recent neutrino oscillation
experiments~\cite{SK-Atm,Solar,KamLAND}. In this section, we propose
some models in which such an ultralight neutrino (or large mass scale
$M_{\rm eff}$) can be naturally implemented.

%%%%%%%%%%%%%%%%%%%%%%%%%%%%%%%%%%%%%%%%%%%%%%%%%%%%%%%%%%%%%%%%%%%
\subsection{Model for the ultralight neutrino with a broken discrete
symmetry}%%%%%%%%%%%%%%%%%%%%%%%%%%%%%%%%%%%%%%%%%%%%%%%%%%%%%%%%%%
%%%%%%%%%%%%%%%%%%%%%%%%%%%%%%%%%%%%%%%%%%%%%%%%%%%%%%%%%%%%%%%%%%%
\label{SEC-model-FN}

Here, we show an explicit model based on a Froggatt-Nielsen (FN)
mechanism~\cite{FN}, in which the required large mass hierarchy is
naturally obtained. We adopt a discrete $Z_6$ as the FN symmetry instead
of a continuous ${\rm U}(1)_{\rm FN}$. We see that the discrete symmetry
is crucial to produce the required large mass hierarchy in the neutrino
sector. To see this, let us first recall the case of the continuous
${\rm U}(1)_{\rm FN}$, which is described in \SEC{SEC-FN}. In that case,
the mass matrix of the neutrino in \EQ{EQ-mnu-FN} gives rise to a mild
hierarchy
%%%
\begin{eqnarray} 
 \mnu{1}:\mnu{2}:\mnu{3} 
  \sim
  \varepsilon^2 :1:1
  =
  {\cal O}(10^{-2}) : {\cal O}(1): {\cal O}(1)
  \,,
\end{eqnarray}
%%%
and hence it cannot explain the large hierarchy required in the present
scenario. To change the above point, we suppose that the broken FN
symmetry is not a ${\rm U}(1)_{\rm FN}$ but a discrete symmetry $Z_n$
with $n = 2 d$.  Then, the mass matrix for the right-handed neutrino
$M_R$ changes into the following form;
%%%
\begin{eqnarray}
 M_R =
  M_0
  \left(
   \begin{array}{lll}
    \widetilde{\xi_{11}}
     & \widetilde{\xi_{12}}\,\varepsilon^{c+d} 
     & \widetilde{\xi_{13}}\,\varepsilon^{b+d}
     \\
    \widetilde{\xi_{12}}\,\varepsilon^{c+d}
     & \widetilde{\xi_{22}}\,\varepsilon^{2 c}
     & \widetilde{\xi_{23}}\,\varepsilon^{b+c}
     \\
    \widetilde{\xi_{13}}\,\varepsilon^{b+d}
     & \widetilde{\xi_{23}}\,\varepsilon^{b+c}
     & \widetilde{\xi_{33}}\,\varepsilon^{2 b}
   \end{array}
   \right)\,.
   \label{matMR}
\end{eqnarray}
%%%
Here, we have assumed $d ( = n/2) > c \ge b \ge 0$. Notice that the
Majorana mass for $N_1$ is no longer suppressed by the power of
$\varepsilon$, which is a basic point to yield an extremely small
neutrino mass $\mnu{1}$. Although the mass matrix $M_R$ is modified, the
structure of the neutrino mass matrix looks similar to the original one:
%%%
\begin{eqnarray}
 \label{EQ-mnu-Z6}
 m_{\nu}  &=&
 m_D^T \frac{1}{M_R} m_D
 \nonumber \\
 &=&
  \frac{\varepsilon^{2a} \vev{H_u}^2}{M_0}
  \left(
   \begin{array}{ccc}
    \varepsilon & 0 & 0
     \\
    0 & 1 & 0
     \\
    0 & 0 & 1
   \end{array}
   \right)
   \left(
    \,\{\widetilde{y_{ij}}\}\,
    \right)^T
    \left(
     \begin{array}{ccc}
      \widetilde{\xi_{11}}\,\varepsilon^{-2d}
       & \widetilde{\xi_{12}}
       & \widetilde{\xi_{13}}
       \\
      \widetilde{\xi_{12}}
       & \widetilde{\xi_{22}}
       & \widetilde{\xi_{23}}
       \\
      \widetilde{\xi_{13}}
       & \widetilde{\xi_{23}}
       & \widetilde{\xi_{33}}
     \end{array}
     \right)^{-1}
     \left(
      \,\{\widetilde{y_{ij}}\}\,
      \right)
      \left(
       \begin{array}{ccc}
	\varepsilon & 0 & 0
	 \\
	0 & 1 & 0
	 \\
	0 & 0 & 1
       \end{array}
       \right)
       \nonumber \\
 &=&
  \frac{\varepsilon^{2a} \vev{H_u}^2}{M_0}
  \left(
   \begin{array}{ccc}
    {\cal O}(\varepsilon^2)
     & {\cal O}(\varepsilon)
     & {\cal O}(\varepsilon)
     \\
    {\cal O}(\varepsilon) 
     & {\cal O}(1) 
     & {\cal O}(1)
     \\
    {\cal O}(\varepsilon)
     & {\cal O}(1)
     & {\cal O}(1)
   \end{array}
   \right)\,.
\end{eqnarray}
%%%
However, in spite of the large components of the mass matrix, we see
that one of the mass eigenvalue of this mass matrix strongly suppressed
as $\sim \varepsilon^{2(1+d)}$. (This suppression is also understood
directly by taking the determinant of the above mass matrix.) Thus, the
required large mass hierarchy between $\mnu{1}$ and $\mnu{2,3}$ can be
naturally obtained by taking $d=3$ ($Z_6$).

Notice that the neutrino mass matrix \EQ{EQ-mnu-Z6} can still lead to a
large $\nu_{\mu}$--$\nu_{\tau}$ mixing angle. The mass hierarchy of the
charged leptons also remains unchanged.\footnote{This is because the
charges of the Yukawa couplings of the charged leptons are smaller than
$n=6$, $Q(\overline{E}_i)+Q(L_j) \le(a+3) < 6$.}  Hence, this model
preserves all the desirable features in the continuous ${\rm U}(1)_{\rm
FN}$ model, explaining the mass of the ultralight neutrino $\mnu{1}$ as
well.

%%%%%%%%%%%%%%%%%%%%%%%%%%%%%%%%%%%%%%%%%%%%%%%%%%%%%%%%%%%%
\begin{figure}[t]%%%%%%%%%%%%%%%%%%%%%%%%%%%%%%%%%%%%%%%%%%%
%%%%%%%%%%%%%%%%%%%%%%%%%%%%%%%%%%%%%%%%%%%%%%%%%%%%%%%%%%%%
  \psfrag{r}[r][l][1.0][90]{$r = \displaystyle\frac{\delta m^2_{\rm sol}}
  {\delta m^2_{\rm atm}}$}
  \psfrag{mn1}[t][b][1.0][0]{$\mnu{1}\,\,[\EV ~]$}
  \centerline{ \scalebox{1.0}{\includegraphics{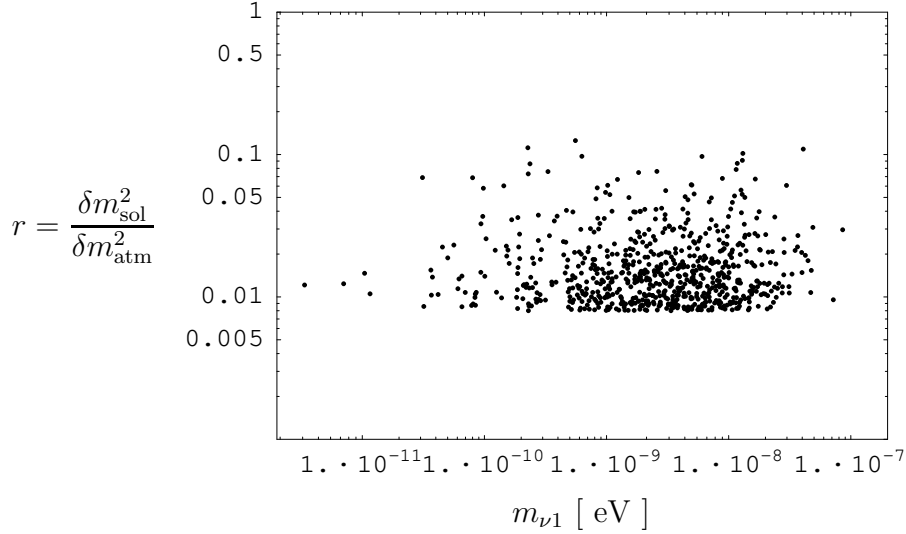}} }
  \caption{The plot for $r = \delta m^2_{\rm sol} /\delta m^2_{\rm atm}$
    and $\mnu{1}$ in a Froggatt-Nielsen model with a discrete $Z_6$
    symmetry.}
  \label{FIG-mnu1}
%%%%%%%%%%%%%%%%%%%%%%%%%%%%%%%%%%%%%%%%%%%%%%%%%%%%%%%%%%%%
\end{figure}%%%%%%%%%%%%%%%%%%%%%%%%%%%%%%%%%%%%%%%%%%%%%%%%
%%%%%%%%%%%%%%%%%%%%%%%%%%%%%%%%%%%%%%%%%%%%%%%%%%%%%%%%%%%%

To demonstrate our point, we randomly generate ${\cal O}(1)$ couplings
$\widetilde{y_{ij}}$ and $\widetilde{\xi_{ij}}$.  Namely we calculate
the mass matrix for neutrinos, taking the magnitudes of the couplings
$\widetilde{y_{ij}}$ and $\widetilde{\xi_{ij}}$ to be in a range $0.5$
-- $1.5$ and their phases to be $0$ -- $2\pi$.  We also take
$\varepsilon = 0.05$ -- $0.1$ randomly.\footnote{A similar calculation
was done in Ref.~\cite{SatoYana-paper}, where they adopted the
continuous U$(1)_{\rm FN}$ model.}  From the obtained neutrino mass
matrix, we calculated the mass eigenstates $\mnu{i}$ and the mixing
matrix $U_{\alpha i}$,\footnote{The matrix $U_{\alpha i}$ is defined
in \SEC{SEC-LAfromN}.} and then required the parameters $\delta
m^2_{\rm sol}\equiv \mnu{3}^2- \mnu{2}^2$, $\delta m^2_{\rm atm} =
\mnu{2}^2- \mnu{1}^2$, $\sin^2 2\theta_{\rm atm}\equiv 4|U_{\mu
3}|^2\left(1 - |U_{\mu 3}|^2 \right)$, and $\tan^2\theta_{\rm
sol}\equiv |U_{e2}/U_{e1}|^2$ to be consistent with the data of
neutrino-oscillation experiments~\cite{SK-Atm,Solar,KamLAND}. Here, we
take large-angle MSW solution to the solar neutrino deficits, since it
is the only solution allowed by the recently announced KamLAND
results~\cite{KamLAND}, and adopt the following
constraints~\cite{SK-Atm,PostKam}:
%%%
\begin{eqnarray}
  &&0.9\le \sin^2 2\theta_{\rm atm} \le 1.0
  \,,
  \nonumber\\
  && r\equiv \delta m^2_{\rm sol}/\delta m^2_{\rm atm}
  =
  0.008 - 0.15
  \nonumber\\
  &&
  \tan^2\theta_{\rm sol}
  =
  0.2 - 0.9
  \,.
\end{eqnarray}
%%%
We also required
%%%
\begin{eqnarray}
 |U_{e3}| < 0.15\,,
\end{eqnarray}
%%%
to satisfy CHOOZ limit~\cite{CHOOZ}.  \FIG{FIG-mnu1} shows the obtained
mass of the lightest neutrino, $\mnu{1}$. We can see that an ultralight
neutrino of mass $\mnu{1}\simeq (0.1$--$3) \times 10^{-9}$ eV is
naturally obtained.

%%%%%%%%%%%%%%%%%%%%%%%%%%%%%%%%%%%%%%%%%%%%%%%%%%%%%%%%%%%%%%%%%%%
\subsection{Large effective masses for right-handed neutrinos}%%%%%
%%%%%%%%%%%%%%%%%%%%%%%%%%%%%%%%%%%%%%%%%%%%%%%%%%%%%%%%%%%%%%%%%%%
\label{SEC-model-PQ}

In \SEC{SEC-model-FN}, we have shown a model in which an ultralight
neutrino with mass $\mnu{1}\simeq (0.1$--$3) \times 10^{-9}$ can be
naturally implemented. Here, we discuss another possibility. The point
is that the large effective mass scale $M_{\rm eff}$ is required {\it
only during the leptogenesis}, not necessarily at the true
vacuum. Therefore, if the effective masses of the right-handed neutrinos
in the early universe are dynamical values and they are different from
the values in the true vacuum, we might naturally obtain a large
effective mass scale $M_{\rm eff}$ during the leptogenesis.

Suppose that the right-handed neutrino masses $M_i$ are given by a
vacuum expectation value of some field $X$ with a superpotential
\begin{eqnarray}
 W = \frac{1}{2}\xi_i  X\,\, N_i N_i
 \,,
\end{eqnarray}
which gives $M_i = \xi_i\vev{X}$. If $\vev{X}$ takes a large value
during the leptogenesis, therefore, $M_i$ can be much larger than the
values in the true vacuum. To demonstrate this point, let us consider
that the field $X$ is responsible to the Peccei-Quinn symmetry
breaking~\cite{PQ}. Then, as we will see, there automatically exists a
flat direction containing the $X$ field. The $X$ field can have a large
value as $X \simeq M_G$ during the inflation whereas it has a vacuum
expectation value of $\vev{X} \simeq F_a$ in the true vacuum. Here, the
Peccei-Quinn breaking scale $F_a$ is constrained by laboratory
experiments, astrophysics, and cosmology as $F_a \simeq
10^{10}$--$10^{12}\GEV$~\cite{axion-Kim}.

We consider the following superpotential for the Peccei-Quinn symmetry
breaking sector:
\begin{eqnarray}
 \label{EQ-PQ-breaking}
 W &=& \lambda Y ( X \overline{X} - F_a^2 )
 \,,
\end{eqnarray}
where $\lambda$ is a coupling constant, and $Y$, $X$, and $\overline{X}$
are supermultiplets which are singlets under the standard-model gauge
groups, which have $0$, $+1$, and $-1$ Peccei-Quinn charges,
respectively.  From \EQ{EQ-PQ-breaking} we see that there exists a flat
direction $X \overline{X} = F_a^2$. We parameterize this direction by
the scalar field $\sigma$, which is called ``saxion''.  This saxion
$\sigma$ receives a soft SUSY breaking mass of $m_\sigma \simeq
m_{3/2}$, and the $X$ and $\overline X$ have vacuum expectation values
of the order of $F_a$.\footnote{We assume here that the soft SUSY
breaking masses for $X$ and $\overline{X}$ are almost the same, i.e.,
$m_X \simeq m_{\overline X}$ ($\simeq m_{3/2}$).}

Along this flat direction the $X$ field can have a large value during
the inflation, if the additional SUSY breaking effects during the
inflation induce a negative mass squared for $X$ (see
\SEC{SEC-Initial}). The initial value of $X$ is then expected to be $X
\simeq M_G$, since the supergravity effects prevent the $X$ field from
running over $M_G$ [see \EQ{EQ-SUGRA}], while the initial value of the
$\overline{X}$ field is given by $\overline{X} \simeq F_a^2/M_G$.

In order to fix the mass scale $M_{\rm eff}$ at the true vacuum, let us
take the mass of the lightest neutrino to be $\mnu{1} = 1\times
10^{-4}\EV$ for a representative value, with which the hierarchy of the
neutrino masses becomes mild. Then the scale $M_{\rm eff}$ is given by
%%%
\begin{eqnarray}
 M_{\rm eff} 
  &\simeq&
  3 \times 10^{17}  \GEV
  \left( \frac{ 10^{-4} \EV }{ \mnu{1} } \right)
  \left( \frac{ \vev{X} }{ F_a } \right)
  \,.
\end{eqnarray}
Here, we have used $M_{\rm eff}\propto \vev{X}$ and $M_{\rm eff} =
\vev{H_u}^2/\mnu{1}$ at the true vacuum $\vev{X} \simeq F_a$. Although
this scale $M_{\rm eff}$ at the true vacuum is too low to produce the
enough baryon asymmetry (see \FIG{FIG-BA-LHu}), the large initial value
of $X$, $X_0\simeq M_G$ gives rise to
%%%
\begin{eqnarray}
 \label{EQ-Meff-init}
  M_{\rm eff}
  =
  7 \times 10^{25}  \GEV
  \left( \frac{ 10^{-4} \EV}{ {m_\nu}_1 } \right)
  \left( \frac{ 10^{10} \GEV }{ F_a } \right)
  \left( \frac{ X_0 }{ M_G } \right)
  \,.
\end{eqnarray}
For such a high scale $M_{\rm eff}$, the evolution of the $\phi$ field
is completely free from the thermal effects and the early oscillation
can be avoided, even if the reheating temperature $T_R$ is as high as
$T_R\simeq 10^8\GEV$.\footnote{This can be seen by calculating the
oscillation time $H_{\rm osc}$ in \EQ{EQ-Hosc-final} with $M_{\rm
eff}\simeq 10^{26}\GEV$.}

The baryon asymmetry is then obtained by using \EQS{EQ-nBs-final} and
(\ref{EQ-Meff-init}) as
%%%
\begin{eqnarray}
 \label{EQ-nBs-PQ}
 \frac{n_B}{s} &=& 
  4 \times 10^{-5}
  \,
  \delta_{\rm ph}
  \,
  \left( \frac{ T_R }{10^8 \GEV } \right)
  \left( \frac{ 10^{-4} \EV }{ {m_\nu}_1 } \right)
  \left( \frac{ 10^{10} \GEV }{ F_a } \right)
  \left( \frac{ X_0 }{ M_G } \right)
  \,, 
\end{eqnarray}
where we have used $H_{\rm osc}\simeq m_{\phi}\simeq m_{3/2}|a_m|$.
This shows that too large amount of the lepton asymmetry is
produced. However, this asymmetry is sufficiently diluted, since there
exists substantial entropy production by the decay of the saxion
$\sigma$.

The saxion begins the coherent oscillation with the initial amplitude
$X_0 \simeq M_G$ at $H \simeq m_\sigma \simeq m_{3/2}$, just after
the production of the lepton asymmetry ends.  The oscillation energy at
that time is given by $\rho_\sigma \simeq m_\sigma^2 X_0^2 /2$.  When
the reheating process of the inflation completes, the ratio of
$\rho_\sigma$ to the energy density $\rho_{\rm rad}$ of the radiation of
the universe is estimated as
\begin{eqnarray}
 \frac{ \rho_\sigma }{ \rho_{\rm rad} }
  \simeq
  \frac{ 1 }{ 6 }
  \left( \frac{ X_0 }{ M_G }\right)^2
  \,.
\end{eqnarray}
Notice that $\rho_\sigma$ decreases at the rate $R^{-3}$ as the universe
expands, while $\rho_{\rm rad}$ decreases as $R^{-4}$.  Therefore, the
oscillation energy of the saxion dominates the energy of the universe
soon after the reheating process completes.  Here, it is useful to take
the ratio of $\rho_\sigma$ to the entropy density $s$,
%%%
\begin{eqnarray}
 \frac{\rho_\sigma}{s}
  \simeq
  \frac{1}{8}
  T_R
  \left( \frac{ X_0 }{ M_G }\right)^2
  \,,
\end{eqnarray}
%%%
which takes a constant value until the saxion decays.

This energy density of the saxion is transferred into the thermal bath
when it decays. Here, it should be noted that the saxion might decay
dominantly into two axions. If it is the case, the extra energy of the
axion at the BBN epoch raises the Hubble expansion of the universe,
which leads to overproduction of $^4$He.  To avoid this difficulty, the
branching ratio of the saxion decay into two axion should be smaller
than about 0.1.  Here, we simply assume this is the case.\footnote{ This
is realized when $m_X^2 \simeq m_{\overline{X}}^2$.} Then, the saxion
decays dominantly into two gluons with the decay rate
%%%
\begin{eqnarray}
 \Gamma (\sigma \rightarrow 2g )
  =
  \frac{ \alpha_s^2 }{ 32 \pi^3 }
  \frac{ m_\sigma^3 }{ F_a^2 }
  \,.
\end{eqnarray}
%%%
Through this decay the universe is ``reheated'' again, and its reheating
temperature $T_\sigma$ is estimated as
\begin{eqnarray}
 \label{EQ-Tsigma}
  T_\sigma =
  10  \GEV
  \left( \frac{ m_\sigma}{1 \TEV}\right)^{3/2}
  \left( \frac{ 10^{10} \GEV}{ F_a } \right)
  \,.
\end{eqnarray}
Notice that the saxion decay takes place far before the beginning of the
Big-Bang Nucleosynthesis, and hence is cosmologically harmless.

The saxion decay increases the entropy of the universe by the rate
%%%
\begin{eqnarray}
 \label{EQ-entropy-PQ}
 \Delta
  \equiv
  \frac{s_{\rm after}}{s_{\rm before}}
  &=&
  \frac{\rho_\sigma}{s_{\rm before}}
  \times
  \frac{s_{\rm after}}{\rho_\sigma}
  \nonumber\\
 &=&
  \frac{1}{8}
  T_R
  \left( \frac{ X_0 }{ M_G }\right)^2
  \times
  \frac{4}{3 T_\sigma}
  \nonumber \\
 &=&
  2 \times 10^6 
  \left( \frac{ T_R }{10^8 \GEV } \right)
  \left( \frac{ 1 \TEV }{ m_\sigma } \right)^{3/2}
  \left( \frac{ F_a }{ 10^{10} \GEV } \right)
  \left( \frac{ X_0 }{ M_G} \right)^2
  \,.
\end{eqnarray}
%%%
Because of this entropy production by the saxion decay, the primordial
baryon asymmetry \EQ{EQ-nBs-PQ} is also diluted by the rate $\Delta$.
Then the present baryon asymmetry is given by
\begin{eqnarray}
 \frac{ n_B }{s}
  &=&
  0.2 \times 10^{-10}
  \left( \frac{ 10^{-4} \EV }{ \mnu{1} } \right)
  \left( \frac{ 10^{10} \GEV }{ F_a } \right)^2
  \left( \frac{ m_\sigma }{ 1 \TEV } \right)^{3/2}
  \left( \frac{ M_G }{ X_0 } \right)
  \delta_{\rm ph}
  \,.
\end{eqnarray}
Notice that the present baryon asymmetry is independent of the reheating
temperature $T_R$, while it depends on the Peccei-Quinn scale $F_a$. We
see that the desired baryon asymmetry is obtained even with the lightest
neutrino mass of $\mnu{1} \simeq 10^{-4}\EV$ for $F_a \simeq
10^{10}\GEV$.

Finally, we comment on the cosmological consequences of this model.

{\bf (i)} The entropy production by the saxion in \EQ{EQ-entropy-PQ}
ensures that we are free from the cosmological gravitino problems, since
the number density of gravitinos produced at the reheating process is
diluted by the rate $\Delta$.

{\bf (ii)} This model does not suffer from the cosmological problem of
the ``axino'', which is a fermionic superpartner of the axion.  The
axinos are produced in the reheating process by the thermal scatterings
and may lead to a cosmological difficulty~\cite{axino-problem}.
However, in this model, the interaction of axino at the reheating epoch
is suppressed by $X_0 \simeq M_G$, not by $F_a$, and hence the
production of axino is less effective.  Furthermore, the entropy
production by the saxion decay dilutes the axino abundance. Thus, the
axino becomes completely cosmologically harmless in this model.

{\bf (iii)} The reheating temperature of the saxion decay \EQ{EQ-Tsigma}
is high enough to thermalize the lightest SUSY particles (LSPs) if the
mass of the LSP is less than $200\GEV$, and hence the stable LSP can be
a dominant component of the dark matter.

\clearpage
%%%%%%%%%%%%%%%%%%%%%%%%%%%%%%%%%%%%%%%%%%%%%%%%%%%%%%%%%%%%%%%%%%%
\section{Effects of the gauged ${\rm U}(1)_{B-L}$ symmetry}%%%%%%%%%%
%%%%%%%%%%%%%%%%%%%%%%%%%%%%%%%%%%%%%%%%%%%%%%%%%%%%%%%%%%%%%%%%%%%
\label{SEC-withB-L}

In \SEC{SEC-LHu}, we found a very interesting aspect of the leptogenesis
via $L H_u$ flat direction, the ``reheating temperature independence of
the cosmological baryon asymmetry.'' The resultant baryon asymmetry is
almost determined by the mass of the lightest neutrino (and the
gravitino mass), and hence we can predict the mass of the lightest
neutrino $\mnu{1}$ as $\mnu{1}\simeq 10^{-9}\EV$ from the observed
baryon asymmetry.

In this section, we discuss the effects of gauging the ${\rm
U}(1)_{B-L}$ symmetry on the leptogenesis via the $L H_u$ flat
direction. As mentioned in \SEC{SEC-1-1}, the absence of ${\rm
U}(1)_{B-L}$ gauge anomaly automatically requires three families of
right-handed neutrinos, and the breaking of $B-L$ symmetry naturally
gives rise to the heavy Majorana masses of these right-handed neutrinos, 
which explains the tiny neutrino masses via the seesaw mechanism.

Below the $B-L$ breaking scale, the behavior of the scalar potential of
the $\phi$ field does not change much from the one discussed in
\SEC{SEC-LHu}. Therefore, if the $B-L$ breaking scale $v$ is larger than
the initial value of $\phi$ field obtained in \SEC{SEC-Initial},
%%%
\begin{eqnarray}
 \label{EQ-F-stop}
  v \gsim \sqrt{M_{\rm eff} H_{\rm inf}}
  \,,
\end{eqnarray}
%%%
the amplitude of the $\phi$ field is always smaller than the $B-L$
breaking scale, and the dynamics of the $\phi$ field is basically the
same as the one discussed in \SEC{SEC-LHu}. Hence, the resultant baryon
asymmetry reduces to the one obtained in \SEC{SEC-resultinLHu} [see
\EQ{EQ-nBs-final} and \FIG{FIG-LHu-numerical}]. We will call this case
``$F$-term stopping case.''

On the other hand, if $v\lsim \sqrt{M_{\rm eff}H_{\rm inf}}$, the
$\phi$ field develops its initial value as large as the $B-L$ breaking
scale. At this scale, the potential of the flat direction $\phi$ can
be lifted by the effect of the ${\rm U}(1)_{B-L}$
$D$-term~\cite{FHY-ADBL}.  (We denote this case by ``$D$-term stopping
case.'') As we shall show in this section, the dynamics of the $\phi$
field is drastically changed in this $D$-term stopping case. Actually,
the obtained baryon asymmetry is much larger than the case without a
${\rm U}(1)_{B-L}$ symmetry, and hence than the $F$-term stopping
case.

It is interesting that the resultant baryon asymmetry in the $D$-term
stopping case linearly depends on the reheating temperatures of
inflation, but it is completely independent of the gravitino mass. This
is totally an opposite situation to the result obtained in
\SEC{SEC-LHu}, in which the final baryon asymmetry is almost independent
of the reheating temperatures of inflation and linearly depends on the
gravitino mass [see \EQ{EQ-nBs-final} and \FIG{FIG-LHu-numerical}]. This
``gravitino-mass independence of the cosmological baryon asymmetry''
provides us a great advantage in gauge-mediated SUSY breaking
scenarios~\cite{GMSB}.

\subsection{The model}

In order to discuss the effect of the ${\rm U}(1)_{B-L}$ gauge symmetry,
we should investigate the $B-L$ breaking sector, which gives the masses
of the right-handed neutrinos. To demonstrate our point, we use the
following superpotential:
\begin{eqnarray}
 \label{EQ-super-BL}
  W
  =
  h N L H_u
  +
  \frac{1}{2}\xi S N N
  +
  \lambda X \left(S\bar{S} - v^2\right)
  +
  \mu H_{u}H_{d}
  \,,
\end{eqnarray}
where, $h$, $\xi$, $\lambda$ are coupling constants and we assume
$\lambda={\cal{O}}(1)$.  $X$, $S$ and $\bar{S}$ are singlets under the
MSSM gauge groups and they carry the ${\rm U}(1)_{B-L}$ charges of $0$,
$-2$ and $+2$, respectively. $v$ is the the breaking scale of the ${\rm
U}(1)_{B-L}$ symmetry. Here and hereafter, we take the couplings $h$,
$\xi$, $\lambda$ and $\mu$, $v$ to be real, by field redefinitions.  We
will omit the family indices, since the ``effectively flat'' direction
relevant for the present baryon asymmetry is again given by the flattest
flat direction, $L = L_1$.

The shape of the superpotential in \EQ{EQ-super-BL} is the same as
that of the SUSY hybrid inflation discussed in
\SEC{SEC-Hybrid-1}.\footnote{The superpotential in \EQ{EQ-super-BL} is
also similar to the one in the model in \SEC{SEC-model-PQ}. However,
the model adopted here is different from that, since the ${\rm U}(1)$
symmetry here is gauged, while in \SEC{SEC-model-PQ} the Peccei-Quinn
${\rm U}(1)$ symmetry is global.}  Actually, If the $X$ field in
\EQ{EQ-super-BL} develops a large expectation value and the energy
density of the universe is dominated by the vacuum energy $V\simeq
\lambda^2 v^4$, hybrid inflation takes place. Here, however, we assume
that the $B-L$ breaking sector in \EQ{EQ-super-BL} is not directly
related to the inflation, and discuss the effects of this ${\rm
U}(1)_{B-L}$ on the dynamics of the $L H_u$ flat direction.

As we have shown in \SEC{SEC-Initial}, the K\"ahler potential must have
non-minimal couplings of the $\phi$ field to the inflaton $\chi$, since
otherwise the $\phi$ field gets a large positive mass term of the order
of the Hubble parameter and it is driven exponentially towards the
origin during inflation~\cite{DRT}. We assume that there are also
non-minimal couplings of other fields to the inflaton in the K\"ahler
potential:
%%%
\begin{eqnarray}
 K &=& 
  \sum_Y Y^{\dagger}Y + \chi^{\dagger}\chi
  \nonumber\\
 &+&
  \sum_Y
  \left(
   \frac{c_Y}{M_G}
   \chi
   Y^{\dagger}Y
   + {\rm H.c.}
   \right)
   +
   \sum_Y
   \frac{b_{Y}}{M_G^2}
   \chi^{\dagger}\chi
   Y^{\dagger}Y
   +
   \cdots
   \,,
\end{eqnarray}
%%%
where $c_Y$ and $b_Y$ are complex and real couplings of order unity,
respectively, and $Y$ denotes $L$, $H_u$, $H_d$, $X$, $S$, $\bar{S}$,
and $N$. Then, the full scalar potential relevant to the flat direction
field $L = H_u = \phi/\sqrt{2}$ is given by~\footnote{Here, we take
$H_d\simeq 0$ assuming a positive Hubble-order mass term for the $H_d$
field. This is necessary in order to avoid a contamination of $H_d$ to
the $LH_u$ flat direction. We have checked both analytically and
numerically that the contamination of the $H_d$ field becomes relevant
only after the $H\lsim \mu$, and hence the following discussion is not
affected much by the $H_d$ contamination.}
%%%
\begin{eqnarray}
 \label{EQ-full-V}
 V
 &=&
 \frac{1}{2}g^2
 \left(
  -2 |S|^2 +2|\bar{S}|^2 +|N|^2 -\frac{1}{2}|\phi|^2
  \right)^2
  \nonumber\\
 &+&
  \frac{1}{2}\mu^2 |\phi|^2
  +\left|\frac{1}{2}h \phi^2 + \xi S N \right|^2
  +h^2|N \phi|^2
  \nonumber\\
 &+&
  \left|
   \frac{1}{2}\xi N^2 +\lambda X \bar{S}
   \right|^2
   +\lambda^2 \left|S \bar{S}-v^2 \right|^2 
   +\lambda^2| X S|^2
   \nonumber\\
 &+&
  3 H^2
  \left(\sum_Y(1-b'_Y)|Y|^2\right)
  -\sqrt{3}H \left(\sum_Y c_Y Y W_Y+{\rm H.c.}\right)
  +V_{\rm SB}
  \,,
\end{eqnarray}
%%%
where $b_Y'\equiv b_Y-|c_Y|^2$ and $W_Y\equiv \partial W / \partial Y$,
and we redefined the phase of $c_Y$.  $g={\cal{O}}(1)$ is the gauge
coupling constant of the ${\rm U}(1)_{B-L}$. $V_{\rm SB}$ represents
SUSY breaking terms in the true vacuum.  All fields denote the scalar
components of the corresponding superfields. Here, we omit the potential
coming from thermal effects, which will be discussed in
\SEC{SEC-BL-condition}.

The scalar potential in \EQ{EQ-full-V} is so complicated that it seems
very difficult to solve the dynamics of the relevant
fields. Fortunately, however, we only need to know the shape around the
minimum of this potential, if the curvatures around the bottom of the
potential are as large as the Hubble parameter during inflation. This is
because all the scalar fields, which have masses as large as the Hubble
parameter, settle down at the bottom of the potential during inflation
and trace this potential minimum in the subsequent evolution. Therefore,
we first find out the relations to minimize the potential in
\EQ{EQ-full-V}.  Though this is also a very hard task because of the
complexity of the potential, one can find out the minimum of the
potential at least when $|\phi|\lsim v$ and $H\lsim v$ are satisfied.
{}From \EQ{EQ-full-V}, we find the approximate minimum of $F$-terms for
$|\phi|,\, H\lsim v$,
%%%
\begin{eqnarray}
 \bar{S}\simeq \frac{v^2}{S}\,,
  \quad 
  X\simeq 0\,,
  \quad 
  N\simeq -\frac{h \phi^2}{2\xi S}\,.
\label{EQ-minima}
\end{eqnarray}
%%%
Notice that the curvatures associated with these three relations are of
the order of the $B-L$ breaking scale $v$, and hence these relations
hold as far as $H\lsim v$. By using these relations, we can reduce the
scalar potential in \EQ{EQ-full-V} to the following one:
%%%
\begin{eqnarray}
 \label{EQ-VBL-prefinal}
 V
  &\simeq&
  \frac{1}{2}g^2
  \left(
   -2|S|^2
   +2\frac{v^4}{|S|^2}
   +\frac{h^2}{4 \xi^2}\frac{|\phi|^4}{|S|^2}
   -\frac{1}{2}|\phi|^2
   \right)^2
   +
   \frac{h^4}{4\xi^2}\frac{|\phi|^6}{|S|^2}
   +
   \frac{h^4}{64\xi^2}\frac{|\phi|^8}{|S|^4}
   \nonumber\\
 &-&
  H^2|\phi|^2
  +3(1-b_S')H^2|S|^2
  +3(1-b_{\bar{S}}')H^2\frac{v^4}{|S|^2}
  +3(1-b_N')H^2\frac{h^2}{4\xi^2}\frac{|\phi|^4}{|S|^2}
  \nonumber\\
 &+&
  \frac{\sqrt{3}}{2}H 
  \left(
   c_{\phi}'\frac{h^2}{\xi}\frac{\phi^4}{S}+{\rm H.c.}
   \right)
   +\frac{1}{2}\mu^2|\phi|^2
   +V_{\rm SB}
   \,,
\end{eqnarray}
where $c_{\phi}'\equiv c_{\phi}-(1/4)c_S$. Here and hereafter, we assume 
$3(1-b_\phi')\simeq -1$ for simplicity. (This corresponds to $c_H\simeq
1$ in the previous analysis. See \EQ{EQ-VH}.)

One may wonder why the ``flat'' direction field $\phi$ can develop a
large expectation value in spite of the presence of the ${\rm
U}(1)_{B-L}$ $D$-term.  This is because the $S$ field shifts and absorbs
the $D$-term potential:
%%%
\begin{eqnarray}
 |S|^2
  &\simeq&
  -\frac{1}{8}|\phi|^2
  +\sqrt{v^4+\left(\frac{1}{64}+\frac{h^2}{8\xi^2}\right)|\phi|^4}
  \,.
\label{EQ-S-phi}
\end{eqnarray}
%%%
The curvature associated with this relation is also of the order of the
$B-L$ breaking scale. Then we get the following effective potential:
%%%
\begin{eqnarray}
 V
  &\simeq&
  V_{\rm SB}
  +
  \frac{1}{2}\mu^2|\phi|^2
  +
  \frac{\sqrt{3}}{2}H
  \left(c_{\phi}'\frac{h^2}{\xi}\frac{\phi^4}{S}+ {\rm H.c.}\right)
  +\frac{h^4}{4\xi^2}\frac{|\phi|^6}{|S|^2}
  +{\cal O}
  \left(\frac{|\phi|^8}{|S|^4}\right)
  \nonumber\\
 &-&
  H^2|\phi|^2
  +3(1-b_S')H^2|S|^2
  +3(1-b_{\bar{S}}')H^2  \frac{v^4}{|S|^2}
  +3(1-b_N')H^2\frac{h^2}{4\xi^2}\frac{|\phi|^4}{|S|^2}
  \,.
  \nonumber\\
  \label{EQ-Vfinal}
\end{eqnarray}
%%%

In the following discussion, we assume $H\lsim v$ and that the relations
in \EQ{EQ-minima} and in \EQ{EQ-S-phi} are satisfied during the
leptogenesis works. Although the approximation we have made here is
crude, we will justify by a numerical calculation that the effective
potential in \EQ{EQ-Vfinal} is valid at least for the following
discussion.

As we will see later, the third term in \EQ{EQ-Vfinal},
\begin{eqnarray}
\label{EQ-HA}
 V_{\rm phase}
  &=&
  \frac{\sqrt{3}}{2}
  H
  \left(
   c_{\phi}'\frac{h^2}{\xi}
   \frac{\phi^4}{S} + {\rm H.c.}
   \right)
   \,,
\end{eqnarray}
which depends on the phase of $\phi$, plays a crucial role in the
$D$-term stopping case.  We will call this term ``Hubble $A$-term.'' 
(This corresponds to the Hubble $A$-term in \EQ{EQ-VH} in the previous
analysis.)

\subsection{Initial amplitude and initial phase}

In this subsection, we discuss the evolution of the $\phi$ field during
the inflation. 

\subsubsection{initial amplitude}

First, we discuss the amplitude of the $\phi$ field. Let us first
investigate the potential in the range $|\phi|\lsim v$.  In this range,
the $S$ field in \EQ{EQ-S-phi} is expanded into the following form:
%%%
\begin{eqnarray}
 |S|^2
  &\simeq&
  v^2
  -\frac{1}{8}|\phi|^2
  +{\cal O}
  \left(\frac{|\phi|^4}{v^2}\right)\,,
  \qquad {\rm for} \quad |\phi|\lsim v
  \,.
  \label{EQ-s-expand}
\end{eqnarray}
%%%
By substituting \EQ{EQ-s-expand} into \EQ{EQ-Vfinal}, we get the
effective potential for $\phi \lsim v$ in the following form:
%%%
\begin{eqnarray}
 V
  &\simeq& 
  V_{\rm SB}
  +\frac{1}{2}\mu^2|\phi|^2
  +\frac{\sqrt{3}}{2}\frac{H}{M_{\rm eff}}
  \left( c_{\phi}'\phi^4+{\rm H.c.} \right)
  +\frac{|\phi|^6}{4M_{\rm eff}^2}
  \nonumber\\
 &-&
  H^2|\phi|^2
  +\frac{3}{8}\left(b_S'-b_{\bar{S}}'\right)
  H^2|\phi|^2
  +\cdots
  \,,
\label{EQ-eVbv}
\end{eqnarray}
%%%
where the ellipsis denotes higher order terms in $|\phi|^2/v^2$.  Here,
$M_{\rm eff}\equiv \xi v/h^2$ corresponds to the mass scale which was
denoted by the same symbol in \SEC{SEC-LHu}. (The mass of the
right-handed neutrino in the present model is given by $\xi v$.) Recall
that it is related to the neutrino mass as $\mnu{1}=\vev{H_u}^2/M_{\rm
eff}$.

The scalar potential in \EQ{EQ-eVbv} is nothing but the potential
obtained in \EQ{EQ-Vtotal} in the previous analysis, with some
modifications.\footnote{As we mentioned, the potential terms coming from
thermal effects will be discussed in \SEC{SEC-BL-condition}.} However,
there is an important difference in the Hubble-induced SUSY breaking
mass terms. Since there appears an extra Hubble mass term, we need the
following condition in order that the $\phi$ field develops a large
expectation value during the inflation:
%%%
\begin{eqnarray}
 b_S'-b_{\bar{S}}'\lsim \frac{8}{3}
  \,.
\end{eqnarray}
%%%
If this is not satisfied, the $\phi$ field is driven toward the origin
during the inflation, and the leptogenesis cannot work.  In the
following discussion, we assume that the above condition is
satisfied. Then the $\phi$ field has a negative Hubble mass term at
least in the range of $|\phi|\lsim v$.

Now let us discuss the scale where the flat direction is lifted.  If the
balance point between the negative Hubble mass term and the $F$-term
potential $|\phi|^6/(4M_{\rm eff}^2)$ in \EQ{EQ-eVbv} is below the $B-L$
breaking scale $v$, the $\phi$ field is stopped at this balance point,
%%%
\begin{eqnarray}
 \label{EQ-Fst}
  |\phi|\simeq \sqrt{M_{\rm eff}H_{\rm inf}}< v
  \,.
\end{eqnarray}
%%%
This is the condition for the ``$F$-term stopping case.'' In this case,
the evolution of the $\phi$ field and the amount of the generated baryon
asymmetry result in the same conclusions as discussed in \SEC{SEC-LHu}.

On the other hand, if the condition
%%%
\begin{eqnarray}
 \sqrt{M_{\rm eff}H_{\rm inf}}\gsim v
  \,,
\end{eqnarray}
%%%
is satisfied, the $\phi$ field can develop its expectation value as
large as $v$. In this case, we can stop the $\phi$ field below the
balance point $\sqrt{M_{\rm eff}H_{\rm inf}}$ by using the $D$-term
potential. This is very crucial to enhance the final baryon asymmetry in
the $D$-term stopping case. To see this, let us discuss what happens if
the $\phi$ field develops its value as large as the $B-L$ breaking scale
$v$. At this scale (i.e., $|\phi|\simeq v$), the expansion of the $S$
field given in \EQ{EQ-s-expand} becomes invalid and above this scale we
must use another expansion of the $S$ field as follows:
%%%
\begin{eqnarray}
 \label{EQ-s-expand2}
 |S|^2
  &\simeq&
  4\frac{v^4}{|\phi|^2}
  +\frac{h^2}{2\xi^2}|\phi|^2
  +
  \left[
   {\cal{O}}\left( \frac{v^4}{|\phi|^4} \right)
   +
   {\cal{O}}\left( \frac{h^2}{\xi^2} \right)
   \right]^2
   |\phi|^2
   \qquad {\rm for}
   \quad |\phi|\gsim v
   \,.
\end{eqnarray}
%%%
One might wonder whether the above expansion is reliable, since
\EQ{EQ-S-phi} is based on \EQ{EQ-minima}, which may not be valid for
$|\phi|\gg v$. Here, we first derive the conditions to fix the $\phi$
field at the scale $v$ during the inflation, assuming this expansion is
effectively valid at least for $|\phi|\sim v$. We will justify later the
validity of the obtained conditions by numerical calculations.

By substituting the expansion \EQ{EQ-s-expand2} into \EQ{EQ-Vfinal}, we
obtain the following Hubble-induced mass term:\footnote{ Here, we assume
$h^2/\xi^2\ll 1$. If this is not the case, we must include the
Hubble mass term coming from the coupling of the right-handed Majorana
neutrino to the inflaton, the last term in \EQ{EQ-Vfinal}.  }
%%%
\begin{eqnarray}
 \delta V\simeq
  \left(-1+\frac{3}{4}(1-b_{\bar{S}}')\right)
  H^2|\phi|^2
  +{\cal{O}}
  \left(\frac{v^4}{|\phi|^2}H^2\right)
  \qquad {\rm for}
  \quad  |\phi|\gsim v
  \,.
\label{eVav}
\end{eqnarray} 
%%%
Therefore, the $\phi$ field can have a positive Hubble mass term for
$|\phi|\gsim v$, if
%%%
\begin{eqnarray}
 b_{\bar{S}}'\lsim -\frac{1}{3}
  \,.
\end{eqnarray}
%%%
Therefore, if this condition is satisfied, the $\phi$ field cannot
develop its expectation value above the scale $v$, and hence it is fixed
at the $B-L$ breaking scale $v$ during the inflation.

To summarize, the $\phi$ field, and hence $S$ and $\bar{S}$ fields as
well, are stopped at the $B-L$ breaking scale $v$ during the inflation,
if the following conditions are satisfied,
%%%
\begin{eqnarray}
 b_S'-b_{\bar{S}}'\lsim \frac{8}{3}
  \,,
  \qquad
  \sqrt{M_{\rm eff} H_{\rm inf}}\gsim v
  \,,
  \qquad 
  b_{\bar{S}}'\lsim -\frac{1}{3}
  \,.
\label{EQ-Dstop}
\end{eqnarray}
%%%
These are the conditions for the $D$-term stopping case.

%%%%%%%%%%%%%%%%%%%%%%%%%%%%%%%%%%%%%%%%%%%%%%%%%%%%%%%%%%%%
\begin{figure}%%%%%%%%%%%%%%%%%%%%%%%%%%%%%%%%%%%%%%%%%%%%%%
%%%%%%%%%%%%%%%%%%%%%%%%%%%%%%%%%%%%%%%%%%%%%%%%%%%%%%%%%%%%
 \centerline{ {\psfig{figure=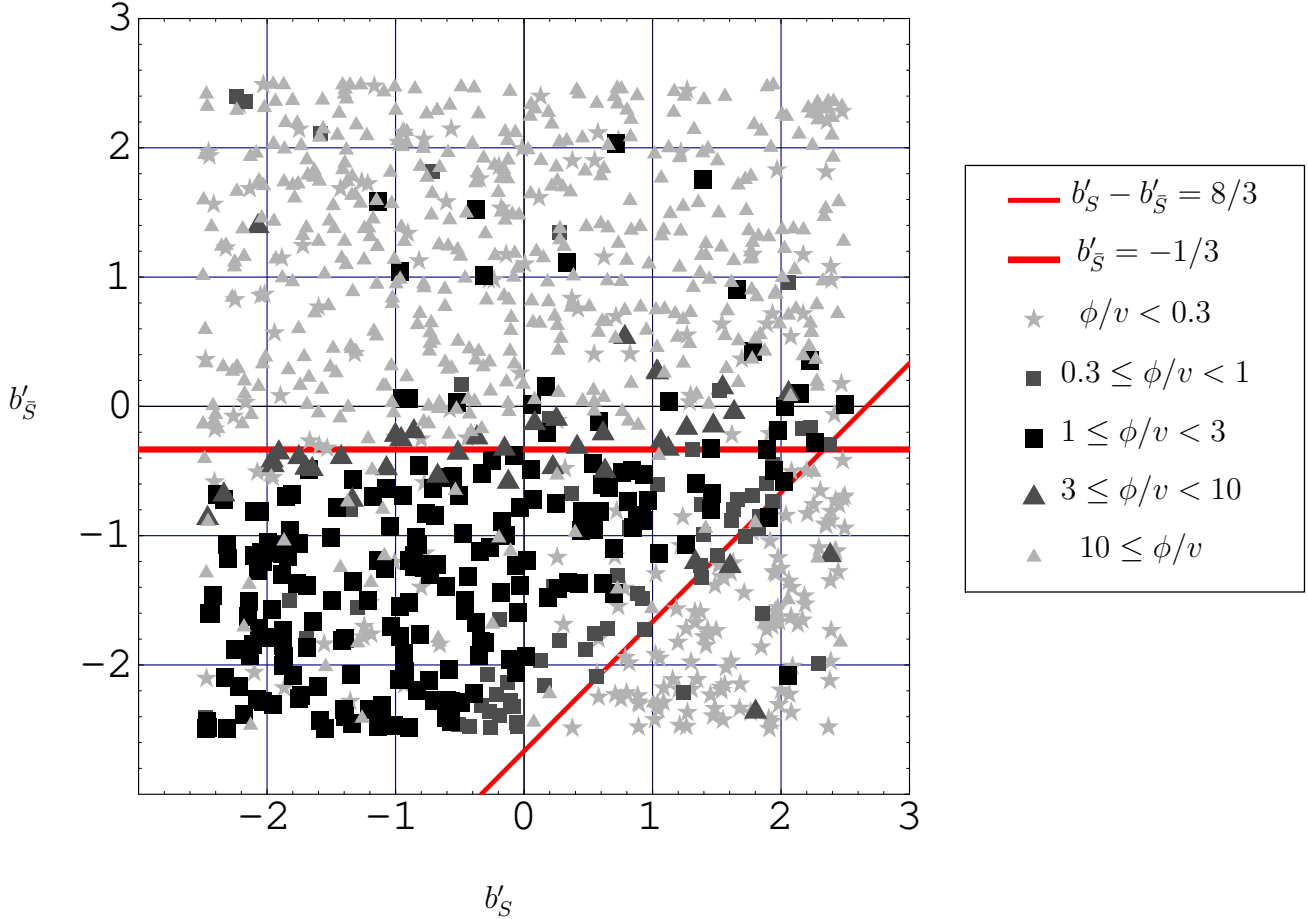,height=12cm}} }
 %%%%%%%%%%%%%%%%%%%%%%%%%
 \begin{picture}(0,0)%%%%%
  %%%%%%%%%%%%%%%%%%%%%%%%%
  \put(-30,190){$b_{\bar{S}}'$}
  \put(150,1){$b_S'$}
  \put(372,268){$b_S'-b_{\bar{S}}'=8/3$}
  \put(374,246){$b_{\bar{S}}'=-1/3$}
  \put(375,223){$\phi/v<0.3$}
  \put(368,200){$0.3\leq\phi/v<1$}
  \put(368,178){$1\leq\phi/v<3$}
  \put(368,156){$3\leq\phi/v<10$} 
  \put(375,134){$10\leq\phi/v$}
 %%%%%%%%%%%%%%%%%%%%%%%%%
 \end{picture}%%%%%%%%%%%%
 %%%%%%%%%%%%%%%%%%%%%%%%%
 \\ \caption{The amplitudes of the $\phi$ field fixed during inflation
 which are determined by numerical calculations. In this calculation, we
 have used the full scalar potential in \EQ{EQ-full-V} to follow the
 evolution of the relevant fields.  Here, we have assumed that $ H_{\rm
 inf}/v=0.1$, $3(1-b_{\phi}')=-1$, $ h=10^{-4}$, and $g = \xi = \lambda
 =1$, and have randomly generated other coupling constants $b_Y$,
 $|c_Y|$ in the range $-2.5\leq b_Y\leq 2.5$ and $0\leq |c_Y|\leq 2.5$,
 respectively.  Various symbols denote the $|\phi|$ to $v$ ratio at the
 end of inflation.  We see that the $\phi$ field is, in fact, stopped at
 the $B-L$ breaking scale $v$ if the conditions in \EQ{EQ-Dstop} are
 satisfied.  }
%%%%%%%%%%%%%%%%%%%%%%%%%%%%%%%%%%%%%%%%%%%%%%%%%%%%%%%%%%%%
 \label{FIG-phi-B-L}%%%%%%%%%%%%%%%%%%%%%%%%%%%%%%%%%%%%%%%%
\end{figure}%%%%%%%%%%%%%%%%%%%%%%%%%%%%%%%%%%%%%%%%%%%%%%%%
%%%%%%%%%%%%%%%%%%%%%%%%%%%%%%%%%%%%%%%%%%%%%%%%%%%%%%%%%%%%

In order to check the reliability of the conditions in \EQ{EQ-Dstop},
(especially that of the last condition $b_{\bar{S}}'\lsim - 1/3$,) we
have numerically solved the coupled equations of motions
%%%
\begin{eqnarray}
 \ddot{Y}+3H\dot{Y}+\frac{\partial V}{\partial Y^*} = 0
  \,,
\end{eqnarray}
%%%
for the relevant fields $Y = \phi$, $N$, $S$, $\bar{S}$ and $X$, using
the full scalar potential in \EQ{EQ-full-V}. We show the result in
\FIG{FIG-phi-B-L}, where the amplitude of the $\phi$ field at the end of
the inflation is plotted in $b'_S$--$b'_{\bar{S}}$ plane. Here, we have
taken $M_{\rm eff} = \xi v / h^2 = 10^8 v$ and $H_{\rm inf} = 0.1 v$,
and hence the second condition $ \sqrt{M_{\rm eff} H_{\rm inf}}\,\,\gsim
v$ is satisfied.

It is found from \FIG{FIG-phi-B-L} that the conditions in \EQ{EQ-Dstop}
well explain the result of this numerical calculation. Actually, for
$b_S'-b_{\bar{S}}' \gsim 8/3$ (positive Hubble mass term for
$|\phi|\lsim v$), the amplitude is damped as $|\phi|\ll v$, while for
$b_{\bar{S}}'\gsim -1/3$ (negative Hubble mass term for $|\phi|\gsim
v$), the $\phi$ field overshoots the $B-L$ breaking scale $v$.

Thus, the $\phi$ field can stop at the $B-L$ breaking scale $v$ if the
conditions $b_S'-b_{\bar{S}}' \lsim 8/3$ and $b_{\bar{S}}'\lsim - 1/3$
are satisfied. In fact, we see from \FIG{FIG-phi-B-L} that the amplitude
of the $\phi$ field at the end of the inflation lies in the range
%%%
\begin{eqnarray}
 |\phi|\simeq (1 - 3)\times v
  \,,
\end{eqnarray}
%%%
in most of the parameter space where the conditions in \EQ{EQ-Dstop} are
satisfied.

\subsubsection{initial phase}

Now let us turn to discuss the phase of the $\phi$ field at the end of
the inflation, which is crucial to estimate the resultant lepton
asymmetry.

First, in the $F$-term stopping case, the amplitude of the $\phi$ field
is fixed at $\sqrt{M_{\rm eff}H_{\rm inf}}$. An important point here is
that the curvature around the valley of the Hubble $A$-term potential in
\EQ{EQ-HA} (which we denote by the symbol $m^2_{\rm{phase}}$) is of the
order of the Hubble parameter in this case. This can be seen from the
following relation:
\begin{eqnarray}
 m^2_{\rm{phase}}
  &\simeq&
  \left|
   \frac{\partial^2}{\partial \phi^2}V_{\rm phase}
   \right|
   \nonumber\\
 &\simeq&
  \frac{H_{\rm inf}}{M_{\rm eff}}|\phi|^2
  \simeq H_{\rm inf}^2
  \,.
\end{eqnarray}
Therefore, as discussed in \SEC{SEC-LHu}, the phase of the $\phi$ field
settles down at the bottom of the valley of the $A$-term potential
during the inflation, and hence the Hubble $A$-term cannot supply a
torque to rotate the $\phi$ field.  In this case, the relevant torque
for the $\phi$ field only comes from the ordinary $A$-term potential
proportional to the gravitino mass. As stressed in
\SEC{SEC-resultinLHu}, this leads to a suppression of the baryon
asymmetry in gauge-mediation models with a small gravitino mass.

However, if the conditions in \EQ{EQ-Dstop} are satisfied, namely, in
the $D$-term stopping case, the curvature of the potential along the
phase direction is smaller than the Hubble parameter during the
inflation:
\begin{eqnarray}
 m^2_{\rm{phase}}
  &\simeq&
  \left|
   \frac{\partial^2}{\partial \phi^2}V_{\rm phase}
   \right|
   \nonumber\\
 &\simeq&
  \frac{H_{\rm inf}}{M_{\rm eff}} |\phi|^2
  \simeq H_{\rm inf}^2
  \left(\frac{|\phi|}{\sqrt{M_{\rm eff} H_{\rm inf}}}\right)^2
  < H_{\rm inf}^2
  \,.
\end{eqnarray}
In this case, therefore, there is no reason to expect that the $\phi$
field sits down at the bottom of the valley of the $A$-term potential in
\EQ{EQ-HA} during the inflation. In other words, unless there is an
accidental fine tuning of the initial phase of the $\phi$ field, it is
generally displaced from the bottom of the valley of this $A$-term
potential when the inflation ends.  

Therefore, the $A$-term potential kicks the $\phi$ field along the phase
direction when the Hubble parameter becomes comparable to the curvature
along the phase direction $m^2_{\rm{phase}}$.  As we will show in the
next subsection, this phase rotation caused by the Hubble $A$-term is
very crucial to enhance the final baryon asymmetry.

\subsection{Lepton asymmetry}

Now let us discuss the dynamics of the $\phi$ field and calculate the
resultant lepton asymmetry in the present model. In the $F$-term
stopping case, the amplitude of the $\phi$ field is always smaller than
the $B-L$ breaking scale $v$ [see \EQ{EQ-Fst}] and the evolution of the
$\phi$ field is determined by the effective potential in \EQ{EQ-eVbv}
plus thermal effects, which we have investigated in \SEC{SEC-LHu}. Thus,
the resultant baryon asymmetry is given by \EQ{EQ-nBs-final} (or
\FIG{FIG-LHu-numerical}).

We turn to the $D$-term stopping case. As discussed in the previous
subsection, the amplitude of the $\phi$ field at the end of the
inflation is given by
%%%
\begin{eqnarray}
 |\phi| \simeq v
  \,,
\end{eqnarray}
and the phase of the $\phi$ field is generally displaced from the bottom
of the valley of the Hubble $A$-term potential.

After the end of the inflation, the field value of the $\phi$ field
takes an almost constant value for a while. Then the Hubble $A$-term in
\EQ{EQ-HA} kicks the $\phi$ field along the phase direction when the
Hubble parameter $H$ becomes comparable with the curvature:
%%%
\begin{eqnarray}
 H^2 \simeq m_{\rm phase}^2 
  \left(
   \simeq 
   \frac{H}{M_{\rm eff}} v^2
   \right)
   \,.
\end{eqnarray}
%%%
Thus, the Hubble parameter at this time $H_{\rm ph}$ is given by
%%%
\begin{eqnarray}
 H_{\rm ph}\simeq \frac{v^2}{M_{\rm eff}}
  \,.
\end{eqnarray}
%%%
At this time, the phase of the $\phi$ field, $\arg(\phi)$, starts to
oscillate around the bottom of the valley of the Hubble $A$-term
potential.  At the same time, the amplitude of the $\phi$ field,
$|\phi|$, starts to decrease according to the following equation:
%%%
\begin{eqnarray}
 |\phi|\simeq \sqrt{M_{\rm eff} H}
  \,,
\end{eqnarray}
%%%
which is the balance point between the negative Hubble mass term and the 
$|\phi|^6 /(4 M_{\rm eff}^2)$ potential.

Since the phase of the $\phi$ field starts its oscillation at $H=H_{\rm
ph}$, it has already had a large acceleration along the {\it phase}
direction before it starts coherent oscillation along the {\it radius}
direction around the origin when $H=H_{\rm osc}$.  This $H_{\rm osc}$ is
determined by thermal effects, soft SUSY breaking mass term and
$\mu$-term in the same way as in \SEC{SEC-LHu}.  Here, we have assumed
that $H_{\rm ph} > H_{\rm osc}$, which we will discuss in
\SEC{SEC-BL-condition}.

\begin{figure}%%%%%  FIGURE Amp and Phase  %%%%%%%%%%%%%%%%%
%%%%%%%%%%%%%%%%%%%%%%%%%%%%%%%%%%%%%%%%%%%%%%%%%%%%%%%%%%%%
 \centerline{ {\psfig{figure=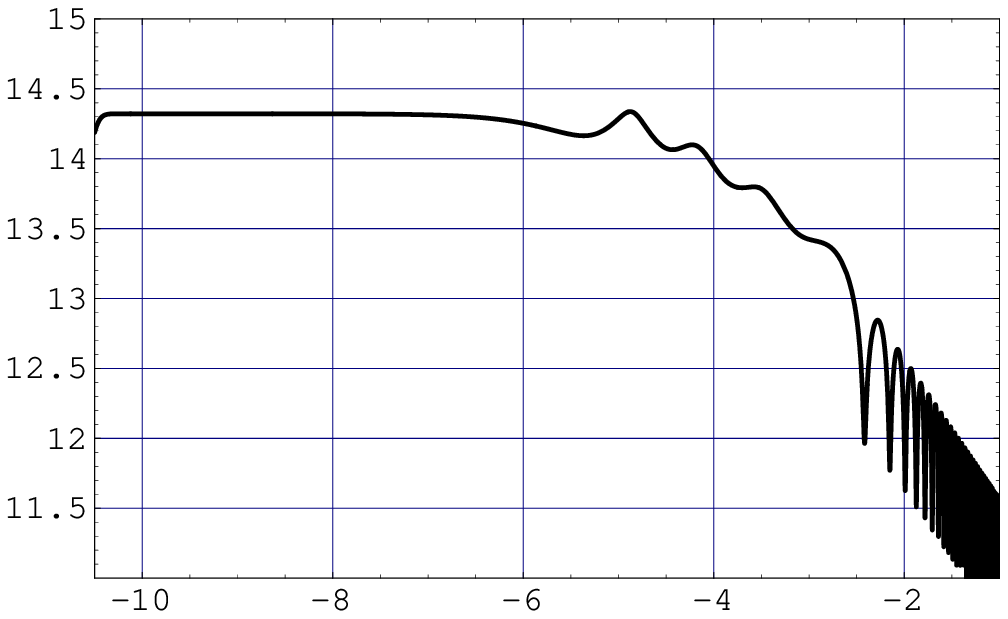,height=6cm}} }
 %%%%%%%%%%%%%%%%%%%%%%%%%
 \begin{picture}(0,0)%%%%%
 %%%%%%%%%%%%%%%%%%%%%%%%%
  \put(10,100){$\log_{10}\left(\Frac{|\phi|}{\GEV}\right)$}
  \put(190, 1){$\log_{10}\left(\Frac{t}{\GEV^{-1}}\right)$}
 %%%%%%%%%%%%%%%%%%%%%%%%%
 \end{picture}%%%%%%%%%%%%
 %%%%%%%%%%%%%%%%%%%%%%%%%
 \\ \caption{The evolution of the $\phi$ field estimated by a numerical
 calculation. We assume that $M_{\rm eff}=3\times 10^{23}\GEV$ (\i.e.,
 $\mnu{1}=10^{-10}\EV$), $v=10^{14}\GEV$, $H_{\rm inf}=10^{12}\GEV$,
 $m_\phi =10^3\GEV$. We see that the $\phi$ field starts to move at
 $H=H_{\rm ph} \simeq 3\times 10^4\GEV$ according to the equation;
 $|\phi|\simeq\sqrt{M_{\rm eff}H}$. Here, $t=2/(3H)$ is the cosmic time.
 The $\phi$ field starts to oscillate around the origin at $H_{\rm
 osc}\simeq m_{\phi}$ since we neglect the thermal effects in this
 calculation.}  
%%%%%%%%%%%%%%%%%%%%%%%%%%%%%%%%%%%%%%%%%%%%%%%%%%%%%%%%%%%%
 \label{FIG-Amp}%%%%%%%%%%%%%%%%%%%%%%%%%%%%%%%%%%%%%%%%%%%%
%%%%%%%%%%%%%%%%%%%%%%%%%%%%%%%%%%%%%%%%%%%%%%%%%%%%%%%%%%%%
 \centerline{ {\psfig{figure=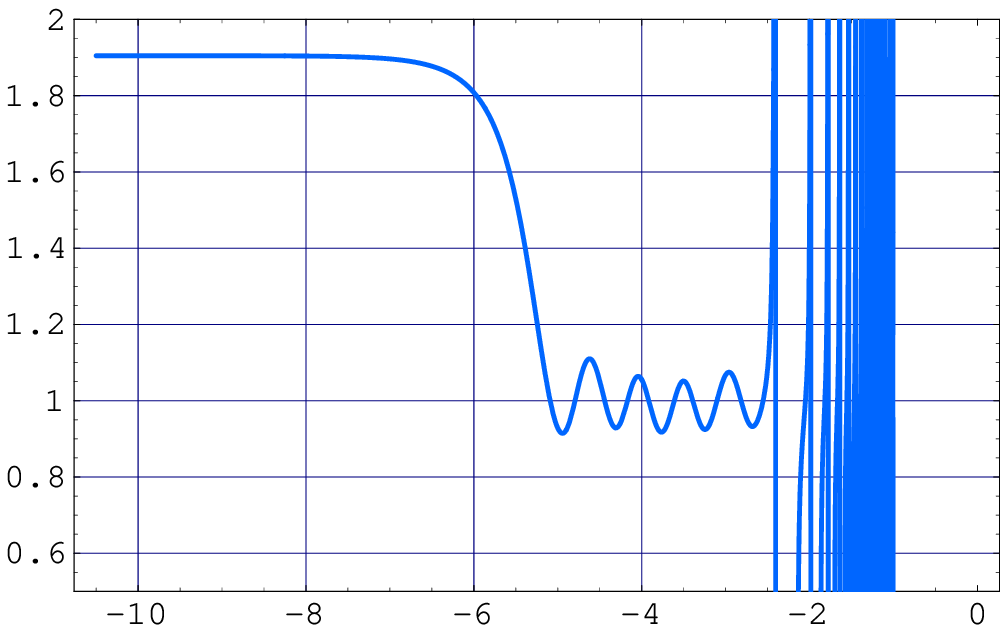,height=6cm}} }
 %%%%%%%%%%%%%%%%%%%%%%%%%
 \begin{picture}(0,0)%%%%%
 %%%%%%%%%%%%%%%%%%%%%%%%%
  \put(40,100){$\Frac{\arg(\phi)}{\pi/4}$}
  \put(190,1){$\log_{10}\left(\Frac{t}{\GEV^{-1}}\right)$}
 %%%%%%%%%%%%%%%%%%%%%%%%%
 \end{picture}%%%%%%%%%%%%
 %%%%%%%%%%%%%%%%%%%%%%%%%
 \\ \caption{ The evolution of the phase of the $\phi$ field estimated
 by a numerical calculation.  The parameters used in this figure are the
 same as in \FIG{FIG-Amp}.  Here, we also defined that $c_{\phi}'$ is
 real, and hence the valleys coming from the Hubble $A$-term lie along
 $\arg(\phi)=\pi/4$, $3\pi/4$, $5 \pi/4$, $7\pi/4$.  We see that the
 phase of the $\phi$ field starts to oscillate around the bottom of the
 valley at $H\simeq H_{\rm ph}$.  } 
%%%%%%%%%%%%%%%%%%%%%%%%%%%%%%%%%%%%%%%%%%%%%%%%%%%%%%%%%%%%
 \label{FIG-phase}%%%%%%%%%%%%%%%%%%%%%%%%%%%%%%%%%%%%%%%%%%
\end{figure}

In \FIG{FIG-Amp} and \FIG{FIG-phase}, we show the evolution of the
amplitude and the phase of the $\phi$ field obtained by numerical
calculations, respectively. In these figures, $t$ is the cosmic time in
the matter dominated universe $t=2/(3H)$.  We see from these figures
that the evolution of the $\phi$ field is well explained by above
arguments.

%%%%%%%%%%%%%%%%%%%%%%%%%%%%%%%%%%%%%%%%%%%%%%%%%%%%%%%%%%%%
%%%%%%%%%%%%%%%%%%%%%%%%%%%%%%%%%%%%%%%%%%%%%%%%
%%%%%%%%%%%%%%%%%%%%%%%%%%%%%%%%%%%%%%%%%%%%%%%%%%%%%%%%%%%%

We are now at the point to estimate the baryon asymmetry. {}From the
equation of motion (\ref{EQ-EOM}), the evolution of the lepton asymmetry
$n_L = (i/2)(\dot{\phi}^* \phi - \phi^* \dot{\phi})$ is given by
%%%
\begin{eqnarray}
 \dot{n}_L + 3 H n_L 
  &=&
  2\sqrt{3}\frac{H}{M_{\rm eff}}
  {\rm Im}\left(c_{\phi}'\phi^4\right)
  \,,
\end{eqnarray}
%%%
where we have used $|S|\simeq v$ for $H\lsim H_{\rm ph}$.  This can be
easily integrated and we obtain the lepton number at time $t$ as
\begin{eqnarray}
 \label{EQ-BL-nL-t}
  \left[ R^3 n_L \right] (t)
  &\simeq& 
  \frac{2\sqrt{3}}{M_{\rm eff}} 
  \int^t dt' 
  R^3 
  H
  \,
  {\rm Im}
  \left(
   c_{\phi}' \phi^4 
   \right) 
   \,.
\end{eqnarray}
Before the $\phi$'s oscillation, i.e., for $H_{\rm osc}\lsim H\lsim
H_{\rm ph}$, the amplitude of the $\phi$ field decreases as
$|\phi|\simeq \sqrt{M_{\rm eff}H}$ and the sign of ${\rm Im}
(c_{\phi}'\phi^4)$ in \EQ{EQ-BL-nL-t} changes with $1/H$ time
scale. (See \FIG{FIG-Amp} and \FIG{FIG-phase}.)  Therefore, we see from
\EQ{EQ-BL-nL-t} that the total lepton number oscillates with $1/H$ time
scale with almost constant amplitude, since $R^3 H |\phi^4| \propto H
\propto t^{-1}$ in this period.  

Then, the total lepton number is fixed when the $\phi$ field starts its
coherent oscillation around the origin at $H=H_{\rm osc}$, since the
amplitude of the $\phi$ field decreases as fast as $|\phi|\propto H$ and
the integrand of \EQ{EQ-BL-nL-t} decreases as $\sim t^{-3}$ after the
oscillation.  The lepton number density at time $t_{\rm osc}=2/(3H_{\rm
osc})$ is then given by
%%%
\begin{eqnarray}
 n_L (t_{\rm osc})
  &\simeq&
  \frac{2\sqrt{3}}{M_{\rm eff}}
  H_{\rm osc}
  {\rm Im}
  \left[
   c_{\phi}' \phi^4 (t_{\rm osc})
   \right]
   t_{\rm osc}
   \nonumber\\
 &\simeq&
  \frac{4}{\sqrt{3}}|c_{\phi}'|M_{\rm eff} H_{\rm osc}^2\delta_{\rm ph}
  \,,
\end{eqnarray}
%%%
where $\delta_{\rm ph}\simeq \sin(4\arg\phi + \arg c_{\phi}')$ is the
effective $CP$-phase. By using \EQ{EQ-nLtR-nLtosc}, the final lepton
asymmetry is given by
%%%
\begin{eqnarray}
 \label{EQ-BL-nLs}
  \frac{n_L}{s}
  \simeq
  \frac{M_{\rm eff} T_R}{\sqrt{3} M_G^2}
  |c_{\phi}'|
  \delta_{\rm ph}
  \,.
\end{eqnarray}
%%%
Here, we have used the fact that the $\phi$ field starts its oscillation
before the reheating process completes, i.e., $H_{\rm osc} >
\Gamma_\chi$, which was verified in \SEC{SEC-resultinLHu}.  Notice that
the resultant lepton asymmetry is independent of the time when the
$\phi$ field starts the oscillation around the origin, $H_{\rm
osc}$.\footnote{We have also confirmed by numerical calculations that
the final lepton asymmetry $n_L / s$ is independent of the oscillation
time $H_{\rm osc}$.} This is because the total lepton number itself
oscillates with a {\it constant} amplitude for $H_{\rm osc} \lsim H
\lsim H_{\rm ph}$, as discussed just above. Thus, the order of the
lepton asymmetry is already fixed just after $H\simeq H_{\rm ph}$,
before the oscillation time $H_{\rm osc}$.

By using the relation between the lepton and baryon asymmetry in the
presence of the sphaleron effect, we obtain
%%%
\begin{eqnarray}
 \label{EQ-BL-nBs-final}
 \frac{n_B}{s}
  &\simeq&
  0.35 \times \frac{n_L}{s}
  \nonumber\\
 &\simeq&
  1\times 10^{-10}
  \left(
   \frac{T_R}{10^8\GEV}
   \right)
   \left(
    \frac{10^{-6}\EV}{\mnu{1}}
    \right)^{-1}
    |c_{\phi}'|
    \delta_{\rm ph}
    \,.
\end{eqnarray}
%%%

Here, we give several comments on the obtained baryon asymmetry. First,
by comparing \EQ{EQ-BL-nLs} and \EQ{EQ-nLs-final}, we can see that the
lepton asymmetry in the $D$-term stopping case is enhanced compared
with the $F$-term stopping case (and the case without ${\rm
U}(1)_{B-L}$), since it is no longer suppressed by the oscillation time
$H_{\rm osc}$.

Next, as stressed before, one sees that the resultant baryon asymmetry
is independent of the gravitino mass $m_{3/2}$. Therefore, we can obtain
an enough baryon asymmetry even with a very light gravitino. This is the
crucial point in this $D$-term stopping case. Here, we have to take care
of the gravitino problem. For example, if the mass of the lightest
neutrino is as small as $\mnu{1}\simeq 10^{-10}\EV$, the present baryon
asymmetry is explained with a low reheating temperature $T_R\simeq
10^4\GEV$ [see \EQ{EQ-BL-nBs-final}]. In such a low reheating
temperature, we are free from the overproduction of gravitinos even in
gauge-mediated SUSY breaking scenarios with the small gravitino mass
$m_{3/2}={\cal{O}}(100\KEV)$~\cite{Gprob-GMSB}.

On the other hand, if the mass of the gravitino is large enough, say,
$m_{3/2}\simeq 1\TEV$, we can avoid the cosmological gravitino problem
even if the reheating temperature is rather high as $T_R\simeq 10^9\GEV$
(see \SEC{SEC-grav}). In such a case, we see from \EQ{EQ-BL-nBs-final}
that the mass of the lightest neutrino $\mnu{1}\simeq 10^{-4}\EV$ is
small enough to explain the present baryon asymmetry.

%%%%%%%%%%%%%%%%%%%%%%%%%%%%%%%%%%%%%%%%%%%%%%%%%%%%%%%%%%%%
\begin{figure}[t]%%  FIGURE  nBs           %%%%%%%%%%%%%%%%%
%%%%%%%%%%%%%%%%%%%%%%%%%%%%%%%%%%%%%%%%%%%%%%%%%%%%%%%%%%%%
 \centerline{ {\psfig{figure=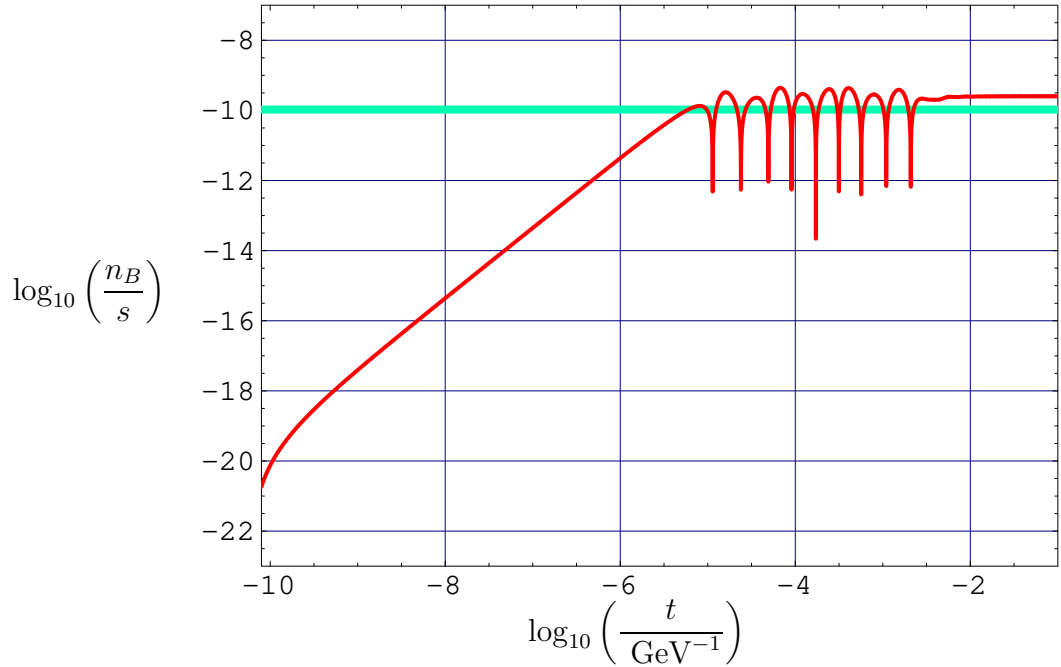,height=8cm}} }
 %%%%%%%%%%%%%%%%%%%%%%%%%
 \begin{picture}(0,0)%%%%%
 %%%%%%%%%%%%%%%%%%%%%%%%%
  \put(-5,130){$\log_{10}\left( \Frac{n_B}{s} \right)$}
  \put(190,1){$\log_{10}\left( \Frac{t}{\GEV^{-1}} \right)$}
 %%%%%%%%%%%%%%%%%%%%%%%%%
 \end{picture}%%%%%%%%%%%%
 %%%%%%%%%%%%%%%%%%%%%%%%%
\\
\caption{The evolution of the baryon asymmetry $n_B/s$ estimated by a
numerical calculation. Parameters used in this figure are the same as in
\FIG{FIG-Amp} and \FIG{FIG-phase}. Here, we also take $T_R=10^4\GEV$.
The thick line denotes the analytic estimation in \EQ{EQ-BL-nBs-final}.
We can see that the asymmetry begins to oscillate with constant
amplitude at $H\simeq H_{\rm ph}$ and it is fixed when the $\phi$ field
starts the oscillations around the origin. The analytical estimation of
the baryon asymmetry agrees well with the obtained value by the
numerical calculation.}
%%%%%%%%%%%%%%%%%%%%%%%%%%%%%%%%%%%%%%%%%%%%%%%%%%%%%%%%%%%%
 \label{FIG-BL-nBs}%%%%%%%%%%%%%%%%%%%%%%%%%%%%%%%%%%%%%%%%%
\end{figure}%%%%%%%%%%%%%%%%%%%%%%%%%%%%%%%%%%%%%%%%%%%%%%%%
%%%%%%%%%%%%%%%%%%%%%%%%%%%%%%%%%%%%%%%%%%%%%%%%%%%%%%%%%%%%

Finally, we show the evolution of the baryon asymmetry $n_B/s$ obtained
by a numerical calculation in \FIG{FIG-BL-nBs}. {}From this figure we
see that the asymmetry begins to oscillate with almost constant
amplitude at $H=H_{\rm ph}$ and it is fixed at the time when the $\phi$
field starts to oscillate around the origin, $H = H_{\rm osc}$. We also
see that the amount of generated baryon asymmetry agrees well with our
analytic estimation in \EQ{EQ-BL-nBs-final}. We confirm from this figure
that the resultant asymmetry is well explained by the arguments
described in this subsection.

\subsection{Conditions to avoid an early oscillation}
\label{SEC-BL-condition}

So far, we have neglected the thermal effects and assumed that there is
no early oscillation, i.e., $H_{\rm ph} > H_{\rm osc}$. If the $\phi$
field starts its oscillation before $H = H_{\rm ph}$, i.e., if $H_{\rm
osc} < H_{\rm ph}$, the resultant baryon asymmetry is strongly
suppressed. (In this case, the resultant baryon asymmetry is inversely
proportional to $H_{\rm osc}^2$.) Thus, we verify the assumption $H_{\rm
ph} > H_{\rm osc}$ here.

First, in order to avoid the early oscillation caused by the soft SUSY
breaking mass $m_{\phi}$ and the $\mu$-term, we need
%%%
\begin{eqnarray}
 \label{EQ-1st-cond}
  m_{\phi}\,,\,\,\mu < H_{\rm ph}\simeq \frac{v^2}{M_{\rm eff}}
  \,.
\end{eqnarray}

Next, the thermal mass term $c_k f_k^2 T^2$ discussed in
\SEC{SEC-LHu-evol} may cause the early oscillation. This can be avoided
if the following condition is satisfied:
%%%
\begin{eqnarray}
 \label{EQ-2nd-cond}
  H^2 > \sum_{f_k|\phi| < T} c_k f_k^2 T^2
  \,,
  \qquad
  {\rm for}
  \quad
  H > H_{\rm ph}
  \,.
\end{eqnarray}

Finally, we need the following third condition to avoid the early
oscillation caused by the thermal log term $T^4\ln (|\phi|^2)$, which
was also explained in \SEC{SEC-LHu-evol}:
%%%
\begin{eqnarray}
 \label{EQ-3rd-cond}
  H^2 > \frac{a_g \alpha_S^2 T^4}{v^2}
  \,,
  \qquad
  {\rm for}
  \quad
  H > H_{\rm ph}
  \,.
\end{eqnarray}
Here, we have used $|\phi|\simeq v$ for $H > H_{\rm ph}$.

From the second (\ref{EQ-2nd-cond}) and third (\ref{EQ-3rd-cond})
conditions, the reheating temperature is bounded from above
as~\footnote{At first sight, \EQ{EQ-2ndplus3rd} seems to be obtained
just by rewriting the relation $H_{\rm osc} < H_{\rm ph}$ by using the
formula of $H_{\rm osc}$ in \EQ{EQ-Hosc-final}. However, it is slightly
different, since $|\phi|$ takes a constant value $v$ for $H\gsim H_{\rm
ph}$ in the present case, while $|\phi|\simeq \sqrt{M_{\rm eff} H}$ in
the previous case.}
%%%
\begin{eqnarray}
 \label{EQ-2ndplus3rd}
 T_R < \min
  \left[
   T_{R\,i}\,,\,\,
   \frac{v^2}{a_{g}^{1/2}\alpha_S M_G^{1/2} M_{\rm eff}^{1/2}}
   \right]
   \,,
\end{eqnarray}
where
\begin{equation}
T_{R\,i}
 =
 \max
 \left(
  \frac{f_i v^{3/2}}
  {c_i^{1/4} M_G^{1/2}}
  \,,\,\,
  \frac{v^3}
  {c_i f_i^2 M_G^{1/2} M_{\rm eff}^{3/2}}
  \right)
  \,.
\end{equation}

\begin{figure}%%%%%  FIGURE  allowed       %%%%%%%%%%%%%%%%%
%%%%%%%%%%%%%%%%%%%%%%%%%%%%%%%%%%%%%%%%%%%%%%%%%%%%%%%%%%%%
 \centerline{ {\psfig{figure=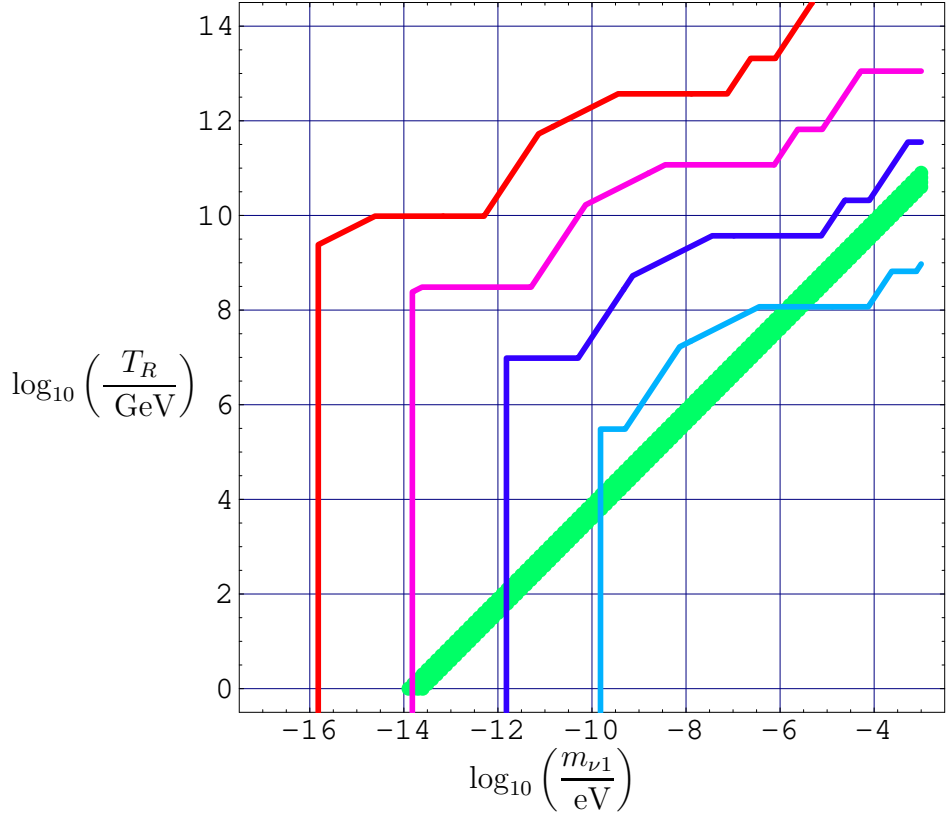,height=10cm}} }
 %%%%%%%%%%%%%%%%%%%%%%%%%
 \begin{picture}(0,0)%%%%%
 %%%%%%%%%%%%%%%%%%%%%%%%%
  \put(17,150){$\log_{10}\left(\Frac{T_R}{\GEV}\right)$}
  \put(190,1){$\log_{10}\left(\Frac{\mnu{1}}{\EV}\right)$}
 %%%%%%%%%%%%%%%%%%%%%%%%%
 \end{picture}%%%%%%%%%%%%
 %%%%%%%%%%%%%%%%%%%%%%%%%
\\
 \caption{The plot of the parameter regions which are free from the
 early oscillation in $\mnu{1}-T_R$ plane.  The regions below the four
 solid lines are free from the early oscillation. These four lines
 correspond to the breaking scale of the ${\rm U}(1)_{B-L}$, $v=10^{16}$,
 $10^{15}$, $10^{14}$, $10^{13}\GEV$ from left to right,
 respectively. The shaded region corresponds to the present baryon
 asymmetry, $n_B/s\simeq (0.4-1)\times 10^{-10}$. We have taken
 $\delta_{\rm ph}=1$ and $|c_{\phi}'|=1$ in this figure. We see that if
 we take the breaking scale of the ${\rm U}(1)_{B-L}$ as $v\gsim 10^{14}\GEV$,
 the early oscillation of the $\phi$ field $H_{\rm osc}>H_{\rm ph}$ can
 be avoided in most region of parameter space.}
%
%%%%%%%%%%%%%%%%%%%%%%%%%%%%%%%%%%%%%%%%%%%%%%%%%%%%%%%%%%%%
 \label{FIG-BL-allowed}%%%%%%%%%%%%%%%%%%%%%%%%%%%%%%%%%%%%%
\end{figure}

We show the allowed regions which are free from the early oscillation in
\FIG{FIG-BL-allowed}, taking account the conditions in \EQS{EQ-1st-cond}
and (\ref{EQ-2ndplus3rd}). {}From this figure, we see that the early
oscillations can be easily avoided, especially when the $B-L$ breaking
scale satisfies $v\gsim 10^{14}\GEV$.
%%%%%%%%%%%%%%%%%%%%%%%%%%%%%%%%%%%%%%%%%%%%%%%%%%%%%%%%%%%%
%%%%%%%%%%%%%%%%%%%%%%%%%%%%%%%%%%%%%%%%%%%%%%%%
%%%%%%%%%%%%%%%%%%%%%%%%%%%%%%%%%%%%%%%%%%%%%%%%%%%%%%%%%%%%

\clearpage
%%%%%%%%%%%%%%%%%%%%%%%%%%%%%%%%%%%%%%%%%%%%%%%%%%%%%%%%%%%%%%%%%%%
\section{The neutrino-less double-beta decay}%%%%%%%%%%%%%%%%%%%%%%%
%%%%%%%%%%%%%%%%%%%%%%%%%%%%%%%%%%%%%%%%%%%%%%%%%%%%%%%%%%%%%%%%%%%
\label{SEC-double-beta}

In \SEC{SEC-LHu}, we have investigated the leptogenesis via $L H_u$
flat direction in detail and have found that the produced baryon
asymmetry is determined almost independently of the reheating
temperature $T_R$, because of the early oscillation of the flat
direction field $\phi$ caused by thermal effects. As shown there, an
ultralight neutrino $\mnu{1}\simeq 10^{-9}\EV$ is required to explain
the empirical baryon asymmetry $n_B/s\simeq 10^{-10}$, except for the
case where the model in \SEC{SEC-model-PQ} is adopted.\footnote{See
also the last comment in \SEC{SEC-resultinLHu}.} By introducing a
gauged ${\rm U}(1)_{B-L}$ symmetry, the early oscillation of the flat
direction due to the thermal effect can be avoided
(\SEC{SEC-withB-L}), but small neutrino mass $\mnu{1}\ll 10^{-4}\EV$
is still necessary if we require low enough reheating temperature
$T_R\lsim 10^9\GEV$ to avoid the cosmological gravitino problem in a
wide range of gravitino mass (see \FIG{FIG-BL-allowed}). In this
section we show that such a tiny mass of the lightest neutrino mass
$\mnu{1}\ll 10^{-4}\EV$ have an interesting implication on a
low-energy experiment; the neutrino-less double beta
($0\nu\beta\beta$) decay, which will be a strong evidence for the
lepton number violation if observed.

The crucial parameter to determine the $0\nu\beta\beta$ decay rate is
the $\nu_e$--$\nu_e$ component of the neutrino mass matrix, which is
given by
%%%
\begin{eqnarray}
 m_{ee}\equiv \sum_i U_{ei}^2 \mnu{i}
  \,,
\end{eqnarray}
%%%
where the mixing matrix $U_{\alpha i}$ is defined in \SEC{SEC-LAfromN}.
If the mass of the lightest neutrino is actually so small as $\mnu{1}\ll
10^{-4}\EV$, the contribution from $\mnu{1}$ to $m_{ee}$ can be
neglected as long as we consider an accessible $0\nu\beta\beta$ decay
rate, which corresponds to $|m_{ee}|\gsim 10^{-3}\EV$. Thus, the
parameter $|m_{ee}|$ is written in terms of masses and mixings of two
other neutrinos as
%%%
\begin{eqnarray}
 |m_{ee}| = |U_{e2}^2 \mnu{2} + U_{e3}^2 \mnu{3}| 
  \,,
\end{eqnarray}
%%%
which immediately leads to the following upper and lower
bounds:
\begin{eqnarray}
  |m_{ee}|_{\max} &=& |U_{e2}|^2 \mnu{2} + |U_{e3}|^2 \mnu{3}\,,
  \nonumber
  \\
  |m_{ee}|_{\min} &=& \left|\, |U_{e2}|^2 \mnu{2} - |U_{e3}|^2 \mnu{3} \,\right|\,.  
  \label{EQ-mee-bounds}
\end{eqnarray}
In general, the mass pattern of the neutrinos can be classified into
two cases: the normal hierarchy and the inverted hierarchy. Let us
consider these two cases in turn.

%%%%%%%%%%%%%%%%%%%%%%%%%%%%%%%%%%%%%%%%%%%%%%%%%%%%%%%%%%%%
%%%%%%%%%%%%%%%%%%%%%%%%%%%%%%%%%%%%%%%%%%%%%%%%%%%%%%%%%%%%
%%%%%%%%%%%%%%%%%%%%%%%%%%%%%%%%%%%%%%%%%%%%%%%%%%%%%%%%%%%%
\subsubsection*{normal mass hierarchy}
%%%%%%%%%%%%%%%%%%%%%%%%%%%%%%%%%%%%%%%%%%%%%%%%%%%%%%%%%%%%
%%%%%%%%%%%%%%%%%%%%%%%%%%%%%%%%%%%%%%%%%%%%%%%%%%%%%%%%%%%%
%%%%%%%%%%%%%%%%%%%%%%%%%%%%%%%%%%%%%%%%%%%%%%%%%%%%%%%%%%%%

In the case of normal mass hierarchy, the parameters of atmospheric
and solar neutrino oscillations are given by
\begin{equation}
  \delta m_{\rm atm}^2 = \mnu{3}^2 - \mnu{2}^2\,,
  \qquad
  \delta m_{\rm sol}^2= \mnu{2}^2\,,
  \qquad
  \tan^2\theta_{\rm sol} \equiv |U_{e2}/U_{e1}|^2
  \,,
\end{equation}
and the mixing
angle constrained by CHOOZ experiment~\cite{CHOOZ} corresponds to
$\sin\theta_{\rm chooz} = |U_{e3}|$, which is severely bounded as
$\sin\theta_{\rm chooz} < 0.15$. In terms of these oscillation
parameters, \EQS{EQ-mee-bounds} become
\begin{eqnarray}
  |m_{ee}|_{\max} 
  &=& 
  c_{\rm chooz}^2 s_{\rm sol}^2 \sqrt{\delta m_{\rm sol}^2}
  +
  s_{\rm chooz}^2 \sqrt{\delta m_{\rm atm}^2 + \delta m_{\rm sol}^2}
  \,,
  \nonumber
  \\
  |m_{ee}|_{\min} 
  &=& 
  \left|\,
  c_{\rm chooz}^2 s_{\rm sol}^2 \sqrt{\delta m_{\rm sol}^2}
  -
  s_{\rm chooz}^2 \sqrt{\delta m_{\rm atm}^2 + \delta m_{\rm sol}^2}
  \,\right|
  \,,
\end{eqnarray}
where $c_X^2\equiv \cos^2\theta_X$ and $s_X^2\equiv \sin^2\theta_X$.
Here, we can safely neglect the contribution of $\delta m_{\rm sol}^2$
in the second term $s_{\rm chooz}^2 \sqrt{\delta m_{\rm atm}^2 +
\delta m_{\rm sol}^2}$, since it changes $|m_{ee}|$ at most
$\Delta|m_{ee}| < s_{\rm chooz}^2 \delta m_{\rm sol}^2 / (2
\sqrt{\delta m_{\rm atm}^2}) < 10^{-4}\EV$. 

\begin{figure}[t!]%%%       FIGURE  Mee     %%%%%%%%%%%%%%%%%
%%%%%%%%%%%%%%%%%%%%%%%%%%%%%%%%%%%%%%%%%%%%%%%%%%%%%%%%%%%%
  \psfrag{xxx}[c][c][0.6][0]
	 {$\sin^2 \theta_{\rm sol} 
	   \sqrt{\Frac
	     { \delta m_{\rm sol}^2 }
	     { \delta m_{\rm atm}^2 }}$}
	 \psfrag{mee}[c][c][0.6][-90]
		{$\Frac
		  { |m_{ee}| }
		  { \sqrt{\delta m_{\rm atm}^2} }
		  $}
		\psfrag{chooz015}[r][r][0.5][0]{$\sin\theta_{\rm chooz} =\ 0.15$}
		\psfrag{chooz010}[r][r][0.5][0]{$0.10$}
		\psfrag{chooz005}[r][r][0.5][0]{$0.05$}
		\psfrag{chooz000}[r][r][0.5][0]{$0$}
		\centerline{ \scalebox{1.8}{\includegraphics{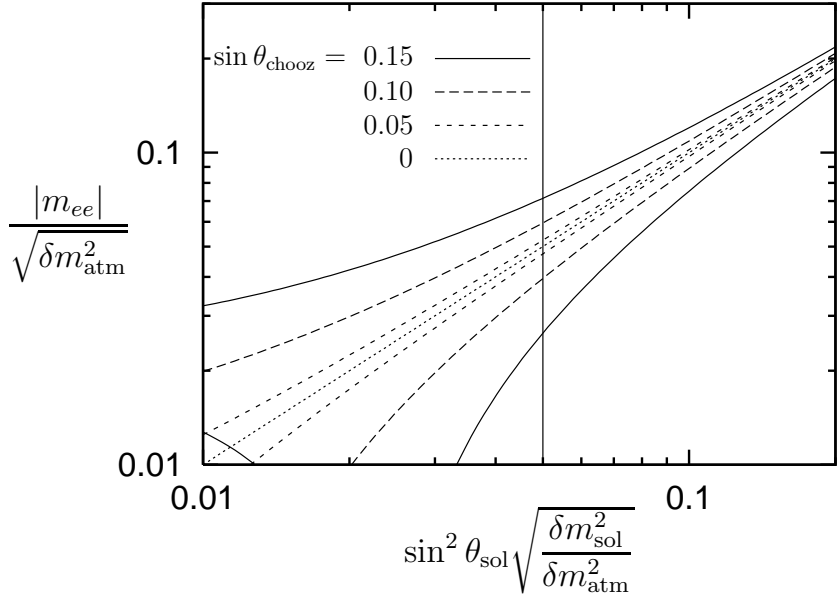}}}
		\vspace{0.7cm}
		\caption{The predicted value of $|m_{ee}|$ in units of
		  $\sqrt{\delta m_{\rm atm}^2}$, in the case of normal
		  mass hierarchy. The lines represent the upper and
		  lower values of $|m_{ee}|$ for $\sin\theta_{\rm
		  chooz} = 0.15$, $0.10$, $0.05$ and $0$. The vertical
		  line corresponds to the set of best fit values,
		  $\sin^2\theta_{\rm sol}\simeq 0.3$, $\delta m_{\rm sol}^2
		  \simeq 7\times 10^{-5}\EV^2$~\cite{PostKam} and $\delta
		  m_{\rm atm}^2 = 2.5\times 10^{-3}\EV^2$~\cite{SK-Atm}.}
%%%%%%%%%%%%%%%%%%%%%%%%%%%%%%%%%%%%%%%%%%%%%%%%%%%%%%%%%%%%
 \label{FIG-0nbb-normal}%%%%%%%%%%%%%%%%%%%%%%%%%%%%%%%%%%%%
\end{figure}

Then, the ratio of the mass parameter $|m_{ee}|$ to the mass scale of
atmospheric neutrino oscillation $\sqrt{\delta m_{\rm atm}^2}$ is
given by
\begin{eqnarray}
  \frac
      { |m_{ee}|_{\max} }
      { \sqrt{\delta m_{\rm atm}^2} }
      &=& 
      c_{\rm chooz}^2 
      s_{\rm sol}^2 
      \sqrt{\frac
	{ \delta m_{\rm sol}^2 }
	{ \delta m_{\rm atm}^2 }
      }
      +
      s_{\rm chooz}^2
      \,,
      \nonumber
      \\
      \frac{|m_{ee}|_{\min}}{\sqrt{ \delta m_{\rm atm}^2}}
      &=& 
      \left|\,
      c_{\rm chooz}^2 
      s_{\rm sol}^2 
      \sqrt{\frac
	{ \delta m_{\rm sol}^2 }
	{ \delta m_{\rm atm}^2 }
      }
      -
      s_{\rm chooz}^2
      \,\right|
      \,.
\end{eqnarray}
\FIG{FIG-0nbb-normal} shows the upper and lower bounds on $|m_{ee}|$
in units of $\sqrt{ \delta m_{\rm atm}^2}$, for $\sin\theta_{\rm
chooz} = 0.15$, $0.10$, $0.05$ and $0$.  As can be seen in
\FIG{FIG-0nbb-normal}, the mass parameter $|m_{ee}|$ is predicted in a
narrow range. We stress that this is because of the absence of the
contribution from the lightest neutrino, i.e., $\mnu{1}\ll
10^{-4}\EV$.  It is quite interesting to see that the predicted
$|m_{ee}|$ can be as large as $|m_{ee}|\simeq 0.001$--$0.01\EV$, which
could be accessible at future $0\nu\beta\beta$ decay
experiments~\cite{nu0bb}.  Furthermore, it is also found from
\FIG{FIG-0nbb-normal} that, if the $\sin\theta_{\rm chooz}$ becomes
more constrained by future experiments, $|m_{ee}|$ is predicted in a
much narrower range.

%%%%%%%%%%%%%%%%%%%%%%%%%%%%%%%%%%%%%%%%%%%%%%%%%%%%%%%%%%%%
%%%%%%%%%%%%%%%%%%%%%%%%%%%%%%%%%%%%%%%%%%%%%%%%
%%%%%%%%%%%%%%%%%%%%%%%%%%%%%%%%%%%%%%%%%%%%%%%%%%%%%%%%%%%%
%%%%%%%%%%%%%%%%%%%%%%%%%%%%%%%%%%%%%%%%%%%%%%%%%%%%%%%%%%%%
\begin{figure}[h!]%%%       FIGURE  Mee     %%%%%%%%%%%%%%%%%
%%%%%%%%%%%%%%%%%%%%%%%%%%%%%%%%%%%%%%%%%%%%%%%%%%%%%%%%%%%%
  \psfrag{xxx}[c][c][0.6][0]{$\tan^2 \theta_{\rm sol}$}
  \psfrag{mee}[c][c][0.6][-90]
	 {$\Frac
	   { |m_{ee}| }
	   { \sqrt{\delta m_{\rm atm}^2} }
	   $}
	 \centerline{ \scalebox{1.8}{\includegraphics{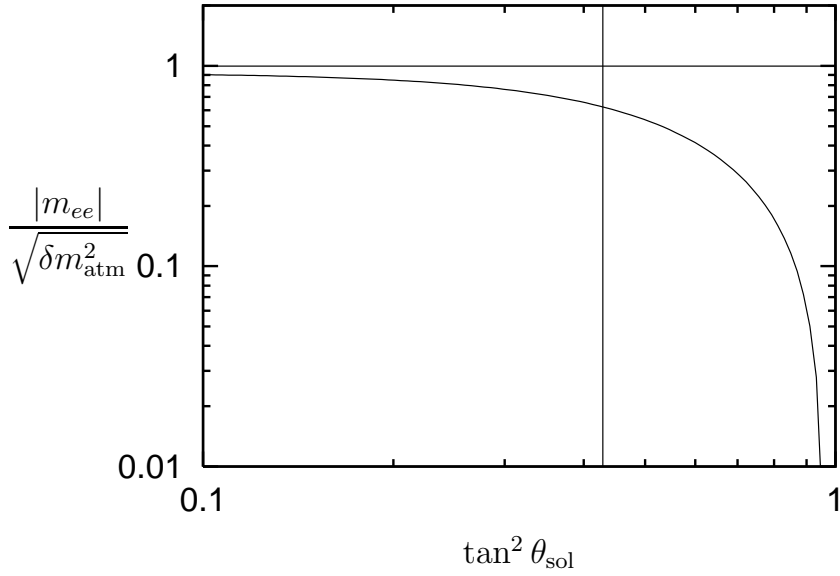}}}
	 \caption{The predicted value of $|m_{ee}|$ in units of
	   $\sqrt{\delta m_{\rm atm}^2}$, in the case of inverted mass
	   hierarchy. The lines represent the upper and lower values
	   of $|m_{ee}|$. The vertical line corresponds to the best
	   fit value, $\sin^2\theta_{\rm sol}\simeq 0.3$~\cite{PostKam}.}
	   %%%%%%%%%%%%%%%%%%%%%%%%%%%%%%%%%%%%%%%%%%%%%%%%%%%%%%%%%%%%
	 \label{FIG-0nbb-inverted}%%%%%%%%%%%%%%%%%%%%%%%%%%%%%%%%%%%%
\end{figure}%%%%%%%%%%%%%%%%%%%%%%%%%%%%%%%%%%%%%%%%%%%%%%%%
%%%%%%%%%%%%%%%%%%%%%%%%%%%%%%%%%%%%%%%%%%%%%%%%%%%%%%%%%%%%

%%%%%%%%%%%%%%%%%%%%%%%%%%%%%%%%%%%%%%%%%%%%%%%%%%%%%%%%%%%%
%%%%%%%%%%%%%%%%%%%%%%%%%%%%%%%%%%%%%%%%%%%%%%%%%%%%%%%%%%%%
%%%%%%%%%%%%%%%%%%%%%%%%%%%%%%%%%%%%%%%%%%%%%%%%%%%%%%%%%%%%
\subsubsection*{inverted hierarchy}
%%%%%%%%%%%%%%%%%%%%%%%%%%%%%%%%%%%%%%%%%%%%%%%%%%%%%%%%%%%%
%%%%%%%%%%%%%%%%%%%%%%%%%%%%%%%%%%%%%%%%%%%%%%%%%%%%%%%%%%%%
%%%%%%%%%%%%%%%%%%%%%%%%%%%%%%%%%%%%%%%%%%%%%%%%%%%%%%%%%%%%

Now let us turn to discuss the case of inverted mass hierarchy, where
the parameters of atmospheric and solar neutrino oscillations are
given by\footnote{The convention adopted here is different from the
one in Ref.~\cite{FHY-ADL-2}, where $\mnu{i}$ are defined as $\mnu{3}
< \mnu{1} < \mnu{2}$ for inverted hierarchy.}
\begin{equation}
  \delta m_{\rm atm}^2 = \mnu{3}^2\,,
  \qquad 
  \delta m_{\rm sol}^2 = \mnu{3}^2 - \mnu{2}^2\,,
  \qquad
  \tan^2\theta_{\rm sol} \equiv |U_{e3}/U_{e2}|^2\,.
\end{equation}
The mixing angle constrained by CHOOZ experiment~\cite{CHOOZ} now
corresponds to $\sin\theta_{\rm chooz} = |U_{e1}|$. Then from
\EQS{EQ-mee-bounds} we obtain
\begin{eqnarray}
  \frac
      { |m_{ee}|_{\max} }
      { \sqrt{\delta m_{\rm atm}^2} }
      &=& 
      c_{\rm chooz}^2
      \left[
	c_{\rm sol}^2
	\sqrt{
	  1-
	  \frac
	  { \delta m_{\rm sol}^2 }
	  { \delta m_{\rm atm}^2 }
	}
	+
	s_{\rm sol}^2 	
	\right]
      \,,
      \nonumber
      \\
      \frac
	  { |m_{ee}|_{\min} }
	  { \sqrt{\delta m_{\rm atm}^2} }
	  &=& 
	  c_{\rm chooz}^2
	  \left|
	  c_{\rm sol}^2
	  \sqrt{
	    1-
	    \frac
		{ \delta m_{\rm sol}^2 }
		{ \delta m_{\rm atm}^2 }
	  }
	  -
	  s_{\rm sol}^2 	
	  \right|
	    \,.
\end{eqnarray}
In \FIG{FIG-0nbb-inverted}, we show the upper and lower bounds on
$|m_{ee}|$ in units of $\sqrt{ \delta m_{\rm atm}^2}$, for given
$\tan^2\theta_{\rm sol}$. Here, we have used $\sin\theta_{\rm
chooz}=0$ for simplicity, but $|m_{ee}|$ changes (reduces) only 2\%
even for $\sin\theta_{\rm chooz}=0.15$. As for the ratio
$r\equiv\delta m_{\rm sol}^2/\delta m_{\rm atm}^2$, we have adopted $r
= 0.008$--$0.15$~\cite{SK-Atm,PostKam}. As seen from the figure, the
$|m_{ee}|$ is restricted in a very narrow range such as
$|m_{ee}|\simeq (0.01$--$0.05)\EV\times (\sqrt{\delta m_{\rm
atm}^2}/0.05\EV)$ for $\tan^2\theta_{\rm sol} < 0.7$, which is indeed
in the reach of the future $0\nu\beta\beta$ decay experiments~\cite{nu0bb}.

\vspace{1cm}

The prediction of $|m_{ee}|$ discussed in this section is a generic
consequence of a hierarchical neutrino masses $\mnu{1}\ll \mnu{2}$,
$\mnu{3}$. Thus the $0\nu\beta\beta$ decay experiments, combined with
neutrino oscillation experiments, can provide a consistency test for
the tiny mass of the lightest neutrino $\mnu{1}\ll 10^{-4}\EV$
required in the leptogenesis via $L H_u$ flat direction.

\clearpage
%%%%%%%%%%%%%%%%%%%%%%%%%%%%%%%%%%%%%%%%%%%%%%%%%%%%%%%%%%%%%%%%%%%
% Change the label of section to *.A.  %%%%%%%%%%%%%%%%%%%%%%%%%%%%
%%%%%%%%%%%%%%%%%%%%%%%%%%%%%%%%%%%%%%%%%%%%%%%%%%%%%%%%%%%%%%%%%%%
\renewcommand{\thesection}{\thechapter.A}%%%%%%%%%%%%%%%%%%%%%%%%%%
%%%%%%%%%%%%%%%%%%%%%%%%%%%%%%%%%%%%%%%%%%%%%%%%%%%%%%%%%%%%%%%%%%%
\section{$D$-flat condition for the $L H_u$ flat direction}%%%%%%%%
%%%%%%%%%%%%%%%%%%%%%%%%%%%%%%%%%%%%%%%%%%%%%%%%%%%%%%%%%%%%%%%%%%%
\label{SEC-App-Dflat}

In this appendix, we discuss the $D$-flat condition along the $L H_u$
direction.  Let us take the number of generations of the lepton doublets
$L_i$ to be $N$ in general. Then, there are $(2 + 2 N)$ complex scalar
fields in the $L$-$H_u$ sector.
%%%
\begin{eqnarray}
 \widetilde{L_i} &=&
  \left(
   \begin{array}{c}
    \widetilde{\nu_L}_i\\
    \widetilde{e_L}_i
   \end{array}
   \right)
   \qquad
   i = 1\cdots N
   \,,
   \nonumber\\
 H_u &=& 
  \left(
   \begin{array}{c}
    H_u^+\\
    H_u^0
   \end{array}
   \right)
   \,.
\end{eqnarray}
First, the freedom to make ${\rm SU}(2)_L$ gauge transformation allows
us to rotate away a possible vacuum expectation value of one of the weak
isospin components of one of the scalar fields. Thus, we can take $H_u^+
= 0$ without loss of generality. Then, vanishing $D$-term potential
requires the following relations:
%%%
\begin{eqnarray}
 - |\widetilde{\nu_L}_i|^2 
  - |\widetilde{e_L}_i|^2
  + |H_u^0|^2
  &=&
  0
  \qquad
  {\rm for\,\,U}(1)_Y
  \,,
  \\
 &&\nonumber\\
  \widetilde{\nu_L}_i^*
  \widetilde{e_L}_i
  +
  \widetilde{e_L}_i^*
  \widetilde{\nu_L}_i
  &=&
  0
  \nonumber\\
  \widetilde{\nu_L}_i^*
  \widetilde{e_L}_i
  -
  \widetilde{e_L}_i^*
  \widetilde{\nu_L}_i
  &=&
  0
  \qquad
  {\rm for\,\,SU}(2)_L
  \,,
  \nonumber\\
  |\widetilde{\nu_L}_i|^2
  -
  |\widetilde{e_L}_i|^2
  -
  |H_u^0|^2
  &=&
  0
\end{eqnarray}
%%%
where the summations over $i$ are suppressed for simplicity. {}From
these equations, the $D$-flat condition is given by
%%%
\begin{eqnarray}
 \label{EQ-D-flat}
 \sum_{i=1}^N
  |\widetilde{\nu_L}_i|^2
  =
  |H_u^0|^2
  \,.
\end{eqnarray}
It is still possible to rotate away one of the phases by using the ${\rm
U}(1)_Y$ gauge transformation (or third component of the ${\rm SU}(2)_L$
gauge transformations\footnote{Notice that one ${\rm U}(1)$ group
remains unbroken by the condensation of the flat direction
field.}). Therefore, this flat ``direction'' is parameterized by [$(N+1)$
complex fields] $-$ [one real constraint \EQ{EQ-D-flat}] $-$ [one phase]
$=2N$ real fields.  In the absence of effective superpotential
\EQ{EQ-LHu-start}, this ``direction'' would be completely flat in the
SUSY limit. Furthermore, it would be isotropic for $N$ directions,
$\widetilde{\nu_L}_1,\cdots,\widetilde{\nu_L}_N$. However, due to the
scalar potential induced by the operator \EQ{EQ-LHu-start}, the
potential is lifted and modified. Then, the flattest direction among the
above field space is parameterized by one scalar field $\phi$, which
corresponds to the first family $i = 1$.

%%%%%%%%%%%%%%%%%%%%%%%%%%%%%%%%%%%%%%%%%%%%%%%%%%%%%%%%%%%%%%%%%%%
% Change again the label of the section to *.number. %%%%%%%%%%%%%%
%%%%%%%%%%%%%%%%%%%%%%%%%%%%%%%%%%%%%%%%%%%%%%%%%%%%%%%%%%%%%%%%%%%
\renewcommand{\thesection}{\thechapter.\arabic{section}}%%%%%%%%%%%
%%%%%%%%%%%%%%%%%%%%%%%%%%%%%%%%%%%%%%%%%%%%%%%%%%%%%%%%%%%%%%%%%%%

%%%%%%%%%%%%%%%%%%%%%%%%%%%%%%
%\include{Chap-Conc}%%%%%%%%%%%
%%%%%%%%%%%%%%%%%%%%%%%%%%%%%%%%%%%%%%%%%%%%%%%%%%%%%%%%%%%%%%%%%%%
%%%%%%%%%%%%%%%%%%%%%%%%%%%%%%%%%%%%%%%%%%%%%%%%%%%%%%%%%%%%%%%%%%%
%%%%%%%%%%%%%%%%%%%%%%%%%%%%%%%%%%%%%%%%%%%%%%%%%%%%%%%%%%%%%%%%%%%
\chapter{Conclusions and discussion}%%%%%%%%%%%%%%%%%%%%%%%%%%%%%%%
%%%%%%%%%%%%%%%%%%%%%%%%%%%%%%%%%%%%%%%%%%%%%%%%%%%%%%%%%%%%%%%%%%%
%%%%%%%%%%%%%%%%%%%%%%%%%%%%%%%%%%%%%%%%%%%%%%%%%%%%%%%%%%%%%%%%%%%
%%%%%%%%%%%%%%%%%%%%%%%%%%%%%%%%%%%%%%%%%%%%%%%%%%%%%%%%%%%%%%%%%%%
\label{SEC-Conc}

%%%%%%%%%%%%%%%%%%%%%%%%
\begin{table}%%%%%%%
 %%%%%%%%%%%%%%%%%%%%%%%
 \begin{center}
  \begin{tabular}{|c|cl|}
   \hline
   & \multicolumn{2}{|c|}{}
   \\
   & \multicolumn{2}{|c|}{produced baryon asymmetry}
   \\
   & \multicolumn{2}{|c|}{}
   \\
   \hline \hline
   &&
   \\
   thermal$^{~(1)}$
   & $\Frac{n_B}{s} $
   & $\simeq 0.35\times \kappa\,\Frac{\epsilon_1}{240}$
   \\
   &
   & $\simeq 0.3\times 10^{-10}
   \left(\Frac{\kappa}{0.1}\right)
   \left(\Frac{M_1}{10^9\GEV}\right)
   \cdot\left(\Frac{\mnu{3}}{0.05\EV}\right)
   \delta_{\rm eff}
   $
   \\
   &&
   \\
   \hline
   &&
   \\ 
   inflaton decay$^{~(2)~(a)}$
   & $\Frac{n_B}{s} $
   & $\simeq 0.35\times \Frac{3}{2}\epsilon_1 B_r \Frac{T_R}{m_\chi}$
   \\
   &
   & $\simeq 0.5\times 10^{-10}B_r
   \left(\Frac{T_R}{10^6\GEV}\right)
   \left(\Frac{2\,M_1}{m_\chi}\right)
   \cdot\left(\Frac{\mnu{3}}{0.05\EV}\right)
   \delta_{\rm eff}$
   \\
   &&
   \\
   \hline
   &&
   \\
   $\widetilde{N}$ dominant$^{~(3)}$
   & $\Frac{n_B}{s} $
   & $\simeq 0.35\times \Frac{3}{4}\epsilon_1 \Frac{T_{N_1}}{M_1}$
   \\
   &
   & $\simeq 0.5\times 10^{-10}\left(\Frac{T_{N_1}}{10^6\GEV}\right)
   \cdot\left(\Frac{\mnu{3}}{0.05\EV}\right)
   \delta_{\rm eff}$
   \\
   &&
   \\ 
   \hline \hline
   &&
   \\
   $L H_u$\qquad
   & $\Frac{n_B}{s} $
   & $\simeq 0.35\times
   \Frac{M_{\rm eff}T_R}{12 M_G^2}
   \left(
   \Frac{m_{3/2}|a_m|}{H_{\rm osc}}
   \right)
   \delta_{\rm ph}$
   \\
   flat direction$^{~(4)~(b)}$
   &&
   \\
   && $\to\,\,\mnu{1}\simeq (0.1$--$3)\times 10^{-9}\EV^{~(c)}$
   \,\,for $T_R\gsim 10^5\GEV$
   \\
   &&  \quad(See \FIG{FIG-LHu-numerical}.)
   \\
   &&
   \\
   \hline
   &&
   \\
   $+$ gauged ${\rm U}(1)_{B-L}$
   & $\Frac{n_B}{s} $
   & $\simeq 0.35\times 
   \Frac{M_{\rm eff}T_R}{\sqrt{3}M_G^2}|c'_\phi|\delta_{\rm ph}$
   \\
   \,$D$-term stopping$^{~(5)}$
   &
   & $\simeq 1\times 10^{-10}
   \left(\Frac{T_R}{10^8\GEV}\right)
   \left(\Frac{10^{-6}\EV}{\mnu{1}}\right)^{-1}
   |c'_\phi|\delta_{\rm ph}$
   \\
   &&
   \\
   \hline
  \end{tabular}
 \end{center}
 \caption{Summary of produced baryon asymmetries in various
 leptogenesis mechanisms. Necessary conditions are: (1) $T_R >
 M_1$. (2) $M_1 > T_R$ and $m_\chi > 2 M_1$. (3) $M_1 > T_{N_1}$ and
 $T_R\gg T_{N_1}$ (for $|\sneu_i|\simeq M_G$). (4) negative Hubble
 mass for $\phi$. (5) See \EQ{EQ-Dstop} and \FIG{FIG-BL-allowed}. Some
 comments are in order: (a) Perturbative decay is assumed. For
 preheating, see \SEC{SEC-preheating}. (b) $H_{\rm inf} > H_{\rm osc}$
 is assumed. (See the last comment in \SEC{SEC-resultinLHu}.) (c) This
 is not the case if effective mass for the right-handed neutrinos are
 very large during leptogenesis (\SEC{SEC-model-PQ}). }
 %%%%%%%%%%%%%%%%%%%%%%%
 \label{Table-Summary}%%
 %%%%%%%%%%%%%%%%%%%%%%%
\end{table}%%%%%%%%%%%%%
%%%%%%%%%%%%%%%%%%%%%%%%

In this thesis, we have investigated various leptogenesis mechanisms in
the framework of SUSY. We have also discussed some aspects of those
scenarios, such as constraints from cosmological gravitino problems and
possible experimental signatures. In particular, we have made essential
improvements compared with earlier works in the following points:
\begin{itemize}

 \item In the leptogenesis scenario where the lepton asymmetry is
       created by the decay of right-handed sneutrino $\widetilde{N}$
       having dominated the early universe (\SEC{SEC-Ntilde}), the
       dilution of the gravitino due to the entropy production by the
       $\widetilde{N}$'s decay has been discussed. As a result, it has
       been shown that the cosmological gravitino problem is drastically
       ameliorated.

 \item In the leptogenesis via the $L H_u$ flat direction
       (Chapter~\ref{Chap-LHu}), all the relevant thermal effects has
       been taken into account, and detailed analyses on the dynamics of
       the flat direction field has been done both by analytic and
       numerical calculations.

\end{itemize}

In this chapter, we summarize the aspects of various leptogenesis
mechanisms investigated in this thesis. The produced baryon asymmetry in
each mechanism is well summarized in Table~\ref{Table-Summary}.

First of all, we see that in all the mechanisms the neutrino mass plays
a very crucial role to determine the resultant baryon asymmetry. In the
mechanisms where lepton asymmetry is produced by the asymmetric decay of
the right-handed (s)neutrino (Chapter~\ref{Chap-Ndecay}), the asymmetry
parameter $\epsilon_1$ is proportional to the mass of the heaviest
neutrino $\mnu{3}$ for fixed effective $CP$-violating phase $\delta_{\rm
eff}$. We see that the information from the atmospheric neutrino
oscillation~\cite{SK-Atm}, which leads to $\mnu{3}\simeq 0.05\EV$ as
long as masses of the light neutrinos are not degenerate, is very
important to calculate the final baryon asymmetry.  On the other hand,
when the lepton asymmetry is produced via the $L H_u$ flat direction
(Chapter~\ref{Chap-LHu}), the mass of the lightest neutrino $\mnu{1}$ is
the crucial parameter.

It is interesting to see that the dependences on the neutrino masses are
opposite in those two scenarios. As for the $\epsilon_1$ parameter of the
right-handed neutrino decay, larger Yukawa coupling is favored to make
$\epsilon_1$ larger. When we normalize the Yukawa couplings by neutrino
masses (\SEC{SEC-LAfromN}), $\epsilon_1$ turns out to be proportional to
$\mnu{3}$ for fixed $\delta_{\rm eff}$. On the other hand, for
leptogenesis via $L H_u$ flat direction, flatter direction is favored to
enhance the final lepton asymmetry. This is because, the flatter the
potential is, the larger the amplitude of $\phi$ becomes, and the larger 
the final lepton asymmetry is produced. Thus in this case the relevant
neutrino mass is the lightest one, $\mnu{1}$.

Now let us comment on each mechanism in turn.

\subsubsection*{$\bullet$ thermal production of the right-handed (s)neutrino}

The simplest one is the thermal production of the right-handed
(s)neutrino (\SEC{SEC-thermal}). The attractive point in this scenario
is nothing but the fact that the production of the right-handed
(s)neutrino requires no assumption besides high enough reheating
temperature $T_R > M_1$. The production occurs just by ordinary thermal
scatterings.

On the other hand, since this scenario requires high enough reheating
temperature, cosmological gravitino problems should be considered
seriously. One solution is the stable (LSP) gravitino with mass
$m_{3/2}\sim 100\GEV$. In this case, one has to take care of the decays
of the next-to-lightest SUSY particle (NLSP) after the big-bang
nucleosynthesis~\cite{Gprob-GMSB,LSPgrav}, which gives severe
constraints on the lifetime and the abundance of the NLSP at the
decay. Since both of the lifetime and the abundance of the NLSP is
determined by the properties of the NLSP (and the gravitino mass), they
might be going to be well calculated if SUSY particles are discovered
and their properties are studied in future collider experiments. This is
quite an interesting possibility.

If the gravitino is not the LSP, cosmological gravitino problem is
relatively severe.  Therefore, it will be interesting and might be
necessary to determine a definite lower bound on the reheating
temperature $T_R$ at which the leptogenesis from thermally produced
right-handed (s)neutrino works marginally. This requires to see what
happens if the reheating temperature is just around the mass of the
right-handed neutrino, $T_R\sim M_1$.

\subsubsection*{$\bullet$ leptogenesis in inflaton decay}

The advantage of this mechanism is that the production of the
right-handed (s)neutrino is possible for $T_R < M_1$. Thus the
cosmological gravitino problem can be avoided in a wider range of the
gravitino mass. Actually, we have shown that the enough lepton
asymmetry can be produced even with a reheating temperature $T_R\simeq
10^6\GEV$.  On the other hand, the produced baryon asymmetry crucially
depends on the physics of the inflation. In particular, we have taken
the branching ratio $B_r(\chi \to N_1 N_1)$ to be $B_r\simeq 1$. In
this sense the first one of the hybrid inflation models we have
discussed in \SEC{SEC-Hybrid-1}, which has Yukawa couplings between
the inflaton and right-handed neutrinos, is attractive since it
automatically gives rise to $B_r\simeq 1$.

We should mention that the current and future satellite experiments on
anisotropies of the cosmic microwave background
radiation~\cite{MAP-Planck} will give precious information about the
inflation physics, especially the spectrum index $n_s$. Thus we will
be able to distinguish which inflation model is favored by using
future data.

By the way, from Table~\ref{Table-Summary}, it is found that the formula
of the baryon asymmetry from inflaton decay reproduces that of the
$\widetilde{N}$-dominant universe if we take $B_r = 1$, $T_R = T_{N_1}$
and $2 M_1 = m_\chi$. This corresponds to the case where the
right-handed sneutrino plays a role of the inflaton
itself~\cite{MSYY-1,MSYY-2}.

\subsubsection*{$\bullet$ $\widetilde{N}$-dominant universe}

The resultant baryon asymmetry in this scenario is expressed in a very
simple way, as given in Table~\ref{Table-Summary}. It is determined only
by the decay temperature $T_{N_1}$ for fixed $\mnu{3}$ and $\delta_{\rm
eff}$. This is because the $\sneu$'s decay produces dominant component
of the radiation as well as the lepton asymmetry at once.

The most attractive feature of this scenario is the dilution of the
gravitino. In fact, as we have shown, cosmological gravitino problems
can be avoided even for $T_R\gg 10^{11}\GEV$, which is very special
among other baryogenesis scenarios.

It is also quite an interesting possibility that the cosmic density
perturbation coming from the $\sneu$'s fluctuation might be detected
at future experiments~\cite{MM-Ntilde}.

\subsubsection*{$\bullet$ $L H_u$ flat direction}

We have performed a detailed analysis on this scenario, including all
the relevant thermal effects. In spite of the complicated potential and
dynamics of the flat direction field $\phi$, it was shown that the
result is very simple and attractive. The baryon asymmetry is determined
almost only by the mass of the lightest neutrino $\mnu{1}$. Remarkably,
we can estimate (or predict) the $\mnu{1}$ from the empirical baryon
asymmetry $n_B/s\simeq (0.4$--$1)\times 10^{-10}$ as
\begin{eqnarray}
 \mnu{1}\simeq (0.1-3)\times 10^{-9}\EV
  \,,
\end{eqnarray}
almost
independently of the reheating temperature for $T_R\gsim 10^5\GEV$. It
is amazing that the baryogenesis which happened in the very early
universe is directly related to the neutrino mass.

\subsubsection*{$\bullet$ $L H_u$ flat direction $+$ gauged ${\rm
U}(1)_{B-L}$, $D$-term stopping case}

When we introduce an gauged ${\rm U}(1)_{B-L}$ symmetry, the dynamics of
the $L H_u$ flat direction field $\phi$ can be modified, if certain
conditions are satisfied ($D$-term stopping case). Then the resultant
baryon asymmetry is enhanced compared with the case with $F$-term
stopping case, or the case without a gauged ${\rm U}(1)_{B-L}$ symmetry. 
Although the ``reheating temperature independence'' disappears in this
case (and $\mnu{1}$ is no longer predicted), this scenario has an
advantage in the case of gauge-mediated SUSY breaking (GMSB)
models~\cite{GMSB}, compared to the previous case. The point is that the
phase rotational motion of the $\phi$ field is provided not by the usual
SUSY-breaking $A$-term which is proportional to the gravitino mass, but
by a Hubble-induced $A$-term. Thus this mechanism can work even with a
very light gravitino as in the GMSB models.

\subsubsection*{$\bullet$ The neutrinoless double beta decay}

We have also shown that a very light mass of the lightest neutrino
$\mnu{1}\ll 10^{-4}\EV$, which is suggested by the above leptogenesis
mechanisms via $L H_u$ flat direction, leads to a high predictability
of the rate of the neutrino-less double beta ($0\nu\beta\beta$)
decay. Although it is highly difficult to determine the mass of the
lightest neutrino directly, future experiments on $0\nu\beta\beta$
decay and neutrino oscillations will provide us precious information
by which we can test the consistency of the very small lightest
neutrino mass $\mnu{1}\ll 10^{-4}\EV$, as shown in
\FIG{FIG-0nbb-normal} and \ref{FIG-0nbb-inverted}.

%%%%%%%%%%%%%%%%%%%%%%%%%%%%%%%%%%%%%%%%%%%%%%%%%%%%%%%%%%%%%%%%%%%
\newpage%%%%%%%%%%%%%%%%%%%%%%%%%%%%%%%%%%%%%%%%%%%%%%%%%%%%%%%%%%%
\section*{Acknowledgement}%%%%%%%%%%%%%%%%%%%%%%%%%%%%%%%%%%%%%%%%%
%%%%%%%%%%%%%%%%%%%%%%%%%%%%%%%%%%%%%%%%%%%%%%%%%%%%%%%%%%%%%%%%%%%

\hspace{\parindent}I am very grateful to my advisor T.~Yanagida for
various instructive suggestions, stimulating discussions and
collaborations, hearty encouragements and continuous support.

I would like to thank T.~Asaka, M.~Fujii, M.~Kawasaki and H.~Murayama,
who are collaborators in various parts of this thesis, for fruitful
discussions and collaborations.

I would also like to thank T.~Watari for helpful discussions, and
K.~Ichikawa for helping me maintain the computer.

Finally, I thank all the members of particle physics theory group at
University of Tokyo for their hospitality.

%%%%%%%%%%%%%%%%%%%%%%%%%%%%%%
\appendix%%%%%%%%%%%%%%%%%%%%%
%%%%%%%%%%%%%%%%%%%%%%%%%%%%%%
%\include{Chap-App}%%%%%%%%%%%%
%%%%%%%%%%%%%%%%%%%%%%%%%%%%%%%%%%%%%%%%%%%%%%%%%%%%%%%%%%%%%%%%%%%
%%%%%%%%%%%%%%%%%%%%%%%%%%%%%%%%%%%%%%%%%%%%%%%%%%%%%%%%%%%%%%%%%%%
%%%%%%%%%%%%%%%%%%%%%%%%%%%%%%%%%%%%%%%%%%%%%%%%%%%%%%%%%%%%%%%%%%%
\chapter{Notations}%%%%%%%%%%%%%%%%%%%%%%%%%%%%%%%%%%%%%%%%%%%%%%%%
%%%%%%%%%%%%%%%%%%%%%%%%%%%%%%%%%%%%%%%%%%%%%%%%%%%%%%%%%%%%%%%%%%%
%%%%%%%%%%%%%%%%%%%%%%%%%%%%%%%%%%%%%%%%%%%%%%%%%%%%%%%%%%%%%%%%%%%
%%%%%%%%%%%%%%%%%%%%%%%%%%%%%%%%%%%%%%%%%%%%%%%%%%%%%%%%%%%%%%%%%%%
\label{SEC-Notation}

We use natural units $c= h/(2\pi)=1$, where $c$ is the speed of light
(in vacuum) and $h$ is the Planck constant. We also take the Boltzmann
constant $k_B$ to be unity, which leads to $1\EV = 11605 K$.

$M_G \equiv (8\pi G)^{-1/2} = 2.4\times 10^{18}\GEV$ denotes the reduced 
Planck scale, where $G$ is the Newton constant.

Throughout this thesis, except for generic discussions in
\SEC{SEC-LHu-overview}, \ref{SEC-LHu}, \ref{SEC-double-beta} and
\ref{SEC-App-Dflat}, we assume the existence of the right-handed
neutrinos, which have superpotentials:
%%%
\begin{eqnarray}
 W = \frac{1}{2} M_i N_i N_i + h_{i\alpha} N_i L_{\alpha} H_u
  \,,
\end{eqnarray}
%%%
where $N_i$ ($i = 1,2,3$), $L_{\alpha}$ ($\alpha = e, \mu, \tau$) and
$H_u$ denote the supermultiplets of the heavy right-handed neutrinos,
lepton doublets and the Higgs doublet which couples to up-type quarks,
respectively. 

We often use a basis where the mass matrix of the light neutrinos
becomes diagonal, where the Yukawa couplings are given by
$\widehat{h}_{ij}$. (The left subscript $i$ denotes the family of the
heavy right-handed neutrinos $N_i$, while the right subscript $j$
denotes the family of the light neutrinos $\nu_j$ in mass
eigenstates. See \SEC{SEC-LAfromN} and \EQ{EQ-rtd-Ykw}.)

As for the parameters related to the cosmology, see next
Appendix~\ref{SEC-App-SCInf}. 

%%%%%%%%%%%%%%%%%%%%%%%%%%%%%%%%%%%%%%%%%%%%%%%%%%%%%%%%%%%%%%%%%%%
%%%%%%%%%%%%%%%%%%%%%%%%%%%%%%%%%%%%%%%%%%%%%%%%%%%%%%%%%%%%%%%%%%%
%%%%%%%%%%%%%%%%%%%%%%%%%%%%%%%%%%%%%%%%%%%%%%%%%%%%%%%%%%%%%%%%%%%
\chapter{Some notes of standard cosmology}%%%%%%%%%%%%%%%%%%%%%%%%%
%%%%%%%%%%%%%%%%%%%%%%%%%%%%%%%%%%%%%%%%%%%%%%%%%%%%%%%%%%%%%%%%%%%
%%%%%%%%%%%%%%%%%%%%%%%%%%%%%%%%%%%%%%%%%%%%%%%%%%%%%%%%%%%%%%%%%%%
%%%%%%%%%%%%%%%%%%%%%%%%%%%%%%%%%%%%%%%%%%%%%%%%%%%%%%%%%%%%%%%%%%%
\label{SEC-App-SCInf}

In this appendix, we briefly review some ingredients of the standard
cosmology, in order to introduce our notations and relevant
equations. For more details, see, for example, Ref.~\cite{KT}.

%%%%%%%%%%%%%%%%%%%%%%%%%%%%%%%%%%%%%%%%%%%%%%%%%%%%%%%%%%%%%%%%%%%
\section{Friedmann-Robertson-Walker universe}%%%%%%%%%%%%%%%%%%%%%%
%%%%%%%%%%%%%%%%%%%%%%%%%%%%%%%%%%%%%%%%%%%%%%%%%%%%%%%%%%%%%%%%%%%

The standard cosmology is based on the following homogeneous and
isotropic metric (Robertson-Walker metric):
%%%
\begin{eqnarray}
 ds^2 = dt^2 - R(t)^2
  \left\{
   \frac{dr^2}{1 - k r^2}
   + r^2
   \left(
    d \theta^2
    +
    \sin^2\theta\,d\phi^2
    \right)
     \right\}
      \,,
\end{eqnarray}
%%%
where $R(t)$ is the ``scale factor'' of the expanding universe, and $t$
is the cosmic time. $(t, r, \theta, \phi)$ is called  the comoving
coordinates. {}From the above metric and the perfect-fluid form of the
energy-momentum tensor $T_{\mu\nu} = {\rm diag}(\rho, p, p, p)$
($\rho(t)$ and $p(t)$ denote the energy density and the pressure,
respectively), the Einstein equations lead to the following two
equations:
%%%
\begin{eqnarray}
 \left(
  \frac{\dot{R}}{R}
  \right)^2
  +
  \frac{k}{R^2}
  &=&
  \frac{8\pi G}{3}
  \rho
  \label{EQ-Friedman-org}
  \,,
  \\
  \frac{d}{dt}
  \left(
   \rho R^3
   \right)
   &=&
   -p
   \frac{d}{dt}
   \left(
    R^3
    \right)
    \label{EQ-rho-p}
    \,,
\end{eqnarray}
where the overdot denotes the derivative with cosmic time $t$, and $G$
is the Newton constant.

The second equation~(\ref{EQ-rho-p}), which corresponds to the
conservation of the energy and momentum $T^{\mu\nu}_{\quad ;\nu}=0$,
describes the dilution of the energy density due to the expansion:
\begin{eqnarray}
 \label{EQ-dilutions}
  {\rm radiation} 
  &(p_{\rm rad} = \frac{1}{3}\rho_{\rm rad})&
  \rho_{\rm rad}\propto R^{-4}
  \,,
  \nonumber\\
 {\rm matter}
  &(p_{\rm mat} = 0)&
  \rho_{\rm mat}\propto R^{-3}
  \,,
  \nonumber\\
 {\rm vacuum}
  &(p_{\rm vac} = -\rho_{\rm vac})&
  \rho_{\rm vac} = const
  \,.
\end{eqnarray}
Notice that the above relations hold for each component of the energy
density $\rho_X$, as long as the energy-momentum conservation
$T^{\mu\nu}_{X\,\,\, ;\nu} = 0$ is satisfied separately for each $X$.

The other equation~(\ref{EQ-Friedman-org}), called Friedmann equation,
describes the expansion of the universe.  Notice that the curvature term
$k/R^2$ decreases with time more slowly than the right-hand side of
\EQ{EQ-Friedman-org} as long as $\rho$ is dominated by matter
($\rho_{\rm mat}$) or radiation ($\rho_{\rm rad}$) [see
\EQ{EQ-dilutions}]. Therefore, it can be safely neglected as far as we
consider the early universe.

The Hubble parameter is defined as
%%%
\begin{eqnarray}
 H \equiv \frac{\dot{R}}{R}
  \,,
\end{eqnarray}
%%%
which means how fast the universe is expanding at a given stage of its
expansion. In terms of this Hubble parameter $H$, reduced Planck mass
$M_G\equiv (8\pi G)^{-1/2}=2.4\times 10^{18}\GEV$, critical energy
density $\rho_{\rm crit}\equiv 3 M_G^2 H^2$ and density parameter
$\Omega\equiv\rho/\rho_{\rm crit}$, Eq.~(\ref{EQ-Friedman-org}) is
rewritten as follows:
\begin{eqnarray}
 \Omega
  =
  \frac{\rho}{3 M_G^2 H^2}
  = 1 + \frac{k}{R^2 H^2}
  \,.
\end{eqnarray}
Neglecting the curvature term, usually we use the following form of the
Friedmann equation:
\begin{eqnarray}
 \label{EQ-Friedmann}
 \rho = 3 M_G^2 H^2
  \,.
\end{eqnarray}

The present universe ($t=t_0$) observed is as follows~\cite{PDB}. The
present Hubble expansion rate is $H_0 = 100\,h\,[{\rm km}\,{\rm
sec}^{-1}\,{\rm Mpc}^{-1}]$, where $h \simeq 0.7$, and hence critical
energy density is $\rho_{\rm crit}(t_0)=3 M_G^2 H_0^2 = 1.9\,h^2\times
10^{-29}\,[{\rm g}\,{\rm cm}^{-3}]$. In terms of density parameter
$\Omega_X \equiv \rho_X / \rho_{\rm crit}$, total energy density is
given by $\Omega(t_0)\simeq 1$, which includes radiation $\Omega_{\rm
rad}(t_0) = (2.47 + 0.56 N_{\nu}^{\rm eff})\, h^{-2} \times 10^{-5}$
($N_{\nu}^{\rm eff}$ is the number of the generations of neutrinos which
are still relativistic today), matter $\Omega_{\rm mat}(t_0)\simeq 0.3$
(including baryon $\Omega_B(t_0)\simeq 0.02 h^{-2}$) and ``dark energy''
$\Omega_{\rm vac}(t_0)\simeq 0.7$.\footnote{The notation ``$\Omega_{\rm
vac}$'' might not be correct, since the ``dark energy'' density could be
a dynamical value (like ``quintessence''), not a cosmological constant.}
For more details and the errors included in these values, see
Ref.~\cite{PDB}. The nature of the cold dark matter (dominant component
of the $\Omega_{\rm mat}$) as well as that of the ``dark energy'' are
big puzzles in cosmology, which is beyond the scope of this
appendix. (As for the cold dark matter, however, supersymmetry provides
a good candidate, the lightest SUSY particle (LSP). See also
Sec.~\ref{SEC-Nlep-grav}.)

{}From the above ratio of the present energy densities, it is found
that the early universe was dominated by radiation (for $R \lsim
10^{-4} R(t_0)$), and subsequently dominated by
matter. Eqs.~(\ref{EQ-dilutions}) and (\ref{EQ-Friedmann}) lead to the
following familiar relations for each epoch.\footnote{If we include
the change of $g_*(T)$ with time, which is defined below, the relation
in \EQ{EQ-R-H-t} is slightly modified.}
\begin{eqnarray}
 \label{EQ-R-H-t}
 {\rm radiation\!\!-\!\!dominated\,\,\,universe}
  &
  R\propto t^{1/2}
  \,,
  &
  H = \frac{1}{2t}
  \,,
  \nonumber
  \\
 {\rm matter\!\!-\!\!dominated\,\,\,universe}
  &
  R\propto t^{2/3}
  \,,
  &
  H = \frac{2}{3t}
  \,.
\end{eqnarray}

%%%%%%%%%%%%%%%%%%%%%%%%%%%%%%%%%%%%%%%%%%%%%%%%%%%%%%%%%%%%%%%%%%%
\section{Thermodynamics}%%%%%%%%%%%%%%%%%%%%%%%%%%%%%%%%%%%%%%%%%%%
%%%%%%%%%%%%%%%%%%%%%%%%%%%%%%%%%%%%%%%%%%%%%%%%%%%%%%%%%%%%%%%%%%%
\label{SEC-App-Thermo}

Here, we briefly review some basic thermodynamics in the
radiation-dominated early universe, in which many particles are
relativistic and in thermal equilibrium. First of all, the equilibrium
density of particles of type $i$ with momenta in a range $d^3 p$
centered on ${\bf p}$ is given by
%%%
\begin{eqnarray}
 g_i\frac{d^3 p}{2\pi^3}
  f_i({\bf p})
  \,,
\end{eqnarray}
where $g_i$ is the number of degrees of freedom and $f_i({\bf p})$ is
the Fermi-Dirac or Bose-Einstein distribution function:
%%%
\begin{eqnarray}
 f_i({\bf p}) = \frac{1}{\exp\left(\Frac{E_i-\mu_i}{T}\right)\pm 1}
  \,.
\end{eqnarray}
Here, $E_i$ is the energy $E_i\equiv \sqrt{{\bf p}^2 + m_i^2}$, $\mu_i$
is the chemical potential of the particle $i$, and the plus (minus) sign
is for fermions (bosons). The number density $n_i$, energy density
$\rho_i$ and pressure $p_i$ of particle $i$ are then given by the
following equations:
%%%
\begin{eqnarray}
 n_i 
  &=&
  \frac{g_i}{2\pi^3}
  \int
  f_i({\bf p})
  d^3 p
  \,,
  \label{EQ-number-density}
  \\
 \rho_i 
  &=&
  \frac{g_i}{2\pi^3}
  \int
  E_i
  f_i({\bf p})
  d^3 p
  \,,
  \\
 p_i 
  &=&
  \frac{g_i}{2\pi^3}
  \int
  \frac{{\bf p}^2}{3 E_i}
  f_i({\bf p})
  d^3 p
  \,.
\end{eqnarray}
%%%
%%%%%%%%%%%%%%%%%%%%%%%%
\begin{table}[t]%%%%%%%%
%%%%%%%%%%%%%%%%%%%%%%%%
 \begin{center}
  \begin{tabular}{|c|c||c|}
   \hline
   \multicolumn{2}{|c||}{$T\gg m_i$}
   & $T\ll m_i$
   \\
   \hline
   fermion & boson & 
   \\
   \cline{1-2}
   $n_i = \Frac{3}{4}   g_i
   \left(\Frac{\zeta(3)}{\pi^2}\right)
   T^3$
   &
   $n_i =   g_i
   \left(\Frac{\zeta(3)}{\pi^2}\right)
   T^3$
   &
   $n_i =    g_i
   \left(\Frac{m_i T}{2\pi}\right)^{3/2}
   \exp\left(-\Frac{m_i}{T}\right)$
   \\
   $\rho_i = \Frac{7}{8}   g_i
   \left(\Frac{\pi^2}{30}\right)
   T^4$
   &
   $\rho_i =   g_i
   \left(\Frac{\pi^2}{30}\right)
   T^4$
   &
   $\rho_i = m_i n_i$
   \\
   $p_i =
   \Frac{1}{3}\rho_i$
   &
   $p_i =
   \Frac{1}{3}\rho_i$
   &
   $p_i = T n_i
   \,\,(\ll \rho_i)$
   \\ \hline
  \end{tabular}
 \end{center}
 \caption{The number density $n_i$, energy density $\rho_i$ and pressure
 $p_i$ of the particle $i$, which is thermal equilibrium, in the limits
 of $T\gg m_i$ and $T\ll m_i$. We have assumed $|\mu_i|\ll T$ and
 $|\mu_i|< m_i$.}  \label{Table-nprho}
%%%%%%%%%%%%%%%%%%%%%%%%
\end{table}%%%%%%%%%%%%%
%%%%%%%%%%%%%%%%%%%%%%%%
In Table~\ref{Table-nprho}, we show these quantities for the
relativistic ($T\gg m_i$) and non-relativistic ($T\ll m_i$)
limits. Here, we have assumed $|\mu_i|\ll T$ and no Bose-Einstein
condensation ($|\mu_i|<m_i$).

Because the energy density of a non-relativistic particle is
exponentially suppressed compared with the relativistic one, the total
energy density of the radiation $\rho_{\rm rad}$ is given by the
following simple form:
%%%
\begin{eqnarray}
 \label{EQ-rho-rad}
 \rho_{\rm rad} = \frac{\pi^2}{30}g_*(T) T^4
  \,,
\end{eqnarray}
where
\begin{eqnarray}
 \label{EQ-gstar-def}
  g_*(T)\equiv
  \sum_{\begin{array}{c}
   m_i\ll T\\
	 i = {\rm boson}
	\end{array}}
  \!\!\!\!\!\!
  g_i
  \,\,\,
  +
  \,\,\,
  \frac{7}{8}
  \!\!\!\!\!
  \sum_{\begin{array}{c}
   m_j\ll T\\
	 j = {\rm fermion}
	\end{array}}
  \!\!\!\!\!\!
  g_j
  \,\,.
\end{eqnarray}
If there are particles which have different temperatures from that of
the photon $T$, another factor $(T_i/T)^4$ should be multiplied in the
above expression. (For example, at $T\ll \MEV$, neutrinos have
temperature $T_{\nu} = (4/11)^{1/3} T$ for $m_\nu \ll T_\nu$.)

Notice that all leptogenesis mechanisms discussed in this thesis work at
temperatures far above the electroweak scale $T\gg 1\TEV$, where all the
MSSM particles are expected to be in thermal equilibrium. In this case,
we obtain
\begin{eqnarray}
 \label{EQ-gstarMSSM}
  g_* = 228.75
  \qquad
  {\rm for}
  \quad
  {\rm MSSM}
  \,.
\end{eqnarray}

In the expanding universe, it is convenient to introduce the entropy
density $s$, which is defined by
%%%
\begin{eqnarray}
 \label{EQ-entropy}
 s
  &\equiv& \frac{\rho + p}{T}
  \nonumber\\
 &=& \frac{4}{3\,T}\rho
  =\frac{2 \pi^2}{45}g_*(T) T^3
  \,.
\end{eqnarray}
%%%
(Again, in the presence of particles with different temperatures, a
factor of $(T_i/T)^3$ is multiplied in \EQ{EQ-entropy}. In this case,
the $g_*$ in \EQ{EQ-entropy} becomes slightly different from the $g_*$
in \EQ{EQ-rho-rad}.) Notice that the entropy per comoving volume $s
R^3$ is conserved as far as no entropy production takes place. Thus it
is quite convenient to take the ratio $n_X/s$ when we discuss some
number density $n_X$. For example, if some $X$-number is conserved,
the ratio of the $X$-number density to the entropy density takes a
constant value
%%%
\begin{eqnarray}
 \frac{n_X}{s} = const
  \,,
\end{eqnarray}
%%%
as long as there is no entropy production, since both $n_X$ and $s$
scales as $R^{-3}$ as the universe expands. As another example, if the
$X$-particle is in thermal equilibrium and relativistic ($T\gg m_X$),
the ratio is given by
%%%
\begin{eqnarray}
 \label{EQ-nXs-thermal}
  \frac{n_X^{\rm eq}}{s} = \frac{45\zeta (3)}{2\pi^4}
  \frac{g_X}{g_*(T)}
  \quad
  \left(
   \times \frac{3}{4}\quad {\rm for\,\,fermion}
   \right)
   \,,
\end{eqnarray}
%%%
where the temperature (or time) dependence only comes from $g_*(T)$.

Before closing this section, we calculate the relations between the
particle number asymmetry $n_i^{(+)} - n_i^{(-)}$ and the particle's
chemical potential $\mu_i$, which can be obtained by integrating
\EQ{EQ-number-density}.  In order to calculate the asymmetry, it is
necessary to calculate higher order terms than those in
Table~\ref{Table-nprho}. The results are given by
%%%
\begin{eqnarray}
 n_i^{(+)} - n_i^{(-)}
  &=&
  \frac{1}{6}g_i T^3
  \left[
  \left(\frac{\mu_i}{T}\right)
  + \cdots
  \right]
  \qquad {\rm for\quad fermion}
  \,,
  \nonumber\\
 n_i^{(+)} - n_i^{(-)}
  &=&
  \frac{1}{3}g_i T^3
  \left[
  \left(\frac{\mu_i}{T}\right)
  + \cdots
  \right]
  \qquad {\rm for\quad boson}
  \,,
  \label{EQ-chemical}
\end{eqnarray}
%%%
where ellipses denote higher order terms in the expansions of $m_i/T$
and $\mu_i/T$.  Here, we have assumed no Bose-Einstein condensation
$|\mu_i|< m_i$ for boson, and relativistic limit $m_i\ll
T$. \EQ{EQ-chemical} is used in \SEC{SEC-sphaleron}, in deriving the
relation between the baryon and lepton asymmetries in the presence of
sphaleron effect. In this case, the chemical potentials are of order
$|\mu|\sim 10^{-10}\,\,T$, and hence the above assumptions are
justified.

%%%%%%%%%%%%%%%%%%%%%%%%%%%%%%

%%%%%%%%%%%%%%%%%%%%%%%%%%%%%%%%%%%%%%%%%%%%%%%%%%%%%%%%%%%%%%%%%%%
%\include{Reference}%%%%%%%%%%%%%%%%%%%%%%%%%%%%%%%%%%%%%%%%%%%%%%%%
%%%%%%%%%%%%%%%%%%%%%%%%%%%%%%%%%%%%%%%%%%%%%%%%%%%%%%%%%%%%%%%%%%%
%%%%%%%%%%%%%%%%%%%%%%%%%%%%%%%%%%%%%%%%%%%%%%%%%%%%%%%%%%%%%%%%%%%
%%%%%%%%%%%%%%%%%%%%%%%%%%%%%%%%%%%%%%%%%%%%%%
%%%%%%%%%%%%%%%%%%%%%%%%%%%%%%%%%%%%%%%%%%%%%%%%%%%%%%%%%%%%%%%%%%%

%%%%%%%%%%%%%%%%%%%%%%%%%%%%%
%%%%%%%%%%%%%%%%%%%%%%%%%%%%%
%%%%%%%%%%%%%%%%%%%%%%%%%%%%%
\end{document}